\pdfoutput=1
\documentclass[11pt]{article}
\usepackage[utf8]{inputenc}
\usepackage{mathtools}
\RequirePackage[utf8]{inputenc}

\RequirePackage{amsmath}
\RequirePackage{amssymb}
\RequirePackage{graphicx,color}
\RequirePackage{amsthm}
\RequirePackage{mathrsfs}
\RequirePackage{bm}

\RequirePackage[boxsize=0.5em,aligntableaux=center]{ytableau}

\RequirePackage[labelformat=simple]{subcaption}

\RequirePackage{ifpdf}

\newcommand{\bC}{\mathbb{C}}

\newcommand{\bP}{\mathbb{P}}

\newcommand{\bZ}{\mathbb{Z}}

\newcommand{\cA}{\mathcal{A}}
\newcommand{\cB}{\mathcal{B}}
\newcommand{\cC}{\mathcal{C}}

\newcommand{\cF}{\mathscr{F}}

\newcommand{\cL}{\mathcal{L}}
\newcommand{\cM}{\mathcal{M}}

\newcommand{\cO}{\mathcal{O}}

\newcommand{\cS}{\mathcal{S}}

\newcommand{\Spin}{\mathrm{Spin}}

\newcommand{\dsz}[2]{\bigl\langle#1,#2\bigr\rangle}

\newcommand{\SO}{SO}

\makeatletter

\newcommand*\wthelper[2]{%
        \hbox{\dimen@\accentfontxheight#1%
                \accentfontxheight#11.2\dimen@
                $\m@th#1\widetilde{#2}$%
                \accentfontxheight#1\dimen@
        }%
}

\newcommand*\accentfontxheight[1]{%
        \fontdimen5\ifx#1\displaystyle
                \textfont
        \else\ifx#1\textstyle
                \textfont
        \else\ifx#1\scriptstyle
                \scriptfont
        \else
                \scriptscriptfont
        \fi\fi\fi3
}
\makeatother

\usepackage{epsfig,verbatim}
\usepackage{amsmath, amssymb, graphics}
\usepackage{slashed}            
\usepackage{bbm}                
\usepackage{units}              
\usepackage{xspace}             
\usepackage{enumerate}          
\usepackage{soul}
\usepackage{upgreek}
\usepackage{tikz}
\usetikzlibrary{shapes,arrows}
\usetikzlibrary{arrows.meta,
                calc, chains,
                quotes,
                positioning,
                shapes.geometric}
                
\usepackage{packages/jheppub}

\newcommand{\mathsym}[1]{{}}

\newcommand{\rmi}{\mathrm{i}}

\newcommand{\beq}{\begin{equation}}  \newcommand{\eeq}{\end{equation}}
\def\beqa{\begin{eqnarray}}
\def\eeqa{\end{eqnarray}}

\newcommand{\eq}[1]{(\ref{#1})}

\renewcommand\){\right)}
\renewcommand\[{\left[}
\renewcommand\]{\right]}

\renewcommand{\Re}{\operatorname{Re}}
\renewcommand{\Im}{\operatorname{Im}}

\newcommand{\ba}{\begin{eqnarray}}
\newcommand{\ea}{\end{eqnarray}}

\newcommand{\be}{\begin{equation}}
\newcommand{\ee}{\end{equation}}
\newcommand{\bea}{\begin{eqnarray}} 
\newcommand{\eea}{\end{eqnarray}}
\newcommand{\ft}[2]{{\textstyle\frac{#1}{#2}}}

\def\pd{\ensuremath{\partial}}

\def\Re{\mathop{\rm Re}\nolimits}
\def\Im{\mathop{\rm Im}\nolimits}

\def\rme{{\mathrm e}}
\def\rmi{{\mathrm i}}

\newsavebox{\uuunit}
\sbox{\uuunit}
    {\setlength{\unitlength}{0.825em}
     \begin{picture}(0.6,0.7)
        \thinlines
        \put(0,0){\line(1,0){0.5}}
        \put(0.15,0){\line(0,1){0.7}}
        \put(0.35,0){\line(0,1){0.8}}
       \multiput(0.3,0.8)(-0.04,-0.02){12}{\rule{0.5pt}{0.5pt}}
     \end {picture}}

\def\cA{{\cal A}} \def\cB{{\cal B}} \def\cC{{\cal C}} \def\cD{{\cal D}}
  \def\cF{{\cal F}}  
   \def\cK{{\cal K}} \def\cL{{\cal L}}
 \def\cM{{\cal M}}  \def\cO{{\cal O}} 
  \def\cR{{\cal R}}

 \def\cS{{\cal S}}   \def\cV{{\cal V}}

\def\atH0{|_{H_{0}}}
\def\AtH0{\bigg|_{H_{0}}}

\csname @addtoreset\endcsname{equation}{section}

\title{\centering Nothing is certain in string compactifications}

\author[\heartsuit]{Iñaki García Etxebarria,}
\author[\clubsuit]{Miguel Montero,}
\author[\diamondsuit]{Kepa Sousa,}
\author[\clubsuit]{and Irene Valenzuela}

\affiliation[\heartsuit]{Department of Mathematical Sciences, Durham University,\\
Durham, DH1 3LE, United Kingdom}
\affiliation[\diamondsuit]{Institute of Theoretical Physics, Charles University,\\ 
V Holes\v{o}vi\v{c}k\'ach 2, 18000  Prague 8, Czech Republic}
\affiliation[\clubsuit]{Jefferson Physical Laboratory, Harvard University,\\
Cambridge, MA 02138, USA}

\emailAdd{inaki.garcia-etxebarria@durham.ac.uk}
\emailAdd{mmontero@g.harvard.edu}
\emailAdd{kepa.sousa@gmail.com}
\emailAdd{ivalenzuela@g.harvard.edu}

\abstract{
A bubble of nothing is a spacetime instability where a compact dimension collapses. After nucleation, it expands at the speed of light, leaving ``nothing'' behind.
We argue that the topological and dynamical mechanisms which could protect a compactification against decay to nothing seem to be absent in string compactifications once supersymmetry is broken. The topological obstruction lies in a bordism group and, surprisingly, it can disappear 
even for a SUSY-compatible spin structure.
As a proof of principle, we construct an explicit bubble of nothing  for a $T^3$ with completely periodic (SUSY-compatible) spin structure in an Einstein dilaton Gauss-Bonnet theory, which arises  in the low-energy limit of certain heterotic and type II flux compactifications. 
Without the topological protection, supersymmetric compactifications are purely stabilized by a Coleman-deLuccia mechanism, which relies on a certain local energy condition. This is violated in our example by the nonsupersymmetric GB term. In the presence of fluxes this energy condition gets modified and its violation might be related to the Weak Gravity Conjecture.

We expect that our techniques can be used to construct a plethora of new bubbles of nothing in any setup where the low-energy bordism group vanishes, including type II compactifications on $CY_3$, AdS flux compactifications on 5-manifolds, and M-theory on 7-manifolds. 
This lends further evidence to the conjecture that any non-supersymmetric vacuum of quantum gravity is ultimately unstable.}
\begin{document}
\hypersetup{pageanchor=false}
\makeatletter
\let\old@fpheader\@fpheader

\makeatother

\maketitle

\hypersetup{pageanchor=true}
\section{Introduction}\label{sec:intro}

It is known that non-supersymmetric vacua typically exhibit instabilities, either at the perturbative or non-perturbative level. In fact, not a single exactly stable non-supersymmetric string theory vacuum is known to us. But can we ensure that this is a necessary implication of breaking supersymmetry? Is it consistent to have a non-supersymmetric stable vacuum? In \cite{Ooguri:2016pdq,Freivogel:2016qwc} it was conjectured that any non-supersymmetric vacuum of a consistent theory of quantum gravity is indeed unstable. The conjecture is motivated by the Weak Gravity Conjecture \cite{ArkaniHamed:2006dz} in the case in which the effective theory arises upon compactifying a higher dimensional theory and the vacuum is supported by fluxes, i.e, non-vanishing gauge field strengths in the compactified dimensions. But the decay mode provided by the Weak Gravity Conjecture relies on the presence of these fluxes and seems insufficient to guarantee the instability of any non-supersymmetric vacuum. The quest for some universal instability that can be described without referring to the specific ingredients of the compactification space is the question that drives the present work.

Perhaps the best candidate for such a universal instability whenever there are extra dimensions is the bubble of nothing. Witten \cite{Witten:1981gj} showed that the Kaluza-Klein vacuum of a circle compactification is non-perturbatively unstable to decay to nothing. In other words, there is a perfectly well defined solution to the Einstein's equations that has zero energy, just like the vacuum, but which describes a hole in space that simply pops up and starts expanding at the speed of light, eventually eating up the whole space-time. Geometrically, the compactified circle shrinks to zero size at the wall of the bubble, but the solution is smooth from the higher dimensional point of view. It is known, though, that this solution is forbidden if there are fermions with supersymmetric preserving (periodic) boundary conditions on the internal circle. Hence, even if supersymmetry is broken at some energy scale, the vacuum will be topologically protected against this instability as long as there are fermions with the right boundary conditions. Therefore, one might be tempted to take the view that the bubble of nothing is just a quirk of some particular solutions that is not really relevant or generic, since it can be dealt with via topological changes that are invisible at low energies. However,  as we will see, nothing really matters\footnote{Since the dawn of this project, it has been our intention to use ``Nothing really matters'' as the title of the manuscript. However, we were title-scooped by the interesting paper \cite{Dibitetto:2020csn}, also about (a different kind of) bubbles of nothing. The search for an alternate title was hard but we tried our best. The reader must judge if we came close to the high bar we set. Among the second-runners we have ``Nothing is real in string theory'', ``Nothing comes for free in string compactifications'', ``Nothing can surprise us'' or ``Nothing is final in string theory''.}\ldots \, .

Actually, a counterexample to this idea was already presented in  \cite{Blanco-Pillado:2016xvf}. There  it was shown that the nonsupersymmetric Kaluza-Klein vacuum endowed with a Wilson line  may still decay to nothing even if the fermions exhibit supersymmetry-preserving boundary conditions. 
 In that setting the coupling between the Wilson line and the fermions renders the decay  topologically unobstructed, and the stability of supersymmetric compactifications is instead dynamically enforced.  In the spirit of finding a decay channel as generic as possible, in this paper we will show that bubbles of nothing compatible supersymmetric boundary conditions are far more general than the scenario discussed in \cite{Blanco-Pillado:2016xvf}, and do not rely on specific ingredients such as Wilson lines or fluxes. For the first time in the literature we will explicitly construct bubbles of nothing compatible with supersymmetric boundary conditions, and which do not require an ad hoc gauge coupling for the fermions.
  This opens up a new type of decay mode that might be universally present even if supersymmetry is only broken at low energies. 

In order to determine if there is a topological obstruction to construct a bubble of nothing, one needs to study whether the internal compactification space can be smoothly shrunk to zero size. In mathematical terms, this occurs whenever the compactification space is bordant to a point, i.e. it belongs to the trivial class of the relevant bordism group. This is denoted as $\Omega_d$, where $d$ is the dimension of the internal manifold. Let us consider that the effective theory contains fermions such that the manifold supports a spin structure. The relevant bordism group is called $\Omega_d^{\text{Spin}}$, and these groups have already been classified in the literature for any $d$ (see for instance \cite{10.2307/1970690}). For a one dimensional manifold (the circle), one has $\Omega_1^{\text{Spin}}=\mathbb{Z}_2$, implying that there are two different classes corresponding to the two choices of boundary conditions for the fermions: periodic or antiperiodic. Only the associated to antiperiodic boundary conditions will allow for bubbles of nothing, as expected. The same occurs for two-dimensional manifolds. However, the situation changes for dimension larger than three. Interestingly, $\Omega^{\text{Spin}}_3=0$, implying that \emph{any} three-dimensional manifold can be topologically shrunk to a point, including the one consistent with periodic supersymmetric preserving boundary conditions! In other words, there is no topological obstruction to construct a bubble of nothing in effective field theories with three extra dimensions. Moreover, the topological obstruction is also absent when there are six and seven extra dimensions since $\Omega^{\text{Spin}}_6=\Omega^{\text{Spin}}_7=0$. This can have important implications for four dimensional effective field theories arising from string theory compactifications of type II, heterotic or M-theory, since they always involve a six or seven compactification manifold.

We should also remark that it has been recently conjectured \cite{McNamara:2019rup} that a consistent theory of quantum gravity must include sufficient ingredients to guarantee that $\Omega^{QG}=0$ for any dimension bigger than two. Otherwise, one can argue that the theory will contain some conserved global charge which would be inconsistent with the well known swampland criteria of not having global symmetries in quantum gravity \cite{Abbott:1989jw,Banks:1988yz,Coleman:1989zu,Kallosh:1995hi,Susskind:1995da,Banks:2010zn,Beem:2014zpa,Harlow:2018jwu,Harlow:2018tng}. If this conjecture holds, it implies that some sort of bubbles of nothing are always topologically allowed in any compactification. However, as we will see, this is not enough to argue for a universal vacuum instability yet, as one needs to study the dynamics of the bubble and check that it will indeed expand eating up the whole space-time. 

If the topological obstruction for these bubbles of nothing is absent, what can protect then a vacuum from decaying? The first thing that probably comes to your mind is supersymmetry.  Indeed, when considering  the non-perturbative stability of false vacua \cite{Coleman:1977py,Callan:1977pt}, if  supersymmetry is unbroken the decay rate will be zero, as the euclidean action of the instanton  associated to the nucleation of a ``true vacuum'' bubble  will diverge \cite{Coleman:1980aw,Cvetic:1992dc,Cvetic:1992st}. 
Similarly, in the case of the Kaluza-Klein compactification of \cite{Blanco-Pillado:2016xvf}, where no topological protection is present,  the Coleman-DeLuccia mechanism was also shown to prevent the decay to nothing in the absence of supersymmetry breaking.
This result motivated Blanco-Pillado et al. to conjecture that this form of  dynamical  suppression is the generic mechanism  enforcing the stability of topologically unprotected supersymmetric compactifications.

 However, the dynamical protection might disappear whenever supersymmetry is broken. One of the goals of this paper is to understand under what circumstances this indeed occurs. The answer is that we need to either break explicitly supersymmetry or, if we want to preserve some covariantly constant spinor and only break supersymmetry spontaneously at lower energies,  a certain energy condition needs to be violated. In the absence of further ingredients that modify the spin connection, the energy condition that needs to be violated is known as the Dominant Energy Condition, as already implied by the Positive Energy Theorem \cite{Schon:1979rg,Witten:1981mf}. This energy condition is  just true for some classical systems, and is often violated by quantum effects, higher derivative corrections, or in the presence of fluxes. Since there is no other principle upholding it that we are aware of, we would expect the condition to be false in nonsupersymmetric string compactifications. We will find this is indeed the case in examples, but we believe the story is general. Thus, the picture one gets is that a vacuum can be in principle be insured against decay either by topology or dynamics, but the first does not happen in quantum gravity and the second only takes place whenever there is SUSY. Thus, in the end, every non-supersymmetric vacuum should decay.

Before getting too deep in these ideas, and for the sake of concreteness, in this paper we will focus on the more modest goal of understanding in detail the decay to nothing of a vacuum $\mathbb{M}_{D-3}\times T^3/\Gamma$ with $D\geq 6$. As a proof of principle for the existence of these new types of bubbles of nothing, we are going to explicitly construct the bubble for an effective field theory involving only Einstein
 gravity with quadratic curvature terms, and a dilaton in lower dimensions. Recall that $\Omega_3^{\text{Spin}}=0$, implying that the bubble of nothing can be constructed completely within the framework of the ($D$-dimensional) low-energy effective field theory, without the need of invoking exotic UV ingredients. This will make easier to construct smooth solutions such that the semi-classical  description of the decay is justified. As a supersymmetry breaking source, the theory includes a Gauss-Bonnet higher derivative term, which will indeed violate the dominant energy condition, allowing us to construct bubble solutions with  a non-vanishing vacuum decay rate.

Since we carry out our analysis in a particular effective field theory coupled to Einstein's gravity, we need to make sure we are not in the Swampland. Otherwise, the bubble solutions we find might just be an artifact caused by the lack of consistent UV completion. We will dispel doubts on this point by showing that the effective theory under consideration with the Gauss-Bonnet term can be embedded in string theory compactifications, as well as discuss the potential impact on string phenomenology.

Our explicit construction for the bubble of nothing of $T^3$ allows us to resolve a puzzle posed in \cite{Acharya:2019mcu}. In that reference, Acharya analyzed the same question we are interested in --- to what extent is it possible to have a stable, non-supersymmetric vacuum. This naturally leads one to consider a Ricci-flat compact space (so that one can solve  Einstein's equations) with no covariant spinors (so that there is no supersymmetry). A nice class of examples are $T^3$ quotients $T^3/\Gamma$ where $\Gamma$ is a fixed-point free discrete isometry of $T^3$. As discussed in \cite{PFAFFLE2000367,Acharya:2019mcu}, there are 28 classes of quotients, including spin structures. 27 of them do not admit any covariantly constant spinors. 26 of these 27 classes descend from a parent $T^3$ with antiperiodic boundary conditions along one of the cycles, and this allows for a suitable quotient of Witten's bubble of nothing to act as a bubble of nothing for the quotient as well. Thus, out of the 28 classes, 1 is supersymmetric and stable, 26 have known bubbles of nothing, but there is one left (class $G3$ in \cite{Acharya:2019mcu}) for which no bubble of nothing was known. Our techniques allow us to close the gap and explicitly construct a bubble of nothing for this last class. Topologically, it is an elliptic fibration with an $E_6$ singularity. Thus, all non-supersymmetric quotients of $T^3$ admit bubble of nothing instabilities. Regarding the geometry of these bounce solutions,  all the quotients of the Witten's bubble presented in \cite{Acharya:2019mcu} contained orbifold singularities. Here we will also prove that these geometries can be regularised,  and we will construct   the explicit smooth instanton solutions mediating these decays.

Finally, it is worth mentioning that  the techniques employed in the present paper can also be applied in other contexts. For instance, the family of elliptic fibrations characterizing our solutions includes  the K3 manifold, and thus our methods can be used to obtain smooth and approximately Calabi-Yau metrics  for the K3 surface, as done in \cite{gross2000,Kachru:2018van}. While our approach is similar in spirit to  \cite{gross2000}, we use a different approximation scheme to theirs. Actually, our method (also alternative to  \cite{Kachru:2018van}) allows to obtain systematically higher order corrections to the  metrics of \cite{gross2000}. Furthermore,  our construction  provides a detailed characterisation of the warping  induced by higher derivative terms and fluxes in these geometries (see e.g. Appendix A, where we extend our results to an AdS compactification on $T^3$ with fluxes). Therefore, it is straightforward to adapt  our results  to obtain an explicit geometric description of   flux compactifications on a   warped K3 manifold.

\subsection{Reading guide}
We have organized our work as follows:\begin{itemize}
\item  Section \ref{sec:rev}, we discuss general background on bubbles of nothing, as well as obstructions to their existence related to topology and the Positive Energy Theorem.
\item  Section \ref{sec:nuts} contains the core result of our paper succinctly summarized: We have explicitly constructed a bubble of nothing for a $T^3$ with supersymmetry-preserving boundary conditions in an Einstein-dilaton model with higher-derivative terms, and given the decay rate explicitly.
\item Section \ref{sec:det} discusses in detail how the effective action and ansatz that we use allows us to evade the topological and dynamical constraints.
\item Section \ref{sec:detailedBON} is the core of the paper, where an explicit metric for the bubble solution is constructed in layers where different approximations are used. In the near bubble region (layer II) a mix of exact and perturbative solutions are used, while far from the bubble core (layer I) Einstein's equations are solved numerically. We discuss appropriate matching of boundary conditions across layers and compute decay rates.

\item Section \ref{sec:phys} contains a simple stringy embedding of our bubble, as well as miscellanea regarding generalizations of positive energy theorems, including fluxes, and a discussion of the implications of our results for String Phenomenology and the relation to Swampland constraints.

\item We finish with our conclusions in Section \ref{sec:conclus} as well as some technical details and generalizations relegated to Appendices.
\end{itemize}

A very minimalistic reading of our paper would contain Sections \ref{sec:rev} and \ref{sec:nuts}. We have written the paper in such a way that the reader can get a very good idea of our work by reading only these two sections (so only 16 pages!). From them on, there are several possibilities. Sections  \ref{sec:det} and \ref{sec:detailedBON} are most important for a reader interested in the explicit construction of our bubble of nothing and the GR/field-theory aspects of the model. By contrast, Section \ref{sec:phys}  is more on the stringy side of things, including also generalizations of the topological and dynamical obstructions in the presence of fluxes. These can be read separately to a large extent, though of course some interdependence is unavoidable. 

\section{Bubbles of nothing}\label{sec:rev}

We will begin with reviewing what bubbles of nothing are, and what are the necessary conditions for these euclidean solutions to exist and yield a non-perturbatively instability of the vacuum. We will distinguish between a topological and a dynamical obstruction, and show how the topological obstruction is absent for some higher dimensional compactification spaces.

\subsection{Review: Bubble of nothing}
\label{sec:WittenBON}

As its name suggests, a bubble of nothing represents a semiclassical non-perturbative decay mode from the vacuum to nothing, i.e. the vacuum annihilates. The bubble yields a hole in space-time which grows at the speed of light, and leads to the end of space-time from the point of view of a four
dimensional observer.

The first construction of a bubble of nothing (BON) was done by Witten in \cite{Witten:1981gj}, as an instability of the Kaluza-Klein (KK) vacuum. Let us consider a KK circle compactification of a five dimensional theory to four dimensions, so the space-time is $\mathbb{M}_4\times S^1$. 

The instanton solution (also called \emph{bounce}) can be constructed by starting from the euclidean version of the Schwarzschild spacetime,
\beq
ds^2_{5}=r^2 d\Omega_3^2+ \frac{dr^2}{1-\cR^2/r^2}+R_{\text{kk}}^2\left(1-\frac{\cR^2}{r^2}\right)d\theta^2, \label{Wittenbounce}
\eeq
where $\theta\in [0,2 \pi)$ is the periodic coordinate on the circle $S^1$ with radius $R_{\text{kk}}$, and $\cR$ is the size of the bubble at the time of nucleation. We will denote this spacetime by $\cM_5$.

The bounce solution asymptotes to the euclidean KK vacuum when $r\rightarrow \infty$. In order to get the endpoint of the vacuum decay, we need to analytically continue the euclidean solution back to Minkowski signature along a new appropriate time variable.  The false vacuum decays then into the Lorentzian space which coincides with this bounce solution at $t=0$.  In this case, if we write the line element on the three sphere as
\be
d\Omega_3 = d\chi^2 + \sin^2\chi d\Omega_2^2, \qquad \text{with} \qquad \chi \in [0, \pi),
\ee
the plane $\chi=\pi/2$ can play the role of $t=0$, so by replacing $\chi\rightarrow \pi/2 +i\psi$ we get the Minkowski signature solution
\beq
ds^2_{5}=-r^2d\psi^2+\frac{dr^2}{1-\cR^2/r^2}+r^2\cosh^2\psi d\Omega^2_2 +R_{\text{kk}}^2\left(1-\frac{\cR^2}{r^2}\right)d\theta^2. \label{metric}
\eeq
 At large $r$ this solution  approaches to the vacuum of  $\mathbb{M}_4\times S^1$, as can be seen rewriting  the line element in terms of the coordinates $x=r\cosh\psi, t=r\sinh\psi$ 
 \beq
ds^2_{5}\underset{r\to \infty}{\approx}-dt^2+dx^2+x^2d\Omega_2^2+R_{\text{kk}}^2d\theta^2.\label{asflat}
\eeq
However, the coordinates $r$ and $\psi$ do not span all of Minkowski space. From the point of view of a four dimensional observer, the full space corresponds to Minkowski space where the region $x^2-t^2<\cR^2$ has been removed. The wall of the bubble then corresponds to the frontier of the four-dimensional space-time, and grows with time as
 \beq
 x_{\text{bubble}}(t)=\sqrt{\cR^2+t^2},
 \eeq
 In particular,  we can see now that the bubble radius at $t=0$ is given by the parameter $\cR =x_{\text{bubble}}(0)$. 
 The size of the collapsing $S^1$, which we will denote by $C(r)$, is given by
 \beq
C(r)=R_{\text{kk}}\sqrt{1-\cR^2/r^2},
 \eeq
 so it approaches $R_{\text{kk}}$ at large $r$ and shrinks to zero size at the bubble surface, located at $r=\cR$.  As shown in \cite{Witten:1981gj}, the condition  $\cR=R_{\text{kk}}$ needs to be imposed\footnote{As shown in \cite{Blanco-Pillado:2016xvf} this condition may be relaxed in more general scenarios, where additional interactions may provide a mechanism to regularise the conical singularity. We will also encounter this situation below when considering the resolution of the orbifold singularities in the  bounce solutions   of \cite{Acharya:2019mcu}.} to avoid the presence of a conical singularity at the bubble surface, thus ensuring  that the full spacetime is non-singular and geodesically complete. Requiring that the bounce geometry is smooth is essential for the semiclassical description of the decay to be accurate. Indeed, if the spacetime curvature is not everywhere well below the Planck scale  we would  need to have some knowledge of the UV physics to describe the decay, but  nevertheless the existence of a singular bounce solution may still indicate the presence of a non-perturbative instability. 
 
   The euclidean BON solution  \eqref{Wittenbounce} can also be rewritten in a different  gauge, more convenient for the computations below,  as follows
 \be
ds^2_{5} =\cR^2W(\rho)^2 d\Omega_3^2+ d\rho^2+C^2(\rho)d\theta^2, 
\label{eq:wittenBON2}
 \ee
 where the new radial coordinate takes values in $\rho \in [0,\infty)$,  and with the bubble  located at $\rho=0$. Here the metric profile functions are defined  by  the equations
 \be
W' = \cR^{-1}\sqrt{1- W^{-2}}, \qquad C(\rho) = R_{\text{kk}} \cR \, W'(\rho), \qquad \text{and} \qquad W(0)=1,
 \ee
and it is immediate to check that the line element in \eqref{Wittenbounce}  can be recovered with the change of variables  $r(\rho) = \cR W(\rho)$.  
Then, the three-sphere $S^3$ defined by  $\rho=0$ represents  the bubble world-volume, which back in  Minkowskian signature turns into a  $dS_ 3$, that is, the expanding bubble surface.   

 Many works have studied different aspects of these bubble instabilities in different setups, including the context of flux compactifications \cite{Yang:2009wz,BlancoPillado:2010df,BlancoPillado:2010et,BlancoPillado:2011me,Brown:2010mf,Brown:2011gt,Blanco-Pillado:2016xvf}, and    in string theory \cite{Fabinger:2000jd,Dine:2004uw,Horowitz:2007pr,deAlwis:2013gka,Ooguri:2017njy,Acharya:2019mcu} (see also \cite{Dibitetto:2020csn,Brown:2014rka}). However, many of these constructions are a slight generalization of Witten's bubble in which a circle from an extra dimension shrinks to zero size. Regarding  scenarios with a more complicated compact space, the only  explicit smooth solutions which are known  describe the collapse of spherical compactifications, as in \cite{Yang:2009wz,BlancoPillado:2010et,BlancoPillado:2011me,Brown:2010mf,Brown:2011gt}, and the more recent construction \cite{Ooguri:2017njy} where the internal manifold  is a homogeneous space with a fibered  two-sphere that collapses. In this sense, other singular bounce geometries with interesting topologies  are those of  \cite{Acharya:2019mcu} and \cite{Horowitz:2007pr}.

A very important caveat is that the bubble of nothing \eqref{Wittenbounce} is only topologically compatible with antiperiodic boundary conditions of the fermions on the circle.  This can be seen as follows:  Since in the bubble of nothing the KK circle shrinks to a point, topologically the spacetime   is a three-sphere $S^3$ times a disk $\cD$. The KK circle far away from the core of the bubble can be identified with the boundary of the disk. If the theory has fermions, then we need to define fermions on a disk. A two-dimensional disk looks like $\mathbb{R}^2$, so we can define fermions in the usual way. But then, the most salient feature of fermions is that they flip sign under a $2\pi$ rotation. This $2\pi$ rotation on the disk amounts to a translation on the boundary $S^1$; as a result, fermions must have antiperiodic boundary conditions in the decaying vacuum.

Therefore either the theory is non-supersymmetric already in high dimensions, or there is explicit supersymmetry breaking coming from Scherk-Schwarz (antiperiodic) boundary conditions on the circle. This can lead to the misleading conclusion that vacua with spontaneously broken supersymmetry are topologically protected against bubbles of nothing. One of the goals of this paper is to show that this statement is incorrect, and we can have more general bubbles of nothing that are compatible with a supersymmetric spin structure. What will protect susy vacua from decaying will not be a topological but a dynamical obstruction, as we will explain in the following. 

\subsection{Topological obstruction\label{sec:top_obs}}

In the previous Subsection we saw that whether or not a bubble exists depends crucially on the spin structure. In absence of e.g. extra $U(1)$'s which might provide Wilson lines along the circle (see \cite{Blanco-Pillado:2016xvf}), the spin structure cannot be deformed continuously, so it provides a topological obstruction to the existence of the bubble. 

As usual, topological obstructions are particularly interesting, since they are extremely robust. Suppose one takes a compactification on $S^1$ with periodic boundary conditions, so that a bubble cannot appear. Even if one deforms the effective field theory in an arbitrary way (for instance, breaking supersymmetry either explicitly or spontaneously), the spin structure cannot change and the bubble of nothing still does not exist. One can always imagine there is some deep UV domain wall, out of reach of the effective field theory, that can change the spin structure (see \cite{Garcia-Etxebarria:2015ota,McNamara:2019rup}, or keep on reading), but this is certainly impossible using low-energy physics only. 

We thus have two mechanisms that ensure the absence of a bubble of nothing: the topological obstruction related to spin structures, and supersymmetry, which ensures stability of the vacuum. Although they coincide for Witten's bubble, they are actually logically independent, as we will see momentarily.

The topological obstruction to the existence of bubbles admits a natural mathematical description via bordisms, generalizing the picture near the end of the last Subsection.  From a topological point of view, all that one needs to construct Witten's bubble of nothing is to be able to ``fill up'' the interior of the $S^1$; the resulting disk $\cD$ ``interpolates'' smoothly between the $S^1$ and ``nothing''. For instance, if we describe Witten's bounce by the line element \eqref{eq:wittenBON2}, then the disc $\cD$ is the manifold parametrised by $\rho$ and $\theta$, and the complete instanton spacetime is the warped product $\cM_5 \cong \cD \times_W S^3$. In addition, when the theory contains fermions, one also needs to be able to extend the spin structure on $S^1$ to the spin structure on the disk.

This picture can be readily generalized to the case with an arbitrary space-time dimension $D$, and where the $S^1$ is replaced by a generic compactification manifold $\cC_d$, of any dimension $d$, with a given spin structure. Then the potentially decaying vacuum will be  of the form    $\mathbb{M}_{D-d} \times \cC_d$.  A bubble of nothing for this compactification requires the existence of  a $d+1$-dimensional manifold $\cB_{d+1}$ with $\cC_d=\partial \cB_{d+1}$, such that the spin structure on $\cC_d$ extends to $\cB_{d+1}$. Then, as we will describe in detail in Section  \ref{sec:BONansatz}, the appropriate generalisation of the euclidean BON  spacetime is a warped product of the manifold $\cB_{d+1}$,  and a sphere  $S^{D-1-d}$ associated to the bubble world-volume,    so that  $\cM_{\text{BON}} \cong \cB_{d+1} \times_W S^{D-1-d}$.

\begin{figure}
\begin{center}
\includegraphics[width=0.55\textwidth]{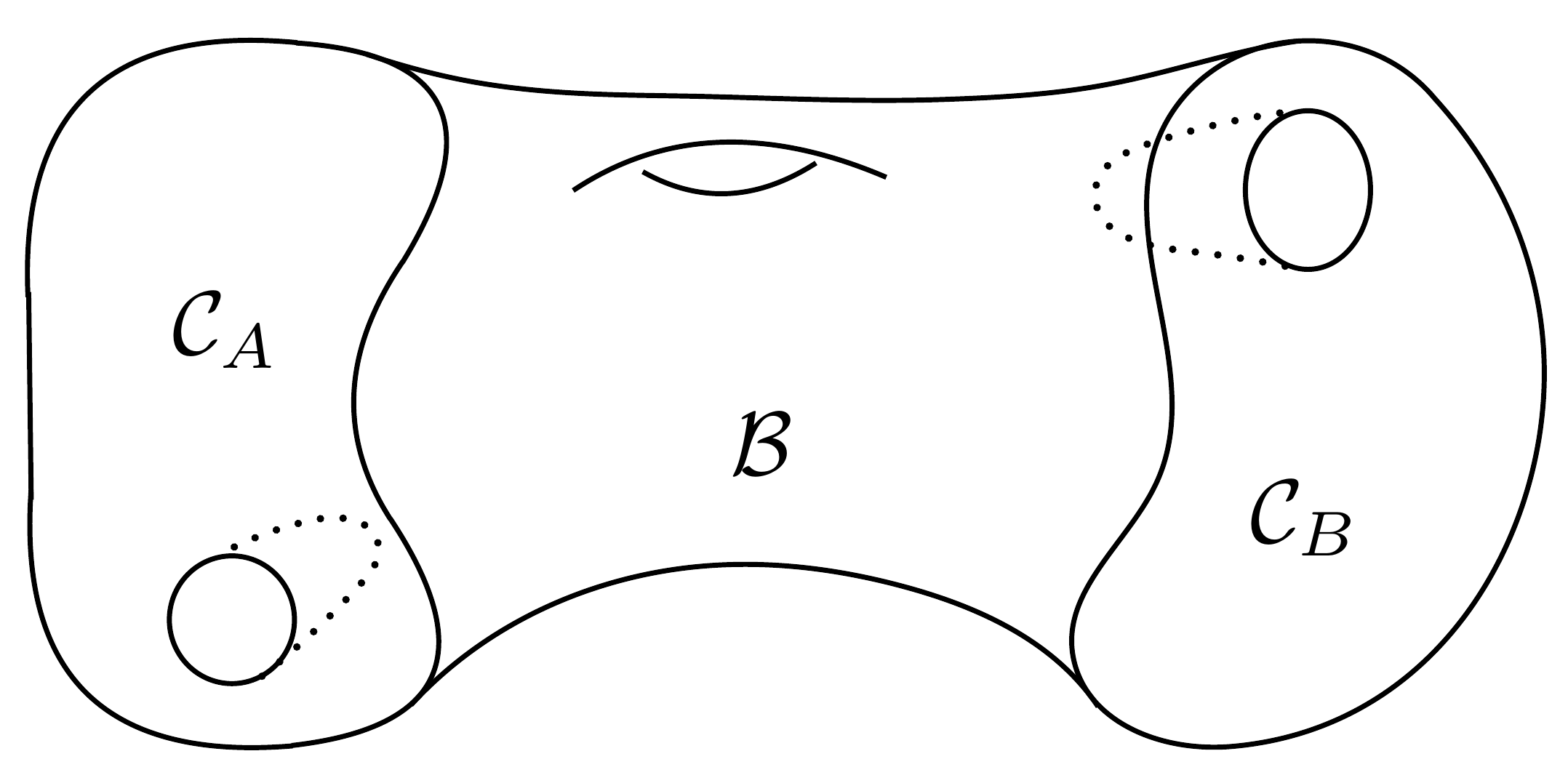}\end{center}
\caption{Two $d$-dimensional manifolds $\mathcal{C}_A$ and $\mathcal{C}_B$ are equivalent in bordism if together they form the boundary of a $(d+1)$-dimensional manifold $\mathcal{B}$.}
\label{f2}
\end{figure}

In general, such a $\cB_{d+1}$ may not exist. Mathematicians have given a full answer to the question of when does and when it doesn't, via bordism groups  \cite{10.2307/1970690}. Bordism is an equivalence relation between $d$-dimensional manifolds: $\cC^{A}_d$ and $\cC^{B}_d$ are equivalent if there is a manifold $\cB_{d+1}$ of one dimension higher such that $\partial \cB_{d+1}=\cC^{A}_d\cup\cC^{B}_d$ (see Figure \ref{f2}). Equivalence classes of manifolds defined in this way have a natural (abelian) group structure, where the group operation is to take the equivalence class of the disjoint union of manifolds\footnote{This can be replaced by a connected sum, as the two notions are equivalent under bordism.}, and  the trivial element is the class represented by any manifold which is a boundary.   If the manifolds $\cC^{A}_d,\cC_{d}^B$ carry any extra structure (such as an orientation, spin structure, or gauge bundle), we can also demand that this structure extends to $\cB_{d+1}$, leading to the notion of twisted bordism groups. The one we are interested in is the $d$-dimensional spin bordism group $\Omega_d^{\text{Spin}}$.  Then, there will be no  topological obstruction to a bubble of nothing for a given compact space  $\cC_d$, i.e. there is a manifold  $\cB_{d+1}$ such that $\pd \cB_{d+1} \cong \cC_d$,  when $\cC_d$ belongs to the trivial class in $\Omega_d^{\text{Spin}}$. We shall refer to the corresponding manifold $\cB_{d+1}$ as a \emph{nulbordism} or a  \emph{bordism} for $\cC_d$.

Let us now revisit Witten's bubble in this more formal language. In this case the compact manifold is the circe $\cC_1 \cong S^1$ supplemented with a given choice of  boundary conditions for the fermions.  The mathematical fact that protects supersymmetric
compactifications on $S^1$ from the decay to nothing is that the one-dimensional $\Spin$
bordism, $\Omega_1^{\text{Spin}}=\mathbb{Z}_2$,  has a non-trivial element. The trivial class corresponds to the circle with antiperiodic (susy-breaking) boundary conditions, and so that $\cB_2\cong \cD$ is topologically a disc. The nontrivial class is generated precisely by an $S^1$ with a periodic (i.e. susy-preserving)  spin structure.  So this generator is not the boundary of any manifold, and
in particular there is no spin structure on the disk $\cD$ that gives rise
to the periodic spin structure on the boundary $S^1$.

The same story persists at degree two: $\Omega^\Spin_2=\bZ_2$, and
the non-trivial generator can be taken to be $T^2$ with the fully
periodic $\Spin$ structure (notice that antiperiodic boundary conditions along any one-cycle would allow us to use the one-dimensional nulbordism and write $T^2$ as the boundary of $\cD\times S^1$). Again, $T^2$ compactifications seem to
be topologically protected. 

The situation changes drastically in three dimensions. Here we have that
\begin{equation}
  \Omega_3^\Spin = 0\, . \label{booooom}
\end{equation}
This tells us that there is no obstruction to constructing the bordism
to nothing of $T^3$, even if we choose the supersymmetry preserving
boundary conditions!  We emphasize that the same is true for any $T^{d\geq 3}$,
since (topologically) we can always\footnote{We are working at the
  level of topology, so we can always deform the torus to the
  factorized case.} deform to the product $T^3\times T^{d-3}$ and
construct a bordism of the first factor. We will come back to this point in Section \ref{sec:detailedTopology}.  

We cannot refrain from stressing again that \eqref{booooom} means that supersymmetry and topological protection are two distinct mechanisms to ensure stability against bubbles of nothing, and that it is possible to have either without the other! 

What about higher dimensions? The spin bordism groups through degree 10 are\footnote{See \cite{Garcia-Etxebarria:2018ajm,McNamara:2019rup} for an extended and more general tables of bordism groups.}
\begin{equation}
  \label{eq:bordism-table}
  \def\arraystretch{1.5}
  \arraycolsep=4pt
  \begin{array}{c|ccccccccccc}
    d & 0 & 1 & 2 & 3 & 4 & 5 & 6 & 7 & 8 & 9 & 10 \\
    \hline
    \Omega^\text{Spin}_d & \bZ & \bZ_2 & \bZ_2 & 0 & \bZ & 0 & 0 & 0 &
                                                                      2\bZ
                                          & 2\bZ_2 & 3\bZ_2
  \end{array}
\end{equation}

In particular we also have that $\Omega_d^{\text{Spin}}=0$ if $d=6,7$. This is very interesting as they are precisely the relevant groups for compactifications of 10 dimensional string theory and M-theory to four dimensions. We will comment more on this in Section \ref{sec:phys}.

Let us finally remark that it was recently conjectured in \cite{McNamara:2019rup}  that $\Omega_d^{QG}=0$ for any consistent theory of Quantum Gravity. The reasoning goes as follows: if this cobordism group is not trivial, different equivalence classes can be associated to different conserved global charges that imply the presence of an exact global $(D-d-1)$-form symmetry, where $D$ is the space-time dimension. This would be inconsistent with the swampland criterion requiring the absence of global symmetries\footnote{Exact global symmetries are commonly believed to be inconsistent with quantum gravity. Strong evidence has been given in the context of AdS/CFT \cite{Harlow:2018jwu,Harlow:2018tng} and perturbative string theory \cite{Banks:1988yz}.}  in quantum gravity \cite{Palti:2019pca}. Therefore, a consistent theory of quantum gravity must contain the necessary defects that guarantee triviality of the cobordism classes. We have seen that for $d=3$ it is enough to consider a spin structure to get $\Omega_d^{\text{Spin}}=0$ while in other cases additional structures might be needed (see  \cite{McNamara:2019rup}  for more details). We can see that an immediate consequence of this conjecture is that there is no longer any topological obstruction to construct bubbles of nothing in any consistent quantum theory of gravity. Notice, though, that in some cases one might need to include UV stringy defects that prevent us from constructing smooth solutions within the supergravity approximation. Hence, we will restrict our attention to  $\Omega_3^{\text{Spin}}$ from now on and construct an explicit smooth solution for this case.

\subsection{Dynamical obstruction: The Positive Energy Theorem}\label{sec:pet}
In spite of \eqref{booooom}, we know that a pure $T^3$ compactification with periodic boundary conditions must somehow be a stable vacuum in Einstein's gravity, at least in less than 12 dimensions. This is because Einstein's gravity is a consistent truncation of supergravity, and a $T^3$ compactification preserves supersymmetry. A vacuum preserving any supercharge must necessarily be stable, since the supercharge can be written as a boundary integral of the supercurrent \cite{Deser:1977hu,Witten:1981mf}. 

One might think that this supersymmetric protection against decay is due to some delicate supersymmetric cancellation that will disappear as soon as SUSY is broken, even slightly. This would mean that on general grounds we should expect  bubble of nothing instabilities generically whenever SUSY is broken. Alas, at the classical level, this is not the case; the dynamical protection against decay is robustly built-in in Einstein's equations themselves, and is a consequence of the Positive Energy Theorem \cite{Witten:1981mf} and its generalization \cite{2004CMaPh.244..335D,Dai_2005}, which covers cases including compactifications. See also \cite{Hertog:2003xg,Hertog:2003ru} for attempts to construct negative energy solutions in string compactifications, which end up being obstructed by the PET. 

These theorems guarantee, under certain assumptions which we list momentarily, that the ADM mass of any spacetime that asymptotes to $\mathbb{M}_{D-d}\times \cC_{d}$, where $\cC_d$ is some compact manifold, is bounded below by zero and that the only solutions that have exactly zero mass is $\mathbb{M}_{D-d}\times \cC_{d}$ itself \footnote{
There are two slightly different theorems to consider.  In \cite{2004CMaPh.244..335D}, it is proven that whenever the Weak Energy Condition holds, any valid initial condition to Einstein's equations with vanishing time derivatives for the gravitational field must have $m_{\text{ADM}}\geq0$ with equality only for $M^{D-d}\times\mathcal{C}_d$. In \cite{Dai_2005}, the assumption on the time derivatives is dropped if one replaces the WEC by the Dominant Energy Condition, but the proof of unicity of the  $m_{\text{ADM}}=0$ solution is lost unless the asymptotic manifold $\mathcal{C}_d$ is Riemann-flat. Since in this paper we construct bubbles of nothing for $T^3$ quotients, we are in this last case, and that is why throughout the paper we phrase the discussion in terms of the DEC. For more general compactifications, it would be more appropriate to use the first theorem in \cite{2004CMaPh.244..335D}, and restrict to time-symmetric initial conditions. Most of the discussion we have in this paper regarding the DEC applies to WEC as well.}. 

A bubble of nothing spacetime is an euclidean solution to the equations of motion, and when restricted to the $t=0$ slice it is an asymptotically flat solution, as explained around \eqref{asflat} for the particular case of the KK bubble. This solution in fact has vanishing ADM mass, as it must be the case for any vacuum decay channel due to energy conservation. Since the Positive Energy Theorem (PET) forbids this, we conclude that the vacuum is dynamically protected against decay via bubbles of nothing whenever the assumptions of the PET hold.

So it all boils down to what these assumptions are and how easily can be broken. Suppose we are interested in a particular $D$-dimensional manifold $\cM_D$ that asymptotes to $\mathbb{M}_{D-d}\times \cC_{d}$.  The Positive Energy Theorem of \cite{2004CMaPh.244..335D} guarantees that any solution of this kind to Einstein's equations (with matter) on $\cM_D$ not identical to $\mathbb{M}_{D-d}\times \cC_{d}$, will have a positive ADM mass as long as
\begin{enumerate}
\item[\bf 1.] $\cM_D$ admits a $\text{Spin}$ structure, with an asymptotically covariantly constant spinor.
\item[\bf 2.] The matter in the theory satisfies the Dominant Energy Condition:
\begin{equation} -T^M_{\phantom{M}N} k^N\quad\text{is  causal and future-pointing}\label{dec}\end{equation}
whenever the vector $k^N$ is also causal and future-pointing, $M,N=\{0,\ldots,D-1\}$.
\end{enumerate}
The first condition is topological in nature, and it implies that $\cM_D$ itself admits covariantly constant spinors. This will always be the case in supersymmetric compactifications, and indeed, Witten's proof of the PET was inspired by these. 

A compactification $\mathbb{M}_{D-3}\times T^3$ with periodic boundary conditions on $T^3$ admits covariantly constant spinors;  therefore, the presence of a bubble solution with vanishing ADM mass depends on whether the second condition is violated. As long as the DEC applies, we will not be able to construct a bubble of nothing, even  if  supersymmetry is explicitly broken and regardless of the  absence of a topological protection. From the point of view of the semiclassical decay, we expect the stability to be enforced  via the Coleman-DeLuccia mechanism (dynamically), as it does to  prevent the non-perturbative decay  of supersymmetric vacua \cite{Coleman:1980aw,Cvetic:1992st}, and as it  has also been observed to obstruct the decay to nothing in    \cite{Blanco-Pillado:2016xvf}.  That is, in the absence of DEC violating sources the critical radius of the bubble and its euclidean action  should diverge,  so that the decay rate  vanishes.

It is amusing that, although  there is no topological obstruction for the decay to nothing in the sense of the previous Subsection, the PET can still protect the vacuum $\mathbb{M}_{D-3}\times T^3$ from decaying. This is in contrast to the $S^1$ case with antiperiodic boundary conditions, where there is neither topological obstruction (because we are in the trivial class in $\Omega_1^{\text{Spin}}$), nor the PET applies since the first condition is not satisfied (no covariantly constant spinors at infinity), as illustrated by Witten's bubble of nothing.

 To sum up, there can only be a bubble of nothing if there is no topological obstruction and the PET does not apply. Checking that the PET does not apply requires in turn checking a local condition (the DEC) and a global one (existence of asymptotically covariantly constant spinors). This state of affairs is illustrated schematically in Figure \ref{fig:flow0}.
\tikzstyle{decision} = [rectangle, draw, 
    text width=6em, text badly centered, node distance=4cm, inner sep=4pt]
\tikzstyle{block} = [rectangle, draw, 
    text width=5em, text centered, rounded corners, minimum height=4em]
\tikzstyle{line} = [draw,-{Latex[length=2mm,width=2mm]}]
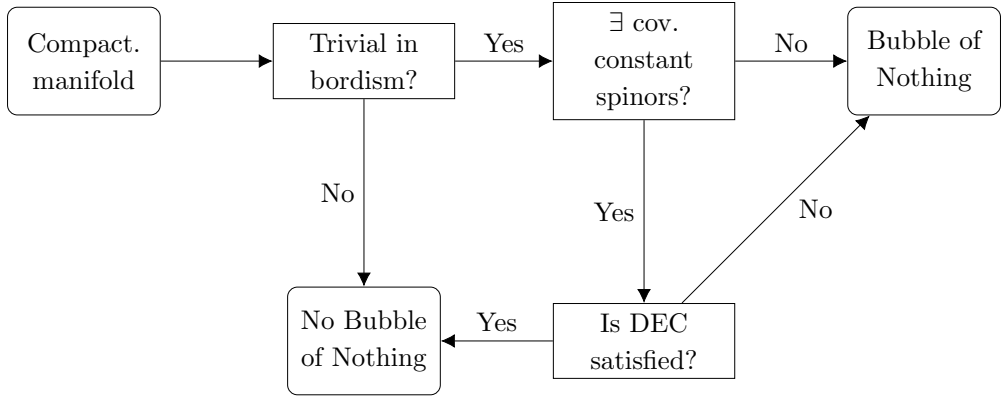
\begin{figure}
\begin{center}
\resizebox{.85\textwidth}{!}{\begin{tikzpicture}[node distance = 4cm, auto]
    \node [block] (init) {Compact. manifold};
    \node [decision, right of=init] (topobs) {Trivial in bordism?};
    \node [decision, right of=topobs] (cccs) {$\exists$ cov. constant spinors?};
    \node [decision, below of=cccs] (ec) {Is DEC satisfied?};
    \node [block, below of=topobs](nobon){No Bubble of Nothing};
    \node [block,right of=cccs](bon){Bubble of Nothing};
    
    \path [line] (init) -- (topobs);
    \path [line]  (topobs) --node[anchor=south] {Yes} (cccs);
        \path [line]  (topobs) --node[anchor=east] {No} (nobon);
        \path [line]  (cccs) --node[anchor=south] {No} (bon);
        \path [line]  (cccs) --node[anchor=east] {Yes} (ec);
         \path [line]  (ec) --node[anchor=south] {Yes} (nobon);
          \path [line]  (ec) -- node[anchor=west] { \, No} (bon);
\end{tikzpicture}}\end{center}

\caption{Flowchart illustrating when can one get a bubble of nothing. Given a compactification manifold $\mathcal{C}$, one first checks that there is no topological obstruction (that the manifold is trivial in bordism). Assuming this is the case, one must make sure there are either no covariantly constant spinors in the compactification manifold, or that the relevant energy condition is violated. If either of these happens, there \emph{can} be a bubble of nothing. As we will see in the paper, our expectation is that if it can be there, it will be.}
\label{fig:flow0}
\end{figure}

So what about breaking the second condition? At first sight, breaking the Dominant Energy Condition seems like a bad idea, since it can lead to traversable wormholes  and time machines (see e.g. \cite{Morris:1988tu,Curiel:2014zba}). However, while these pathological objects require a violation of the DEC, the converse is not true; the DEC is violated (although by tiny amounts) by quantum effects such as Casimir energies \cite{Curiel:2014zba}, false vacua (in the Coleman-DeLuccia sense \cite{Coleman:1977py}), and just about in any AdS vacuum. So it is probably safe to say that while writing down a random DEC-violating theory is not allowed, some violations are. 

In this paper, we will study how both assumptions in the theorem can be weakened in a reasonable way. We will find that both can be broken naturally in string theory, and correspond to different ways to break supersymmetry; breaking the first condition corresponds to compactification on a manifold which admits no covariantly constant spinors, which will always break supersymmetry; while the second depends on the matter content and higher derivative corrections of the EFT. To give an example of the latter, we will write down in the next Section a concrete model that violates the DEC by including a higher derivative correction proportional to the Gauss-Bonet term, and construct explicit bubble of nothing solutions to it. In Section \ref{sec:phys} we will provide an string embedding of the model into heterotic string theory on $T^4$ and its type IIB dual.

The assumptions in Witten's proof of the PET are closely related to each other.  As we show in Section \ref{sec:phys}, it is possible to modify the proof of the PET to work with e.g. a $\text{Spin}^c$ instead of a Spin structure, which then leads to a different energy condition. For instance, the results of \cite{Blanco-Pillado:2016xvf} can be understood in this way. Indeed, the fermions in the model considered there are charged under a $U(1)$ gauge field, and thus the relevant fermionic structure is precisely $\text{Spin}^c$.  Since $\Omega_1^{\text{Spin}^c}=0$, there is no topological obstruction whatsoever to the existence of bubbles of nothing in a theory with charged fermions.  In particular, a $S^1$ with periodic boundary conditions is the boundary of a disk with flux. Regarding the dynamical obstruction, this compactification admits asymptotically covariantly constant  \emph{charged} spinors\footnote{Consider an $S^2$ with flux. The index theorem says that the Dirac equation has a single zero mode, the restriction of which to each hemisphere provides the desired asymptotically covariantly constant spinor, after a suitable conformal transformation (which maps zero modes to zero modes since the massless Dirac equation is conformally invariant).}. But the model in \cite{Blanco-Pillado:2016xvf} violates the modified energy condition for the $\text{Spin}^c$ PET (a BPS bound), except in the supersymmetric limit. This is why there is a bubble of nothing. Note that the model in \cite{Blanco-Pillado:2016xvf} always satisfies the ordinary DEC. This modified energy theorem was also used in \cite{Gibbons:1982jg} to show that the mass of any charged black hole solution is above extremality. The general picture is that one has several slightly different versions of the PET, with slightly different assumptions; as long as one of these applies, we will have no bubble of nothing. We will discuss this in more detail in Section \ref{sec:phys}.

\section{Our model in a nutshell}\label{sec:nuts}
The main goal of this paper is to learn to what extent can the obstructions discussed in Section \ref{sec:rev} be lifted in reasonable setups when supersymmetry is broken and, ultimately, to what extent is a vacuum necessarily unstable whenever SUSY is broken. 

To do this, we would need to show one has bubbles of nothing whenever the relevant bordism group vanishes and there is no local energy condition preventing the decay. We comment on this briefly in Section \ref{sec:phys}, but we do not have a general construction. Instead, we will focus on a concrete class of compactifications $\mathbb{M}_{D-3} \times \cC_3$, which illustrate what we believe are general features,  where the internal manifold is a three-torus or quotients of it by free actions  $\cC_3\cong T^3/\Gamma$, with arbitrary spin structure. In doing so, we provide an example of a more convoluted bubble of nothing that is not simply described by a shrinking circle or a sphere, while at the same time being able to do explicit calculations. We are not aware of similar constructions in the literature. In this Section we introduce our model and briefly present our results.

\subsection{Topology of the solutions}
\label{sec:topology-in-a-nutshell}

\begin{figure}
\begin{center}
\includegraphics[width=0.35\textwidth]{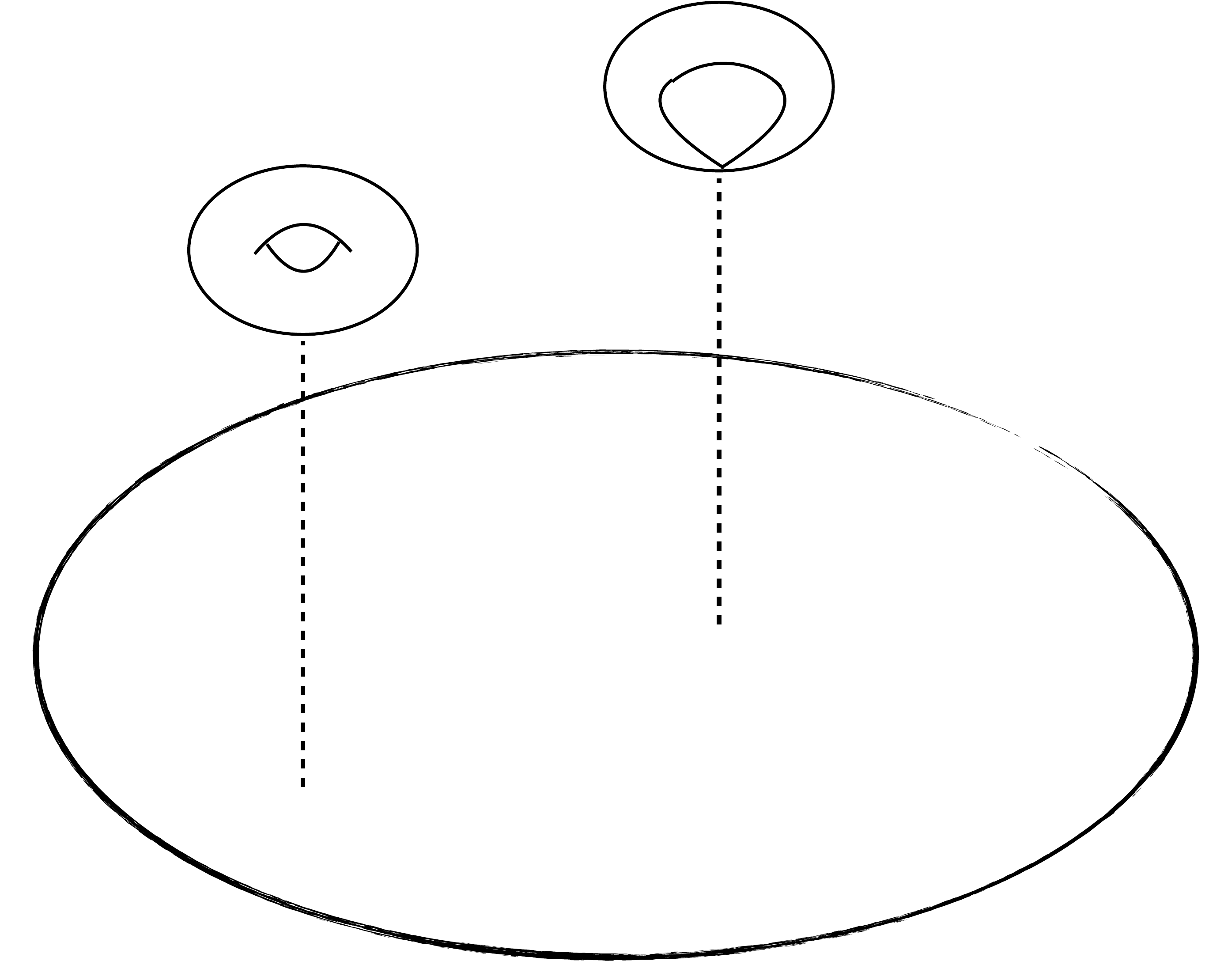}\end{center}
\caption{Schematic representation of the Weierstrass fibration over a disk \eq{eq:Weierstrass}: There is a $T^2$, which can pinch off at a discrete set  of points. The bubbles we will consider in this paper all share this general topological structure.}
\label{f1}
\end{figure}

We will start discussing the compact space  $\cC_3\cong T^3$ with supersymmetry-preserving (periodic) boundary conditions. As discussed in Section \ref{sec:rev}, the fact that $\Omega_3^{\text{Spin}}=0$ tells us that there is a spin four-manifold $\cB_4$ such that $\partial \cB_4=T^3$. This manifold is a candidate for constructing a $T^3$ bubble of nothing, but what is it? The precise answer can be found in pg. 524 of \cite{scorpan2005wild}, and we discuss it in more detail later on, but we will give the idea first. Let us regard $T^3=T^2\times S^1$ as a trivial fibration of a $T^2$ over a circle, and then introduce a disk $\cD$ such that $\partial \cD=S^1$. If one could extend the $T^2$ fibration and its spin structure on the boundary 
over the whole disk, the total space of such fibration would give the desired $\cB_4 \cong \cD \times T^2$. It turns out that one can do this, with the caveat that the fiber $T^2$ must pinch off in a discrete set of points inside the disk. This behavior might be familiar from elliptic fibrations in F-theory compactifications \cite{Weigand:2010wm,Weigand:2018rez} and indeed, that's what $\cB_4$ is: an elliptic fibration over $\mathbb{C}$ (a conformal rescaling of our disk $\cD$), described by a Weierstrass model

\begin{equation}
  \label{eq:Weierstrass}
  y^2 = x^3 + f(u)x + g(u),
\end{equation}
parametrized by the coordinate $u$. All three coordinates $x,y,u$ take values in $\mathbb{C}$. This configuration is illustrated in Figure \ref{f1}. These fibrations have been studied extensively \cite{Weigand:2018rez}, and in complex codimension one, they are completely classified. The number of degenerations, or pinchings of the fibration, is controlled by the zeroes of $f$ and $g$, and their vanishing degree. The total number of degenerations is the degree of the discriminant polynomial $\Delta=4f^3+27g^2$. To construct a nulbordism for $T^3$ with periodic spin structure, we need $\Delta$ to have degree 12.\footnote{Proofs of all these statements can be found in Section \ref{sec:detailedTopology}.} If the vanishing degree of $f$ or $g$ at a point is low enough (for instance, if all the zeroes are isolated), the total space of the fibration is smooth, even if the torus fiber itself becomes singular. Actually, from a geometrical point of view,  isolated degenerations can be described locally as Taub-NUT points, that is, Kaluza-Klein monopoles \cite{Gross:1983hb,PhysRevLett.51.87}.
 So we just need to have all 12 degenerations separate from each other and we have a smooth $\cB_4$.

There is another description of $\cB_4$ that might be more familiar. The boundary of $\cB_4$ is $T^3$, so we can take two copies of $\cB_4$, reverse orientation, and glue them along their common $T^3$ boundary. The resulting compact manifold is a K3, since it has by construction an elliptic fibration with 24 degenerations and a $\mathbb{P}^1$ base (the result of gluing the two $\cD$'s of each copy of $\cB_4$).  Thus, $\cB_4$ can be described as ``half a K3''. This particular decomposition of K3 comes up in discussions of the ``stable degeneration limit'' \cite{Aspinwall:1997ye}.

Let us now consider compactifications on the quotients of tori $\cC_3 \cong T^3/\Gamma$ by a non-trivial freely acting discrete symmetry $\Gamma$.
In particular will focus on the set of examples  given by the six classes of compact orientable manifolds admitting a (Riemann)-flat metric; a discussion can be found in \cite{PFAFFLE2000367,Acharya:2019mcu}.  In the above example, $T^3$ was written as a trivial torus fibration over $S^1$, but the idea works in the same way if we have a more general (nonsingular) torus fibration over $S^1$.  
All these manifolds $T^3/\Gamma$ are $T^2$ fibrations over $S^1$, where the $T^2$ comes back to itself up to an $PSL(2,\mathbb{Z})$ action. These manifolds are all spin, and taking into account the spin structure, there are 28 different possibilities. All of them admit nulbordisms in terms of a Weierstrass fibration \eqref{eq:Weierstrass}, though the total degree of $\Delta$ changes. 

These 28 classes are interesting because 27 of them do not admit covariantly constant spinors; they break necessarily all supersymmetry, and so they would be nice candidates for Minkowski nonsusy vacua at weak coupling\footnote{One expects quantum effects to introduce a running potential, but as long as this running is towards weak coupling, these are perfectly well-defined solutions.}. Reference \cite{Acharya:2019mcu} was able to construct bubbles of nothing in 26 out of 27 of these cases, showing that they are nonperturbatively unstable. The bubbles constructed there are products $\cD \times T^2$, with a trivial disk fibration\footnote{Reference \cite{Acharya:2019mcu} constructed these bubbles by taking a quotient of Witten's bubble of nothing that has fixed points. As a result, the bubbles in that reference actually contain orbifold singularities where the geometry is not smooth. These are the kind of mild singularity we can often ignore in string theory, but strictly speaking, these bubbles are not solutions to the GR equations of motion. Instead, wee can construct smooth bubbles for all 28 classes; we do so in Section \ref{sec:detailedBON}}. Our bubbles become the orbifold bubbles of \cite{Acharya:2019mcu} in a certain limit. We have constructed nulbordisms using Weierstrass fibrations  \eq{eq:Weierstrass} for all 27 cases; below, we will discuss explicitly the bubble for class $G3$, the only one left out in \cite{Acharya:2019mcu}. The only difference with the $T^3$ case is that the degree of $\Delta$ is 8 instead of 12.

\subsection{The EFT model}
\label{sec:NUTSmodel}

As explained in Section \ref{sec:rev}, constructing a topological manifold $\cB_4$ is only half the story; we also need to find a metric on it that asymptotes quickly enough to the flat metric on $\mathbb{M}_{D-3}\times T^3/\Gamma$. And here, a general obstruction is provided by the Positive Energy Theorem (PET); as long as the solution admits covariantly constant spinors at infinity and the DEC holds, there will be no bubble of nothing. 

For the 27 quotients of $T^3$ without covariantly constant spinors, the PET provides no obstruction\footnote{In Section \ref{sec:phys}, we will discuss some variations of the Positive Energy Theorem that could apply to these scenarios, but there is no obstruction in the end.}. But for $T^3$, it shows that one will not have a bubble unless the DEC is violated. Even in this case it is a challenge to construct an actual solution to the euclidean equations of motion representing a bubble of nothing, and this is what we will accomplish in this paper.

We will now write down a low-energy EFT that violates the DEC, in which we will construct the bubbles. The model involves  the spacetime metric $g_{MN}$,  an anti-symmetric tensor $B_{MN}$, and  a  dilaton field $\phi$, with the spacetime indices $M,N$ running in $0,\ldots,D-1$. The corresponding action (written in the string-frame\footnote{The action in Einstein frame is obtained with a  conformal scaling of the metric $g_{MN} = \rme^{\frac{4}{D-2} (\phi-\phi_0)}g^E_{MN}$. See eq. (15.12) in reference \cite{Ortin:2015hya}.}) has the form
\be
\boxed{S_{s} = -\frac{g^2_s}{16 \pi G_{D}} \int_{\cM_D} d^Dx \sqrt{-g} \rme^{-2 \phi }\Big[ R + 4 (\nabla \phi)^2-\ft1{12} H^2+\ft18 \alpha R^2_{GB}\Big],}
\label{eq:action}
\ee
where   $H_{MNP} = 3 \pd_{[M} B_{NP]}$ is the field strength of $B_{MN}$, and $G_{D}$ is  the $D$-dimensional Newton's constant.
When the parameter $\alpha$ is set to zero, the model can be identified with  the NSNS sector in the low-energy description of superstring and bosonic string theories. In that case,    $g_s = \rme^{\phi_\infty}$ represents the string coupling, which is determined by  the  expectation value of the dilaton, $\phi_\infty$.  
It can be checked explicitly that DEC is satisfied when $\alpha=0$, what makes sense since this is a consistent truncation of a supersymmetric theory, and we know there are no bubbles of nothing anyway.
Therefore, all the fun comes when we turn on the last term in the action \eqref{eq:action}, which is the dimensionally extended Gauss-Bonnet invariant 
\be
R^2_{GB} = R^2 - 4 R_{MN} R^{MN} + R_{MNPQ} R^{MNPQ}. \label{gbt}
\ee
 On a four-dimensional manifold $\cM_4$, \eqref{gbt} is topological, and its integral gives the Euler characteristic $\chi(\cM_4)$
 \be
 \int_{\cM_4} R^2_{GB} = 32 \pi ^2\,  \chi(\cM_4).
\ee
On higher dimensions, the term is no longer topological but it still is special in that it gives rise to second-order equations of motion for the metric (the corresponding theories are called Lovelock \cite{Padmanabhan:2013xyr}), thus avoiding the ghosts associated to the Ostrogradski instability. 

Turning on this deformation (and nothing else) breaks supersymmetry and the DEC. We have included it as a means to break supersymmetry explicitly in a controlled way, with the coupling constant $\alpha$ acting as a deformation parameter which controls the scale of supersymmetry breaking.  Although this  supersymmetry breaking mechanism might look contrived at first, it has a number of properties which will allow us to find explicit solutions in this theory.

  On the one hand,  we are studying the decay of a toroidal compactification, which is a flat geometry,  and therefore after deforming  the theory with the term $R_{GB}^2$ the compactification will still be a solution to the Euler-Lagrange equations. That would not be the case, for example, if we tried  to the deform the theory including a cosmological constant. 

 On the other hand, we will consider the $R_{GB}^2$ term as a small (perturbative) deformation of the theory, using   a vacuum solution to the Einstein's equations as  background geometry. In that situation, to leading order in perturbation theory, the net effect of such deformation is a warping of the bordism geometry, what simplifies  considerably the analysis of the Euler-Lagrange equations.

Furthermore, this deformation $R_{GB}^2$ can also be motivated in string theory. This quadratic higher derivative correction appears both for bosonic and heterotic strings as leading order $\alpha'$ corrections \cite{Gross:1986mw,Metsaev:1987zx,Tseytlin:1995bi}, in  M-theory  upon compactification\footnote{For more discussions about these terms on  M-theory see \cite{Duff:1995wd,Vafa:1995fj,Bachas:1999um,Gukov:1999ya,Green:1997di}, and in flux compactifications \cite{Gukov:1999ya,Becker:1996gj}.}  on $K3$ to  $D=7$ \cite{Duff:1995wd}, in type IIA compactified in $K3$ to\footnote{This is expected from the heterotic/type IIA duality in $D=6$ \cite{Witten:1995ex}.} $D=6$ \cite{Antoniadis:1997eg} and in orientifold compactifications of type IIB (and their type I duals) \cite{Tseytlin:1995bi}. In the particular  case of  superstring theories,  supersymmetry  requires additional terms to be included in the action together with the quadratic curvature terms \cite{Peeters:2000qj,Gukov:1999ya}.  We will describe the string theory embedding of our model in more detail  in Section \ref{sec:phys} and provide an explicit embedding of the action \eqref{eq:action} with $D=6$ as a toroidal compactification of heterotic string theory. 

It is also important to notice that only $\alpha>0$ is a physical deformation; the other sign leads to trouble with unitarity along the lines of \cite{Cheung:2016wjt}, and naked singularities \cite{Boulware:1985wk}.
This is consistent with the fact that in all situations where this quadratic deformation arises in a string theory compactification to flat space its  coefficient is positive $\alpha>0$ \cite{Metsaev:1987zx}. This is also consistent with the connection between the Weak Gravity  Conjecture and higher derivative corrections (see e.g. \cite{Cheung:2018cwt,Hamada:2018dde,Andriolo:2018lvp}), though this depends on additional higher-derivative terms. In any case, this particular deformation should only be taken as an example that allows us to construct an explicit solution, but there could many other supersymmetry breaking mechanisms that yield a finite rate for the bubble. Our goal in this paper is simply to provide an example as a proof of principle for the presence of these new types of bubbles of nothing.

\subsection{Main result: new bubbles of nothing}
\label{sec:NUTSresult}

The main technical result of our paper is that, when the Gauss-Bonnet coupling  $\alpha>0$ is turned on, there is a bubble of nothing mediating the decay  of the compactification  $\mathbb{M}_{D-3}\times T^3$, which has  the topology described above. 

Furthermore,  we will also construct  \emph{smooth}  instantons mediating the decay to nothing of the  27  non-supersymmetric compactifications $\mathbb{M}_{D-3} \times T^3/\Gamma$ in  \cite{PFAFFLE2000367,Acharya:2019mcu}, including the missing case where the compact space is $G3$ \cite{Acharya:2019mcu}. The  BON instantons for this family of non-supersymmetric compactifications exist, and have a finite decay rate, even in case $\alpha=0$.  To construct these instatons we have used  a combination of perturbation theory, space-time matching techniques and numerical methods, so the specific details of the solution  are rather involved.  Here we will only  summarise the general properties of these BON solutions, and we will discuss them at length  in Sections \ref{sec:det} and \ref{sec:detailedBON}.

 The general form of the instanton solutions mediating these decays can be characterised by the following $\SO(D-4)$ symmetric ansatz 
\be
ds^2_{\text{BON}} = W^2(y) \cR^2 d\Omega^2_{D-4} + h_{\alpha\beta}^\cB(y) d y^\alpha d y^\beta, \qquad \phi = \phi(y), \qquad  B_{MN}=0,
\ee
which in particular represents a warp product euclidean spacetime of the form $\cM_D \cong \cB_4 \times_W S^{D-4}$. Here $\cR$ is   the bubble nucleation radius, and  $h_{\alpha \beta}^\cB$ is the metric on the manifold $\cB_4$, which is parametrised by the coordinates $y^\alpha$, $\alpha= 1,\ldots, 4$.
  Interestingly,  the bounce solutions of this family describe a \emph{multi-centered bubbles of nothing}, with the various bubble cores located at the $N\le12$ points on $\cB_4$ where the $T^2$ fibre degenerates. As we mentioned above, each degeneration point carries a unit of Taub-NUT charge, and thus they can be locally described as KK monopoles.   These are the first bubbles of nothing of this kind to ever appear in the literature.
  
  Far from the KK monopoles the  bordism geometry has the form  $\cB_4 \to \mathbb{R} \times T^3/\Gamma$, the total spacetime approaches the euclidean vacuum $\mathbb{R}^{D-4} \times T^3/\Gamma$, and  the dilaton  its expectation value $\phi \to \phi_\infty$. More specifically, if we parametrise the $\mathbb{R}$ factor of $\cB_4$ with the coordinate $\rho(y)\equiv \cR W(y)$ we find 
\be
ds^2_{\text{BON}} \to \rho^2 d\Omega^2_{D-4} + d\rho^2 +h_{\bar \alpha\bar \beta}^\cC d y^{\bar \alpha} d y^{\bar \beta}, \qquad \phi \to \phi_\infty,
\label{vacuum}
\ee
where $h_{\bar \alpha \bar \beta}^\cC$ is the flat  metric on the compact space $\cC_3\cong T^3/\Gamma$, with coordinates  $y^{\bar \alpha}$, and $\bar \alpha=1,2,3$. 

 Setting aside the difference on the number of degenerations on $\cB_4$, the most important distinction between the BON decay of $T^3$ and the non-supersymmetric compactifications arises when comparing their  decay rates, $\Gamma_{\text{dec}} \sim \rme^{- S_{\text{BON}}}$, where   $S_{\text{BON}}$ is the euclidean BON action.  In the case of the $T^3$ compactification ($N=12$ degenerations) the bubble nucleation radius $\cR$ and the euclidean action, which are computed explicitly in Section \ref{eq:decayRates}, behave as
\be
\cR(\alpha) \propto \Big( \frac{24 \pi^2}{\cV_{T^3}}\alpha\Big)^{-1}, \qquad  S_{\text{BON}}(\alpha) \propto
\Big( \frac{24 \pi^2}{\cV_{T^3}}\alpha\Big)^{-(D-5)},
\label{eq:actionG3}
\ee
where 
$\mathcal{V}_{T^3}$ is the asymptotic volume of the $T^3$ compact space. As we anticipated in Section \ref{sec:pet}, since the compactification has  no  topological protection against the decay to nothing, and in the limit $\alpha\to0$ (where DEC holds) the decay is forbidden by the  Positive Energy Theorem, the stability of the supersymmetric compactification has to be enforced dynamically. Indeed, as we turn off  the Gauss-Bonnet term $\alpha\to 0$, both the bubble nucleation radius and the euclidean action grow unbounded and the decay rate vanishes. In other words, the stability of the supersymmetric compactification is protected via the Coleman-DeLuccia mechanism. This is in agreement with the conjecture made in \cite{Blanco-Pillado:2016xvf}. Conversely, when $\alpha\neq0$ the model violates DEC, and the Positive Energy Theorem can not protect the stability of the compactification (the second condition of the PET does not hold),    so the bubble on nothing instability appears.

Regarding the non-supersymmetric compactifications $T^3/\Gamma$, where $\cB_4$ is a Weierstrass fibration  with $N<12$ degenerations, we find that the bubble nucleation radius and the instanton action remain finite even if  we turn off the Gauss-Bonnet coupling. More specifically, in  the limit $\alpha \to 0$ we find that the radius $\cR$ and the euclidean action are given by
\be
 \cR= \frac{6(D-5)}{(12-N)} R_{\text{kk}}, \qquad \text{and} \qquad S_{\text{BON}} =
  \frac{A_{D-4}}{8 \pi G_{D-3}}\frac{\cR^{D-5}}{2},
\label{eq:G3orbifoldS0}
\ee
with the particular case of   $G3$ corresponding to $N=8$. Here, $R_{\text{kk}}$ is the radius of the base circle when writing the compact space as a $T^2$ fibration over $S^1$. The constant  $G_{D-3}\equiv G_D/\cV_{\cC_3}$ is the $D-3$-dimensional Newton's constant, and $\cV_{\cC_3}$ the volume of the compact space $\cC_3\cong T^3/\Gamma$.    As discussed in \ref{sec:pet}, in the case $\alpha=0$,  the DEC is not violated, but  these compactifications do not admit  covariantly constant spinors and thus  the Positive Energy Theorem provides no protection (the first condition of the PET does not hold). As a consequence  the bubble of nothing instability is present even when $\alpha=0$. As we will show in Section  \ref{eq:decayRates} in these cases the net effect of turning on  the Gauss-Bonnet coupling is to decrease the bubble nucleation radius and, in consequence, to enhance slightly the decay rate. It is also interesting to note that setting $D=7$ and $N=0$ we obtain the  nucleation radius and the action of  Witten's original bubble of nothing\footnote{The bounce action was overestimated by a factor of 2 in \cite{Witten:1981gj}. See e.g. appendix C in \cite{Brown:2014rka}.} \cite{Witten:1981gj},
\be
\cR = R_{\text{kk}}, \qquad S_{\text{BON}} = \frac{\pi R_{\text{kk}}^2}{8 G_4},
\ee
since in this case the $T^2$ fibration is trivial, and we could reduce to five dimensions on the $T^2$ factor, thus recovering Witten's original setup.

In field theory, to know whether a particular solution to the euclidean equations of motion is a bounce (mediates an instability) or an instanton (a harmless nonperturbative contribution to the vacuum energy), it is essential to compute the spectrum of fluctuations around the solution \cite{Coleman:1985rnk}. As will become apparent in Section \ref{sec:det}, we have used linear perturbation theory to construct (part of) these solutions. So already at the classical level our solutions are not exact, and we do not know what the fluctuation spectrum looks like. Furthermore, we have not computed any quantum effects. In gravity, this is a daunting task even for simple setups \cite{Brown:2016nqt}; for ours it seems hopeless. So why should anyone trust our bubbles (or any bubble of nothing solution, in fact)?

The answer is that our approximate bubble solution, when restricted to the $t=0$ slice, provides a valid initial condition (in the sense that it satisfies the Hamiltonian constraint) for time evolution in GR, other than the vacuum, with \emph{zero} ADM mass. With a small deformation, we can actually make it negative, as discussed in Appendix \ref{app:negativeStates}, and in fact, as negative as one wants\footnote{The Hamiltonian constraint is  solved only to first order in perturbation theory, although the existence of negative mass states is robust as any further corrections can only lift the negative mass by a tiny amount.}. Neither quantum effects nor classical instabilities can alter this fact, which clearly shows that the spectrum of the Hamiltonian is unbounded from below. It is energetically favorable for the vacuum to nucleate more and more of these solutions, so the instability is unavoidable. More concretely, in asymptotically AdS quantum gravity, energies below that of the vacuum are incompatible with unitarity bounds in the dual CFT \cite{Minwalla:1997ka}; with Minkowski asymptotics, there can be no unitary S-matrix with negative-energy states if one is to avoid tachyons, because a two-particle state of a positive energy particle and a negative energy one can have spacelike momentum and hence be tachyonic. 

In other words, unless we just demand by hand that all these negative energy states magically decouple from the spectrum, the instability seems unavoidable. The actual decay rate might be different, but, in any case, the action of the actual bounce solution must be equal or \emph{lower} than that of the configuration we start with. This is because we know there is an instability, so there must be one bounce solution. If our solution is not a bounce, it must have two or more negative fluctuation modes (it has at least one, since we can deform it to solutions with lower mass, and this mode is always present). In this case we can just follow the gradient flow of the action in configuration space\footnote{This might take us out of the effective field theory and into configurations like e.g. orbifolds, but this is not a problem since the ``action'' (logarithm of the path integral) should still be well-defined.} until there is just one negative mode (which must be exactly true in the actual bounce solution). So in any case, our expressions provide an \emph{upper bound} on the actual decay rate of the vacuum.

\section{Dynamical and Topological constraints}
\label{sec:det}

Let us begin the explicit construction of our bubble of nothing by presenting the field theory model and discussing in more detail the topological and dynamical obstructions that appear in this particular case, and how to overcome them.

For convenience, we will repeat here the field theory model already outlined in Section \ref{sec:NUTSmodel}.
It describes  the dynamics of  the spacetime metric $g_{MN}$,  an anti-symmetric tensor $B_{MN}$, and  a  dilaton field $\phi$, with the spacetime indices $M,N$ running in $0,\ldots,D-1$. The corresponding action written in the string-frame has the form
\be
S_{s} = -\frac{g^2_s}{16 \pi G_{D}} \int_{\cM_D} d^Dx \sqrt{-g} \rme^{-2 \phi }\Big[ R + 4 (\nabla \phi)^2-\ft1{12} H^2+\ft18 \alpha R^2_{GB}\Big],
\label{eq:actionDetailed}
\ee
where   $H_{MNP} = 3 \pd_{[M} B_{NP]}$ is the field strength of $B_{MN}$, and $G_{D}$ is  the $D$-dimensional Newton's constant. Then, the  Euler-Lagrange equations 
are given by
\bea
   R_{MN} &=& -2 \nabla_M \nabla_N\phi    +\ft1{4}   H_M^{\phantom{M}PQ} H_{NPQ}   -\ft14 \alpha \Big[ R_{MRST} R_N^{\phantom{N}RST} \nonumber \\
 && -2 R_{MSNT} R^{ST} -2 R_M^{\phantom{M}S} R_{NS} + R R_{MN}  \Big],
\label{EinsteinEOM}
\eea
for the metric,  while those of the  dilaton and the two-form read
\be
\nabla_{(D)}^2 \phi-   2(\nabla\phi)^2 =- \ft1{12}  H^2 + \ft{\alpha}{16} R^2_{GB}, \qquad \pd_M (\sqrt{-g} \rme^{-2 \phi} H^{MNP}) =0,
\label{dilatonHeom}
\ee
where $\nabla_{(D)}^2$ is the $D$-dimensional Laplace operator.   Note that the model allows for the consistent truncation of the two-form $B_{MN}$, so in the following we will  set $B_{MN}=0$ to simplify the analysis.

\subsection{Topology of the bubble}\label{sec:top}
\label{sec:detailedTopology}

We now provide a few more details (well known to experts, but
hopefully useful for those not familiar with the construction) of the
the nulbordism $\cB_4$ with boundary $T^3$ that we sketched in
Section~\ref{sec:topology-in-a-nutshell}, as well as the generalization to flat manifolds $T^3/\Gamma$ where $\Gamma$ acts freely on the torus. 

\subsubsection*{The nulbordism for $T^3$}

Consider what physicists call the $dP_9$ surface, and
mathematicians more often call the rational elliptic surface. We
denote it by $Z$. Topologically, it can be obtained by blowing up
$\bP^2$ at 9 generic points. $Z$ can be described as an elliptic
fibration over $\bP^1$, in which the fiber degenerates over 12 points
in the base. (So this is, in a well defined sense
\cite{Donagi:2012ts}, ``half a K3'', since on a K3 we have the
elliptic fiber degenerating over 24 points in the base, as mentioned in Section \ref{sec:topology-in-a-nutshell}. We can
represent the space in Weierstrass form: it is given by the locus
\begin{equation}
  \label{eq:Weierstrass2}
  y^2 = x^3 + f(u)xz^4 + g(u)z^6
\end{equation}
inside the toric variety
\begin{equation}
  \begin{array}{c|ccccc}
    & u_1 & u_2 & x & y & z\\
    \hline
    \bC^*_1 & 1 & 1 & 2 & 3 & 0\\
    \bC^*_2 & 0 & 0 & 2 & 3 & 1
  \end{array}
\end{equation}
which is a $\bP^{2,3,1}$ fibration over $\bP^1$, where $(u_1,u_2)$
parameterize the base, and $(x,y,z)$ parameterize the fiber. For
consistency we need to choose $f$ and $g$ to be homogeneous
polynomials of degree $4$ and $6$ in the base coordinates
$(u_1,u_2)$. As a small sanity check, note that the discriminant
$\Delta=4f^3+27g^2$ of the elliptic fibration is indeed a degree-12
polynomial on the $u_i$, so we indeed have 12 degenerations of the
fiber.

This space is not Calabi-Yau, since we have that
\begin{equation}
  c_1(TZ) = 3\ell - \sum_{i=1}^9 e_i
\end{equation}
with $\ell$ the pullback of the hyperplane on $\bP^2$, and $e_i$ the
exceptional divisors coming from the blow-up. These divisors satisfy
$\ell^2=1$, $\ell\cdot e_i=0$ and $e_i\cdot e_j = -\delta_{ij}$. It is
also not $\Spin$, since $w_2(TZ)=c_1(TZ)$ mod 2, so
$\dsz{w_2}{\ell}=1$ mod 2, for instance.

We can take care of both obstructions at once if we remove from the
space a tubular neighborhood of the Poincare dual of $c_1(TZ)$. In
this particular case this is known to be simply the homology class of
the $T^2$ fiber \cite{Heckman-Looijenga}. So, pick any (open) disk
$D_E$ on the base which does not intersect the discriminant
locus. (Any small enough disk around a generic point in the base will
do.) Denote by $E$ the total space of the torus fibration over $D_E$,
with topology $T^2\times D_E$. We then set $\cB_4=Z - E$. This now has
$w_1(T\cB_4)=0$, and in fact $c_1(T\cB_4)=0$, since we have removed a Poincare
dual to the characteristic class of $Z$.\footnote{This example is an
  instance of the ``log-CY'' construction of \cite{Donagi:2012ts}, and
  somewhat explains why the stable degeneration limit of K3 is built
  out of $dP_9$ surfaces.} 

It remains to be shown that the boundary of $\cB_4$, which has topology
$S^1\times T^2 = T^3$ (as the torus fibration around a generic point
in the base is trivial), has a periodic $\Spin$ structure. We can
proceed by contradiction (see \cite{scorpan2005wild} for an argument
that does not use index theory). Assume that on $\partial \cB_4$ we did
not have a fully periodic structure. This means that there is some
one-cycle $L$ in $\partial \cB_4$ with anti-symmetric boundary conditions
on the fermions. Then we can construct another four manifold $W$ by
``filling in'' $L$. It is clearly the case that the $\Spin$ structure
on $\partial \cB_4$ extends over $W$, so by gluing $W$ to $\cB_4$ we end up
with a smooth $\Spin$ four-manifold $K$. In terms of the curvature,
the signature of $K$ can be computed as
\begin{equation}
  \sigma = \frac{1}{3}\int_K p_1(TK)\, .
\end{equation}
This will receive contributions only from $\cB_4$, so it equals the
signature of $dP_9$, which is 8. From here we learn that
\begin{equation}
  \int_K p_1(TK) = 24\, .
\end{equation}
The index theorem tells us that a Dirac fermion on $K$ would have
\begin{equation}
  n_+ - n_- = \int_K \hat{A}(TK) = -\frac{1}{24} \int_K p_1(TK) = -1
\end{equation}
net zero modes. But in four dimensions the eigenvalues of the Dirac
operator always appear in pairs (see for example appendix~B.3 of
\cite{Witten:2015aba}), so this is a contradiction, and $K$ cannot
exist.\footnote{More generally, the fact that in dimensions $d=8k+4$
  the signature is a multiple of 16 is known as Rokhlin's theorem.}

\subsubsection*{The nulbordism for $G3$}

Let us briefly describe the nulbordism for the $G3$ geometry
introduced in \cite{Acharya:2019mcu}. This geometry can be understood
as a fibration of a $T^2$ over $S^1$, with monodromy of the $T^2$
corresponding to a rotation by $2\pi/3$ of the $T^2$.

This kind of fibration arises in a familiar context in
F-theory.\footnote{We refer the reader unfamiliar with F-theory to the
  nice review \cite{Weigand:2018rez}, which contains background for
  all the statements made here.} Consider an elliptic fibration over a
complex plane, and assume that at a given point of the base one has a
degeneration of Kodaira type $IV^*$ (also known as an $E_6$
degeneration in physics). The $SL(2,\bZ)$ action around the
singularity is of order 3, given by a $2\pi/3$ rotation of the
$T^2$. So the total space of the fibration over a small circle in the
base linking the point where the singularity is located will have the
same topology of $G3$, at least if we ignore the spin structure. The
nulbordism of interest to us can then be constructed as the total
space of the fibration over a small disk in the base centered around
the $IV^*$ degeneration. This configuration can be smoothed
straightforwardly, giving rise to an elliptic fibration degenerating
at 8 points in the base.

We still need to show that the spin structure on the space that we
have just constructed is the one we are after, namely the periodic
one. To see this, recall from \cite{Acharya:2019mcu} that there are
two possible spin structures on the space $G3$: the periodic one that
we want, and a second, antiperiodic one. We can characterize which one
we have by reducing on the torus fiber, and considering the effect of
circling the singularity at the origin three times (since the
geometric monodromy is of order three). For the periodic spin
structure the effect of this rotation will leave fermions invariant,
while under the antiperiodic spin structure the fermions will pick up
a sign. In the Kodaira classification there are precisely two
singularities that give rise to monodromies of order three: they are
the $IV$ and $IV^*$ degenerations. Their monodromies are inverses to
each other, so we can glue a $IV$ singularity to a $IV^*$ singularity
to form a closed manifold without further singularities, the result is
a $dP_9$ surface. This surface does not admit a spin structure, so it
must be the case that the spin structures on the elliptic
three-manifolds surrounding the singularities (both of which are
topologically $G3$, if we ignore the spin structure) are opposite,
otherwise the gluing construction would provide $dP_9$ with a spin
structure. On the other hand, we can bring two $IV$ degenerations
together in order to construct a $IV^*$ degeneration, so it must be
the case that the square of the monodromy action on the fermions
around a $IV$ gives the action on the fermions around a $IV^*$. The
only solution to these constraints is that the $G3$ manifold linking
the $IV^*$ singularity has the periodic spin structure (justifying our
choice above), and the one around a $IV$ degeneration the antiperiodic
one.

\subsubsection*{Nulbordisms for $T^3/\Gamma$} 

The techniques we described above work not only for $G3$, but actually allow us to construct topological nulbordisms for any flat torus quotient $T^3/\Gamma$, with any spin structure. These have been completely classified; see \cite{Acharya:2019mcu} and references therein. There are six possible geometries, labeled $G1,\ldots G6$, each of which admits a different number of spin structures, for a total of 28 cases. All cases except for $G6$ can be understood as a $T^2$ fibered over an $S^1$ with a constant complex structure parameter and a nontrivial $SL(2,\mathbb{Z})$ holonomy. Because the complex structure must remain invariant under the $SL(2,\mathbb{Z})$ transformation, for cases $G3,G4$ and $G5$ the complex structure must be chosen $\tau=i$ or $\tau=\sqrt[3]{-1}$, since these are the only points left invariant by a nontrivial subgroup of $PSL(2,\mathbb{Z})$; for $G1,G2$, any $\tau$ works, which we choose for convenience to be $\tau=i$. All of these admit a nulbordism in terms of a Weierstrass fibration with the type of singularity (depending on spin structure) specified on Table \ref{t1}.   A good reference for this is \cite{Weigand:2018rez}.

\begin{table}[!htb]
\begin{center}
\bgroup
\def\arraystretch{0.8}\begin{tabular}{c|c|c|c}
Class& $\#$ of s.s.&$SL(2,\mathbb{Z})$ act.&Kodaira sing.\\\hline
G1&8&$\left(\begin{array}{cc}1&0\\0&1\end{array}\right)$&--\\\hline
G2&8&$\left(\begin{array}{cc}-1&0\\0&-1\end{array}\right)$&$I_0^*$\\\hline
G3&2&$\left(\begin{array}{cc}0&1\\-1&-1\end{array}\right),\left(\begin{array}{cc}0&1\\-1&-1\end{array}\right)$&$IV,IV^*$\\\hline
G4&4&$\left(\begin{array}{cc}0&-1\\1&0\end{array}\right),\left(\begin{array}{cc}0&1\\-1&0\end{array}\right)$&$III,III^*$\\\hline
G5&2&$\left(\begin{array}{cc}1&-1\\1&0\end{array}\right),\left(\begin{array}{cc}0&1\\-1&1\end{array}\right)$&$II,II^*$
\end{tabular}\egroup\end{center}
\caption{Table listing five of the six flat tori quotient geometries, together with their number of spin structures, action of $SL(2,\mathbb{Z})$ (which also gives the holonomy), and maximal Kodaira singularity type of the corresponding nulbordisms. Only the first class, $T^3$ with a periodic spin structure, is compatible with supersymmetry; it is the only one for which there is no nulbordism and hence no bubble of nothing.}
\label{t1}
\end{table}

The only case left, $G6$, is a quotient of $G2$ by an additional $\mathbb{Z}_2$ action $\omega$ defined as follows: If $w$ is a complex coordinate on $T^2$ and $\theta$ parametrizes the $S^1$, then
\begin{equation} \omega:\, (\theta,w)\,\rightarrow\, \left(-\theta, \frac{1+\tau}{2}+w^*\right).\label{ss100}\end{equation}
Topologically, this is not a $T^2$ fibration over a circle, as the other flat tori are. Rather, this corresponds to a $T^2$ fibration over an interval; the torus becomes a Klein bottle at the endpoints.

The singularity $I_0^*$ corresponding to $G2$ can be deformed to four $I_1$ singularities in a complex-conjugation symmetric way. Then, the action \eq{ss100} can be extended to the whole $G2$ nulbordism, acting by complex conjugating the coordinate on the base and on the fiber as illustrated in \eq{ss100}. The resulting action has no fixed points; thus, the quotient of the Weierstrass fibration also leads to an appropriate nulbordism for $G6$.

\subsection{Geometric ansatz for the bubble}
\label{sec:BONansatz}

In the present Section we will describe the general  features of the BON spacetime that we construct below.

In order to  discuss the semiclassical decay of compactifications of the form  $\mathbb{M}_{D-3} \times T^3/\Gamma$, first we need a characterisation of the corresponding euclidean vacuum geometry, namely $\mathbb{R}_{D-3} \times T^3/\Gamma$. It turns out that a useful description for this space  is given in terms of the euclidean line element \eqref{vacuum}, where the non-compact factor in  is expressed using  spherical coordinates. Back  in  Lorentzian signature this gauge corresponds to a de Sitter slicing of  $\mathbb{M}_{D-3}$. Note that, since  the  geometry is flat, it does indeed represent a solution  solution to the Euler-Lagrange equations (\ref{EinsteinEOM}-\ref{dilatonHeom})  provided the dilaton is set to a constant value $\phi=\phi_\infty$.

 We would like to identify  the  most general euclidean line element for a  BON geometry $\cM_D$  mediating the decay of the a $D-$dimensional   vacuum $\mathbb{R}_{D-3} \times T^3/\Gamma$. 
Since we are interested in instanton solutions, we will require the BON ansatz to be invariant under a   $\mathrm{SO}(D-4)$ symmetry acting on the non-compact factor of the background. Any  line element consistent with this symmetry  can be described as  a warped geometry of the form $\cM_D = \cB_4 \times_W S^{D-4}$.  Furthermore, the manifold  $\cB_4$  needs to be an appropriate nulbordism for the compact space $T^3/\Gamma$.  Then, we find
\be
ds^2 =   W^2(y) \cR^2  d\Omega^2_{D-4}  + h_{\alpha \beta}^\cB(y) dy^\alpha dy^\beta,
\label{gralBONansatz}
\ee
where the coordinates $y^\alpha$, with  
$\alpha=1,\ldots, 4$,  parametrise 
 $\cB_4$.  For later convenience, we have written explicitly the bubble nucleation radius $\cR$, which will have to be determined. This is precisely ansatz  anticipated in Section \ref{sec:NUTSresult}. 
In addition, to be able to solve the Euler-Lagrange equations  we will need the dilaton configuration to have the dependence $\phi = \phi(y)$.

With this ansatz the components of the Ricci tensor read
\bea
R_{\mu\nu}  &=&\big(- W^{-1}\nabla^2 W +  (D-5)W^{-2}[\cR^{-2} -(\nabla W)^2]  \big) g_{\mu\nu} \nonumber \\
R_{\alpha \beta}&=& R_{\alpha \beta}^\cB - (D-4)\, W^{-1}  \nabla_\alpha \nabla_\beta  W,
\label{eq:ricciCurvatureAnsatz}
\eea
where $\mu,\nu=0,\ldots, D-3$ label coordinates on the sphere $S^{D-4}$. In the previous expressions $\nabla$ is the Levi-Civita connection compatible with the metric on the bordism $h_{\alpha\beta} ^\cB$, and  $R^{\cB}_{\alpha \beta}$  the associated Ricci tensor.

In order for the geometry above to represent the decay of the vacuum $\mathbb{R}_{D-3} \times T^3/\Gamma$ we also have to impose appropriate boundary conditions on \eqref{gralBONansatz}.  
Note that  the line element of the euclidean vacuum \eqref{vacuum} is  consistent with the $\SO(D-4)$ symmetry of \eqref{gralBONansatz}, and thus it is appropriate for matching the  form of the bounce spacetime far from the bubble, $\cM_D\to \mathbb{R}^{D-3} \times T^3/\Gamma$.
In this asymptotic regime, where $\cB_4\to \mathbb{R} \times T^3/\Gamma$, it is convenient  to split the local  coordinate system for the bordism   as $y^{\alpha} =\{\rho,y^{\bar \alpha}\}$, where  $\bar \alpha = 2,3,4$ label coordinates on the compact space, and $\rho$ parametrises the non-compact direction transverse to it. Furthermore, we will impose the  gauge conditions $h^\cB_{\rho\rho}=1$ and  $h^{\cB}_{\rho\bar \alpha=0}$. Then, the requirement that the BON configuration approaches the vacuum   \eqref{vacuum} far from the bubble can be equivalently expressed as
\be
\rho \to \infty: \qquad  W(\rho,\bar y) \cR\to \rho, \qquad h_{\bar \alpha \bar \beta}^\cB(\rho,\bar y) \to h^{\mathcal{C}}_{\bar \alpha\bar\beta}(\bar y),\qquad  \phi \to \phi_\infty,
\label{eq:BONbcs1}
\ee
where $h^{\mathcal{C}}_{\bar \alpha\bar\beta}$ is the flat metric on $T^3/\Gamma$.

Moreover, if this instanton is to be identified with a bubble of nothing,  at the bubble location  the geometry should approach that of a $(D-4)$-dimensional sphere \emph{of finite radius} $\cR$, where the bordism  is smoothly seals off: $\cM_D \to S^{D-4} \times \mathbb{R}^4$.  In other words, near the bubble there must exist  a local coordinate system $\{\rho,  y^{\bar \alpha}\}$, with bubble location  at $\rho=0$, such that we have
\be
\rho \to 0: \qquad  W(\rho,\bar y) \to 1, \qquad h_{\alpha \beta}^\cB(\rho,\bar y) \to \delta_{\alpha\beta},
\label{eq:BONbcs2}
\ee
while the  dilaton  approaches a  finite value.

The previous requirements ensure that the instanton  interpolates between the compactification at infinity  and the bubble containing \emph{nothing}. In real spacetime, (switching back to Lorentzian signature), far from the bubble core the geometry is $\mathbb{M}_{D-4}\times T^3/\Gamma$, and  near the bubble the spacetime is of the form  $\text{dS}_{D-4} \times \mathbb{R}^4$.  As in the original Witten's bubble, at $\rho=0$ the deSitter factor $\text{dS}_{D-4}$  represents the world-volume of the bubble surface, which nucleates initially at rest with radius $\cR$, and then begins expanding exponentially fast with expansion rate  $\cR^{-1}$.

\subsection{Dynamical constraint}

\label{sec:constraint}

We have seen in Section \ref{sec:top} that there is no topological obstruction to construct a bubble of nothing in $T^3$ compactifications of \eqref{eq:action}. However, there might be a dynamical obstruction that forbids the bubble to expand and to mediate the vacuum decay. In the present Section we will prove that the corresponding  instanton has a infinite action when the vacuum is supersymmetric, i.e. when $\alpha=0$ in \eqref{eq:action}, and therefore the decay rate is zero, so that the stability of the compactification is guaranteed by a Coleman-DeLuccia type of mechanism. We will also discuss under which conditions it would be possible to evade this dynamical constraint, and then we will show  that the quadratic deformation in the action \eqref{eq:action} with $\alpha\neq 0$ has the required form necessary for the decay to occur with a finite rate.

In order to find the dynamical constraint that forbids the decay in supersymmetric settings, we will begin rewriting  the equations of motion for the specific BON ansatz given above when $\alpha=0$. 
The Einstein's equations on  the $S^{D-4}$ sphere reduce in the \emph{Einstein's frame} to
\be
\left((D-5)W^{-1}\nabla^2 W  -\frac12 (D-6)(D-5) [\cR^{-2} - (\nabla W)^2 ]W^{-2}-\frac12 R_{\cB} \right)g_{\mu\nu}=8\pi G_D T_{\mu\nu},
\label{eq:sphere}
\ee
while the trace of the Einstein's equations for the bordism $\cB_4$ reads
\be
3(D-4) W^{-1} \nabla^2 W -2 (D-4)(D-5) [\cR^{-2} - (\nabla W)^2 ]W^{-2}- R_\cB =8\pi G_D h^{\alpha\beta}_{\cB}T_{\alpha\beta}\ .\label{eq:bordism}
\ee
We can combine these equations to give 
\be
W^{-1}\nabla^2 W =- \frac{16 \pi G_D}{D-2}\, \Big( \frac{2\,  T_{00} }{|g_{00}|}+ \frac{(D-6)}{2(D-4)} h^{\alpha \beta}_\cB T_{\alpha\beta}\Big) +\frac{ R_\cB}{(D-4)}. 
\label{eq:combinedConstraint}
\ee
The previous expression can be integrated on the bordism, and after discarding a vanishing boundary term we find
\be
\label{constraint}
0 \le \int_{\cB_4} \sqrt{h_\cB}(\nabla \log W)^2 = \int_{\cB_4}   \sqrt{h_\cB}\Big[ \frac{ R_\cB}{(D-4)}- \frac{16 \pi G_D}{D-2}\, \Big( \frac{2\, T_{00} }{|g_{00}|}+ \frac{(D-6)}{2(D-4)} h^{\alpha \beta}_\cB T_{\alpha\beta}\Big)\Big]. 
\ee
Therefore, in order to satisfy the inequality we need either an integrated positive scalar curvature or a stress energy tensor satisfying
\be
 \int_{\cB_4}   \sqrt{h_\cB} \Big[ 4 (D-4) |g_{00}|^{-1} T_{00}+ (D-6)h^{\alpha \beta}_\cB T_{\alpha\beta}\Big]< 0.
\label{eq:EnergyMomentumIneq}
\ee
In particular, the contribution of the dilaton to the stress energy momentum  is given by
\beq
\label{Tconstraint}
 T^\phi_{\mu\nu}=-\frac{(\nabla \phi)^2}{4 \pi G_D(D-2)} \, g_{\mu\nu}\ ,\quad T^\phi_{\alpha\beta}=\frac{1}{2\pi G_D(D-2)}\Big(\nabla_\alpha \phi\, \nabla_\beta \phi -\frac12 h_{\alpha\beta}^\cB(\nabla \phi)^2\Big)
 \eeq
 which implies that the specific combination appearing in \eqref{eq:EnergyMomentumIneq} is non-negative. In particular, the condition \eqref{eq:EnergyMomentumIneq} can be related to violating the Dominant Energy Condition as follows.
At every point in spacetime we can always find a local orthonormal frame, $\{e^M_m\}$ with $m=0,\ldots,D-1$, which diagonalises  the energy-momentum tensor  (see e.g.  \cite{poisson_2004}). 

Using  this basis we  define the following future directed time-like vector
\be
v^M = 2 \sqrt{D-2} \, e_0^M + \sqrt{D-6} \,   \sum_i e_i^M,
\ee
where  $i=1,\ldots,4$ labels the basis elements for  the tangent space of the  bordism.  It is now easy to check that the inequality \eqref{eq:EnergyMomentumIneq} can be written as follows
\be
 \int_{\cB_4}   \sqrt{h_\cB}\;  T_{MN} v^M v^N \le 0,
\ee
what can only be satisfied provided $T_{MN} v^M v^N<0$ somewhere,  violating   the Dominant Energy Condition.  

From this we see that provided the DEC holds and  the warp factor is non-vanishing $W>0$,  then $W$ is necessarily a constant for Ricci flat bordisms. Since the boundary conditions \eqref{eq:BONbcs1} cannot be satisfied, we conclude that there  are no bubble of nothing solutions. 
 This nicely matches with the Positive Energy Theorem explained in Section \ref{sec:pet}. Regarding the mechanism of dynamical supression, it can also be proven  that  the only solutions  to the equations of motion  in this setting necessarily have a  $\cR\to \infty$. That is, when $\alpha=0$ and the scalar curvature $R_\cB$ vanishes the line element must be of the form
  \be
ds^2 = dx^\mu dx^\mu  + h^\cB_{\alpha \beta}(y) dy^\alpha dy^\beta,
\label{eq:background}
 \ee
with the metric  $h^\cB_{\alpha\beta}$ being Ricci-flat.
 This  can be seen integrating the (Einstein frame) dilaton equation  \eqref{eq:appDilatonEom} over the bordism\footnote{The $D$-dimensional laplacian $\nabla^2_{(D)}$ and the laplacian $\nabla^2$ on $\cB_4$  coincide when $W$ is constant, since $\nabla_{(D)}^2\phi=\nabla^2 \phi + (D-4)\nabla \phi\nabla \log W$. }, what shows that $\phi$ also needs to be a constant to match the boundary conditions \eqref{eq:BONbcs1}.  As this implies that the  energy momentum tensor  is vanishing, it follows from equation \eqref{eq:sphere} that  $\cR$  needs to be infinite when $\alpha=0$, and from the equations on $\cB_4$,  that   $h^\cB_{\alpha\beta}$ is Ricci-flat.  Then, suppose we have a BON solution with finite nucleation radius for some $\alpha\neq0$, as we approach the limit $\alpha\to 0$ the bubble nucleation radius will grow unbounded and the decay rate will vanish. In other words, as we anticipated at the beginning of this Section, the stability of the supersymmetric compactification ($\alpha=0$) is enforced   by  Coleman-DeLuccia type of mechanism.

\begin{figure}
\begin{center}
\includegraphics[width=0.65\textwidth]{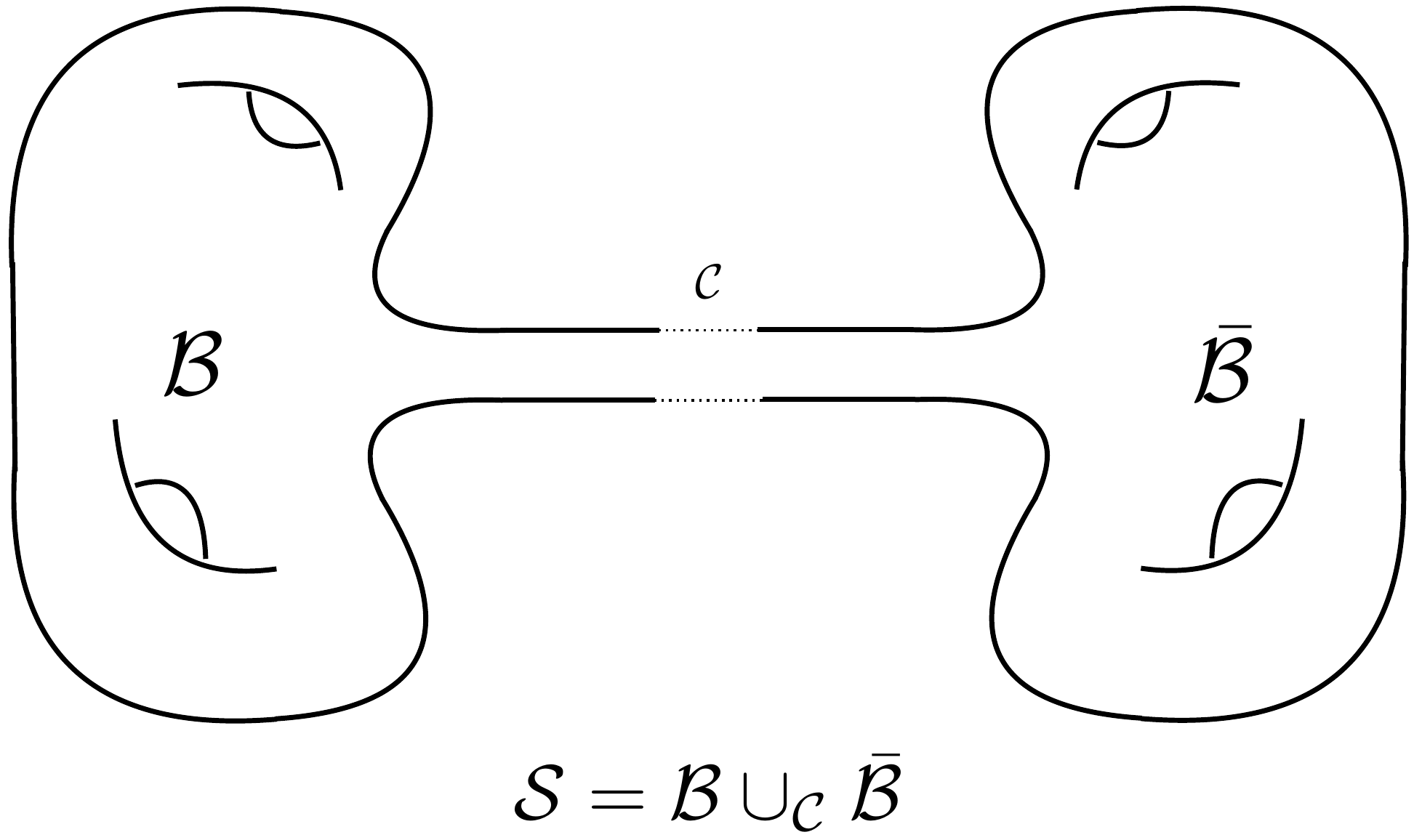}\end{center}
\caption{Starting with a noncompact manifold $\mathcal{B}_4$ with an infinite tube, one can construct an auxiliary compact manifold $\cS$ by taking two copies of $\mathcal{B}_4$, reversing orientation of one copy, and gluing them along their common boundary $\mathcal{C}_3$. $\cS$ is not a complete manifold with respect to the induced metric, but this can be easily fixed as described in the main text.}
\label{f22}
\end{figure}

One could hope to go around this by changing the metric on the bordism so that the total scalar curvature is positive $\int_{\mathcal{B} }R_{\mathcal{B}}>0$, what would allow  to find nontrivial solutions to \eq{constraint}. We will now show that this is impossible. Suppose such a metric existed. Then, one could take two copies of the bordism $\mathcal{B}_4$, reverse the orientation of one of them, and glue them back together, as illustrated in Figure \ref{f22}.  Let us call the compact manifold constructed in this way $\cS\cong \cB^A_4 \cup \cB^B_4 $. The metric on $\mathcal{B}_4$ becomes an incomplete metric on $\cS$, as some points of $\cS$ are at infinite distance from a generic point in $\mathcal{B}_4$. Schematically, we are gluing the two copies of $\mathcal{B}_4$ via an ``infinite throat''. This can be made more explicit as follows: near the boundary of $\mathcal{B}_4$, the bordism metric written in the coordinate system of \eqref{eq:BONbcs1} reads
\begin{equation} h_{\alpha\beta}^{\mathcal{B}}dy^\alpha dy^\beta\rightarrow d\rho^2+h^{\mathcal{C}}_{\bar{\alpha}\bar{\beta}}dy^{\bar{\alpha}} dy^{\bar{\beta}},\label{asympt}\end{equation}
where $\rho$ is the radial coordinate. Making the change of variables $\chi\equiv \pi/2-\arctan(\rho)$, the metric becomes
\begin{equation} h_{\alpha\beta}^{\mathcal{B}}dy^\alpha dy^\beta\sim \frac{d\chi^2}{\sin^4(\chi)}+h^{\mathcal{C}}_{\bar{\alpha}\bar{\beta}}dy^{\bar{\alpha}} dy^{\bar{\beta}}.\end{equation}
The second copy of $\mathcal{B}_4$ can be glued by allowing $\chi$ to take negative values, but the point $\chi=0$ is at infinite distance from any point with $\chi\neq0$. This is easily remedied; deforming the metric to 
\begin{equation} \frac{d\chi^2}{\epsilon(\chi)+\sin^4(\chi)}+h^{\mathcal{C}}_{\bar{\alpha}\bar{\beta}}dy^{\bar{\alpha}} dy^{\bar{\beta}},\end{equation}
where $\epsilon(\chi)$ is a smooth symmetric positive function of compact support located on a small neighbourhood of $\chi=0$, the point $\chi=0$ is now at finite distance. Since we are assuming that $\int_{\mathcal{B}} R_{\mathcal{B}}$ is convergent and positive, the asymptotic region with $\chi\sim0$ must contribute a negligible amount. By taking $\epsilon(\chi)$ small enough,  the sign of the integral $\int_{\mathcal{B}} R_{\mathcal{B}}$ then cannot change. This means we have constructed a complete metric on the compact manifold $\mathcal{\cS}$, with $\int_{\mathcal{\cS}} R_{\mathcal{\cS}} >0$. 

Now, on every smooth compact manifold of dimension $\geq 3$, it is a fact that every metric is conformal to a different metric of constant scalar curvature  (this is known as the solution to the ``Yamabe problem'' in the literature (see e.g. \cite{Rosenberg_manifoldsof,Dai2005} and references therein)). In other words, the metric we have just constructed is conformal $h_{\alpha\beta}^{\cS}= \omega^2 \tilde{h}_{\alpha\beta}^{\cS}$, where $\tilde{h}_{\alpha\beta}^{\cS}$ is a metric of constant scalar curvature $R_0$. One has
\begin{equation} R_{\cS}=\omega^{-2} R_0 +3\nabla^2(\omega^{-2})-18\omega^2\vert\nabla \log\omega\vert^2,\end{equation}
so that the integrated curvature $\int_{\cS}R_{\cS}$ can only be positive if $R_0$ is positive. In other words, because we constructed a metric with positive integrated curvature on $\cS$, this means that a metric of constant positive scalar curvature on $\cS$ also exists. Yet not every manifold admits a metric of everywhere positive scalar curvature. This is a problem which has been exhaustively studied by mathematicians, and a very clear survey of the question can be found in \cite{Rosenberg_manifoldsof}.
In particular, any compact spin four-manifold with nonvanishing integrated Dirac index, $\int_\cS \hat{A}(R)$, does not admit a metric of positive scalar curvature. 

In the particular case for us, where $\mathcal{B}_4$ is $dP_9$ with a hole, the procedure illustrated in Figure \ref{f22} produces a $K3$ manifold. Since the Dirac index on K3 is nonvanishing, this means that it does not admit a metric of positive scalar curvature, and thus, by the above reasoning, our $T^3$ nulbordism $\mathcal{B}_4$ does not admit a metric of integrated positive scalar curvature with the asymptotics \eq{asympt}. 

In dimension bigger than or equal to 5, a stronger statement, know as the trichotomy theorem \cite{Rosenberg_manifoldsof}, implies that a compact manifold which admits a Ricci-flat metric will not admit a metric of positive scalar curvature. Thus, for instance, any higher-dimensional bubble of nothing obtained by e.g. slicing open a CY manifold will not be dynamically allowed in the pure Einstein theory.

The previous discussion gives us a hint of how to deform the model in order to evade this dynamical constraint. It is clear that we need to break supersymmetric in such a way that either the bordism $\cB_4$ adquires a positive scalar curvature or we violate the DEC.  We can indeed show that the deformation given by the Gauss-Bonnet term when $\alpha\neq 0$ evades the dynamical constraint to leading order in perturbation theory by violating the DEC. Let us think of $\alpha$ as a small  perturbation parameter  $\alpha\ll1$. 
To first order in perturbation theory, and taking the geometry \eqref{eq:background} as the background, the Einstein's equations become\footnote{The transformation of the Gauss-Bonnet term needed to switch between the string and Einstein frames can be found in \cite{Carneiro:2004rt}.
 We also used  that $h^\cB_{\alpha\beta}$ is Ricci-flat on the background \eqref{eq:background}, and that its  Riemann tensor  satisfies  $R_{\alpha\gamma\delta\kappa} R_{\beta}^{\phantom{\beta}\gamma\delta\kappa}= \ft14 R^2_{GB} h^\cB_{\alpha\beta}$, since $\cB_4$ is four dimensional \cite{DeWitt:1965jb}.}
\be
G_{\mu\nu} = 8 \pi G_D T_{\mu\nu}^\phi + \ft{1}{16}\alpha \rme^{-\frac{4}{D-2}\phi_\infty} R_{GB}^2 \, g_{\mu\nu} + \cO(\alpha^2), \qquad G_{\alpha \beta}= 8 \pi G_D T_{\alpha\beta}^\phi+ \cO(\alpha^2), 
\ee
where the correction linear in $\alpha$  is evaluated on \eqref{eq:background}. 
This implies an additional contribution to the right hand side of \eqref{constraint} given by
\beq
0< \frac{\alpha}{4(D-2)}   \rme^{-\frac{4}{D-2}\phi_\infty}  \, \int_{\cB_4}  \sqrt{h_\cB}  R_{GB}^2 = \frac{96 \pi^2  \alpha}{(D-2)}   \rme^{-\frac{4}{D-2}\phi_\infty}  \ +\cO(\alpha^2),
\eeq
which indeed satisfies \eqref{eq:EnergyMomentumIneq}, i.e. it violates the DEC. Therefore, the Gauss-Bonnet term evades the dynamical constraint and allows in principle for finite action bubbles of nothing. In the next Section, we will explicitly construct such solutions. 

The other possibility to allow the bubble solutions, without violating DEC, is that the fermionic structure on the boundary is such that the bordism admits a  metric of positive scalar curvature $R_\cB>0$. This would be the case of the Witten BON, and all the 27 cases compactifications on tori quotients $T^3/\Gamma$ where the action $\Gamma$ is non-trivial.  

\section{Detailed construction of the BON solution}

\label{sec:detailedBON}

In this Section, we will describe in great detail how to explicitly construct the bubble of nothing for compactifications on a three-dimensional manifold, either the three-torus  or quotients thereof $T^3/\Gamma$ in the particular theory presented in Sections \ref{sec:nuts} and \ref{sec:det}.  

\subsection{The strategy}

To find BON solutions representing the decay to nothing of a vacuum $\mathbb{M}_{D-3} \times T^3/\Gamma$ we will  solve the set of equations  \eqref{EinsteinEOM} and \eqref{dilatonHeom} using the generic BON ansatz \eqref{gralBONansatz}, while requiring the field configurations to be subject to the boundary conditions \eqref{eq:BONbcs1} and \eqref{eq:BONbcs2}. These boundary conditions ensure that the BON spacetime interpolates between the bubble containing ``nothing'' and  the decaying vacuum at infinity:
\be
\text{bubble core geometry:}\quad S^{D-4} \times \mathbb{R}^{4}\quad \longrightarrow \quad \text{euclidean vacuum:}\quad  \mathbb{R}^{D-3} \times T^3/\Gamma.
\label{eq:BONdefect}
\ee
For the instanton solution to admit the interpretation of a semi-classical decay, we will restrict ourselves to smooth solutions.  More specifically, we will require the spacetime curvature to be everywhere well below the Planck scale, so that the quantum-gravitational effects are  suppressed, and  the semi-classical description of the decay is justified.  In addition, in order  to construct the solutions we will resort to a number of approximations which, on the one hand  will render  the problem tractable, and on the other hand will allow us to have a  faithful characterisation of our BON solutions.  A good understanding of the BON configurations will help us to keep the approximations under control, and guarantee the validity of the solutions that we obtain.

\subsubsection{Approximation scheme.}
\label{sec:appScheme}
As we anticipated in the previous Section, the main simplification  will be to regard $\alpha$ as a small parameter (in appropriate units), and the use of perturbative methods to construct the solutions.  Provided that this condition is satisfied, the spacetime  geometry of the  BON solution becomes specially neat, in that the two main  length-scales characterising these solutions are nicely separated, and the equations of motion  can be studied independently on each of these scales. 

The  natural length-scales that appear in the BON geometry are  the Kaluza-Klein scale $\ell_{\text{kk}}$, associated to the compact space $\cC_3 \cong T^3/\Gamma$, 
 together with the scale of supersymmetry  breaking  $\ell_{\text{ssb}}$. The supersymmetry breaking effects become irrelevant at length-scales larger  than $\ell_{\text{ssb}}$ (energies smaller than $\ell^{-1}_{\text{ssb}}$).  Thus,  in particular, in the case of the $T^3$ compactification where the breaking of supersymmetry is   induced by the Gauss-Bonnet term, we have
 \be
\ell_{\text{kk}}\sim (\cV_{\cC})^{1/3}, \qquad \ell_{\text{ssb}}\equiv(\alpha/\cV_{\cC})^{-1}
 \ee
 with the precise definition of $\ell_{\text{kk}}$  given in \eqref{eq:scalesKKF}, and where $\cV_{\cC}$ is the volume of the compact space $\cC_3$ in the vacuum\footnote{Note that $\alpha$ has dimensions of squared length.}.  For the compactifications on $T^3/\Gamma$, where supersymmetry is already broken by the boundary conditions of the fermions,   the parameter $\ell_{\text{ssb}}$ will be related instead  to this  supersymmetry breaking mechanism. For simplicity, in the discussion that follows just consider the case  where $\ell_{\text{ssb}}$ is controlled by the parameter $\alpha$, but the argument is identical in the other case\footnote{For compactifications $T^3/\Gamma$ the separation of scales $\ell_{\text{kk}}\ll\ell_{\text{ssb}}$ will be achieved considering a manifold $\cB_4$ in a particular degenerate/Large Volume   limit.}.

\begin{figure}[t]
\centering \includegraphics[width=1\textwidth]{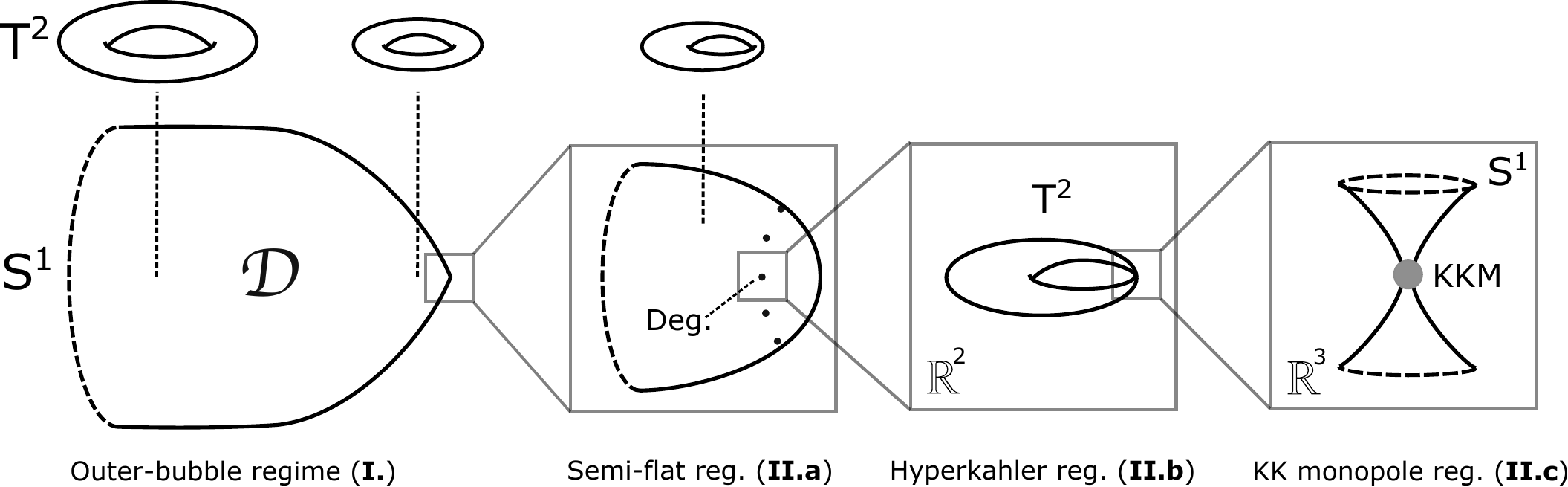}
\caption{Layered structure of the fibration $\cB_4\cong \cD \times T^2$ for the $T^3$ BON. From left to right the diagram displays the different regimes of the manifold $\cB_4$: the outer-bubble regime ({\bf I.}), whose asymptotic boundary matches the compact space $T^3$, and where only the $T^3$ volume is dynamical; the semi-flat regime  ({\bf II.a}), valid away from the degeneration points (\emph{Deg.}),  and where the complex structure of the $T^2$ fibre becomes dynamical; the hyperk\"ahler regime ({\bf II.b}) describing the neighbourhood of the degenerations; and the KK monopoles (\emph{KKM})  ({\bf II.c}) describing the BON cores,  where the compact space is smoothly sealed off, and the metric is locally  $\mathbb{R}^4$.}
  \label{fig:T3diagram}
\end{figure}

When $\alpha$ is small  the BON spacetime has two well differentiated regimes (see Fig. \ref{fig:T3diagram}):
\begin{enumerate}
\item[\bf{I.}]   the {\bf outer-bubble regime}, associated to the  scale $\ell_{\text{ssb}}\gg \ell_{\text{kk}}$, is the outermost layer of the BON geometry which asymptotes at infinity to the vacuum $\mathbb{R}^{D-3} \times T^3/\Gamma$. In this region the effect of KK modes is exponentially suppressed, and thus  the induced metric on the $T^3/\Gamma$ is approximately be flat. Our approximation here consists in assuming   the metric on the compact space to be \emph{exactly flat} (we neglect the KK modes), leading to a behaviour which closely resembles  the original Witten's  BON.
\item[\bf{II.}]An { \bf inner-bubble regime}, describing the features of the geometry on the Kaluza-Klein scale  $\ell_{\text{kk}}$.  
\end{enumerate}
In the inner-bubble region is  where the spacetime exhibits the topology characterised by the Weierstrass model \eqref{eq:Weierstrass} (with $N\le12$ degenerations), and thus where the ``fermionic knot'' is undone.  As we describe below, in this regime the bordism $\cB_4$ is well described by a non-compact  conformally Calabi-Yau manifold, which becomes   
 exactly  Calabi-Yau,  when the deformation is turned off, $\alpha \to 0$.  Therefore, assuming $\alpha\ll1$, we will treat this  inner regime perturbatively, using as a background the Calabi-Yau geometries given by \eqref{eq:Weierstrass}, and then considering the effect of including the  Gauss-Bonnet term (i.e. the warping) as a small deformation  \cite{Banks:1988rj}.

   In addition, in order to have a good control over the background geometry ($\alpha=0$), we will consider the case where the compact space $T^3/\Gamma$ is close to a degenerate limit. Namely, regarding the compact space  as $T^2$ fibration over $S^1$, for simplicity we will discuss the situation when the volume of the $T^2$ fibre is small, $\ell_{\text{kk}} \gg \ell_{\text{fibre}} \sim (\cV_{T^2})^{1/2}$ (see definition \eqref{eq:scalesKKF}). In this limit the inner-bubble region attains a particularly clear structure, and displays three distinct regimes (see Fig. \ref{fig:T3diagram}):
 \begin{enumerate}
\item[{\bf II.a}] {\bf Semi-flat regime.} This is the outermost layer within the inner-bubble region, and provides a transition to the outer-bubble regime.   Almost everywhere in the inner-bubble region the induced metric on the $T^2$ fibre is exponentially close to   flat \cite{gross2000,Kachru:2018van}. Therefore, in this layer we will describe the  BON  with a \emph{semi-flat geometry}  \cite{Greene:1989ya}, which  assumes  the induced metric on the fibre to be to be exactly flat (neglects KK modes from the $T^2$ fibre). 
\end{enumerate}
When discussing the geometry of  the semi-flat region we will encounter  a second natural length-scale, $\ell_{\text{sf}}$ defined in \eqref{eq:lsfDef},  controlling the distance between the $N$  points where the fibre pinches off. For convenience we will work in the regime defined by $\ell_{\text{fibre}}/\ell_{\text{sf}}\ll1$, where the degenerations are well  separated from each other.  

In a small neighbourhood  of the degenerations  (of size  $\sim\ell_{\text{fibre}}\ll \ell_{\text{kk}}$)   the semi-flat description fails \cite{gross2000}.   There the radius $\ell_{\text{fibre}}^{(2)}$ of one of the cycles of the $T^2$ fibre becomes large, leading to a partial  decompactification, while the other one shrinks  (keeping the fibre  volume constant), so that $\ell_{\text{fibre}}^{(1)}\ll\ell_{\text{fibre}}^{(2)}$. In this region of the BON spacetime the KK modes associated to the growing cycle can no longer be neglected and the semi-flat description becomes inadequate. Here the geometry is still well characterised in terms of a hyperk\"ahler metric:
\begin{enumerate}
\item[{\bf II.b}] {\bf Hyperk\"ahler regime.}  This is the intermediate layer of the inner-bubble region, and describes the local spacetime around an \emph{isolated} degeneration point. The metric here  is  approximately  that of a self-dual Taub-NUT space \cite{Gross:1983hb,PhysRevLett.51.87,Hawking:1976jb},  \emph{a KK monopole}, with one of the directions (other than the standard KK circle)   compactified on a\footnote{This geometry  also appears in the literature under the name of the  \emph{Ooguri-Vafa metric} \cite{Ooguri:1996me}.} $S^1$  \cite{Myers:1986rx} (see also \cite{Ortin:2015hya}). In other words, the base manifold of the KK monopole spacetime  is $S^1\times \mathbb{R}^2$.
\end{enumerate}

Finally, near the Taub-NUT point at the core of the region where the $T^2$ fibre degenerates,  the additional $S^1$ identification of the KK monopole spacetime can be ignored. In this limit  we have the
\begin{enumerate}
\item[{\bf II.c}] {\bf KK monopole regime,} which represents the core of the BON spacetime. Here the geometry approaches that of an isolated KK monopole with  $\mathbb{R}^3$   base manifold.
 \end{enumerate}

Summarising, this neat layered  structure of the BON spacetime will arise as long as our approximation scheme holds:
\be
\ell_{\text{Planck}}^{(D)} \quad \ll\quad   \ell_{\text{fibre}}  \quad \ll \quad \ell_{\text{sf}}\quad \sim\quad \ell_{\text{kk}} \quad \ll \quad \ell_{\text{ssb}}
\label{eq:appScheme}
\ee
The first inequality is required for the semiclassical treatment to be appropriate, and  as we shall see below, for consistency  we will also  require that the compact space $T^3/\Gamma$ is in a Large Volume (LV) regime. This approximation scheme   will allow us to discuss the geometry on each of these spacetime regions independently, and to ensure that the obtained solution can be interpreted as an instanton mediating a semiclassical decay.

To have a clear geometric picture of the BON configuration it is useful to regard it as a ``defect'' interpolating between the bubble core and vacuum  geometries \eqref{eq:BONdefect}. In this sense, one can think of the  spacetime as undergoing a series of geometric transitions as we ``zoom out''  from one of Taub-NUT points (the KK monopoles), towards the vacuum at infinity. Then,  the spacetime regions above can be associated  these transitions as follows
\bea
 S^{D-4} \times \mathbb{R}^{4}\quad &\xrightarrow{\text{{\bf II.c}}\rightarrow\, \text{{\bf II.b}}}& \quad  
 S^{D-4} \times \mathbb{R}^{3} \times S^1 \quad \xrightarrow{\text{{\bf II.b}}\rightarrow\, \text{{\bf II.a}}} \quad
  S^{D-4} \times \mathbb{R}^{2} \times T^2 
  \nonumber \\ \quad &\xrightarrow{\text{{\bf II.a}}\rightarrow\, \text{{\bf I}}}& \quad 
  S^{D-4} \times \mathbb{R} \times T^3/\Gamma \quad \xrightarrow{\text{{\bf I}}\rightarrow\, \mathcal{C}} \quad
  \mathbb{R}^{D-3} \times T^3/\Gamma.
\label{eq:transitionSequence}
\eea

As we said above,  the Taub-NUT points in regime ${\bf II.c}$ can be seen as the core of the BON geometry, since there the compact space pinches off, and the geometry is   smoothly sealed off. At these points the sphere $S^{D-4}$ represents the world-volume of the bubble surface. It is worth noting that, since there is a bubble core  associated to each of  the various Taub-NUT points, this spacetime actually represents a multi-centered BON.

Then, as we zoom away from the bubble core towards larger length scales, $\ell_{\text{fibre}}^{(1)}\to\ell_{\text{fibre}}^{(2)}\to \ell_{\text{kk}}$, the compact nature of each of the three cycles of the $T ^3/\Gamma$ becomes  apparent in a sequence of  steps, each  associated to one of the three regimes of the inner-bubble region, respectively ${\bf II.c}$,  ${\bf II.b}$ and  ${\bf II.a}$.   
In the outer-bubble regime, {\bf I.}, the  directions along the sphere $S^{D-4}$ (with radius $\sim \ell_{\text{ssb}}$) combine with the non-compact direction of the bordism $\cB_4$ to give the $D-3$ dimensional euclidean space. Simultaneously,  as we  move towards infinity in this outermost layer, the volume of the  compact space $\cV_{\cC}$  grows,  attaining an asymptotic value which  matches that of the  decaying vacuum $T^3/\Gamma$. This picture of the layered structure of the BON geometry,  together with the requirement that the scales \eqref{eq:appScheme} are well separated, are  the basis  for the approximations we will make  in the semi-flat  and outer-bubble regions where we will neglect, respectively, the effects of the KK modes of the $T^2$ fibre and on the full compact space $T^3/\Gamma$. 

Finally, a delicate  issue in the perturbative construction of our solutions is that of zero-modes. The  Calabi-Yau geometries that we use as background for the perturbative expansion have large moduli spaces, with associated massless excitations (zero-modes). A generic deformation of the theory might turn this flat directions into runaways,  what would make impossible the construction of the BON solution. However, as we will proof explicitly in Section \ref{sec:zeroModeDec}, this is not the case when the theory is deformed with a Gauss-Bonnet term (see \cite{Banks:1988rj}). In our model the massless excitations decouple from this deformation, and thus remain zero-modes to first order in perturbation theory.

\subsubsection{Gluing method.}
\label{sec:gluingStrategy}

The approximation scheme \eqref{eq:appScheme} will allow us to study  the equations of motion separately for different layers of the BON spacetime and, making use of the appropriate approximate descriptions in each of them, we will be able to obtain local solutions there. Therefore, in order to construct the global BON solution we will need to resort  to spacetime matching techniques, also known as \emph{gluing} \cite{Israel:1966rt} (see \cite{Mars:1993mj} for an extensive discussion).  Indeed, to construct the full solution we will have to glue, on the one hand, the hyperk\"ahler and semi-flat regions (resp.  ${\bf II.b}$  and ${\bf II.a}$), and on the other hand, the semi-flat and outer-bubble regimes,  (resp. $\bf{II.a}$ and ${\bf I.}$). To incorporate  the  KK monopole regime, {\bf II.c}, no gluing will be necessary, as it corresponds to the limiting behaviour of the hyperk\"ahler geometry  of region ${\bf II.b}$ near the Taub-NUT points.

To perform the gluing of the different layers we will follow closely the method in  references \cite{Mars:1993mj,Mars:2005ca,Nolan:2018ozv}, which we briefly summarise here. Given two  $D$-dimensional spacetimes $(\cM^{+},g_{MN}^+)$ and $(\cM^-,g_{MN}^-)$, we can construct a new manifold $\cM \equiv \pd \cM^+\cup \cM^-$ by performing a point-by-point identification of the boundaries,   $\pd \cM^{\pm}$, of the constituent spaces. In practice, this identification is done introducing a one-to-one mapping between the boundaries, $\Phi: \pd \cM^- \rightarrow \cM^+$,  the so called   \emph{gluing diffeomorphism}, so that pairs of points related by this map are regarded as the same point in the total manifold. 
For the new manifold $\cM$ to  constitute a well defined spacetime $(\cM, g_{MN}) \equiv (\cM^+,g_{MN}^+)\cup (\cM^-, g_{MN}^-)$, and to be able to write down the Einstein's equations, the metric tensor $g_{MN}$ must be continuous across the matching boundary. 
This condition is implemented imposing that the induced metrics on $\pd \cM^\pm$   agree, i.e. the first fundamental forms $s_{ab}^\pm$,  where $a,b=\{1,\ldots,D-1\}$ are indices on the tangent space to the hypersurface. Furthermore, we will  require the second fundamental forms on the boundaries $K_{ab}^\pm =\ft12 (\cL_{n^\pm} g^\pm)_{ab}$ also to be equal, with $n^\pm$ being the unit normals to $\pd\cM^\pm$. This additional condition is imposed to avoid the presence of a shell/brane on the matching hypersurface, i.e. a Dirac-delta singularity of the energy-momentum tensor. These two requirements  can be  expressed explicitly in terms of the pull-back map, $\Phi^*$,   as follows
\be
\Phi ^*(s^+)_{ab}  = s_{ab}^-, \qquad \Phi^*(K^+)_{ab} = K_{ab}^-.
\label{eq:matchingConditions}
\ee
It should be noted that these matching conditions are those appropriate to General Relativity, and that in general  they need to be modified when considering theories of modified gravity with quadratic curvature terms. However, the modification of  GR by including a Gauss-Bonnet term is special, in that  the matching conditions remain the same as those in GR \cite{Reina:2015gxa}. 

Since  the construction of our solutions requires a perturbative treatment we will also have to consider the  matching procedure in the context of perturbation theory.  When the geometry involves a  deformation parameter $\epsilon\ll1$,  the matching procedure can be adapted to the perturbative framework by promoting all the geometric quantities, i.e. the metric $g_{MN}^\pm$, the gluing diffeomorphism $\Phi$ and the fundamental forms $s_{ab}^\pm$ and $K_{ab}^\pm$, to be functions of the parameter $\epsilon$, and then solving the matching conditions order by order in $\epsilon$ \cite{Mukohyama:2000ga} (see also \cite{Mars:2005ca}).  

To be more specific, when performing  the gluing procedure below we will encounter multiple perturbative expansions: one associated to the  parameter $\alpha\ll1$, and other three  related to the various approximations made to describe the background geometry. That is,  we neglect    the exponentially suppressed KK modes of the $T^2$ fibre in the semi-flat regime, 
 we ignore the KK modes of the compact space $T^3/\Gamma$ in the  outer-bubble region, 
  and we assume that the $N$ degenerations are far from each other.
  
  Let us discuss first the perturbative expansions associated to the approximate description of the background. The  hyperk\"ahler, semi-flat and outer-bubble geometries that we will use to \emph{locally} characterise the different layers  of the BON become  exact solutions of the equations \eqref{EinsteinEOM} and \eqref{dilatonHeom}  in the limit $\alpha\to0$. Yet,  when combined, they  only provide an (exponentially accurate) approximation to the \emph{global} BON solution, even in the limit $\alpha\to 0$. This becomes evident when trying to  perform the gluing procedure since, to zero order  in $\alpha$,  we will only be able to satisfy the matching conditions  \eqref{eq:matchingConditions} up to small corrections associated to the KK modes, and neglecting the mutual backreaction between the degenerations.
  
  Consider for definiteness the matching between the hyperk\"ahler and semi-flat regimes ({\bf II.b}$\to${\bf II.a}). On the one hand, in the semi-flat region we neglect  the KK modes of the $T^2$ fibre, which  are nevertheless important in the  interior the hyperk\"ahler region. On the other hand,  in the hyperk\"ahler regime we ignore the presence of multiple degenerations.    However,  at the boundary between the two layers (i.e. the gluing hypersurface) the mismatch   is extremely small since the KK modes of the hyperk\"ahler regime are already exponentially suppressed there, and the degenerations are located very far from each other. More explicitly, the condition \eqref{eq:matchingConditions} on the first  fundamental form $s_{ab}$ will have the following schematic form\footnote{A similar expression is found  when considering the matching condition on the second fundamental form.} 
\begin{align*}
0&=\Phi^*(s_{{\bf IIb}})-s_{{\bf IIa}}=\\
& \Phi_0^*(s_{\text{hk}}|_{m=0}) - s_{\text{sf}} \; + \;\sum_{m>0}\rme^{-m\frac{\ell_{\text{kk}}}{\ell_{\text{fibre}}}} \; \Delta s^{(1)}_m 
\;+\;
\;\frac{\ell_{\text{fibre}}}{\ell_{\text{sf}}}  \Delta s^{(2)} 
\;+\;
\frac{\ell_{\text{kk}}}{\ell_{\text{ssb}}}  \;\Delta s^{(3)}|_{m=0} + \ldots .
\end{align*}
Here $\Phi_0$ is the zero-order gluing diffeomorphism, $s_{\text{hk}}$ and $s_{\text{sf}}$ are the first fundamental forms of the matching hypersurface when embedded in the  hyperk\"ahler and semi-flat geometries, $m$ is an integer labelling the   KK modes, and the last term represents  the leading order correction in $\alpha$ (recall that $\ell_{\text{ssb}}^{-1} = \alpha/\cV_{T^3}$). 

In this expression we can easily identify  three perturbative expansions: the KK expansion with parameter $\epsilon^{(1)} =\exp(-\frac{\ell_{\text{kk}}}{\ell_{\text{fibre}}})$, the one associated to the mutual influence of the degenerations, controlled by $\epsilon^{(2)} = \ell_{\text{fibre}}/\ell_{\text{sf}}$, 
and the  $\alpha$-expansion with parameter $\epsilon^{(3)} =\ell_{\text{kk}}/\ell_{\text{ssb}}\propto \alpha$. Choosing the scales so that 
$\epsilon^{(1)}, \epsilon^{(2)}\ll \epsilon^{(3)}<1$ 
we can work consistently  to first order in $\epsilon^{(3)}$ (i.e.  in $\alpha$), and to zero-order in the other two expansions. In other words,  the massive KK modes ($m>0$)  and the interaction between degenerations can be consistently neglected in the matching procedure  to first order in the $\alpha-$expansion. In particular,   working order by order in this parameter, we obtain  the condition
\begin{equation}
\hspace{-2cm}\text{zero-order gluing condition on ({\bf II.b}$\to${\bf II.a}):} \qquad \Phi_0^*(s_{\text{hk}}|_{m=0}) = s_{\text{sf}},
\label{eq:zeroOrder1FF}
\end{equation}
so that to zero  order in $\alpha$ we only have to match the zero-mode of  the hyperk\"ahler metric  with the semi-flat metric. Then, when considering the effect of the perturbation to first order in $\alpha$, we will have to compute the coefficient $\Delta s^{(3)}|_{m=0}$ of the expansion above and require it to be vanishing. Note that this term will again only involve the zero-mode sector of the KK expansion, since higher KK contributions are subleading.

Finally, we comment briefly on the gluing  between the semi-flat and outer-bubble regime  ({\bf II.a}$\to${\bf I.}). In the outer-bubble region we neglect completely the KK modes of the compact manifold $T^3/\Gamma$, but in the semi-flat regime some of these modes are still excited (not those associated to the $T^2$ fibre, which is  assumed to be flat).  As in the previous case, the KK modes of the semi-flat geometry appear  suppressed  at the matching boundary between the two regions, and thus the gluing condition for the first fundamental form is schematically 
\begin{equation*}
0=\Phi^*(s_{\bf IIa})-s_{\bf I} = \Phi_0^*(s_{\text{sf}}|_{m=0}) - s_{\text{out}}  \;+\; \sum_{m>0}\rme^{-m\frac{\ell_{\text{ssb}}}{\ell_{\text{kk}}}} \; \Delta s^{(1)}_m \;+\; \frac{\ell_{\text{kk}}}{\ell_{\text{ssb}}} \;\Delta s^{(2)}|_{m=0} + \ldots. 
\end{equation*}
Here $s_{\text{sf}}$ and $s_{\text{out}}$ are the first fundamental forms of the matching hypersurface  when embedded, respectively, in the semi-flat and in the outer-bubble geometries. Following a similar argument as the one given above, and provided \eqref{eq:appScheme} holds, it is easy to see that we can work consistently to first order in $\alpha$  neglecting the effect of the KK modes. In particular  to zero-order we obtain the condition 
\be
\hspace{-3cm}\text{zero-order gluing condition  on ({\bf II.a}$\to${\bf I.}):} \qquad \Phi_0^*(s_{\text{sf}}|_{m=0}) = s_{\text{out}}.
\ee
That is, to leading order in $\alpha$, we just  need to require  that the zero-mode KK sector of the semi-flat  geometry to match with the outer-bubble metric. Finally, to first order in $\alpha$  we will have to compute the coefficient $\Delta s^{(2)}|_{m=0}$ on the zero-mode sector of the KK expansion, and impose the relevant conditions on the metric perturbation so that it vanishes. 

A similar reasoning can be followed when considering the matching condition involving the second fundamental form in \eqref{eq:matchingConditions}.

As a final remark, note that the method used here to construct the background ($\alpha=0$) inner-bubble  spacetime is closely related to the approach in \cite{gross2000} to obtain approximate metrics on $K3$ surfaces. However the two works differ in the gluing methods  employed. As explained above, these space-times are to be used as the background of a perturbative expansion, so we will need a good  characterisation of the deviations between the approximate background ($\alpha=0$) metric   and the exact one. The  gluing methods used here \cite{Israel:1966rt,Mars:1993mj,Mars:2005ca}, (standard in the GR literature), can be used to  obtain in a systematic way higher order corrections to the approximate metrics of  \cite{gross2000}, and thus they   also provide a quantitative characterisation of the error made at each particular order\footnote{See also \cite{Kachru:2018van} for a different method to improve systematically the approximations of \cite{gross2000}.}.

\subsection{Background geometry of the  inner-bubble region.}
\label{sec:innerBackground}

In the present Section we will discuss the properties of the spacetime that we will use as background for the perturbative expansion in the inner-bubble layer. As we summarised above, in the limit $\alpha\to0$ we will describe the different layers of the geometry in terms of exact solutions of the equations of motion, which nevertheless can only be glued together in an approximate way.  We begin this Section presenting some general features of these geometries.

Regarding the  BON  mediating the decay of the compactification on $T^3$, from the discussion  about  the dynamical constraint (Section \ref{sec:constraint}), we know that in the limit $\alpha \to 0$ the spacetime geometry must be of the form
\be
ds^2 =dx^\mu dx^\mu + h^\cB_{\alpha \beta}(y) dy^\alpha dy^\beta, \qquad \text{with} \qquad  R^{\cB}_{\alpha \beta} =0,
\label{eq:susyLineElement}
\ee
with the bordism $\cB_4$ being Calabi-Yau, and the dilaton a constant $\phi =\phi_0$. Thus, in particular, the background metric for the inner-bubble region must also be necessarily Calabi-Yau in this limit. For simplicity, in the case of the BON associated to compactifications on $T^3/\Gamma$  we will also take the background geometry of the inner-bubble layer to be Calabi-Yau, although the dynamical constraint does not force us to make this choice. 

Recall,   as we argued in Section \ref{sec:detailedTopology}, that  the topology of the bordism is determined by the Weierstrass model \eqref{eq:Weierstrass} which can be described as a  $T^2$ fibration over a disc $\cD$.  In order to write down a local ansatz for the metric in the inner-bubble region, we first note that any  Calabi-Yau two-fold is also hyperk\"ahler (and self-dual)  \cite{Eguchi:1980jx}.   In addition, we will  assume the metric to be locally consistent with the dimensional reduction over one of the two $S^1$ cycles of the $T^2$ fibre.  Without further simplifications this leads us already to the family of  metrics we will use to describe the hyperk\"ahler regime:

\paragraph{Hyperk\"ahler ansatz.}   If we choose a coordinate system for the bordism $y^\alpha = \{\psi, y^i\}$, $i=\{1,2,3\}$, with  $\psi \sim \psi +1$ parametrising the $S^1$, then the most general line element that we can write on $\cB_4$ has the form \cite{HitchinMonopoles}
\be
ds^2_{\cB}|_{\text{hk}} = h^\cB_{\alpha \beta}(y)|_{\text{hk}}\; dy^\alpha dy^\beta=\rme^{2\varphi_0} \big(V\,  \mathring h_{ij} d y^i d y^j + V^{-1} (d\psi + \cA_i dy^i)^2\big),
\label{eq:HyperkahlerMetric}
\ee
 where the function $V = V(y)>0$, the one-form $\cA=\cA(y)$ and the three-dimensional metric $\mathring h_{ij}=\mathring h_{ij}(y)$ are independent of $\psi$.  The  overall real constant $\rme^{2\varphi_0}$  is an arbitrary parameter added for later convenience, and which regulates the volume of the bordism. 
With this ansatz, the hyperk\"ahler condition requires the  metric on the base $\mathring h_{ij}$ be flat, while the function $V$ and  the one-form $\cA_i dy^i$ should satisfy
\be
V^{-1}\, \mathring \nabla^2 V =0, \qquad  \text{and} \qquad\pd_i V =  \epsilon_{ijk} \pd_j \cA_k,
\label{eq:selfDualConditions}
\ee
where $\mathring \nabla^2$ is the Laplacian associated to   $\mathring h_{ij}$, $\epsilon_{ijk}$ is the totally antisymmetric Levi-Civita tensor. Since this geometry is Calabi-Yau and has zero Ricci curvature $R_{\alpha\beta}^\cB=0$ \cite{Hawking:1976jb}, together with the ansatz \eqref{eq:susyLineElement},  this geometry corresponds to a exact solution to the equations of motion in the limit $\alpha \to 0$.

The local ansatz that we will use to describe the semi-flat regime can now be easily obtained with one  further simplification: we require the previous metric to be also consistent with the reduction along the second direction of the $T^2$.  In other words, we require the induced geometry in the $T^2$ fibre to be flat.

\paragraph{Semi-flat ansatz.} Explicitly, if we parametrize the second cycle of the $T^2$ by the periodic coordinate $y^1 \sim y^1 +1$, the  semi-flat ansatz is obtained imposing
\be
\mathring h_{11} =1, \qquad \mathring h_{12} =  \mathring h_{13}=0,\qquad \cA_2 = \cA_3=0,  \label{eq:semiFlat1}
\ee
and requiring the non-zero metric components $V$, $\cA_1$, $\mathring h_{22}$, $\mathring h_{23}$ and $\mathring h_{33}$ to be independent of $y^1$. Then,  defining  the complex coordinate $z \equiv y^2 - \rmi y^3$ (which parametrises the disc $\cD$), and the complex field $\tau(z,\bar z)\equiv \cA_1 + \rmi V$, it is straightforward to check that  the hyperk\"ahler conditions \eqref{eq:selfDualConditions}  are simply the Cauchy-Riemann conditions for the complex field on $\tau$ 
\be
\pd_2 V = \pd_3 \cA_1, \qquad \pd_3 V = - \pd_2 \cA_1 \qquad \Longrightarrow \qquad \pd_{\bar z} \tau =0. 
\label{eq:semiFlat2}
\ee
In other words the field $\tau =\tau(z)$, which determines the complex structure of the $T^2$, should be a holomorphic function of $z$.  Moreover, 
the line element \eqref{eq:HyperkahlerMetric} can  be written in the form
\be
ds^2_\cB|_{\text{sf}} =\rme^{2\varphi_0} \Big(\Im(\tau)\, |F(z)|^2 dz d\bar z +\Im(\tau)\, d y^1 d y^1  + \Im(\tau)^{-1} (d\psi + \Re(\tau) dy^1)^2\Big),
\label{eq:semiFlatMetric}
\ee
where $\Im \tau >0$, and  $F=F(z)$  is a holomorphic function determined by the metric components $\{\mathring h_{22},\mathring h_{23}, \mathring h_{33}\}$.  Since this line element is a special case of \eqref{eq:HyperkahlerMetric} it also defines a exact solution to the Euler-Lagrange  equations of our model in the limit $\alpha \to 0$. Note that with this ansatz the volume of the $T^2$ fibre is parametrised by $\varphi_0$, and is  constant over the $z$-plane. 

In the following two Sections we will discuss more in detail the properties of these two  space-times, and how they can be glued together to obtain  an approximate description of the smooth Calabi-Yau associated to the Weierstrass model \eqref{eq:Weierstrass}.

\subsubsection{Layer (II.a): Semi-flat regime.}
\label{sec:semiflatLayer}

To  begin our discussion on the  detailed structure of the inner-bubble region we will consider the semi-flat regime.  As shown in \cite{Greene:1989ya}, the semi-flat description is particularly appropriate to construct geometries consistent with the Weierstrass model \eqref{eq:Weierstrass}, and thus it is in this layer of the BON geometry where the spacetime will exhibit the topology we  discussed in Section \ref{sec:detailedTopology}.

Following \cite{Greene:1989ya}, we first note that in \eqref{eq:semiFlatMetric} the $T^2$ fibre geometry is actually  invariant under the modular group $PSL(2,\mathbb{Z})$, 
with generators
\be
T: \quad \tau \to \tau +1, \qquad \qquad S: \quad  \tau \to -1/\tau.
\label{modularTrans}
\ee
 Therefore, although the complex structure field $\tau$ can take values on the full complex upper-half  plane $\mathbb{H}$, the space of inequivalent toroidal geometries is only given by the fundamental domain $\cF=\mathbb{H}/PSL(2,\mathbb{Z})$, which can be represented by the region where $|\tau|>1$ and  $\Re(\tau) \in [-1/2,1/2)$.  With this at hand, we can  construct semi-flat geometries \eqref{eq:semiFlatMetric}  
 with the required topology making use of the elliptic modular invariant function $j(\tau): \mathbb{C} \rightarrow \cF$, 
  which defines a holomorphic one-to-one mapping between the  full complex plane and the fundamental domain $\cF$. More specifically,  the class of semi-flat geometries which are also solutions of the Weierstrass  model \eqref{eq:Weierstrass}   are those for which the complex structure field $\tau(z)$ satisfies  the ansatz \cite{Weigand:2018rez}
  \be
j(\tau(z)) =12^3 \frac{4 f^3(z)}{4 f^3(z) +27 g^2(z)}, \qquad N\equiv  \deg(4 f^3 +27 g^2) \ge \deg(f^3),
\label{eq:stringAnsatz}
\ee 
where $f=f(z)$ and $g=g(z)$ are holomorphic polynomials on $z$.  
 The condition on the degree of the polynomials  ensures  that the complex structure, and thus the geometry of the two-torus, attains a fixed (finite) value  in the limit $|z|\to\infty$ (the  asymptotic boundary of the bordism $\pd \cB_4$) \cite{Greene:1989ya}.  
 
  The previous ansatz represents a smooth solution to the equations of motion and to \eqref{eq:Weierstrass} everywhere except at the $N$ points $z_a$,  defined by $4f^3(z_a)+27g^2(z_a)=0$ with $a=1,\ldots,N$. At those points  the fibre degenerates and the semi-flat description ceases to be valid.    The appropriate choice for  the function $F(z)$ in \eqref{eq:semiFlatMetric}  is given by \cite{Greene:1989ya}
\be
F(z) =F_0 \, \eta(\tau(z))^2 \prod_{a=1}^N (z- z_a)^{-\frac{1}{12}},
\label{eq:Fz}
\ee
where $\eta(\tau)$ is the Dedekind function\footnote{It is defined by $\eta(\tau) = \rme^{\rmi \pi \tau/12} \prod_{n>1} (1 - \rme^{2 \pi \rmi n \tau})$, and the    the modular group acts on it as follows:   $\eta(\tau+1) = \rme^{\rmi \frac{\pi}{12} } \eta(\tau)$, and 
 $\eta(-1/\tau) = (-\rmi \tau)^{1/2} \eta(\tau)$.} 
  and $F_0$ is an arbitrary  complex number. The Dedekind function $\eta(\tau)$ needs to be included to make the metric on the plane parametrised by $\{z,\bar z\}$ (the first summand in \eqref{eq:semiFlatMetric}) invariant under modular transformations.   As we  will explain  below, the  remaining factor of $F(z)$ ensures that  the  metric on the $z$-plane does not vanish at the degeneration points. 
 
 We will denote by $\cB_{\text{sf}} =\{z \in \mathbb{C}\setminus\{z_a\}, y^1\in \[0,1\), \psi\in \[0,1\)\}$ the spacetime characterised by the semi-flat metric   \eqref{eq:semiFlatMetric}  and the ansatz \eqref{eq:stringAnsatz} away from the degeneration points. This spacetime is to be glued through its boundary $|z|\to \infty$ with the outer-bubble regime, and the hyperk\"ahler regime be used to characterise the geometry in a neighbourhood of $z_a$, where the semi-flat approximation breaks down.

\paragraph{Detailed internal structure.} The ansatz \eqref{eq:stringAnsatz} defines a multivalued mapping  between the fundamental domain and the disc, $\tau^{-1}: \cF\to  \mathbb{C}\setminus\{z_a\}$, with monodromy in $PSL(2,\mathbb{Z})$. Actually,  the fundamental domain $\cF$ is mapped $N$ times  to the $z$-plane \cite{Greene:1989ya}, and the different images of $\cF$, denoted  $\cB_\text{sf}^a$, are \emph{glued} together  across  their boundaries in the sense of \eqref{eq:matchingConditions}, so that $\cB_{\text{sf}} \cong \cup_{a=1}^N \cB_{\text{sf}}^a$. The matching surfaces can be identified as the images of the boundary of fundamental domain $\pd\cF$ on $\mathbb{C}\setminus\{z_a\}$,
which are easily located using that the elliptic modular function  is \emph{real} at $\pd\cF$, with $j(\tau)\in(-\infty,12^3]$.
Since the geometry of the two-torus in adjacent regions $\cM_{\text{sf}}^a$ are  related to each other by modular transformations, 
the matching of the toroidal geometries in different regions of the disc
will require  a non-trivial gluing diffeomorphism. Actually, given to regions $\cB_{\text{sf}}^A$ and $\cB_{\text{sf}}^B$ with the $T^2$ geometry related by a modular transformation $S$ or $T$, the appropriate gluing diffeomorphism    $\Phi_{AB}: \pd \cB_{\text{sf}}^A \to \pd \cB_{\text{sf}}^B$ at the matching surfaces is
\bea
S:&&  \tau^A = -1/\tau_B, \qquad \Longrightarrow \qquad z^A = z^B, \qquad  \psi^B = y_1^A  \; \;  \qquad  \qquad \text{and} \qquad  y_1^B  = - \psi^A,\nonumber \\
T: &&\tau^A = \tau_B +1,  \qquad \Longrightarrow \qquad z^A = z^B, \qquad \psi^B =  \psi^A + y_1^A \qquad \text{and} \qquad  y_1^B = y_1^A.
\label{eq:identifications}
\eea
Note that   the matching of the metric along the directions $\{z,\bar z\}$ is trivial. This is due to the transformation properties of the Dedekind function under $PSL(2,\mathbb{Z})$, which imply that the first summand in \eqref{eq:semiFlatMetric} is completely invariant under the identifications \eqref{eq:identifications}. 

In a generic semi-flat geometry, the non-trivial identifications in the previous gluing procedure might lead to orbifold singularities \cite{Greene:1989ya}. 
These singularities  could arise at points $\{z_\rmi,z_\rho\}$ of the $z$-plane mapped through \eqref{eq:stringAnsatz} to the fixed points of $PSL(2,\mathbb{Z})$ in the fundamental domain, that is, to the configurations $\tau(z_\rmi) =\rmi$ and $\tau(z_\rho) =\rho\equiv \rme^{\frac{2 \pi \rmi}{3}}$.
Indeed,  the field  $\tau$ undergoes a  modular transformation 
 when encircling the points $\{z_\rmi,z_\rho\}$ (see \cite{Ortin:2015hya}), however 
the ansatz \eqref{eq:stringAnsatz} guarantees that  this modular transformation is trivial: when going  (counterclockwise)  around the point $z_\rmi$ (resp. $z_\rho$) the monodromy  is actually $S^{2 p}$ (resp. $(T^{-1}S )^{3p}$), with $p\in\mathbb{Z}$, which is the identity map. Since there is no  need to perform a non-trivial identification in this case,  no orbifold singularity will arise  in the  geometries defined by\footnote{Alternatively, we could also argued the absence of these singularities using the Weierstrass model \eqref{eq:Weierstrass}, which defines a regular geometry provided that we stay away form the degeneration points $z_a$ of the fibre.} \eqref{eq:stringAnsatz}.

 Regarding the degeneration points $z_a$, when we encircle these points counterclockwise the two-torus fibre also undergoes a $T$ monodromy (a Dehn twist), however since the points $z_a$ are already excluded from  $\cB_{\text{sf}}$ due to the failure of the semi-flat description on them, we do not have to discuss the presence of orbifolds there.

\begin{figure}[t]
\centering \hspace{-.5cm}\includegraphics[width=0.45\textwidth]{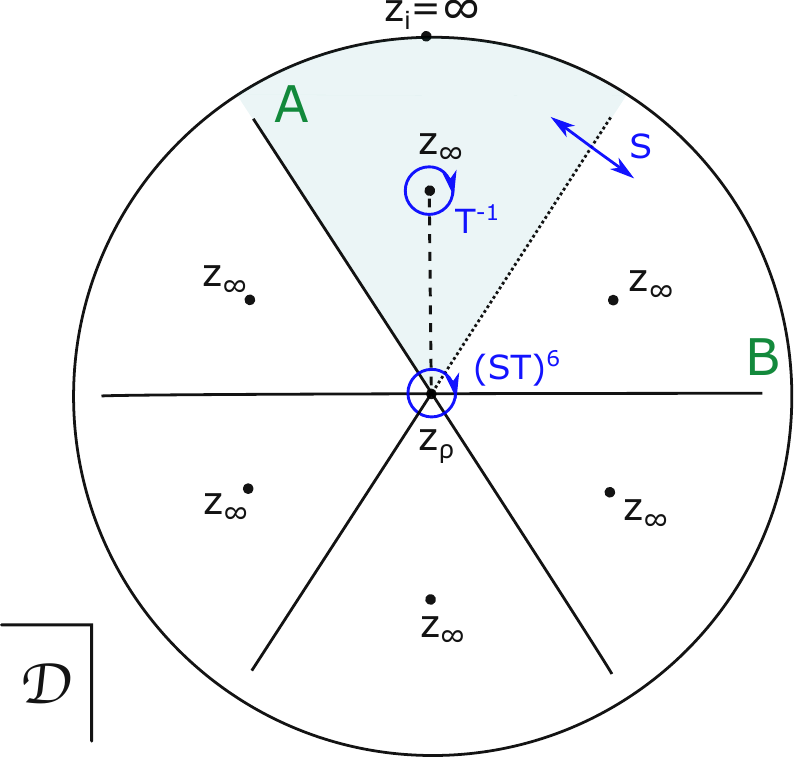}
\caption{Representation of the semi-flat geometry $j_{(\rmi)}$ \eqref{eq:T3ZNsol} with $N=6$, and the monodromy paths. The fundamental domain is mapped six times to the disc $\cD$ via \eqref{eq:T3ZNsol}, and the  shaded region represents a single image of $\cF$. The $T^2$ geometry on adjacent regions (e.g. $A$ and $B$) is related by a $S$ transformation, and the gluing  across the common boundary (dotted line) is done with the diffeomorphism in \eqref{eq:identifications}.  The points  $z_\infty$ represent the degenerations points, and $z_\rmi$, $z_\rho$ are the locations where $\tau$ attains the values  $\tau(z_\rmi)=\rmi$ and  $\tau(z_\rho)=\rho$ respectively  }
  \label{fig:DiscStructure}
\end{figure} 

\paragraph{Explicit semi-flat geometries.} The freedom in the choice of the holomorphic functions $f(z)$ and $g(z)$, together with the arbitrary volume parameters $\rme^{2\varphi_0}$ and  $|F_0|$, allows a large degree of control over the properties of the semi-flat metric. For clarity in the following we will work with a specific choice of the functions $f(z)$ and $g(z)$, although our results can be easily generalised to other geometries.

As we discussed in Section \ref{sec:detailedTopology}, when the compact space is a  quotient  $T^3/\Gamma$ the three-torus must be in one of the two symmetric configurations $\tau=\{\rmi,\rho\}$. While this is not necessary in $T^3$ compactifications, for simplicity  we will  restrict ourselves to semi-flat geometries which asymptote to these symmetric configurations for  $|z| \to \infty$, where the outer-bubble regime (and the asymptotic vacuum) is located. 

Moreover, we will also require the geometry to be invariant under a $\mathbb{Z}_N$ discrete symmetry which rotates the position of the $N$ degeneration  points $z_a$. 
Using that the value of the $j$-function at the  symmetric points is $j(\rmi) = 12^3$ and $ j(\rho) = 0$, we find 
 the following two classes of geometries\footnote{The expression for $j_{(\rmi)}$ is obtained from   \eqref{eq:stringAnsatz} setting  $f = z^p/4^{1/3}$ and $g^2=z_0^{3p}/27$, with  $N=3p$ and  $p\in \mathbb{Z}$. For the ansatz $j_{(\rho)}$  we have to set $g = z^p/\sqrt{27}$ and $f^3 = z_0^{2p}/4$, with $N=2p$ and $p \in \mathbb{Z}$. }\be
 j_{(\rmi)}(\tau)  = 12^3 \frac{ z^N}{z^N - z_0^N}, \quad N\in3\mathbb{Z} ,\qquad\text{and} \qquad  j_{(\rho)}(\tau)  = 12^3 \frac{z_0^N}{z_0^N-z^N},\quad N\in2\mathbb{Z},
\label{eq:T3ZNsol}
\ee
with  $z_0\in\mathbb{C}\setminus\{0\}$, and where $j_{(\rmi)}$ and $j_{(\rho)}$ correspond respectively to the cases  $\tau_\infty = \rmi$, ($z_\rmi=\infty$),  and $\tau_\infty=\rho$ ($z_\rho = \infty$).   It is also easy to check that  for the geometry given by $j_{(\rmi)}$ (resp. $j_{(\rho)}$) the $T^2$ fibre also attains a symmetric configuration at the point  $z_{\rho} =0$  (resp. $z_\rmi=0$), and the  degeneration  points (which are single poles) are  distributed at the $N$ locations 
\be
z_a = z_0 \, \rme^{ \frac{\rmi 2 \pi a}{N}},
\ee
with $a=\{1,\ldots,N\}$. Thus,   the distance between the degenerations  can be tuned changing  $|z_0|$, and their positions rotated varying $\arg(z_0)$. 

\begin{figure}[t]
\centering\hspace{.6cm}\includegraphics[width=0.35\textwidth]{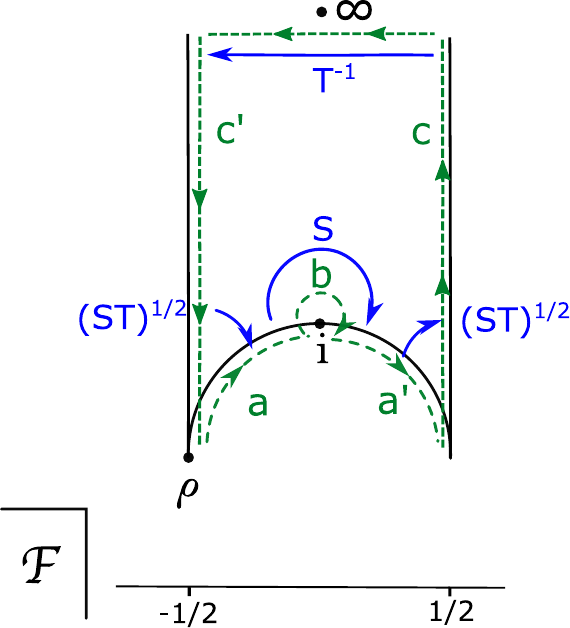} \hspace{1.2cm}
\centering \includegraphics[width=0.47\textwidth]{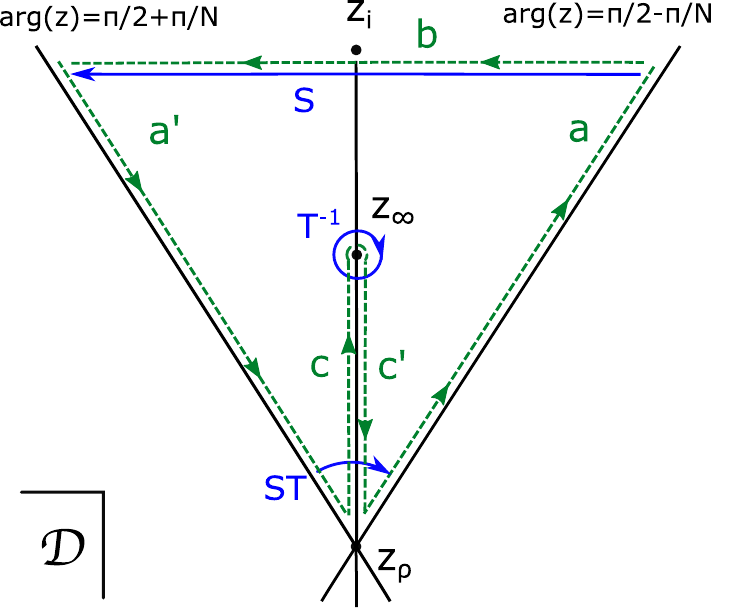}
\caption{An image of the fundamental domain $\cF$ and its boundary (left) on the  disc $\cD$ (right) under the mapping  \eqref{eq:T3ZNsol} with $\arg(z_0)=\pi/2$.  }
  \label{FtoDmap}
\end{figure}

It can be checked that when we encircle counterclockwise the points $z_a$ and $z_\rmi$  the geometry undergoes respectively  monodromies $T$ and $S^N$, and when going around $z_\rho$ it experiences the modular transformation $(T^{-1} S)^N$ in the case of $j_{(\rmi)}$, and $(S T)^N$ for the geometry given by $j_{\rho}$. Finally, the matching surfaces where the $T^2$ fibre  undergoes modular transformations are located at
\bea 
S:&\qquad&\arg(z)= \arg(z_0) + \frac{(2 a +1) \pi}{N} \nonumber\\
T:&\qquad&\arg(z)= \arg(z_0) + \frac{2a \pi}{N}, \qquad   \text{$|z|\le |z_0|$ for $j_{(\rmi)}$, or  $|z|\ge |z_0|$ for $j_{(\rho)}$.} 
\label{eq:semiflatPatches}
\eea
The structure of the geometry associated to the ansatz $j_{(\rmi)}$ is represented in Figs. \ref{fig:DiscStructure} and $\ref{FtoDmap}$ for the case $N=6$.  For simplicity, in all the calculations that follow we will set $\arg z_0=0$.

\paragraph{Asymptotic form of the semi-flat metric $|z| \to \infty$.}

Since the semi-flat spacetime $\cB_{\text{sf}}$ is to be glued through the boundary at $|z|\to \infty$ with the outer-bubble layer,  we need to characterise the behaviour of the geometry in this limit.  The asymptotic form of the  semi-flat metric  \eqref{eq:T3ZNsol}  can be obtained from the following expansions\footnote{Near the  points $z_\rmi$ and $z_\rho$, we have respectively $\tau(z) \approx \rmi +\alpha_\rmi (z-z_\rmi)^{1/2}$ and $\tau(z) \approx \rho +\alpha_\rho (z-z_\rho)^{1/3}$ \cite{Greene:1989ya,Ortin:2015hya} with $\{\alpha_\rmi,\alpha_\rho\}$ complex constants, while $\eta(\tau(z))$ is regular at those points with non-zero first derivative. }
\bea
\tau(z\to \infty) &\to& \tau_\infty + \alpha\, (z_0/z)^{N/p} + \ldots \;,\nonumber \\
|F(z\to \infty)|^2 &\to&  |F_0|^2 |\eta(\tau_\infty)|^4\; |z|^{-N/6}\left(1 + \Re(\beta z_0/z)\right)+\ldots \;,
\label{eq:correctionsTF}
\eea
where $\alpha$ and $\beta$ are some complex  constants, and $p$ is an integer taking  values $2$ and $3$ for the geometries $j_{(\rmi)}$ and $j_{(\rho)}$ respectively. 
Then, to leading order we find
\be
ds^2_{\cB}|_\text{sf} \to   R_{\text{kk}|\theta}^2 |z|^{-\frac{N}{6}}  dz d \bar z + ds^2_{T^2}, \qquad \text{with}\qquad R_{\text{kk}|\theta}^2 \equiv \rme^{2\varphi_0} |F_0|^2 \Im(\tau_\infty)\, |\eta(\tau_\infty)|^4.
\label{eq:StringAsymptotics1}
\ee

Depending on the number of degenerations  we should distinguish two cases, $N<12$ and $N=12$.
  For configurations with  $N<12$ degenerations,  introducing  a polar coordinate system   $\{u, \theta\}$ for the $z$-plane, 
  we can  write the asymptotic form of the line element as follows
\be
ds^2_{\cB}|_\text{sf} \to  d u^2 + u^2\, (1- \ft{N}{12})^2 d\theta^2+ ds^2_{T^2}, \qquad \text{with} \qquad u \equiv \frac{R_{\text{kk}|\theta} |z|^{1-\frac{N}{12}}}{(1-\frac{N}{12})}, \qquad \theta = \arg(z),
\label{eq:asympSF1}
\ee
which describes a geometry of the form $\mathbb{R}^2_\Delta\times T^2/\Gamma$, where $\mathbb{R}^2_\Delta$ is a conical spacetime with deficit angle $\Delta =N\pi/6$. 

 For configurations with $N=12$ degenerations, we need to make  a different  choice for the radial coordinate, namely $u \equiv R_{\text{kk}|\theta} \log |z|$,  which leads  to the asymptotic form of the line element  
\be
ds^2_{\cB}|_{\text{sf}} \to  d u^2 +  R_{\text{kk}|\theta}^2 d\theta^2+ ds^2_{T^2}. \label{eq:asympSF2}
\ee
This metric represents a cylindrical  geometry of the form $\mathbb{R}\times T^3$ (i.e.  $\Delta=2 \pi$), where the 
 radius of the $S^1$ parametrised by $\theta$ is given by  $R_{\text{kk}|\theta}$ \eqref{eq:StringAsymptotics1}, and  the circles parametrised by $\psi$ and $y^1$ have respectively  the following radii
\be
R_{\text{kk}|\psi}^2 =\rme^{2 \varphi_0} \Im(\tau_\infty)^{-1}, \qquad \text{and} \qquad
  R_{\text{kk}|y^1}^2= \rme^{2 \varphi_0} \Im(\tau_\infty)^{-1} |\tau_\infty|^2.
\label{eq:radii}
\ee

Finally, noting that the $T^3$ and fibre volumes are proportional to $\cV_{T^3} \propto \rme^{3 \varphi_0} |F_0|$ and $\cV_{T^2} \propto \rme^{2 \varphi_0}$, we find   that the characteristic length-scales of this geometry are given by
\be
\ell_{\text{kk}} \equiv  \rme^{\varphi_0} |F_0|^{1/3} \qquad\text{and} \qquad \ell_{\text{fibre}}\equiv \rme^{\varphi_0}. \label{eq:scalesKKF}
\ee

\paragraph{Behaviour near the degeneration points.} Although  the semi-flat geometry \eqref{eq:semiFlatMetric} is an exact solution to the equations of motion with $\alpha=0$, it is singular  at the degeneration points  $z_a$, and thus we cannot use it to construct a  BON instanton while working consistently in  the semiclassical regime of quantum gravity. Then, as we mentioned above, to obtain a smooth solution we will excise a small neighbourhood of the degeneration points, and glue there a hyperk\"ahler geometry of the form \eqref{eq:HyperkahlerMetric}. 
In the following paragraphs we will review here the local geometry of the semi-flat metric close to the degeneration points.

 Near a point $z_a$ where the denominator of \eqref{eq:stringAnsatz} has a zero, the behaviour of the complex structure field in \eqref{eq:T3ZNsol} is determined by the relation
\be
j(\tau(z)) \approx \pm \frac{12^3}{N}  \frac{ z_a}{(z-z_a)} + \alpha_N+\cO(|z-z_a|/|z_a|),  
\ee
where the plus sign corresponds to  $j_{(\rmi)}$, and the minus sign to $j_{(\rho)}$, and $\alpha_N$ is a complex constant independent of $z_a$.
Then, using the asymptotic expansion of the $j$-function near\footnote{In the limit $\Im(\tau)\to \infty$ we have $j(\tau)\approx \rme^{-\rmi 2 \pi \tau} + 744 + \cO(\rme^{2\pi \rmi \tau})$. } $\Im \tau \to \infty$, it is easy to check that the complex structure $\tau$ has the well known logarithmic profile 
\be
 \tau(z) \approx -\frac{\rmi }{2 \pi} \log \left[\pm \frac{N}{12^3} \frac{z-z_a}{z_a}\right] +\beta_N \frac{z-z_a}{z_a} + \ldots \, ,
\label{eq:tauNearDeg}
\ee
with $\beta_N$ an $N$-dependent complex parameter. Equivalently, in terms of the function $V(y)$ and the one-form $\cA_i$ which characterise the  hyperk\"ahler metric \eqref{eq:HyperkahlerMetric} we have the leading form
\be
V(z) \approx  \frac{1}{2 \pi} \log \big[12^3 |z_a|/(N |z-z_a|)\big], \qquad \cA_1 \approx \frac{1}{2 \pi} \left(\arg(z-z_a) - \frac{2 \pi a}{N}+ s \pi \right),
\label{eq:imTauDeg}
\ee
where the parameter $s$ takes the values $0$ and $1$ for the solutions  $j_{(\rmi)}$ and  $j_{(\rho)}$ respectively. 

Finally, to write the line element around the points $z_a$ we will also need the local form of the function $F(z)$. Making use of the local expansion of the Dedekind function around\footnote{We use $\eta(\tau) \approx \rme^{\pi \rmi\tau/12} -\rme^{25 \pi \rmi \tau/12}+\ldots$.} 
$\Im(\tau)\to \infty$,  we find
\be
|F(z)|^2 \approx \frac{|F_0|^2}{2 \sqrt{3} |z_a|^{N/6}} \big(1+ \Re\left( \delta_N (z-z_a)/z_a\right) \big) +\ldots .
\label{eq:FnearDeg}
\ee
where $\delta_N$ is a complex constant independent on $z_a$.
Collecting all of these results, we find that near the degeneration points the semi-flat metric  has the form
\bea
\rme^{-2\varphi_0} ds^2_{\cB}|_{\text{sf}} &\to& \frac{|F_0|^2}{4 \pi\sqrt{3} |z_a|^{N/6}}\,   \log (r_0/r)\,  (dr^2 +r^2 d\theta^2 ) \nonumber\\ 
&&+  \frac{1}{2 \pi} \log (r_0/ r)\, d y^1 d y^1  + \frac{2 \pi}{ \log (r_0/r)}  \Big(d\psi + \frac{\theta}{2\pi}  \, dy^1\Big)^2,
\label{eq:degeneratingMetric}
\eea
where $r=|z-z_a|$, $\theta =\arg(z-z_a) - \frac{2 \pi a}{N}+ s \pi$ and $r_0 \equiv 12^3 z_0/N$.
It is straightforward to check that these are singular points of the $D$-dimensional geometry, as it can be checked computing the Gauss-Bonnet invariant 
$R^2_{GB}|_{\text{sf}}\sim (r \log r)^{-4} \to \infty$ in the limit $r\to0$. This behaviour signals the failure of the semi-flat description near the degenerations. 

To understand the geometrical meaning of this divergence, recall that the volume of the $T^2$ fibre is constant over the $z$-plane. Therefore the  profile \eqref{eq:tauNearDeg} for $\tau$ implies that one of the cycles of the fibre shrinks to zero size at $z_a$, while the other one grows unbounded, what
 amounts to a partial decompactification of spacetime.  In addition, in  \eqref{eq:tauNearDeg}  we can see explicitly that as we go around  a  degeneration point the $T^2$ fibre transforms under the action of a $T$ modular transformation, $\tau \to \tau +1$. 

From the previous expressions we can also identify one further length-scale characterising this  layer of the BON spacetime, namely 
\be
\ell_{\text{sf}}^2 \equiv \frac{\rme^{2\varphi_0} |F_0|^2}{2 \sqrt{3} \, z_0^{N/6}} = \frac{\ell_{\text{kk}}^6}{2 \sqrt{3}\, \ell_{\text{fibre}}^4\, z_0^{N/6}}.
\label{eq:lsfDef}
\ee
  At the beginning of  Section \ref{sec:detailedBON}   we mentioned that $\ell_{\text{sf}}$ represents the distance between the degenerations, and yet the parameter $z_0$ controlling their relative position appears in the denominator. This might seem counterintuitive at first sight. However, the numerator contains the actual length-scale $\rme^{\varphi_0} |F_0|$  which determines  proper distances measured by the semi-flat line element \eqref{eq:semiFlatMetric}.  Thus, the proper distance between degenerations can be made arbitrarily large increasing   $\ell_{\text{sf}}$, and regardless of the value  of $z_0$.

\subsubsection{Layer (II.b): Hyperk\"ahler  regime.}
\label{sec:regularisation}

As we discussed above,  near the degeneration points the semi-flat spacetime  \eqref{eq:semiFlatMetric} experiences a partial decompactification, since  the cycle of the $T^2$ fibre parametrised by $y^1$ grows unbounded.

  In such a situation we know  there is an infinite tower of Kaluza-Klein (KK) modes (with dependence on $y^1$) which become light, what suggests that they  may play  an important role in the resolution of the singularity. Recall that in order to find the semi-flat metric \eqref{eq:semiFlatMetric} we started with the hyperk\"ahler ansatz \eqref{eq:HyperkahlerMetric}, and  then imposed consistency 
   with the dimensional reduction along the coordinate $y^1$. Since this condition is equivalent to the truncation of the tower of KK modes with dependence on $y^1$, our argument  suggests that we should lift this constraint near the degeneration points, and try instead to describe the geometry  in terms of the more general hyperk\"ahler ansatz \eqref{eq:HyperkahlerMetric}.  
A simple hyperk\"ahler geometry consistent with the periodicity   $y^1 \sim y^1 + 1$  is characterised by the following harmonic function written in terms of the Gibbons-Hawking ansatz \cite{Hawking:1976jb,Myers:1986rx,Ooguri:1996me,gross2000} 
 \be
V(y^1,\rho) = V_0 + \frac{1}{4 \pi} \sum_{k\in \mathbb{Z}} \frac{1}{\sqrt{(y ^1- k)^2 +\rho^2}} - \frac{1}{2 \pi} \sum_{k>0} \frac{1}{k}.
\label{eq:KKmonopoleRed}
\ee  
where we have introduced the coordinate\footnote{Notice this is a different radial coordinate from the $\rho$ introduced in Section \ref{sec:BONansatz}, which corresponds to the outer layer of the bubble.}  $\rho^2 \equiv (y^2)^2 + (y^3)^2$, and taking the flat metric $\mathring{h}_{ij}$ to be in the canonical form $\delta_{ij}$. The last term  is added to cancel the divergent contribution to $V$ from the locations $y^i = (k,0,0)$ with $k\in\mathbb{Z}$, and $V_0$ is an arbitrary constant. 
The resulting metric  \eqref{eq:HyperkahlerMetric} is  smooth everywhere by construction \cite{Myers:1986rx,Ooguri:1996me} and  clearly has the required discrete isometry along $y^1$.  This space, which we denote by $\cB_{\text{hk}}$, represents a euclidean Taub-NUT geometry, i.e. a  KK monopole, embedded in a $S^1 \times \mathbb{R}^2$ spacetime, that is, with one dimension (other than $\psi$) 
compactified on  a circle.

To clarify the connection between this geometry and the semi-flat metric \eqref{eq:semiFlatMetric}, let us consider the asymptotic $\rho\to \infty$ behaviour of the harmonic function  \eqref{eq:KKmonopoleRed}. For this purpose it is convenient to rewrite the previous expression using  Poisson's summation formula \cite{Ooguri:1996me} 
\be
V(y^1,\rho) = \frac{1}{2 \pi} \log( \rho_0/\rho) +\frac{1}{\pi} \sum_{m>0} K_0(2 \pi m\, \rho) \cos(2 \pi m \, y^1),
\label{eq:poissonKK}
\ee
where $\rho_0 =  2\rme^{2 \pi V_0 -\gamma}$ is determined by the arbitrary constant $V_0$, $\gamma$ is Euler's constant, and $K_0$ is the modified Bessel function of the second kind. Note that the $m=0$ mode in this expansion exhibits at $\rho \to 0$  precisely the same logarithmic divergence as the semi-flat metric near a degeneration point, eq. \eqref{eq:imTauDeg}, but in this case 
the infinite tower of excited KK modes exactly cancels the divergence leading to a smooth geometry.

\paragraph{Asymptotic limit of the hyperk\"ahler geometry.}  Following \cite{Ooguri:1996me,gross2000}, in order to  cure the singular behaviour of the semi-flat metric  \eqref{eq:degeneratingMetric}, we will excise a neighbourhood of the degeneration points of $\cB_{\text{sf}}$ and   glue there the spacetime of the dimensionally reduced KK monopole given by \eqref{eq:poissonKK}. For this purpose we need a more detailed characterisation of the hyperk\"ahler geometry \eqref{eq:poissonKK} in the limit  $\rho\to \infty$.  Using the asymptotic expansion of the modified Bessel function $K_0$ one finds (recall $m>0$)
\be
\rho\to \infty: \qquad K_0(2 \pi m\, \rho) \approx \frac{\rme^{-2 \pi m \, \rho}}{2 \sqrt{m \rho} }  + \ldots \, , 
\label{eq:asympHK}
\ee
that is,  far from the KK monopole the harmonic function  approaches exponentially fast to the limiting behaviour of the semi-flat geometry  \eqref{eq:imTauDeg}. 
Regarding the one-form $\cA_i$, it is convenient to   express its components in a cylindrical coordinate system $\{y^1,\rho,\theta\}$, with $\tan \theta = - y^3/y^2$.

Then, the self-dual conditions  together with \eqref{eq:KKmonopoleRed},  imply that   $\pd_1 A_\rho - \pd_\rho A_1=0$. If we choose the gauge $\cA_\rho=0$,  then the configuration  $\cA_i$
must satisfy the equations
\be
\pd_\rho\cA_1=0,\qquad \pd_\rho \cA_\theta =-\rho\, \pd_{1} V , \qquad  \pd_{1} \cA_\theta -\pd_\theta \cA_1 = \rho\,  \pd_\rho V.
\label{eq:KKselfdual}
\ee
Using the form \eqref{eq:poissonKK}  for the harmonic function $V$ and \eqref{eq:asympHK}, it is easy to see that  away from the KK monopoles the one-form $\cA_i$ 
has the asymptotic behaviour
\be
\cA_\rho =0,\qquad  \cA_\theta\approx -\frac{y^1}{2 \pi}-\frac{\sqrt{\rho}}{2 \pi} \rme^{-2 \pi \rho} \sin(2 \pi y^1), \qquad \cA_1 \approx  -\sqrt{\rho} \rme^{-2 \pi \rho} \cos(2 \pi y^1) (\theta-\theta_0),
\label{eq:asymptoticAKK}
\ee
where we  have kept only the dominant terms in the KK and $ \rho$ expansions, and a specific choice of  integration constants has been made for later convenience. While the asymptotic form of the one-form $\cA_i$ does not match that of  the semi-flat metric \eqref{eq:imTauDeg}, where $\cA_\theta|_{\text{sf}}\approx 0$, we will see below that is still possible to perform the matching of the  spacetimes $\cB_{\text{sf}}$ and $\cB_{\text{hk}}$ with  an appropriate gluing diffeomorphism (see also  \cite{gross2000}).

\paragraph{Gluing  the hyperk\"ahler and semi-flat spacetimes.} In the previous paragraphs we have described the independent semi-flat $\cB_{\text{sf}}$ and hyperk\"aher $\cB_{\text{hk}}$  spacetimes which are to represent respectively   the layers {\bf II.a} and {\bf II.b} of the inner BON geometry in the absence  of the Gauss-Bonnet term, $\alpha \to 0$. We will now perform the matching of these two geometries in the sense of \eqref{eq:matchingConditions} in order to have a complete and well defined inner-bubble  spacetime $\cB_{\bf II}|_{\alpha=0}$. 

First we define the manifold $\cB_{\text{sf}}^*\equiv \cB_{\text{sf}} \setminus\{B_{r^*}(z_a)\}$ as the result of cutting out from the semi-flat space the interior of $N$ balls of radius $r^*$  centered on the degeneration points, $B_{r^*}(z_a)$, with  $|z-z_a|\le r^*$. On the other hand we introduce $N$ copies $\cB_{\text{hk}|\rho^*}^a\equiv \{(\rho,\theta,y^1,\psi)\in \cB_{\text{hk}}\; /\; \rho \le \rho^*\}$ of the region of the  hyperk\"ahler spacetime \eqref{eq:poissonKK}  with radial coordinate bounded by $\rho \le \rho^*$. Then, with this at hand we define the inner-bubble  manifold by
\be
\cB_{\bf II} \equiv \cB_{\text{sf}}^*\cup_{a=1}^N \cB_{\text{hk}}^a.
\label{eq:innerBON}
\ee
Since these two spacetimes cannot be glued exactly we will resort to the perturbative matching methods described in Section \ref{sec:gluingStrategy}. 
We begin defining the small parameter $\epsilon \equiv \ell_{\text{fibre}}/\ell_{\text{kk}} \ll1$, which controls the magnitude of the KK corrections   in the hyperk\"ahler regime. 
Then, consistently with the approximation scheme \eqref{eq:appScheme}, we consider the region of the parameter space where  the length-scales characterising the spacetime geometry,  $\ell_{\text{kk}}$,   $\ell_{\text{fibre}}$ and $\ell_{\text{sf}}$, meet the conditions\footnote{This particular set of  relations between the length-scales has been chosen for simplicity, but more general approximation schemes are also possible.}
\be
\ell_{\text{sf}}=\ell_{\text{kk}} \qquad \Longrightarrow \qquad \frac{\ell_{\text{fibre}}}{\ell_{\text{sf}}}=\epsilon,  \qquad\Longrightarrow \qquad z_0 = \ft{N}{12^3} \hat r_0 \epsilon^{-q}, 
\label{eq:scaling}
\ee
where $q \equiv 24/N\ge2$,  and   $\hat r_0 =\cO(\epsilon^0)$ is  a positive real parameter. To derive the scaling of $z_0$ with $\epsilon$  we have used the definitions \eqref{eq:scalesKKF} and  \eqref{eq:lsfDef}.

    Next, to make the  space $\cB_{\bf II}$  connected we  introduce the gluing diffeomorphisms which identify the common boundaries of the constituent spacetimes.  Let us consider the surgery around one of the degeneration points, $z_a$. The appropriate (leading order) gluing diffeomorphism $\Phi: \pd\cB_{\text{sf}}^* \to \cB_{\text{hk}}^a$ is given by\footnote{Note that the identification is consistent with the periodicity of the coordinates $\theta \sim \theta + 2 \pi k_1$, $y^1 \sim y^1 + k_2$ and $\psi \sim \psi +k_3$, with $k_i\in\mathbb{Z}$.} (see  Section \ref{sec:gluingStrategy})
\be
\Phi:  (r^*, \;\; \theta_-, \; \;y^1_-,\; \;\psi_-) \quad \longrightarrow\quad (\rho^* =  r^*/\epsilon, \;\; \theta_+ =\theta_- , \; \;y^1_+ = y^1_- , \; \; \psi_+ =\psi_- + \frac{\theta_-\,  y^1_-}{2\pi} ),
\label{eq:zeroGluing}
\ee
where the subscripts $``-"$ and $``+"$  refer to coordinates on the boundary hypersurfaces $\pd\cB_{\text{sf}}^*$ and  $\pd \cB_{\text{hk}}^a$ respectively. In particular the coordinates on the semi-flat patch are defined as in eq. \eqref{eq:degeneratingMetric},  and we take $r^* =\cO(\epsilon^0)$. 
To ensure  the continuity of the metric tensor across the matching surface we need to require the induced metrics on the boundaries $\pd\cB_{\text{sf}}^*$ and  $\pd \cB_{\text{hk}}^a$  to agree (left eq. in \eqref{eq:matchingConditions}), and following the perturbative approach of \cite{Mars:2005ca}
we solve the resulting conditions order by order in $\epsilon$. 

To leading order, i.e. from equation eq. \eqref{eq:zeroOrder1FF},   we find that the parameters of the semi-flat and hyperk\"ahler  geometries should satisfy the relation 
\be
\rho_0 = \hat r_0 \epsilon^{-(q+1)} \qquad \Longrightarrow \qquad   \rho_0 = \frac{12^3 |F_0|}{2 \sqrt{3} N}  z_0^{(12-N)/12}. 
\label{eq:gluingParams}
\ee
Recall  that to this order  we neglect completely  the KK modes contribution in $\cB_{\text{hk}}^a$  and the presence of multiple degeneration points in $\cB_{\text{sf}}^*$, and thus the identifications \eqref{eq:zeroGluing} and \eqref{eq:gluingParams} imply that  the
 asymptotic forms of the semi-flat  and the hyperk\"ahler  metric tensors  are identical in a neighbourhood of the  matching boundary. As a consequence,  the requirement that  the there is no shell present on the matching hypersurface (right eq. in \eqref{eq:matchingConditions}) is trivially satisfied. Since there are no  further constraints,  the parameter $r^*$ is left unfixed to leading order in $\epsilon$. 
  
    Summarising, at this point we have already succeeded in constructing the background spacetime $\cB_{\bf II}|_{\alpha=0}$ describing the  inner-bubble region of the bordism $\cB_4$ to zero-order in the deformation, $\alpha\to0$.  The corresponding manifold is defined in terms of the semi-flat $\cB_{\text{sf}}^*$ and hyperk\"ahler  $\cB_{\text{hk}}^a$ spacetimes  via \eqref{eq:innerBON} together with the identifications \eqref{eq:zeroGluing}.  The metric tensor  on $\cB_{\bf II}|_{\alpha=0}$  is determined on the  patch $\cB_{\text{sf}}^*$ by \eqref{eq:semiFlatMetric} and \eqref{eq:T3ZNsol},  on the patches $\cB_{\text{hk}}^a$ by the line element  \eqref{eq:HyperkahlerMetric} given by the harmonic function \eqref{eq:poissonKK} and the parameter \eqref{eq:gluingParams}, and it is continuous  across the matching boundaries to leading order in $\epsilon$. This line element 
    is an \emph{exact solution} of the Euler-Lagrange equations in the interior of $\cB_{\text{sf}}^*$ and $\cB_{\text{hk}}^a$, and it also solves the equations of motion on the matching boundaries to the leading order in $\epsilon$ (with no shells/branes present there).

\paragraph{Validity of the approximations.}

Since the inner-bubble spacetime $\cB_{\bf II}|_{\alpha=0}$ will be used as the background for the perturbative expansion in $\alpha$,  we need to characterise the size of the leading order corrections due to the neglected KK modes  and  the backreaction of the multiple degeneration points. 

As before, we consider the gluing between $\cB_{\text{sf}}^*$ and one of the $N$ copies $\cB_{\text{hk}}^a$ associated to the degeneration at $z_a$. We find that the next-to-leading  correction to the  matching conditions for  the first  fundamental form are:
\begin{align}
0&= r^{-2}_*\rme^{-2 \varphi_0}(\Phi_*(s_{\text{hk}}) - s_{\text{sf}})_{\theta\theta}\hspace{-.5cm} &=\; &\hat \epsilon  \, \Delta s_{\theta\theta}^{(\alpha)}  +  \epsilon^{q} \log\epsilon \;  \Delta s^{(\text{deg})}_{\theta\theta}  + \frac{\rme^{-\frac{2 \pi r^*}{\epsilon}}}{\sqrt{\epsilon}} \Delta s_{\theta\theta}^{(\text{kk})}+\ldots \; , \nonumber \\0&=\hspace{.2cm}\rme^{-2 \varphi_0}(\Phi_*(s_{\text{hk}}) - s_{\text{sf}})_{11} &=\; &  \hat \epsilon \; \Delta s_{11}^{(\alpha)} +\epsilon^q\, \Delta s^{(\text{deg})}_{11}+ \frac{\rme^{-\frac{2 \pi r^*}{\epsilon}}}{\sqrt{\epsilon} \log\epsilon}  \Delta s^{(\text{kk})}_{11}  + \ldots \;, \nonumber\\0&=\hspace{.2cm}\rme^{-2 \varphi_0}(\Phi_*(s_{\text{hk}}) - s_{\text{sf}})_{1\psi}&=\; & \hat \epsilon \; \Delta s_{1\psi}^{(\alpha)} + \frac{\epsilon^q}{\log\epsilon}  \, \Delta s_{1\psi}^{(\text{deg})} + \frac{\rme^{-\frac{2 \pi r^*}{\epsilon}}}{\sqrt{\epsilon} \log \epsilon}  \Delta s^{(\text{kk})}_{1\psi} +\ldots \,, \nonumber\\
0&=\hspace{.2cm}\rme^{-2 \varphi_0}(\Phi_*(s_{\text{hk}}) - s_{\text{sf}})_{\theta\psi}&=\; &  \hat \epsilon\; \Delta s_{\theta\psi}^{(\alpha)} 
+\frac{\rme^{-\frac{2 \pi r^*}{\epsilon}}}{\sqrt{\epsilon} \log \epsilon} \Delta s_{\theta \psi}^{{\text{(kk)}}}+ \ldots \, ,\nonumber \\
0&=\hspace{.2cm}\rme^{-2 \varphi_0}(\Phi_*(s_{\text{hk}}) - s_{\text{sf}})_{\psi\psi}&=\; & \hat \epsilon\; \Delta s_{\psi\psi}^{(\alpha)}+ \frac{\epsilon^q }{(\log\epsilon)^2} \Delta s_{\psi \psi}^{(\text{deg})} + \frac{\sqrt{\epsilon}\, \rme^{-\frac{2 \pi r^*}{\epsilon}} }{(\log \epsilon)^2} \Delta s_{\psi \psi}^{\text{(kk)}}+\ldots \; ,
\label{eq:innerBONmatching}
\end{align}
where we have used the  behaviour of the semi-flat geometry near the degenerations, eqs.  \eqref{eq:FnearDeg} and \eqref{eq:tauNearDeg}, and the asymptotic form of the hyperh\"ahler geometry, eqs. \eqref{eq:asympHK}  and \eqref{eq:asymptoticAKK}.   Here the tensors $\Delta s_{ab}^{(\text{deg})}$ and  $\Delta s_{ab}^{(\text{kk})}$ are functions of $y^1$ and $\theta$, and represent respectively the contribution from distant degeneration points (other than $z_a$) and the massive KK modes. 

For clarity, we have also included the first order correction induced by the  change in the Gauss-Bonnet coupling  $\alpha$, which is assumed to be small  $\alpha =\cO(\hat \epsilon)$, with  $\hat \epsilon\ll1$ being  independent from $\epsilon$.   The associated correction, denoted by  $\Delta s_{ab}^{(\alpha)}$,  will be computed  in Section \ref{sec:nonSusyDeformation}. Recalling that $q\ge2$, and assuming $\epsilon \sim \hat \epsilon$, it is immediate to check that at next-to-leading order it is consistent to take into account only the leading correction in $\alpha$, while neglecting completely  the effects of the KK modes and the presence of multiple 
degenerations. Although we do not show it here, it can also be checked that the matching conditions for the second fundamental form  \eqref{eq:matchingConditions} have a similar structure to \eqref{eq:innerBONmatching}. Thus, to solve the next-to-leading order in perturbation theory it is also sufficient to consider only the dominant corrections in $\alpha$.

From the previous expressions we can also see that the neighbourhood where the semi-flat  approximation becomes inadequate (due to sizeable KK mode effects) has a finite size of order  $r\sim \epsilon$.  On the other hand, these regions should not become too big, since in the semi-flat geometry  the $T^2$ fibre undergoes modular $S$ transformations at the hypersurfaces \eqref{eq:semiflatPatches}, where the role of the $\psi$ and $y^1$ cycles is exchanged. Indeed, the corresponding gluing diffeomorphisms  \eqref{eq:identifications} are only compatible with the matching conditions \eqref{eq:matchingConditions} provided all the massive KK modes of the $T^2$ fibre are exactly zero, so we also need the matching boundary to satisfy  $r_* \lesssim z_0\sim \epsilon^{-q}$. 
Note that these consistency requirements are met in our construction, as we are assuming $r_* \sim \cO(\epsilon^0)$ and $\epsilon\ll1$, which implies
\be
\epsilon \ll r_*  \ll \epsilon^{-q}, \qquad \text{since} \qquad q\ge2. 
\label{eq:activeKKregion}
\ee
In terms of the parameters of the compact space $T^3/\Gamma$ (see \eqref{eq:StringAsymptotics1} and  \eqref{eq:radii}), the limit $\epsilon\to0$ and the scaling \eqref{eq:scaling} imply the relations 
\be
R_{\text{kk}|\theta}^2 \gg R_{\text{kk}|y^1} \cdot R_{\text{kk}|\psi}, \qquad z_0 \sim \left(R_{\text{kk}|\theta} / \sqrt{R_{\text{kk}|y^1} R_{\text{kk}|\psi}}\right)^q\gg1.
\label{eq:semiflatAccurate}
\ee
That is, we are restricting ourselves to a limit of the $T^3/\Gamma$ geometry in which  the cycles of the $T^2$ fibre are much smaller than the base $S^1$, and for consistency we need to ensure that in the  BON instanton all the degenerations are well separated from each other. These two conditions can always be met in the limit $\alpha\to0$, as both the radii, \eqref{eq:StringAsymptotics1} and \eqref{eq:radii}, and the degeneration positions can be freely specified (they are moduli of the background geometry).  Below we will prove  that these parameters remain free moduli when we turn on the Gauss-Bonnet deformation and work to leading order in $\alpha$.

\subsubsection{Bubble core (II.c): the KK monopole regime.}
\label{sec:analyticLimits}

The relations \eqref{eq:semiflatAccurate} we just derived set  bounds on the regime of parameter space where our construction is under control, however the scaling of the overall volume of the compact space $\cV_{\cC}$ is still undetermined. For this purpose we will now discuss the geometry of the bubble core, where the spacetime curvature, and thus the Gauss-Bonnet term is largest. We will show that the requirement that the curvature   remains everywhere below the Planck scale  for small $\epsilon$ determines the right volume scaling.

It is well known that the multi-centered KK monopole solution described by the harmonic function \eqref{eq:KKmonopoleRed} approaches the self-dual Taub-NUT metric near the KK monopole locations, i.e. $\rho\to0$, $y^1 \to k\in\mathbb{Z}$. For definiteness let us consider the $k=0$ image of the Taub-NUT point. Near this position 
the dominant term of the function $V$ in \eqref{eq:KKmonopoleRed}  is the one diverging at  the position  $\{\rho=0, \, y^1=0\}$ and then\footnote{The remaining images of the Taub-NUT point give no contribution at $\{\rho=0, \, y^1=0\}$, as can be checked summing the series \eqref{eq:KKmonopoleRed} at $\rho=0$ \cite{gross2000}: $V|_{\rho=0} = V_0 + \ft{1}{4\pi|y^1|} -\ft{\gamma}{2\pi} - \ft{1}{4\pi}(\Psi(y^1+1) +\Psi(y^1-1))$, where $\Psi$ is the digamma function. }, choosing the gauge $\cA_\rho=0$, we find that the one-form $\cA_i$ is approximately
\be
V(\rho, y^1)|_{\text{core}} \to V_0 + \frac{1}{4 \pi}\frac{1}{\sqrt{\rho^2 +(y^1)^2 }}, \qquad  \cA_\theta|_{\text{core}} \to \frac{-y^1}{4 \pi \sqrt{\rho^2 + (y^1)^2}}, \qquad \cA_1|_{\text{core}} \to 0,
\label{eq:singleKK}
\ee
where we have used the self duality equations \eqref{eq:KKselfdual} to obtain the local behaviour  of $\cA_i$. 
It is straightforward to check that a linear superposition of these solutions located in the array  $\{\rho=0,y^1=k\}$ and  $k\in \mathbb{Z}$ has the right asymptotic behaviour for the one-form  \eqref{eq:asymptoticAKK}, that is, $\cA_\theta \to -y^1/(2 \pi)$. To write the metric in the standard KK monopole form we introduce  the coordinates
\be
R= \sqrt{\rho^2 + (y^1)^2},\qquad \text{and} \qquad  \cos \chi = y^1/R,
\ee
leading to 
\be
\rme^{-2 \varphi_0}d s^2_\cB|_{\text{core}} \to \frac{R  +R_0}{4 \pi R \, R_0}\,  (dR^2 + R^2  d\Omega^2_{(2)}) + \frac{R\, R_0 }{4 \pi (R  +R_0)} \Big(4 \pi d\psi - \cos \chi \,d\theta \Big)^2,
\label{eq:taubNUT}
\ee
where  $d\Omega_{2}^2$ is the two sphere line element. The constant $R_0$ appearing here can be expressed  in terms of the  parameters of the hyperk\"ahler geometry as follows
\be
 R_0 \equiv \frac{1}{4 \pi V_0} = \frac{1}{2} \log \left(\frac{\rme^\gamma \, \hat r_0}{2 \epsilon^{q+1}}\right)^{-1} \ll1.
\label{eq:KKMradius}
\ee
Provided the coordinate  $\psi$ has the periodicity  $\psi \sim \psi +1$, this metric is known to be regular at the origin $R\to0$ (the locus of the  Taub-NUT point), where the full spacetime $\cM_D$  has the local topology  $\cM_D \cong S^{D-4}\times \mathbb{R}^4$. This point can be identified as one of the $N$ bubble cores, where the compact space is smoothly sealed off.

In this region the Gauss-Bonnet invariant can be computed analytically 
\be
R_{GB}^2|_{\text{core}} \to 3 \cdot 2^{7}\cdot \frac{ \pi^2 \, \rme^{-4 \varphi_0} R_0^4  }{ (R_0 + R)^6},
\label{eq:singleKKGB}
\ee
what shows that, since  $R_0\ll1$,  the region of large curvature is very localised  around the KK monopole positions. Actually, in the strict limit $\epsilon\to 0$ we  find that the Gauss-Bonnet term of the background becomes the (curved space) 4-dimensional Dirac-delta function centred at the KK monopole position, i.e.  $R_{GB}^2|_{\text{core}} \approx 32 \pi^2  \; \delta^{(4)}(R)$. However, note that as long as we keep the  value of $\epsilon>0$ finite,    the maximum value of $R_{GB}^2$  (achieved at $R= 0$) is also finite 
\be
R_{GB}^2|_{\text{max}} =3 \cdot 2^{9}\cdot \pi^2 \, \rme^{-4 \varphi_0}  \log \left(\frac{\rme^\gamma \, \hat r_0}{2 \epsilon^{q+1}}\right)^{2}.
\ee  
Recall that for the semi-classical approach to remain valid the spacetime curvature should remain everywhere well below Planck mass. Using for simplicity  Planck  units,  we can achieve this with the choice  
\be
 \rme^{2 \varphi_0} \gtrsim  |\log \epsilon|.
\ee
Note also that the Gauss-Bonnet contribution to the equations of motion appears always multiplied by the parameter $\alpha = \cO(\hat \epsilon )$. Then, the previous scaling  guarantees that the Gauss-Bonnet deformation also  remains small   near the BON core, $\hat \epsilon R_{GB}^2|_{\max}\ll1$. 
Combining this  result with \eqref{eq:scaling}, and assuming $\hat \epsilon \sim \epsilon$ we find
\be
\ell_{\text{Planck}}^{(D)} = \cO(\epsilon^0),\qquad \ell_{\text{fibre}} = \cO(|\log \epsilon|),\qquad \ell_{\text{kk}} \sim \ell_{\text{sf}} =\cO(\epsilon^{-1}),\qquad \ell_{\text{ssb}} = \cO(\epsilon^{-4}),
\label{eq:scalingCheck}
\ee
what proves that our construction is consistent with the approximation scheme \eqref{eq:appScheme}, and that we have parametric control over the approximations that we have made.

Summarising, our  approximations require  the  compact space $T^3/\Gamma$ to be in a degenerate $T^3$/large-volume (degenerate/LV) limit, with a relatively small $T^2$ fibre volume,  but still well above the Planck scale. 
Finally, the degenerations on the bordism $\cB_4$ should also be well separated from each other.

\subsection{Effect of the non-supersymmetric deformation}
\label{sec:nonSusyDeformation}

In the present Section we will consider the effects of turning  on the non-supersymmetric deformation, the Gauss-Bonnet contribution in \eqref{eq:action}, on the inner-bubble region of spacetime. We will  perform a  perturbative analysis  regarding the Gauss-Bonnet coupling $\alpha$ as a small parameter,  $\alpha =\hat \epsilon \hat \alpha $ with $\hat \epsilon\ll1$, using as background geometry the spacetime $\cM_{\bf II}|_{\alpha=0} \cong S^{D-4}\times\cB_{\bf II}|_{\alpha=0}$, where $S^{D-4}$  is a sphere of radius $\cR$, and $\cB_{\bf II}|_{\alpha=0}$ represents the inner-bubble region of the bordism $\cB_4$ which we constructed in the previous Section. 

 For this purpose we promote the metric tensor $g_{MN}$ of the BON ansatz  \eqref{gralBONansatz} (including    $\cR$) and the dilaton  $\phi$
  to be $\hat \epsilon$-dependent quantities, and assume that they admit a Taylor expansion around $\hat \epsilon=0$
\bea
g_{MN}^{\hat \epsilon} &=& g_{MN}|_{\hat \epsilon=0} + \hat \epsilon\, \pd_{\hat \epsilon} g_{MN}|_{\hat \epsilon=0} +\ldots\, , \nonumber \\
\cR^{-1}_{\hat \epsilon} &=& \cR^{-1}|_{\hat \epsilon=0} +\hat \epsilon \pd_{\hat \epsilon} \cR^{-1}|_{\hat \epsilon=0}+\ldots\, , \nonumber \\
\phi_{\hat \epsilon} &=& \phi|_{\hat \epsilon=0} + \hat \epsilon\, \pd_{\hat \epsilon} \phi|_{\hat \epsilon=0} +\ldots \, .
\label{eq:expansions}
\eea
Here, the unperturbed metric  $g_{MN}|_{\hat \epsilon=0}$ is that of  $\cM_{\bf II}$   \eqref{eq:susyLineElement}, and the dilaton value on the  background $\phi|_{\hat \epsilon=0} = \phi_0$ is an arbitrary  constant. 
In particular the warp factor $W$ of the $S^{D-4}$ component of the background  geometry \eqref{eq:susyLineElement}, and the corresponding $S^{D-4}$ radius $\cR^{-1}$ are given by
\be
W|_{\hat \epsilon=0} = 1, \qquad \cR^{-1}|_{\hat \epsilon=0} =0,
\label{eq:zeroOrderR}
\ee
and $h_{\alpha\beta}^\cB|_{\hat \epsilon=0}$ is the metric tensor on the manifold $\cB_{\bf II}|_{\alpha=0}$.
In the following, to ease the notation, we will drop the $``\hat \epsilon=0"$ subscript from background quantities, e.g. $h_{\alpha\beta}^\cB|_{\hat \epsilon=0}\to h_{\alpha \beta}^\cB$, and we will indicate first order perturbations with an script $``1"$
\be
W^{(1)} \equiv  \pd_{\hat \epsilon} W|_{\hat \epsilon=0},
\qquad   \phi^{(1)}\equiv  \pd_{\hat \epsilon} \phi|_{\hat \epsilon=0}, \qquad \text{etc ... \, .} 
\ee
The  perturbation of the  metric on $\cB_{\bf II}$  will  be denoted by $\gamma_{\alpha \beta}\equiv \pd_{\hat \epsilon} h_{\alpha\beta}^\cB|_{\hat \epsilon=0}$.

It is  immediate to write down the  Euler-Lagrange equations for the first order perturbations linearising the equations \eqref{EinsteinEOM} and \eqref{dilatonHeom}. After substituting the expression for the perturbed Ricci curvature of the BON ansatz \eqref{gralBONansatz}, given by eq.  \eqref{eq:ricciCurvatureAnsatz}, the linearised equations for the metric tensor  read\footnote{Here we have  used that in a four dimensional  manifold with metric $h_{\alpha\beta}^\cB$ with zero Ricci tensor, the curvature    satisfies the following relation $R_{\alpha}^{\phantom{\alpha}\delta \gamma \kappa} R_{\beta \delta \gamma \kappa} =\ft14  R_{GB}^2 g_{\alpha\beta}$  \cite{DeWitt:1965jb}.}
\be
\nabla^2 W^{(1)} =0,\qquad    R_{\alpha \beta}^{\cB(1)} =(D-4)  \nabla_\alpha \nabla_\beta  W^{(1)}   -2 \nabla_\alpha \nabla_\beta\phi^{(1)}        -\ft1{16} \hat \alpha  R_{GB}^{2} h^\cB_{\alpha\beta},
\label{eq:linearEOM1}
\ee
while the one of the  dilaton  leads to 
\be
\nabla^2 \phi^{(1)}=  \ft{\hat \alpha}{16} R^{2}_{GB}.
\label{eq:linearEOM2}
\ee
In the previous expressions  $\nabla$ is the Levi-Civita connection compatible with the metric on the unperturbed bordism $h_{\alpha\beta} ^\cB$, and it should be understood that the Gauss-Bonnet term, $R^2_{GB}$, is evaluated on the background metric of $\cB_{\bf II}|_{\alpha=0}$. 

Note that the radius $\cR^{-1}$ of the $S^{D-4}$ component of the geometry is absent from these equations, since it appears quadratically in \eqref{eq:ricciCurvatureAnsatz}, and therefore it becomes relevant only to second order in perturbation theory. To understand this point, first recall that the $\cR$ represents the bubble radius at the time of nucleation which (as we prove below) it is controlled  by the length-scale $\ell_{\text{ssb}}\sim \cR$. Since we are considering  features of the inner-bubble region of the bordism $\cB_{\bf II}$ with natural scales  $\ell_{\text{kk}} \ll \ell_{\text{ssb}}$, the curvature of the $S^{D-4}$ is comparatively very small, $\cR^{-1} \ll \ell_{\text{kk}}$, and thus it can be consistently neglected in the perturbative analysis\footnote{The authors thank J. J. Blanco-Pillado for a  discussion on this point. }.

\subsubsection{Decoupling of the  zero-modes}
\label{sec:zeroModeDec}

In the present Section we will rewrite and simplify the previous set of linearised equations, and we will also discuss the decoupling of the zero-modes (massless deformations of the bordism $\cB_4$) from the non-supersymmetric deformation of the action. This decoupling of the zero-modes is essential for our construction, as it is a requirement for   the existence of solutions to the linearised equations.

To begin the analysis, let us first consider   
the perturbation of the warp factor $W^{(1)}$ on the sphere $S^{D-4}$. The perturbation  $W^{(1)}$  obeys a Laplace equation on the  manifold $\cB_{\bf II}$, whose   boundary at infinity  is topologically a  three-torus quotient,  $\pd \cB_{\bf II} \cong T^3/\Gamma$ (see sec. \ref{sec:semiflatLayer}). As discussed in Section \ref{sec:appScheme},  in order to be able to glue the perturbed inner-bubble geometry to the outer-bubble region, we need the KK modes  associated to the  $\pd \cB_{\bf II}$  directions to decay far from the bubble core.  Since this implies that $W^{(1)}$ should be a constant on  $\pd \cB_{\bf II}$, then  the only non-singular  solutions that we can find to the Laplace equation  are those where $W^{(1)}$ is a constant on $\cB_{\bf II}$. Finally, this constant  can always be set  to zero without loss of generality,  $W^{(1)} =0$, as it can be absorbed with a redefinition of  the  $S ^{D-4}$ radius,  $\cR$.

Next,  to rewrite the equations for the metric perturbations on the manifold  $\cB_{\bf II}$, it is convenient to decompose the perturbation $\gamma_{\alpha\beta}$ into  its trace, which we denote for later convenience by  $8 \varphi^{(1)}$, and its traceless part $\bar \gamma_{\alpha\beta}$. Furthermore, we will fix partially the spacetime gauge imposing the traceless part to be transverse. That is,
\be
\gamma_{\alpha\beta}= \bar \gamma_{\alpha \beta} + 2 \varphi^{(1)} \, h_{\alpha\beta}^\cB, \qquad\text{with}\qquad  \nabla^\alpha \bar \gamma_{\alpha \beta}=0.
\label{eq:TTTdecomposition}
\ee
Comparing this  with  the local metric  ansatz \eqref{eq:HyperkahlerMetric} on the inner-bubble region, it is immediate to associate $\varphi^{(1)}$ with he first order variation of an overall volume factor $\rme^{2 \varphi}$. Thus $\varphi$ can be regarded as a volume modulus with  constant background value $\varphi_0$,  which becomes   spacetime dependent to first order in $\alpha$.
In this gauge, the equations  \eqref{eq:linearEOM1} for the trace  and the transverse traceless components of the  metric perturbation  are simply 
\be
\nabla^2 \Big(\phi^{(1)}  - 3 \varphi^{(1)} \Big) =  - \ft{\hat \alpha}{8} R^2_{GB},
\label{linearGamma2}
\ee
and 
\be
 \nabla^2 \bar \gamma_{\alpha \beta} - 2R^\cB_{\lambda \alpha \lambda' \beta } \bar \gamma^{\lambda \lambda'}=4 \left(\nabla_\alpha \nabla_\beta - \ft14 h_{\alpha \beta}^\cB \nabla^2 \right)\left(
   \phi^{(1)} - \varphi^{(1)} \right),
\label{eq:TTgammaEOM}
\ee
where we have already set $W^{(1)}=0$, and the  Ricci identity has been used to rewrite the left hand side of the last equation (see e.g. \cite{Wald:1984rg}).

The equation for the volume modulus  \eqref{linearGamma2}, when combined with   with the dilaton equation \eqref{eq:linearEOM2}, implies that the difference $\phi^{(1)}-\varphi^{(1)}$ satisfies the Laplace equation on  $\cB_{\bf II}$ and  thus, as in the case of the perturbation $W^{(1)}$, the boundary conditions on $\pd \cB_{\bf II}$ require this combination to be a constant. This constant can be absorbed  by shifting the background values of $\varphi$ and $\phi$, so we find that
the first order perturbations of the dilaton and the volume modulus satisfy the relation $\varphi^{(1)} = \phi^{(1)}$.

Now, if the previous relation is substituted in \eqref{eq:TTgammaEOM}, it can be seen that 
 its right hand side vanishes.  The resulting  equation is the well known \emph{Lichnerowicz equation}, 
\be
   \nabla^2 \bar \gamma_{\alpha \beta} - 2R_{\lambda \alpha \lambda' \beta }^{\cB} \bar \gamma^{\lambda \lambda'}=   0,
    \ee
 whose solutions are  massless deformations of the background geometry, i.e.  moduli of $\cB_4$ \cite{Candelas:1990pi}.  Therefore  $\bar \gamma_{\alpha\beta}$ must be a zero-mode of the inner-bubble background geometry, which can again be absorbed with a redefinition of the background, and thus without loss of generality  we can set  $\bar \gamma_{\alpha\beta}=0$.   We conclude that the  first order variation of the geometry  of $\cB_{\bf II}$ is completely specified by the trace part of the perturbation, $\varphi^{(1)}(y)$, which describes  a \emph{warping} with dependence only on the coordinates on the bordism. 
  
Finally, collecting all of these results we can see  that the resolution of the equations for the first order perturbations,  \eqref{eq:linearEOM1} and \eqref{eq:linearEOM2}, amounts to solving the following Poisson equation on the background geometry of $\cB_{\bf II}$
\be
 \nabla^2 \varphi^{(1)} =\ft{\hat \alpha}{16} R_{GB}^2.
\label{eq:poissonGB}
\ee
By construction, the source term $R_{GB}^2$ is finite everywhere and is  non-zero only in a compact region of the manifold $\cB_4$ (its integral over $\cB_4$ is $32 \pi^2 N$). Then, since we are not  imposing any specific  boundary conditions on the perturbations for now (see Section \ref{sec:outerBON} for clarification),  there are no impediments to finding smooth solutions for \eqref{eq:poissonGB}. 

As anticipated above, all zero-modes of the background geometry   except the volume modulus  \emph{decouple} from the non-supersymmetric deformation, and regarding $\varphi^{(1)}$ there is no run-away potential which could prevent us from constructing the instanton solution.

For later reference we also write here the relation between the perturbation $\varphi^{(1)}$ and the deformation of the Ricci scalar on $\cB_{\bf II}$
\be
R_{\cB}^{(1)} =    - 6 \nabla^2 \varphi^{(1)}.
\label{eq:pertRicciS}
\ee
Thus, the  Gauss-Bonnet term induces a metric deformation leading to a negative scalar curvature on $\cB_4$ ($\alpha>0$), for which there is no topological obstruction. This negative scalar curvature localized at the degenerations of the elliptic fibration is the key physical ingredient that allows us to evade the PET.

\subsubsection{Layer (II.): warped inner-bubble region.}
\label{sec:warpedGeometry}

We will now discuss the warped geometries described by the Poisson equation  \eqref{eq:poissonGB},  when the background is given by the Calabi-Yau geometries discussed in Section \ref{sec:innerBackground}.  We begin with a  characterisation of the perturbation at the  BON core {\bf II.c}, and proceed moving across the layers {\bf II.b} and {\bf II.a} towards the larger scale structure of the instanton solution. For simplicity we will look for solutions where the warp factor has no dependence on the $\psi$ coordinate, i.e. we neglect the associated KK modes, so that  \eqref{eq:poissonGB} on $\cB_{\bf II}$  reduces to the standard  Poisson equation in flat space
\be
\nabla_{(3)}^2 \varphi^{(1)}=  \ft{\hat \alpha}{16} \rme^{2 \varphi_0}\,(V R_{GB}^{2}),
\label{eq:poissonRedKK}
\ee
where $\nabla^2_{(3)}$ is the flat-space Laplace operator in three dimensions, and  the product of $V R_{GB}^2$ is evaluated on the background.

\paragraph{Bubble core ({\bf II.c}).}
As we discussed in Section \ref{sec:analyticLimits}, in this regime the geometry is given by the line element  \eqref{eq:taubNUT} describing a single KK monopole, and the associated Gauss-Bonnet term is  \eqref{eq:singleKKGB}. Then, the warped geometries are described by the solutions to the equation 
\be
 \nabla^2_{(3)} \varphi^{(1)}|_{\text{core}} =  6 \pi \,  \frac{ \rme^{-2 \varphi_0} R_0^3  }{ R (R_0 + R)^5} \, \hat \alpha.
\label{eq:KKMpoisson}
\ee
As the background spacetime of the KK monopole is spherically symmetric we can find solutions where the warp factor depends only on the radial coordinate  $\varphi^{(1)}= \varphi^{(1)}(R)$    \cite{Banks:1988rj}
\be
\varphi^{(1)}|_{\text{core}}= \frac{\pi \rme^{-2 \varphi_0}}{2} \,\hat \alpha\, \frac{(R^3 +2R^2 R_0 -2 R^3_0) R}{ (R+R_0)^3 R_0} + 3 \pi \rme^{-2 \varphi_0} \, \hat \alpha\, \log 12.
\ee
Here one of the integration constants has been fixed requiring regularity at the origin, and the value of the second one has been chosen for later convenience.  
Recall that, as shown in Section \ref{sec:analyticLimits}, the KK monopole radius \eqref{eq:KKMradius} is very small for the class of background geometries that we consider $R_0 \sim |\log(\ell_{\text{fibre}} / \ell_{\text{kk}})|^{-1}\ll1$, (see eq.   \eqref{eq:scalingCheck}).  Then, provided we are interested in the behaviour of the warp factor at radii $R\gg R_0$, the source term  in \eqref{eq:KKMpoisson} can be well approximated by a  Dirac delta at the origin with weight $2 \pi^2 \rme^{-2 \varphi_0}\hat \alpha$,
  while the volume modulus  $\varphi^{(1)}$  takes the Newtonian form
\be
\lim_{R/R_0 \to \infty} \varphi^{(1)}|_{\text{core}}\to\frac{\pi\, \rme^{-2 \varphi_0}}{2}\left(   6 \, \log 12 +  \frac{1}{R_0}  - \frac{1}{R}\right)\, \hat \alpha +\cO(\hat \alpha |\log \epsilon|^{-4}),
\label{eq:pertCoreAssymp}
\ee
where $\rme^{2 \varphi_0} R_0 \sim \cO(\epsilon^0)$, and $\epsilon=\ell_{\text{fibre}} / \ell_{\text{kk}}\ll1$ is the  parameter characterising the degenerate/LV limit of the background. 
Note that this approximation is valid even for moderately small values of the radius $R\lesssim 1$, since   we have  $R_0/R = \cO(|\log \epsilon|^{-1}) \ll1$.

\paragraph{Layer ({\bf II.b}): Hyperk\"ahler regime.} 
Let now us move further way from the KK monopole location, and consider the Poisson equation   \eqref{eq:poissonRedKK} in the whole hyperk\"ahler region which characterises the  neighbourhood of one of the degeneration points, $\cB_{\text{hk}}^a$. 

For the  solutions to the  equation  \eqref{eq:poissonRedKK} to be consistent with the identifications of the background we  should   impose the periodic boundary conditions on the warp factor along the direction $y^1$, that is $\varphi^{(1)}(y^1) = \varphi^{(1)}(y^1+1)$.  Then, we can formally write the solution to  \eqref{eq:poissonRedKK}  using the associated Green's function 
\be
G(w,y^1,w',y^1{}') = \frac{1}{2\pi} \log(|w-w'|)- \frac{1}{\pi}\sum_{m>0 } \cos(2 \pi m (y^1-y^1{}')) K_0(2 \pi m |w-w'|),
\ee
which involves the modified Bessel function  of the second kind $K_0$. Here $w$ represents collectively the coordinates $w = (y^2, y^3)$ on  $\cB_{\text{hk}}^a$, and   $|w| = \rho\le \rho^*$ is the radial coordinate used in Section \ref{sec:regularisation}.
The formal expression for the volume modulus $\varphi^{(1)}$ is
\bea
 \varphi^{(1)}(w,y^1)|_{\text{hk}} &=& \frac{\hat \alpha}{16\pi}  \int_{\cB^{a}_{\text{hk}}} d^2 w' \Big[\ft12  \log\left(|w-w'|\right) (V R_{GB}^2)_0(|w'|) \nonumber\\
&&- \sum_{m>0 } \cos(2 \pi m y^1) K_0(2 \pi m |w-w'|)(V R_{GB}^2)_m(|w'|)\Big] + \varphi^{(1)}_{\text{h}},
\label{eq:greenGamma1}
\eea
where we have expressed the result in  terms of the  Fourier coefficients  of the  source term in \eqref{eq:poissonRedKK}
\be
(V R_{GB})(|w'|)_m = 2 \rme^{2 \varphi_0} \int_0^1 dy^1 \cos(2 \pi m y^1) V R_{GB}^{2}(|w'|,y^1),
\label{eq:RGBkk}
\ee
and $\varphi^{(1)}_{\text{h}}$ is a harmonic function to be determined. 
The  perturbed spacetime $\cB_{\text{hk}}^a$ has to be glued with the semi-flat regime $\cB_{\text{sf}}^*$ across its boundary at $|w|=\rho^*\gg1$, and then we need  to characterise the asymptotic behaviour of the warp factor in the limit $|w|\to \infty$. As we discussed in the previous paragraph, the source term of the Poisson equation \eqref{eq:poissonRedKK} is very localised within a region $|w| \lesssim R_0 \sim \cO(|\log \epsilon|^{-1})$, and it becomes a Dirac delta function located at the KK monopole  centre in the strict degenerate/LV limit of the background,  $\epsilon\to 0$. Therefore, near the boundary $|w|=\rho_*\gg R_0$ the volume modulus  is well approximated by 
\bea
 \varphi^{(1)}(\rho, y^1)|_{\text{hk}}&\approx& 2\pi \hat \alpha \, \rme^{-2 \varphi_0}\left( \ft12 \log(\rho)-\sum_{m>0 } \cos(2 \pi m y^1) K_0(2 \pi m \rho)\right) +\varphi^{(1)}_{\text{h}} \nonumber \\
&\approx& \pi  \, \rme^{-2 \varphi_0}\, \hat \alpha  \log(12^3 \rho/\rho_0) -  \pi  \, \rme^{-2 \varphi_0} \, \hat \alpha \cos(2 \pi y^1) \frac{\rme^{-2 \pi \, \rho}}{\sqrt{\rho}} + \ldots  \; ,
\label{eq:varphi1Assym}
\eea
where in the second step we have fixed an integration constant  comparing the previous expression in the limit $\rho\to0$ with \eqref{eq:pertCoreAssymp}, and using the definitions \eqref{eq:KKMradius} and \eqref{eq:gluingParams}.  Here we can see that, similarly
 to the background geometry, far from the KK monopoles the  corrections to the modulus  $\varphi^{(1)}$ from massive Kaluza-Klein modes are  exponentially suppressed, and only the zero-mode ($m=0$) remains active. Notice also that we have set to a constant the harmonic function   $ \varphi^{(1)}_{\text{h}}$  appearing in \eqref{eq:greenGamma1}
 \be
\varphi^{(1)}_{\text{h}} = -\pi  \, \rme^{-2 \varphi_0}\, \hat \alpha  \log(\rho_0/12^3).
\label{eq:harmonicFunction}
 \ee
As we will see next,  this function is determined by the boundary conditions for $\varphi^{(1)}$ at $|w|=\rho_*$, which in the present case are given by the gluing conditions between the hyperk\"ahler  and the semi-flat regimes.

\paragraph{Layer ({\bf II.a}): semi-flat regime.} We will now leave the neighbourhood of the degeneration points and discuss the behaviour of $\varphi^{(1)}$ in the  semi-flat region of the inner-bubble  geometry, $\cB_{\text{sf}}^*$.  Using  the  metric given by   \eqref{eq:semiFlatMetric} and \eqref{eq:T3ZNsol} as a background, we  look for solutions to the Poisson equation \eqref{eq:poissonGB} requiring that  $\varphi^{(1)}$ has no dependence on $y^1$, that is, we neglect entirely the KK modes associated to the $T^2$ fibre.  Note that this ansatz   avoids a possible conflict with  the identifications \eqref{eq:identifications} on the gluing  hypersurfaces \eqref{eq:semiflatPatches} within $\cB^*_{\text{sf}}$. Actually,  those diffeomorphisms leave invariant the function $\varphi^{(1)} =\varphi^{(1)}(z)$, what ensures that the warp factor is globally defined over the whole semi-flat region. Substituting this  ansatz  into the equation \eqref{eq:poissonGB} we find
\be
 \nabla_{(2)}^2 \varphi^{(1)}|_{\text{sf}} = \ft{\hat \alpha}{16} \rme^{2 \varphi_0} |F|^2 \Im(\tau) R_{GB}^2|_{\text{sf}},
\label{eq:sfPoisson}
\ee
where the operator $\nabla_{(2)}^2$ is the two-dimensional  Laplacian  in flat space, and $R_{GB}|_{\text{sf}}$  is the Gauss-Bonnet invariant  of the metric      \eqref{eq:semiFlatMetric}.

 Let us now estimate the scaling of the source term   with the degenerate/LV parameter of the background, $\epsilon$. The maximum value for the right hand side of \eqref{eq:sfPoisson} is attained near the degeneration points, i.e. at the boundaries $|z-z_a|=r^*$,  where spacetime curvature on $\cB_{\text{sf}}^*$ is largest. Using the local form of the semi-flat geometry \eqref{eq:degeneratingMetric}, \eqref{eq:FnearDeg} and \eqref{eq:imTauDeg}, together with the $\epsilon$ scaling relations \eqref{eq:scaling} and \eqref{eq:scalingCheck}, we obtain
\be
0\le \[ \ft{\hat \alpha}{16} \rme^{2 \varphi_0} |F|^2 \Im(\tau) R_{GB}^2\]_{\text{max}} = \frac{24 \pi \hat \alpha}{\ell_{sf}^2 r_*^4 \log(r_0/r_*)^3} = \cO(\epsilon^2),
\ee  
where $\ell_{\text{sf}}$ was defined in \eqref{eq:lsfDef}. Since we are assuming $\hat \epsilon \sim \epsilon$, we see that the source term of \eqref{eq:sfPoisson}  has to be neglected everywhere  when we work to first order $\hat \epsilon$, and therefore   $\varphi^{(1)}|_{\text{sf}}$ needs to be a harmonic function on the semi-flat layer.

To determine  the function $\varphi^{(1)}|_{\text{sf}}$ we have to provide appropriate boundary conditions. The semi-flat  region $\cB^*_{\text{sf}}$ has $N$ internal boundaries around the degeneration points $z_a$, defined by $|z -z_a|= r^*$, and an external boundary at  $|z| = r_{\text{max}}\gg1$ where the semi-flat region connects with the outer-bubble regime. The boundary conditions that we are seeking will be provided by the gluing conditions \eqref{eq:matchingConditions} between the semi-flat ({\bf II.a}) and hyperk\"ahler ({\bf II.b}) spacetimes at the $N$ internal boundaries, and by the matching of the semi-flat and the outer-bubble region ({\bf I.}) at the external one.   Let consider now   the gluing between the layers ({\bf II.b}) and ({\bf II.a}) of the inner-bubble regime, and leave the matching with the outer region for Section \ref{sec:outerBON}. 
According to the discussion in Section \ref{sec:regularisation}, 
the gluing conditions \eqref{eq:innerBONmatching} require that the first order variations (with respect to  $\hat \epsilon$) of the first,  $s_{ab}^{(1)}$, and second fundamental forms, $K_{ab}^{(1)}$, 
of the matching boundary $\pd \cB_{\text{hk}}^a$ should agree with those of $\pd \cB^*_{\text{sf}}$. Leaving the gluing diffeomorphism \eqref{eq:zeroGluing} unperturbed  
we find the conditions
\begin{align*}
0=&\Delta s^{(\alpha)} =\Phi^*(s_{\text{hk}}^{(1)}) - s_{\text{sf}}^{(1)} =\; 2 \left( \Phi^* (\varphi^{(1)}_{\text{hk}}) - \varphi^{(1)}_{\text{sf}}\right) s^{(0)}_{m=0} \\
0=&\Delta K^{(\alpha)} =  \Phi^*(K_{\text{hk}}^{(1)}) - K_{\text{sf}}^{(1)}=\; \left( \Phi^* (\varphi^{(1)}_{\text{hk}}) - \varphi^{(1)}_{\text{sf}}\right) K^{(0)}_{m=0} + \left( \Phi^* (\nabla_n \varphi^{(1)}_{\text{hk}}) - \nabla_n \varphi^{(1)}_{\text{sf}}\right) s^{(0)}_{m=0}.
\end{align*}
Here the variations $s^{(1)}$ and $K^{(1)}$ have been computed using the results in \cite{Mars:2005ca} together with the form of the first order  metric perturbation \eqref{eq:TTTdecomposition}. The quantities $s^{(0)}_{m=0}$ and $K^{(0)}_{m=0}$ are the KK zero-modes of the  leading order fundamental forms, which we already  proved to satisfy the matching conditions  in Section \ref{sec:regularisation}. The previous equations are equivalent to imposing the continuity of the volume modulus $\varphi^{(1)}$ and its derivative $\nabla_n \varphi^{(1)}$ across the matching hypersurface, where $n$ is the associated unit normal vector\footnote{The unit normal hyperk\"ahler boundary is given by $n^\rho|_{\text{hk}} = \sqrt{2 \pi/\log(\rho_0/\rho_*)} \, \rme^{-\varphi_0}$, and the one corresponding to the semi-flat boundary is $n^r|_{\text{sf}} = (\ell_{\text{fibre}}/\ell_{\text{sf}}) \sqrt{2 \pi/\log(r_0/r^*)} \, \rme^{-\varphi_0}$.}.  Let us consider the following tentative solution which is harmonic in $\pd \cB^*_{\text{sf}}$ 
\be
\varphi^{(1)}|_{\text{sf}}(z)  = \pi \rme^{-2 \varphi_0} \hat \alpha\, \sum_{a}\log(|z-z_a|/z_0),
\label{eq:semiflatWarping}
\ee
and satisfies  $\varphi^{(1)}|_{\text{sf}}(0) =0$.   At the boundary $|z-z_a| = r^*$ near the  degeneration point $z_a$ this function takes the approximate form
\be
|z-z_a| = r^*:\qquad  \varphi^{(1)}|_{\text{sf}}  \approx  \pi \rme^{-2 \varphi_0} \hat \alpha\, \log(N r_*/z_0)  + \cO(\epsilon^q),
\ee
 what can be compared   with \eqref{eq:varphi1Assym} using the gluing diffeomorphism \eqref{eq:zeroGluing}. We find that the zero-mode KK components of  $\varphi^{(1)}|_{\text{hk}}$ and $\nabla_n \varphi^{(1)}|_{\text{hk}}$  match properly  with the function $\varphi^{(1)}|_{\text{sf}}$ and its derivative to leading order in $\epsilon$ (recall that $q\ge2$). Regarding the   
massive KK modes in \eqref{eq:varphi1Assym}, their contributions  are exponentially suppressed by   $\cO(\rme^{2 \pi m r_*/\epsilon})$ factors, and thus  they can  be consistently   neglected.  Therefore, we conclude that the function \eqref{eq:semiflatWarping} is an appropriate  extension of the volume modulus $\varphi^{(1)}$ in the semi-flat layer\footnote{This also proves the consistency of  setting the undetermined harmonic function  $\varphi^{(1)}_h$ of \eqref{eq:greenGamma1} to a constant in \eqref{eq:varphi1Assym}.}.

Summarising, at this point we have  successfully constructed the first order deformation in $\alpha$ of the inner-bubble  spacetime, that is the layer $\cB_{\bf II}$.  The perturbation corresponds to a warping of the background Calabi-Yau geometry on $\cB_{\bf II}$, with  the warp factor  $\exp(2\hat \epsilon\,   \varphi^{(1)})$  given by the expressions \eqref{eq:greenGamma1} and  \eqref{eq:semiflatWarping} on the hyperk\"ahler  ({\bf II.b})  and semi-flat layers ({\bf II.a}) respectively. Finally, the perturbed spacetime $\cB_{\bf II}$  that we just   obtained needs to be glued with the outer-bubble regime across its boundary at $|z|=r_{\text{max}}\gg1$, and thus to finish  this Section we will  analyse the behaviour of the warp factor in the limit  $|z|\to \infty$.  From \eqref{eq:semiflatWarping} it is straightforward  to obtain the  asymptotic  form of $\varphi^{(1)}$  
\be
\lim_{|z| \to \infty}\varphi^{(1)}|_{\bf II} \to     \pi \rme^{-2 \varphi_0} N\, \hat \alpha\, \log(r/z_0)  
-  \pi \rme^{-2 \varphi_0}\, \hat \alpha\, \cos(N\theta) \frac{z_0^N}{r^N} 
+ \ldots\; ,
\label{eq:assympGamma}
\ee
while the derivative along the normal direction to the boundary $\pd \cB_{\bf II}$ reads 
\be
\nabla_n \varphi^{(1)}|_{\pd\cB_{\bf II}} 
= \frac{2 \pi^2 N
}{\cV_{T^3}|_{\pd\cB_{\bf II}}}  \hat \alpha, \qquad \text{with}\qquad   \cV_{T^3}|_{\pd\cB_{\bf II}} = 2 \pi R_{\text{kk}|\theta} \, \rme^{2 \varphi_0}\,   r^{(1-\frac{N}{12})}_{\text{max}},
\label{eq:normalGammaInf}
\ee
where   $n =R_{\text{kk}|\theta}^{-1}\, \pd_{r}$ is the normalised normal vector to the boundary  $\pd\cB_{\bf II}$, with $R_{\text{kk}|\theta}$  defined in \eqref{eq:StringAsymptotics1}, and $z = r \rme^{\rmi \theta}$. As a consistency check, note that this last expression is consistent  with the result of applying Gauss's theorem to the Poisson equation \eqref{eq:poissonGB}.  
This follows from the fact that the source term in \eqref{eq:poissonGB} is proportional the Gauss-Bonnet invariant, and then its  integral over $\cB_{\bf II}$   gives   $2 \pi^2 \, \chi(\cB_{\bf II})$, where Euler characteristic of  $\cB_{\bf II}$ is precisely  the number of degenerations of the fibre, $\chi(\cB_{\bf II})=N$.

\subsection{Layer (I.): the outer-bubble regime.}
\label{sec:outerBON}

In the previous Section we have constructed the inner region of the BON instanton,   that is,  a spacetime $\cM_{\bf II}\cong S^{D-4}\times \cB_{\bf II}$ which solves  to the equations of motion  \eqref{EinsteinEOM} and  \eqref{dilatonHeom}  for small values of perturbation parameter $\alpha$, and where $\cB_{\bf II}$ has  the topology described in Section \ref{sec:detailedTopology}. To make this analysis tractable we have assumed the compact space $\cC\cong T^3/\Gamma$ to be in a degenerate/LV limit.   We will now  discuss  the outermost layer of the bubble geometry, with spacetime denoted by $\cM_{\bf I}$, which interpolates between the inner-bubble region and the asymptotic (euclidean) vacuum $\mathbb{R}^{D-3} \times  T^3/\Gamma$, \eqref{vacuum}. 

As we argued in Section \ref{sec:appScheme}, the massive KK modes of the compact space are expected to be  suppressed in this regime, and thus  we will describe this spacetime region with a solution to the Euler-Lagrange equations where the KK modes are neglected entirely. We will prove the consistency of this approach showing  that the gluing conditions \eqref{eq:matchingConditions} between inner-bubble and outer-bubble regions can still be satisfied  consistently with our approximation scheme.  Below we shall see that the behaviour the spacetime geometry in the outer layer is inherently non-linear due to the boundary conditions \eqref{eq:BONbcs1} of the metric BON ansatz \eqref{gralBONansatz}, and therefore we will have to solve a non-linear set of equations of motion in this layer. Still, as a result of  the $\SO(D-4)$ symmetry of the BON ansatz, and provided we neglect the KK modes on the compact space,  the Euler-Lagrange equations  reduce to a coupled system  ODE's that can be easily  solved using a combination of analytical and numerical methods.

\paragraph{Metric ansatz and Euler-Lagrange equations.} In order to solve the equations of motion we will use the following ansatz consistent with   neglecting  the KK modes on the compact space $T^3/\Gamma$ 
\be
ds^2|_{\cM_{\text{I}}} =W^2(\rho) \cR^2\, d\Omega_{D-4}^2  +d\rho^2 +C^2(\rho)  \,d\theta^2+ \rme^{2 (\varphi-\varphi_{\infty})} d s_{T^2}^2, 
\label{eq:outerAnsatz}
\ee
where $\varphi|_{\bf I} = \varphi|_{\bf I}(\rho)$, and we also take the dilaton to be a function of $\rho$ only,   $\phi|_{\bf I} = \phi|_{\bf I} (\rho)$. In the previous expression $d\Omega_{D-4}^2$ is the line element on the unit $S^{D-4}$ sphere,  $d s_{T^2}^2$ represents the metric on the $T^2$ fibre of the compact space $T^3/\Gamma$ at the vacuum  \eqref{vacuum}, and $\theta$ parametrises the remaining cycle of the three-torus whose  asymptotic radius we denote by $R_{\text{kk}}$. Therefore, the metric and dilaton fields  should approach the following asymptotic configuration  for the BON to meet the boundary conditions \eqref{eq:BONbcs1}
\be
\rho \to \infty: \qquad W( \rho)\to \rho/\cR, \qquad C(\rho)\to R_{\text{kk}},\qquad \varphi(\rho) \to \varphi_\infty, \qquad \text{and}\qquad \phi (\rho)\to\phi_\infty,
\label{eq:outermostBC}
\ee
In Section \ref{sec:warpedGeometry} we showed that when working  in the degenerate/LV limit of the compact space $T^3/\Gamma$,  the contribution of the Gauss-Bonnet term from the curvature on $\cB_4$ can be neglected almost everywhere except on the bubble's multiple cores. Similarly it is easy to check that the contribution to $R_{GB}^2$ from the curvature on the $S^{D-4}$ sphere scales as $\cR^{-4} \sim (\ell_{\text{kk}}/\ell_{\text{ssb}})^{4}\ll1$, and thus it can also be neglected in the equations of motion 
describing the outer-bubble regime.
 We will see below that  this is a very good approximation for the solutions that we find. 
 
 With the ansatz \eqref{eq:outerAnsatz}, and discarding the contribution form the Gauss-Bonnet term, the Euler-Lagrange equation for the dilaton reduces to 
\be
\phi'' + \Big((D-4) \frac{W'}{W}  + \frac{C'}{C}+2\varphi' \Big) \phi'- 2 \phi'{}^2 =0,
\label{eq:dilatonOutermostEqs}
\ee
while the equations  for the  metric profile functions  $W$, $C$  and $\varphi$ are\footnote{The $\rho-\rho$ component of  Einsteins equations also leads to a constraint on the initial conditions at $\rho=\rho_*$ which  can be seen to be  satisfied up to $\cO(\alpha^2)$ corrections, in consistency with the analysis in Section \ref{sec:nonSusyDeformation}.}
\bea
W''   +W'\,\Big(\frac{C'}{C} +2 \varphi' -2 \phi'\Big)- (D-5)W^{-1}(\cR^{-2}- W'{}^2)&=& 0,\nonumber \\
 C''  + (D-4)\frac{ W'}{W} C'+2 C'\varphi' -2 C' \phi' &=&0,  \nonumber \\
\varphi'' + (D-4)\frac{W'}{W}\varphi' +\frac{C'}{C}\varphi' +2 \varphi'{}^2  -2 \phi' \varphi' &=&0.
\label{eq:G3outermostEqs}
\eea
To solve this system of coupled ODE's  we need to specify the boundary conditions for the profile functions  $W(\rho)$, $C(\rho)$, $\varphi(\rho)$  and the dilaton  $\phi(\rho)$ both at infinity $\rho \to \infty$, where they are given by \eqref{eq:outermostBC}, and at the matching surface with the inner-bubble regime, i.e. $\pd \cM_{\bf I}\cong\pd \cM_{\bf II}$.

\paragraph{Gluing with the inner-bubble region.}   From the analysis in the previous Section we know that  far from the degenerations  the  inner-bubble regime $\cM_{\bf II}$ can be described  to first order in $\alpha$ by a field configuration of the following form
\be 
ds^2|_{\bf II} \to \cR^2 d\Omega^2_{D-4} +
\rme^{2 (\varphi-\varphi_0)} \, (du^2 + \hat C^2(u) d \theta^2) +\rme^{2(\varphi- \varphi_{\infty})} ds_{T^2}^2 \ldots \,, 
\label{eq:assympM2}
\ee
with both the volume modulus $\varphi|_{\bf II} \to \varphi|_{\bf II}(u)$ and the dilaton $\phi|_{\bf II}\to \phi|_{\bf II}(u)$ approaching functions of $u$ only, and where   $\hat C(u)$ can be read from   equations \eqref{eq:asympSF1} and \eqref{eq:asympSF2}.    The dots denote higher order corrections in $\alpha$ and $1/u$, which  include massive KK modes of the compact manifold.  
The modulus $\varphi$ and the dilaton $\phi$, are determined by their  perturbative description, with their background values given respectively by $\varphi_0$ and $\phi_0$, and their first order perturbations in terms of the function $\varphi^{(1)}$ in \eqref{eq:assympGamma}  by $\pd_{\hat \epsilon} \phi|_{\hat \epsilon=0}=\pd_{\hat \epsilon}\varphi|_{\hat \epsilon=0} = \varphi^{(1)}$.

 With this parametrisation of $\cM_{\bf II}$ at hand  we  can define the complete BON spacetime manifold $\cM$ as follows  
\be
 \cM \cong \cM_{\bf I}^* \cup\cM_{\bf II}^*,\qquad \cM_{\bf II}^* \cong \cM_{\bf II}\setminus \overline B_{u_*}(0), \qquad \cM_{\bf I}^* \cong \cM_{\bf I}\setminus B_{\rho_*}(0),
\ee
 where the component spacetimes  $\cM_{\bf II}^*$ and $\cM_{\bf I}^*$ are obtained   from $\cM_{\bf II}$ and $\cM_{\bf I}$ cutting out respectively the regions with $u>u_*$ and $\rho<\rho_*$. To make this manifold connected we need to introduce the gluing diffeomorphism $\Phi: \pd \cM^*_{\bf I}  \rightarrow \pd \cM^*_{\bf II}$ which identifies the boundaries of the component spacetimes $\pd \cM_{\bf II}$ ($u=u_*$) and $\pd \cM_{\bf I}$ ($\rho =\rho_*$). Due to the convenient coordinate choices used in \eqref{eq:outerAnsatz} and \eqref{eq:assympM2}, to first order in $\hat \epsilon$  the diffeomorphism $\Phi$ can be taken to be  the trivial map given by $x^\mu|_{\bf II} = x^\mu|_{\bf I}$ and  $y^{\bar \alpha}|_{\bf II}=y^{\bar \alpha}|_{\bf I}$, where $x^\mu$ and $y^{\bar \alpha}$ parametrise respectively  the $S^{D-4}$ and $T^3/\Gamma$  components of the boundaries $\pd \cM_{\bf II} \cong \pd \cM_{\bf I}\cong S^{D-4} \times T^3/\Gamma$. In the next paragraphs we  discuss the gluing conditions \eqref{eq:matchingConditions} that  guarantee the metric tensor to be continuous, and the absence of shells/branes on the matching hypersurface. As we shall see, those equations together with \eqref{eq:outermostBC} provide the necessary conditions to solve  the boundary value problem associated to the system of equations \eqref{eq:dilatonOutermostEqs} and \eqref{eq:G3outermostEqs}. .

\subsubsection{Bubble decay of the \texorpdfstring{$T^3$}{T3} compactification.}
\label{sec:T3BubbleDecay}

We will now focus on the bubble of nothing solution for the vacuum with compact space $T^3$. In this case we have to set  the number of degenerations  on the inner-bubble geometry to $N=12$, so that the bordism $\cB_{\bf I}$ corresponds to a warped ``half-K3 space'', and we can obtain an appropriate ansatz for the outer-bubble regime setting $C(\rho) = \rme^{\varphi-\varphi_\infty} R_{\text{kk}}$ in \eqref{eq:outerAnsatz}.  As the inner bubble region has been constructed using perturbation theory, we will also have to resort to perturbative matching techniques to glue the inner and outer regimes. It is important to stress that, while 
the boundary conditions at the hypersurface separating the two layers are obtained with perturbative methods,   the evolution in the interior of  $\cM_{\bf I}$ described by \eqref{eq:dilatonOutermostEqs} and \eqref{eq:G3outermostEqs} is fully non-linear. More specifically, due to the asymptotic behaviour $\rho\to \infty$ of the function $W(\rho) \to \rho/\cR$ on \eqref{eq:outerAnsatz}, the outer-bubble geometry cannot be regarded as a small perturbation of the background Calabi-Yau geometry used to construct of the inner-bubble layer.

If we neglect the subleading terms  in \eqref{eq:assympM2},  we find that imposing  the metric tensor to be continuous across the matching surface  to first order in $\alpha = \hat \epsilon \hat \alpha$ (left equation in \eqref{eq:matchingConditions})  is equivalent to the conditions (see eqs. \eqref{eq:expansions} and \eqref{eq:zeroOrderR})
\be
  W|_{\bf I} (\rho_*)=1+ \cO(\alpha^2), \qquad \varphi|_{\bf I}(\rho_*) =
 \varphi|_{\bf II}(u_*) +\cO(\alpha^2) ,
\label{G3matching1BIS}
\ee
Furthermore, after identifying  the unit normals  $n|_{\bf II} =\rme^{-\varphi(u_*)} \pd_u$ and $n|_{\bf I} =  \pd_\rho$ to the respective boundaries $\pd\cM_{\bf II}$ and $\pd\cM_{\bf I}$, and using \eqref{eq:normalGammaInf}
 we find that the requirement of having no shell on the matching  boundary is equivalent to 
\be
W'|_{\bf I} (\rho_*)  =\cO(\alpha^2),\qquad 
\varphi'|_{\bf I} (\rho_*) =  \nabla_{n} \varphi|_{\bf II}(u_*) =  \frac{24 \pi^2}{\cV_{T^3}^*}\alpha  + \cO(\alpha^2),
\label{T3matching}
\ee
to first order in $\alpha$, where $\cV_{T^3}^*=\cV_{T^3} \, \rme^{3 (\varphi(u_*) -\varphi_\infty)}$ is the  three-torus volume at the matching point $u=u_*$.
These conditions have to be supplemented with the requirement that the dilaton is smooth, which implies 
\be
 \phi|_{\bf I} (\rho_*)= \phi|_{\bf II}(u_*), \qquad  \phi'(\rho_*)|_{\bf I} = \nabla_{n} \phi|_{\bf II}(u_*) =  \frac{24 \pi^2}{\cV_{T^3}^*}\alpha +  \cO(\alpha^2).
\label{dilatonT3matching}
\ee
To derive the second equation we used that  the first perturbation of the dilaton and that of $\varphi$ are equal, together with \eqref{eq:normalGammaInf}. These initial conditions are consistent with the ansatz $\phi=\varphi - \varphi_\infty+\phi_{\infty}$, which reduces the set of equations \eqref{eq:dilatonOutermostEqs} and \eqref{eq:G3outermostEqs} to the equivalent system
\be
\varphi'' + (D-4) \frac{W'}{W}  \varphi' +  \varphi'{}^2 =0,\qquad
\frac{W''}{W}    - (D-5)\frac{(\cR^{-2}- W'{}^2)}{W^2} +\frac{W'}{W}\,\varphi'= 0.
\label{eq:reducedSystem}
\ee
\begin{figure}[t]
\centering \hspace{-.5cm}\includegraphics[width=0.55\textwidth]{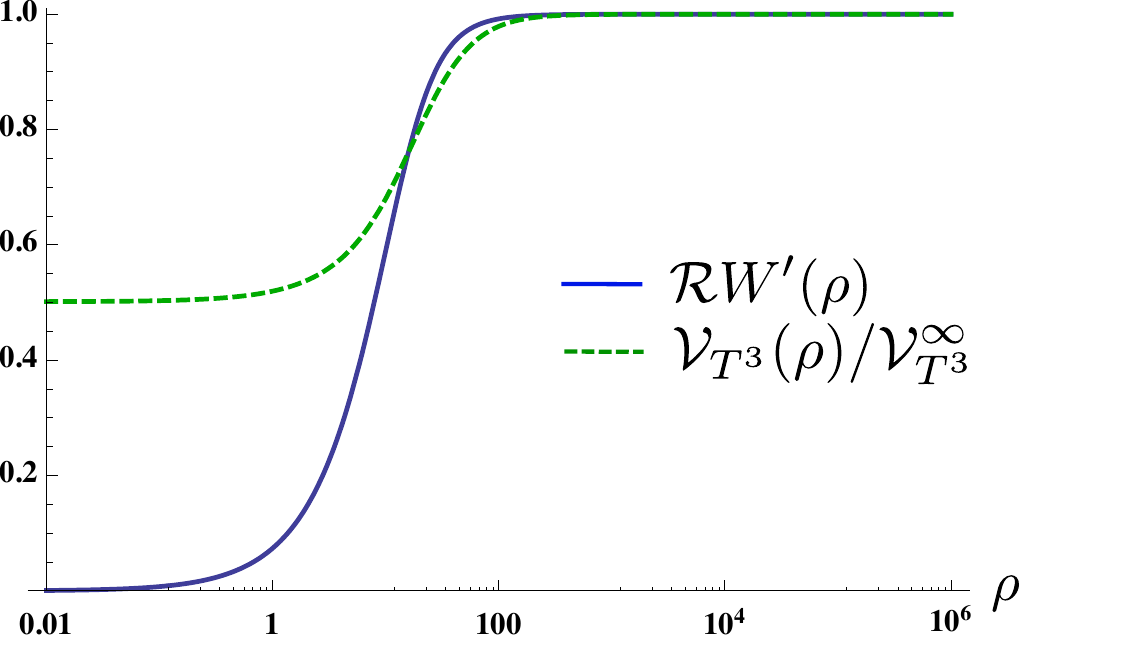} 
\caption{Outer-bubble regime of the $T^3$ BON with deformation parameter $\frac{24 \pi^2}{\cV_{T^3}}\alpha = 0.005$. The solid line is related to   the warp factor on the sphere,  $\cR W'(\rho)$, and the dashed line represents the volume of the $T^3$ compact space, $\cV_{T^3}(\rho)/\cV_{T^3}^\infty$, with the radial coordinate $\rho$ displayed in logarithmic scale. The  bubble nucleation radius is   $\cR = 32.2$, and the three-torus volume at the boundary of the inner-bubble regime is given by  $\cV_{T^3}^*= 0.5 \times \cV_{T^3}^\infty$, where  $\cV_{T^3}^\infty$ is the asymptotic $T^3$ volume, denoted simply by $\cV_{T^3}$ in the text.}   \label{T3BON}
\end{figure} 
Note that the initial value $\varphi(u_*)=\varphi_0+\cO(\alpha)$ and the BON radius $\cR$ are \emph{a priori} free parameters, which can be  varied arbitrarily by changing the background inner-bubble geometry. However, as we shall see, the Euler-Lagrange   equations and the boundary conditions  determine the relation between these two quantities.

 To solve the resulting boundary value problem it is convenient to use the 
the following scaling symmetry satisfied by the equations of motion
\be
\cR \to \lambda^{-1}\, \cR,\qquad  W(\rho) \to   W(\lambda \rho),\qquad \varphi(\rho) \to \varphi(\lambda \rho), \qquad\text{with}\qquad  \lambda \in \mathbb{R}^+,
\label{eq:rescaling1}
\ee
 which also acts  on    the fields derivatives as follows
\be
W'(\rho) \to  \lambda\, W'(\lambda \rho),\qquad \varphi'(\rho) \to \lambda \, \varphi'(\lambda \rho).
\label{eq:rescaling2}
\ee
Therefore, given a solution to the boundary value problem with a specific value for  $\alpha$, for example $\alpha=\mathring \alpha\equiv\cV_{T^3}^*/(24 \pi^2)$,  it is possible to  construct solutions for arbitrary (but small) values of $\alpha$ using that
\be
\cR|_{\alpha}(\rho) =  (\mathring\alpha/\alpha) \, \cR|_{\mathring\alpha} \qquad W|_{\alpha}(\rho) =W|_{\mathring\alpha}(\alpha  \rho/\mathring\alpha) , \qquad  \varphi|_{\alpha}(\rho) = \varphi|_{\mathring\alpha}(\alpha\rho/\mathring \alpha), 
\label{eq:alphaScaling}
\ee
and with $\rho_*(\alpha) = (\mathring \alpha/\alpha) \rho_*(\mathring \alpha)$,
as can be checked comparing \eqref{eq:rescaling2} with \eqref{T3matching} and \eqref{dilatonT3matching}. Without further computations we can already see from \eqref{G3matching1BIS} that the bubble radius behaves as
\be
 \cR = \mathring \cR  \,\rme^{-3\Delta \varphi}  \left(\frac{24 \pi^2}{\cV_{T^3}}\alpha\right)^{-1},
\label{eq:BONradiusScaling}
\ee
where $\mathring \cR$, which is independent of $\alpha$, is a function of $\Delta \varphi$   that needs to be determined. Then, as $\Delta \varphi$ is unaffected by the scaling relation \eqref{eq:rescaling1}, we find that the nucleation radius diverges as we turn off the Gauss-Bonnet coupling, $\alpha \to 0$.  This precisely the behaviour  anticipated in Section \ref{sec:constraint},  when we  discussed  the dynamical constraint.

To find the unknown coefficient in the previous expression we solve numerically the system of equations \eqref{eq:reducedSystem} subject to the boundary conditions \eqref{G3matching1BIS} and \eqref{T3matching}  at $\rho=\rho_*$ (which  we set at $\rho_*=0$), together with \eqref{eq:outermostBC} at $\rho\to \infty$. 
\begin{figure}[t]
\centering \hspace{-.5cm}\includegraphics[width=0.55\textwidth]{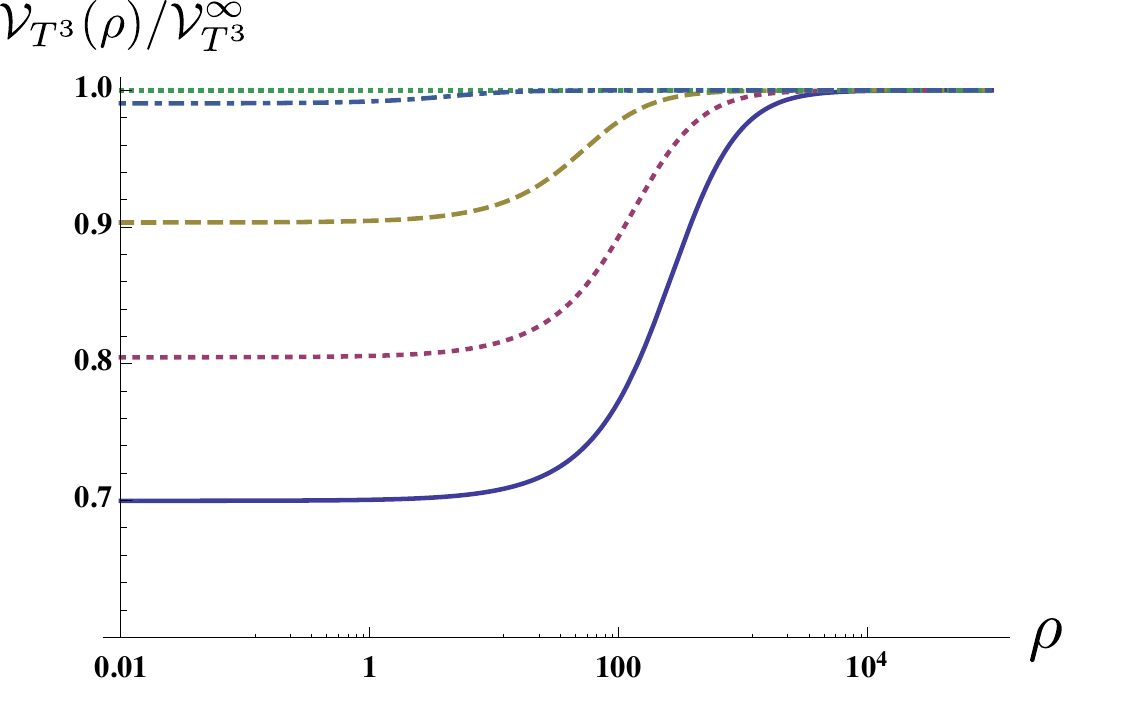} 
\caption{Dependence of the outer-bubble  geometry on the value of the $T^3$ volume  at the matching point, $\cV_{T^3}^* = \cV_{T^3}^\infty \, \rme^{-\Delta \varphi}$. The plot displays the $T^3$ volume $\cV_{T^3}/\cV_{T^3}^\infty$ as a function of the radial coordinate $\rho$ for  a fixed parameter $\frac{24 \pi^2}{\cV_{T^3}}\alpha = 0.005$, and varying values of $\cV_{T^3}^*/\cV_{T^3}^{\infty}= \{0.7,0.8,0.9,0.99\}$ (from bottom to top). The  corresponding radii are respectively  $\cR = \{455.4,\, 231.5,\, 95.1,\, 8.1\}$.
 } 
  \label{T3BONradii}
\end{figure} 
The result of this numerical computation for $D=7$ is shown in Figure \ref{T3BON} for the case $\frac{24 \pi^2}{\cV_{T^3}}\alpha =0.02$ and $\cV_{T^3}^*/\cV_{T^3}= 0.5$, where we find a BON nucleation radius  $\cR = 32.2$.  In this solution we have  estimated the magnitude of the Gauss-Bonnet term, and it is everywhere smaller than $\frac{6 \pi^2}{\cV_{T^3}}\alpha R^2_{GB} \lesssim 10^{-14}$, what justifies neglecting it in the equations of motion.  Qualitatively similar solutions can also be found for other dimensions  $D=6,\ldots,10$. We have also found that  the total growth  of the volume modulus  $\Delta \varphi$,  remains a free parameter of the BON spacetime.  This quantity characterises the ratio between the asymptotic $T^3$ volume $\cV_{T^3}$ and its value at the boundary between the inner and outer regimes $\cV_{T^3}^*$, and our numerical solutions show a one to one correspondence with    the BON nucleation radius (see Fig. \ref{T3BONradii}).

To understand better  the  dependence of the nucleation radius on the quantity $\Delta \varphi$ let us  consider the regime  $\Delta \varphi\to \infty$.  For this purpose, it is convenient to note that the reduced system of equations \eqref{eq:reducedSystem} admits the following first  integral 
\be
W'(\rho) =  \cR^{-1} \left(1- W^{-(D-5)} \right)^{1/2}, \qquad \varphi(\rho) = \varphi_\infty + \log(\cR \, W'(\rho)).
\label{eq:T3limitDinf}
\ee
Imposing the boundary condition $W(0)=1$, we find that in the limit $\rho/\cR \to 0$ the solution of these equations  has the expansion
\be
W|_{\bf I} (\rho) = 1 + \frac{(D-5)\rho^2}{4\cR^2}+\ldots, \qquad \varphi|_{\bf I}(\rho) = \varphi_\infty +\log\left( \frac{(D-5)\rho}{2\cR} \right)+\ldots \;.
\label{eq:T3expansion}
\ee
Then, in order to meet the matching conditions for the volume modulus in \eqref{G3matching1BIS} and \eqref{T3matching}, it is straightforward to see that we need so set the matching boundary at a point $\rho=\rho_*\ll\cR$, given by
\be
\rho_*^{-1} = \frac{24 \pi^2 \alpha}{\cV_{T^3}} \rme^{3 \Delta \varphi}\ll1, \qquad \text{and}\qquad \Delta \varphi  = -\log \left(\frac{(D-5) \rho_*}{2 \cR}\right)\gg1.
\ee
This in turn implies that the  nucleation radius (in string frame) is 
\be
\lim_{\Delta \varphi\to \infty}\cR= \frac{(D-5)\rme^{-2 \Delta \varphi}}{2} \left(\frac{24 \pi^2 \alpha}{\cV_{T^3}}\right)^{-1},
\label{eq:T3RadiusDlarge}
\ee
which satisfies $\cR\gg \rho_*\gg1$, justifying the use of the expansion \eqref{eq:T3expansion}.  In addition we can now also check that  the gluing  condition in  \eqref{G3matching1BIS} and \eqref{T3matching} for the warp factor $W(\rho)$ are also satisfied, and particular we have $W'(\rho_*) = \cO(\cR^{-1} \rme^{-\Delta \varphi})\ll \cO(\cR^{-1})$.

Although in the present  analysis $\Delta \varphi$ (i.e. the ratio  $\cV_{T^3}^*/\cV_{T^3}$) remains a free parameter of the BON geometry, higher order $\cO(\alpha^2)$ corrections will fix this value in general, and possibly also other moduli such as the locations of the degenerations. To illustrate this point in appendix \ref{app:fluxBON} we consider a slightly more complicated model than 
\eqref{eq:action} with additional ingredients which fix the $T^3$ volume at the vacuum. In that scenario we show that both $\Delta \varphi$ and the  BON radius $\cR$ must also  attain a specific value on the BON configuration which is determined by the higher order corrections.

To finish this discussion, it only remains to justify that the active KK modes of the semi-flat layer  are consistently neglected in \eqref{eq:assympM2}. In order to find the magnitude of the leading corrections we need an estimate the position  $u=u_*$ of the boundary of the inner-bubble layer \eqref{eq:assympM2}. Comparing the line elements \eqref{eq:assympM2} and \eqref{eq:outerAnsatz} we find that, 
 up to subleading $\alpha$ corrections, the radial coordinates $u$ and $\rho$ can be identified consistently with the gluing conditions. Therefore, from here and \eqref{eq:alphaScaling} we can see that the value of $u$ at the  matching hypersurface scales as $u_* \sim \alpha^{-1} \sim \epsilon^{-1} = \ell_{\text{kk}}/\ell_{\text{ssb}}$.  
 Then, expressing the leading KK corrections of the semi-flat regime, \eqref{eq:correctionsTF} and  \eqref{eq:assympGamma}, in terms of the variable $u$ (defined in \eqref{eq:asympSF2}),  we find that the error made in the gluing  as a result of neglecting the massive KK modes scales as\footnote{A similar expression can be found for the second fundamental forms $K|_{\bf I}$ and $K|_{\bf II}$.}
\be
(\Phi_*(s_{\bf II}) - s_{\bf I})= \cO(\rme^{ - p \, \ell_{\text{ssb}}/\ell_{\text{kk}}}) \sim \rme^{ - p/ \epsilon} \ll \epsilon, 
\ee
  where  $p$  is a positive real number. With this at hand, we  conclude that the contributions from massive KK modes are indeed subleading, and that it was justified ignoring them in   \eqref{eq:assympM2}. This completes our construction of the BON instanton mediating the decay of the compactification on $T^3$.

\subsubsection{Bubble decay for the \texorpdfstring{$G3$}{G3} compactification.}
\label{sec:G3decay}

We will now discuss the bubble of nothing decay of the vacuum where the compact space is a torus quotient $\cC_3 \cong T^3/\Gamma$. For definiteness concentrate in the vacuum with  $T^3/\Gamma \cong G3$, but the results are also applicable to a generic choice of $\Gamma$. 

In the case of the $G3$ compactification, the appropriate manifold to describe the inner-bubble region is  a warped  non-compact Calabi-Yau manifold of the class discussed in Section \ref{sec:detailedBON} with $N=8$ degenerations. The  matching conditions between the inner and outer bubble regimes are identical to the $T^3$ case for the metric function $W(\rho)$, the volume modulus $\varphi(\rho)$ and the dilaton,  that is, the equations \eqref{G3matching1BIS}, \eqref{T3matching} and \eqref{dilatonT3matching}. These conditions should be supplemented with the gluing constraints for the metric function $C(\rho)$ appearing in \eqref{eq:outerAnsatz}, which we find to be
\bea
C|_{\bf I}(\rho_*) &=& u_*\Big(1 - \frac{\Delta}{2 \pi}\Big)\Big(1+ \varphi^{(1)}|_{\bf II}(u_*)\Big)+\cO(\alpha^2), \nonumber \\
 C'|_{\bf I}(\rho_*) &=& \Big(1 - \frac{\Delta}{2 \pi}\Big)+ u_*\Big(1 - \frac{\Delta}{2 \pi}\Big) \nabla_n \varphi|_{\bf II}(u_*) + \cO(\alpha^2),
\label{eq:Cgluing}
\eea
where $\Delta=N \pi/6$ is the deficit angle in the geometry \eqref{eq:asympSF1}.
Then, as for the $T^3$ compactification, in the $G3$ case it is also consistent to use the  ansatz $\phi = \varphi - \varphi_\infty + \phi_\infty$ in the equations  \eqref{eq:G3outermostEqs}, which now reduce to  
\bea
W''   +W'\,\frac{C'}{C} - (D-5)W^{-1}(\cR^{-2}- W'{}^2)&=& 0,\\
 C''  + (D-4)\frac{ W'}{W} C'&=&0,  \\
\varphi'' + (D-4)\frac{W'}{W}\varphi' +\frac{C'}{C}\varphi'  &=&0.
\label{eq:G3system}
\eea
Let us first consider  the limit $\alpha\to0$, where the  volume modulus can be set to its asymptotic value $\varphi=\varphi_\infty$.  The   equations for $W(\rho)$ and $C(\rho)$ admit the following first integral
\be
 W'(\rho) =  \cR^{-1} \left(1- W^{-(D-5)} \right)^{1/2}, \qquad C(\rho) = R_{\text{kk}} \cR \, W'(\rho)
\label{eq:firstIntegral}
\ee
where we will impose the boundary condition  $W(0) = 1$. Actually, as we saw in Section \ref{sec:WittenBON}, this is just  the standard Witten's BON instanton in disguise. In this case the Witten's solution is embedded in an asymptotically flat spacetime with $D-3$ non-compact directions, where $\theta$ parametrises the collapsing  $S^1$ in \eqref{eq:outerAnsatz}, and with two inert  extra dimensions compactified in a two-torus. In the limit $\rho/\cR\to0$  the solution to \eqref{eq:firstIntegral} admits the expansion
\be
W(\rho) = 1 + \frac{(D-5)\rho^2}{2 \cR^2} + \ldots, \qquad C(\rho) = R_{\text{kk}} \frac{(D-5)}{2 \cR} \rho+\ldots \, .
\ee
Assuming for now that the nucleation radius scales as $\cR=\cO(\epsilon^{-q})$ with $q\in \mathbb{R}^+$, we can satisfy all the matching conditions  for $W(\rho)$ in \eqref{G3matching1BIS} and \eqref{T3matching} to order $\cO(\epsilon^{2q})$  setting the matching boundary at a point $\rho^*=\cO(1)\ll\cR$. Furthermore, the conditions \eqref{eq:Cgluing} for the metric function $C(\rho)$  can also be solved  setting   $u_* = \rho_*$, and  imposing the relation 
 \be
 \cR = R_{\text{kk}} \frac{(D-5)}{2(1-\frac{\Delta}{2 \pi})},\qquad \text{with} \qquad \frac{\Delta}{2 \pi} = \frac{2}{3}.
\label{eq:G3radius}
 \ee
 The consistency of this construction requires that  the asymptotic radius of the collapsing cycle also scales as  $R_{\text{kk}}= \cO(\epsilon^{-q})$. Therefore, as we anticipated in Section \ref{sec:constraint}, the bubble nucleation radius is finite even though we have turned off the Gauss-Bonnet coupling.

For  small non-zero values of  $\alpha$ we obtain a  deformation of the Witten's BON spacetime  \eqref{eq:firstIntegral}. In particular the matching constraints for the metric functions $W(\rho)$ and $C(\rho)$ can be satisfied to order $\cO(\alpha)$  setting the positions of  the gluing boundaries as  $\rho_* = u_* (1 + \varphi|_{\bf II}(u^*))$, and with an appropriate modification of the deficit angle $\Delta$ in the formula for $\cR$  \eqref{eq:G3radius} 
\be
\frac{\Delta}{2 \pi} = \frac{N}{12} - \frac{24 \pi^2 \alpha}{\cV_{G3}}  \rme^{2 \Delta \varphi}R_{\text{kk}} + \cO(\alpha^2), 
\label{eq:deformedDeficit}
\ee
where $\cV_{G3}$ is the volume of the $G3$ compact space at the vacuum, and $\Delta \varphi$ determines  the volume growth of the $T^2$ fibre in  the outer bubble region, $\cV_{T^2}(\rho_*)/\cV_{T^2} = \rme^{-2\Delta \varphi}$.
This implies that the effect of the Gauss-Bonnet term is to reduce the BON radius and, as we shall see below, to increase nucleation rate of these bubbles. Interestingly, as opposed to our discussion on the $T^3$ compactification, in this case we have been able to fully determine the bubble radius even when the Gauss-Bonnet coupling is turned on. The reason for this is that the bubble we just constructed is a deformation of the standard Witten's BON, which is an exact solution to  Einstein's equations in the outer bubble layer.   Thus the nucleation radius was already determined to zero order in $\alpha$.

Regarding the volume modulus $\varphi$,  we can determine its behaviour integrating  the third equation in \eqref{eq:G3system}, with $W$ and $C$ given by Witten's solution  \eqref{eq:firstIntegral}. In particular, we find that the volume growth of the $T^2$ fibre in  the outer bubble region can be obtained using that 
\be
\varphi'(\rho) = \frac{\varphi'(\rho_*) C(\rho_*)}{C(\rho) W(\rho)^{D-4}} \qquad \Longrightarrow \qquad \Delta \varphi =  \varphi'(\rho_*) C(\rho_*) \int _{\rho_*}^\infty \frac{d\rho}{C(\rho) W(\rho)^{D-4}}. 
\ee
Then, it is immediate to find an expression for $\Delta \varphi$ valid in the limit of small $\alpha$. Assuming $|\Delta \varphi|\ll1$ we obtain
\be
\Delta \varphi =\log\big(R_{\text{kk}}/C(\rho_*)\big)\,  \frac{R_{\text{kk}}}{(1-\frac{\Delta}{2 \pi})}\, \frac{24 \pi^2 \alpha}{\cV_{G3}} + \cO(\alpha^2),
\ee
where we have used the equations \eqref{eq:firstIntegral} and \eqref{eq:G3radius}. Note that $|\Delta \varphi|$  can always be made small   tuning conveniently the parameter $\alpha$,  or making the $G3$ volume large.

Before we conclude, to ensure the validity of the  construction, we need to estimate the size of the KK corrections  neglected in \eqref{eq:assympM2}. For simplicity we will just consider the $G3$ case ($N=8$) in the model without the Gauss-Bonnet term, $\alpha=0$, but a similar analysis can be done in a generic situation where the compact space is given by any torus quotient $T^3/\Gamma$. From \eqref{eq:asympSF1} and  \eqref{eq:correctionsTF}, we find that   the leading KK corrections to the gluing constraints  \eqref{eq:matchingConditions}  at the boundary between the inner and outer-bubble regions scale as
\be
(\Phi_*(s_{\bf II}) - s_{\bf I})\sim (z_0/|z|)^p\sim \Big(\frac{R_{\text{kk}}^* \ell_{\text{fibre}}^2}{\ell_{\text{kk}}^3}\Big)^{-3 p } \left(\frac{\ell_{\text{kk}}^6}{\ell_{\text{fibre}}^4\ell_{\text{sf}}^2}\right)^{\frac{3p}{4}}, \qquad p \in \mathbb{R}^+
\ee
and with a similar expression for the second fundamental forms. 
This estimate is written in terms of the length scales defined in  \eqref{eq:scalesKKF} and  \eqref{eq:lsfDef},  and using the size of the collapsing cycle at the matching point  $R_{\text{kk}}^* = u_* (1 - \frac{\Delta}{12})$.  Assuming the approximation scheme summarised by \eqref{eq:scalingCheck}, we find that the error associated to neglecting  the massive KK modes  can be made arbitrarily small if the size the collapsing cycle at $u=u_*$ scales as $R_{\text{kk}}^* \gtrsim \cO(\epsilon^{-5})$. As the position of the gluing surface is arbitrary,  this scaling can always be achieved   setting the matching hypersurface sufficiently far from the degeneration points, e.g.  with $u_* = \cO(\epsilon^{-5})$. Finally,  radius of the collapsing cycle at the vacuum  and  the bubble nucleation size  should scale as $R_{\text{kk}} \sim \cR \sim \cO(\epsilon^{-6})\gg R_{\text{kk}}^*$, what guarantees that  we have parametric control over all the approximations that we have made.

For completeness, to end this Section, we will write down the \emph{outer-bubble} line element  in the case $\alpha=0$,   for a generic  spacetime dimension, and an arbitrary compact space $T^3/\Gamma$. To write the metric given by \eqref{eq:firstIntegral}  in a more familiar  gauge, we use a new radial coordinate defined by $r(\rho) \equiv \cR W(\rho)$, which leads us to
\be
ds^2|_{\bf I} = r^2 d\Omega_{D-4}^2 +  \Big(1 - \frac{\cR^{D-5}}{r^{D-5}}\Big)^{-1} d r^2 + R_{\text{kk}}^2 \Big(1 - \frac{\cR^{D-5}}{r^{D-5} }\Big) d\theta^2 + ds^2_{T^2},\label{G3n}
\ee
with the bubble nucleation radius given by
\be
\cR= R_{\text{kk}} \frac{6 (D-5)}{(12-N)}.
\ee
The line element above can be easily recognised as the euclidean version of a $(D-2)$-dimensional Schwarzschild black hole, with two extra dimensions compactified on a two-torus. It is important to emphasize that this metric  is only appropriate for describing  the outer-bubble regime of the BON spacetime. That is, for values of the radial coordinate
\be
r \ge \cR + \frac{(D-5) \rho^2_*}{2 \cR}  >\cR.
\ee
Actually, even if the $T^2$ fibration on the KK circle was trivial,  the metric  \eqref{G3n} would still have a conical singularity at $r =\cR$, which would  disappear only with the specific choice of parameters $\cR=(D-5)R_{\text{kk}}/2$ (as in the original Witten's bubble). In the present construction,  we have instead cut  out a small
neighbourhood of $r=\cR$, and replaced it with the \emph{smooth} inner-bubble region described in the previous sections.

\subsection{Decay rates}
\label{eq:decayRates}

In this Section we will compute the  bubble nucleation rate per unit world-volume in the $(D-3)$ dimensional non-compact space,  $\Gamma_{\text{dec}}/\cV_{D-3} = A \rme^{- S_{\text{BON}}}$, which   is given in terms of the euclidean BON action $S_{\text{BON}}$  \cite{Coleman:1977py}. 
For the variational problem to be well defined we need to supplement the action \eqref{eq:action} with the Gibbons-Hawking boundary term\footnote{For the sign  conventions see \cite{poisson_2004}. The boundary term in string frame is the same as in Einstein frame (see \cite{Dyer:2008hb,Casadio:2001ff}).} \cite{PhysRevD.15.2752,York:1972sj} 
\be
S = S_s - \frac{\sigma}{8 \pi G_D} \int_{\pd \cM}d \zeta^{(D-1)} \sqrt{s} [\cK-\cK_0],  
\label{eq:boundaryTerm}
\ee
where $s_{ab}$ is the induced metric on the spacetime boundary $\pd \cM_D$, $s = \det(s_{ab})$ and $\zeta^a$  are  coordinates parametrising $\pd \cM_D$, with  $a=\{1,\ldots,D-1\}$. The constant $\sigma= n_M n^M$ is the norm of the outwards pointing normal vector $n^M$ to the boundary, and $\cK = s^{ab}\nabla_{a} n_b$ is the trace of the second fundamental form of $\pd\cM_D$.  For convenience we have also subtracted the value of the boundary term  computed reference spacetime, with $\cK_0$ representing the trace of the corresponding  second fundamental form.   Actually the bulk contribution to the instanton action is zero, and thus $S_{\text{BON}}$ is completely determined by a boundary term.  This is a direct consequence  of  the dilaton equation of motion \eqref{dilatonHeom}, which allows to write the on-shell string frame action \eqref{eq:action}  in the form
\be
 S_{s}|_{\text{BON}} = 
-\frac{g_s^2}{8 \pi G_D} \int_{\cM} d^Dx \sqrt{-g}\; \nabla_{(D)}^2 (\rme^{-2\phi}),
\ee
which is also a boundary term. Then we have the following expression for the full action
\be
S_{\text{BON}} =
 - \frac{1}{8 \pi G_{D-3} \cV_{\cC}} \int_{\pd \cM}d \zeta^{(D-1)} \sqrt{s} [\cK-\cK_0 -2 \nabla_n \phi],
\label{eq:rawBONaction}
\ee
which is expressed in terms of $(D-3)$-dimensional Newton's constant $G_{D-3} \equiv G_D/\cV_{\cC}$, where $\cV_{\cC}$ is the volume of the compact space at the vacuum. 

To compute the elements appearing in the instanton action, first we need to write the metric of the outer-bubble region \eqref{eq:outerAnsatz} in the same gauge as the vacuum \eqref{vacuum}
\be
ds^2 = r^2 d \Omega_{D-4}^2 + H(r)^2 d  r^2 +C(r)^2  d\theta^2 +  \rme^{2(\varphi(r)-\varphi_\infty)} ds^2_{T^2},
\label{eq:WittensGauge}
\ee
where the new radial coordinate is defined by $r(\rho) = \cR\, W(\rho)$, and $H(r) = 1/\cR\, W'|_{\rho(r)}$.
In this gauge the spacetime boundary is defined as the hypersurface at   $r=r_\infty$, after  taking the limit $r_\infty\to \infty$. Then, the trace of the second fundamental form on this hypersurface reads
\be
\cK = \Big[(D-4) \frac{\pd_\rho W}{W} + \frac{\pd_\rho C}{C}+  2\pd_\rho \varphi\Big]_{\rho(r_\infty)},
\label{eq:2FFinfty}
\ee
and the determinant of the first fundamental form  in \eqref{eq:rawBONaction} is 
\be
\sqrt{s} = r^{(D-4)} \omega_{D-4} \sqrt{h_{T^2}} \; C \rme^{2(\varphi-\varphi_\infty)}|_{\rho(r_{\infty})},
\ee
where $\omega_{D-4}$ is the area element of the $D-4$ unit sphere. Next we will  study the asymptotic behaviour  of the BON instantons presented above in the limit $\rho\to \infty$ to find an explicit expression for the two quantities $\cK$ and $\sqrt{s}$.

As we have seen above, the BON instantons  mediating  the decay of compactifications on a  three-torus or its quotients $T^3/\Gamma$ are  all consistent with the ansatz $\phi = \varphi - \varphi_\infty + \phi_\infty$, so we just need to consider system of equations \eqref{eq:G3system} for the outer bubble region.  Integrating the second and third equations of   \eqref{eq:G3system}, we find that the  functions $C(\rho)$ and $\varphi(\rho)$ have the following asymptotic form in the limit $\rho \to \infty$
\bea
C'(\rho) = \frac{C'(\rho_*)}{W^{D-4}}\qquad &\Longrightarrow& \qquad C(\rho) = R_{\text{kk}} - \frac{C'(\rho_*)\cR^{D-4}}{(D-5) \, \rho^{D-5}} + \ldots\, \nonumber \\
 \varphi'(\rho)  \rme^{\varphi(\rho)}=\frac{\varphi'(\rho_*)  \rme^{\varphi(\rho_*)}}{W^{D-4}}\,  \qquad &\Longrightarrow& \qquad \varphi'(\rho)  =\varphi_\infty -\frac{\varphi'(\rho_*)  \rme^{-\Delta \varphi} \cR^{D-4}}{(D-5)\, \rho^{D-5}}+\ldots\, ,
\eea
where we used the boundary conditions \eqref{eq:outermostBC} at infinity, and $\rho_*$ is the position of the boundary between the inner and outer bubble regions. As for the  metric function $W(\rho)$, from the first equation in  \eqref{eq:G3system}  we obtain 
\bea
D=6:&\qquad& \cR W(\rho) =\rho - \rho_0  - \frac{C'(\rho_*)\cR^{D-4}}{(D-5)R_{\text{kk}}}\log \rho +  \ldots \,,\nonumber\\
D>6: &\qquad&\cR W(\rho) =\rho - \rho_0 + \frac{C'(\rho_*)\cR^{D-4}}{(D-6) (D-5)R_{\text{kk}}\, \rho^{(D-6)}} +  \ldots \, .
\label{eq:asymptG}
\eea
Here the parameter $\rho_0$ is an  integration constant  
which can be eliminated with the redefinition $\rho \to \rho+ \rho_0$, and  discarding subleading terms  of the expansion. 
 Substituting the previous results into the expression for the second fundamental form $\cK$ \eqref{eq:2FFinfty} we arrive at
\be
\cK = \frac{(D-4)}{r}  - \frac{C'(\rho_*)\cR^{D-4}}{(D-5) R_{\text{kk}}\, r^{D-4}}  + \frac{2 \varphi'(\rho_*)  \rme^{-\Delta \varphi}}{r^{D-4}},
\ee
where the (divergent) vacuum contribution, which we need to subtract, is  $\cK_0 = \frac{(D-4)}{r}$. Finally the instanton action,   is given by the following formula
\be
S_{BON} = \frac{A_{D-4}}{8 \pi G_{D-3}} \frac{C'(\rho_*)\cR^{D-4}}{R_{\text{kk}}(D-5)} .
\ee
where we have also taken into account  the  contribution from the dilaton in \eqref{eq:rawBONaction}.

\paragraph{BON action for the compactification on $T^3$.} In this case we have $C(\rho)= R_{\text{kk}} \rme^{\varphi(\rho)-\varphi_\infty}$,  which combined with the gluing conditions  \eqref{T3matching} and the relation \eqref{eq:BONradiusScaling} allows to find  $C'(\rho_*)$, and in turn the euclidean BON action
\be
S_{BON} = \frac{A_{D-4}}{8 \pi G_{D-3}} \frac{ \mathring \cR^{D-4} }{(D-5)} \rme^{-(3 D-14) \Delta \varphi} \, \left(\frac{24 \pi^2 \alpha}{\cV_{T^3}}\right)^{-(D-5)}.
\ee
Therefore, regardless of the value of $\Delta \varphi$, when we turn off the Gauss-Bonnet coupling $\alpha\to 0$  the action  grows unbounded, and then the decay rate becomes exponentially suppressed as a result of a   Coleman-deLuccia mechanism. 

In particular, in the regime  $\Delta \varphi\to \infty$ discussed above we find
\be
\Delta \varphi\to \infty:\qquad S_{BON} = \frac{A_{D-4}}{16 \pi G_{D-3}} \cR^{D-5},
\label{eq:T3actions}
\ee
where the radius $\cR$ is  given   by the formula    \eqref{eq:T3RadiusDlarge}.  Here again we can see that that the bubble nucleation rate will be exponentially suppressed when we set to zero the parameter $\alpha$, since the BON radius diverges in this limit.

\paragraph{BON action for the compactification on $T^3/\Gamma$.} In this case expression for $C'(\rho_*)$ can be found from the gluing conditions   \eqref{T3matching} and \eqref{eq:Cgluing}, leading to 
\be
S_{BON} =  
\frac{A_{D-4}}{16 \pi G_{D-3}} \cR^{D-5} \qquad \text{with}  \qquad  \cR= \frac{(D-5)}{2 (1-\frac{\Delta}{2 \pi})} R_{\text{kk}},
\label{eq:G3orbifoldS}
\ee
and $\Delta$ given by \eqref{eq:deformedDeficit}. As anticipated above, in this case the BON action remains finite when we set to zero the Gauss-Bonnet coupling. This is consistent with the fact that, unlike in the case of $T^3$, here we do not need the GB term to violate the dominant energy condition, as the PET is already violated since the manifold does not admit a covariantly constant spinor.  It can also be seen that when $\alpha\neq0$ the deficit angle $\Delta$ decreases slightly, so the net effect of the Gauss-Bonnet term is a  small enhancement of the bubble nucleation rate.

The action and radius of  Witten's original bubble of nothing can be recovered  substituting $D=7$, $N=0$  and $\alpha=0$ in the previous formula
\be
S_{\text{BON}}^{\text{w}} = \frac{\pi R_{\text{kk}}^2}{8 G_4}, \qquad \cR_{\text{w}} = R_{\text{kk}}.
\ee
As an interesting coincidence, we also note that if   we set $N=12$ in the formula for the deficit angle \eqref{eq:deformedDeficit} and then plug it into \eqref{eq:G3orbifoldS} we recover  \eqref{eq:T3actions} (with $\cR$ given by \eqref{eq:T3RadiusDlarge}), the  euclidean action of the bubble mediating decay of the $T^3$ compactification in the limit $\cV_{T^3}/\cV_{T^3}(\rho^*)\to \infty$.

\section{Physical implications and string theory embedding}\label{sec:phys}

The explicit construction of the BON solution for a $T^3$ compactification of the previous Sections is only the beginning towards many more new types of bubbles of nothing that can be present  in non-supersymmetric vacua. In this Section, we will provide an string embedding of our field theory model to show that it is a sensitive solution in a consistent theory of quantum gravity as well as discuss generalizations of these bubbles including fluxes or charged fermions. We will also discuss implications for String Phenomenology and the possibility for these bubbles to be a universal decay mode for any string compactification breaking supersymmetry.

\subsection{String theory embedding\label{sec:string}}

In the previous Sections, we have explicitly constructed a (self-consistent, approximate in a derivative expansion) bubble of nothing solution for $T^3$ with the fully periodic spin structure, in the Einstein Gauss-Bonnet dilaton theory \eq{eq:action}. The construction avoids Witten's positive energy theorem because the relevant energy condition is not satisfied, see Subsection \ref{sec:pet}.

One might worry that the low-energy solution we constructed only exists because somehow we did something pathological: perhaps the action \eq{eq:action} is secretly ill because e.g. it does not satisfy the dominant energy condition classically. We comment on energy conditions in gravity in Subsection \ref{sec:DEC}, but first, here we (hope to) dispel any doubts about the validity of the Einstein Gauss-Bonnet dilaton theory \eq{eq:action} by showing that it is (almost) a consistent truncation of a valid string compactification to six dimensions. We do this in several different ways.

\subsubsection{Heterotic embedding\label{sec:heterotic}}
As starters, consider compactification of heterotic string theory on $T^4$. The tree-level bosonic effective lagrangian of the NS sector of heterotic up to four derivatives in the metric is \cite{Gross:1986mw,Liu:2013dna} 
\begin{equation}\mathcal{L}=e^{-2\phi}\left[R+4(\nabla\phi)^2-\frac{1}{12}H^2-\frac{\alpha'}{4}\text{tr}(F\wedge *F)+\frac{\alpha'}{8}R_{MNRS}(\Omega_+)R^{MNRS}(\Omega_+)\right],\label{L0}\end{equation}
where $\Omega_+$ is the connection with torsion
\begin{equation}\textbf{R}(\Omega_+)=\textbf{R}+\frac12d\mathcal{H}+\mathcal{H}\wedge\mathcal{H},\quad\mathcal{H}^{ab}\equiv H^{ab}_M dx^M.\label{ctr}\end{equation}
The equations of motion corresponding to (\ref{L0}) are 
\begin{eqnarray}
R-4(\nabla \phi)^2+4\Box\phi-\frac1{12}H^2-\frac{\alpha'}{4}\text{tr}(F\wedge *F)+\frac{\alpha'}{8}R_{MNRS}(\Omega_+)R^{MNRS}(\Omega_+)&=&0,\nonumber\\
R_{MN}+2\nabla_M\nabla_N\phi-\frac14H^2_{MN}
-\frac{\alpha'}{4}\text{tr}( F_{MN}^2) 
+\frac{\alpha'}{4}R_{MPRS}(\Omega_+)
R_N{}^{PRS}(\Omega_+)&=&0,\nonumber\\
d(e^{-2\phi}*H)&=&0,\nonumber\\
e^{2\phi}d(e^{-2\phi}*F)+A\wedge*F-*F\wedge A+F\wedge*H&=&0.
\label{eq:HetEOM}
\end{eqnarray}
On top of this, there is the usual heterotic Bianchi identity,
\begin{equation}d H=\frac{\alpha'}{4}\left[\text{tr}(R(\Omega_+)^2)-\text{tr}(F^2)\right].\end{equation}
We will now show how to embed the solution to the Einstein-dilaton-Gauss-Bonnet theory that we constructed in previous Sections in this theory. First, the heterotic action \eq{L0} contains only a Riemann squared term, instead of the full Gauss-Bonnet. However, this will not matter in the following, since the bubble solution we have is constructed by perturbing a Ricci-flat background metric (and in the asymptotic region, where we do not use perturbation theory anymore, the effect of the Gauss-Bonnet term is small). So we may pretend \eq{L0} contains a full Gauss-Bonnet term.

The solution we constructed is a bubble of nothing for a toroidal compactification to three dimensions, $T^3\times\mathbb{R}^3$. We will embed it into the equations of motion \eq{eq:HetEOM} by taking the gauge fields to vanish identically, and taking $H$ to be the solution to the equations
\begin{equation} dH=\frac{\alpha'}{4}\text{tr}(R^2),\quad d(e^{-2\phi}*H)=0.\label{bianchi-1}\end{equation}
These equations always have a solution for any $\phi$, but notice that since $\text{tr}(R^2)\neq0$ in our solution, we are forced to take $H\neq0$. This is unavoidable, caused by the topology of the bordism, and as we will see, intimately related to supersymmetry. 

The $H$ field obtained in this way is linear in $\alpha'$, so when plugged back into the first two equations in \eq{eq:HetEOM} it only gives a $\mathcal{O}(\alpha'^2)$ contribution. The additional terms in \eq{ctr} are also higher order in $\alpha'$. To first order in $\alpha'$, we get the same equations of motion we have discussed in the previous Sections, so the bubble solution seems to have been successfully embedded. However, this result is \emph{too} good! In particular, we seem to have embedded the bubble of nothing into a genuinely supersymmetric compactification. This is impossible, since in a supersymmetric theory there are no negative ADM mass solutions and the only zero solution is the vacuum; this is due to the fact that the energy operator is the square of the supercharge, which is itself a boundary integral of the supercurrents \cite{Witten:1981mf}.

The resolution is that in fact we do not recover asymptotically the $T^3\times\mathbb{R}^3$ compactification; due to the Bianchi identity, we end up with a flux compactification instead, since
\begin{equation}\int_{T^3,r\rightarrow\infty} H= \int_{\cB_4}dH=4 \pi^2 \alpha'  \chi(\cB_4).\end{equation}
Thus, the embedding works, but it automatically turns on a flux at infinity that breaks supersymmetry. In other words, the bubble we constructed embeds in string theory, but it is an instability of a nonsupersymmetric flux compactification to three dimensions. 

Due to the flux we have turned on, the asymptotic compactification is not stable. There will be runaway potentials for the moduli, and also a nonzero vacuum energy (proportional to $(\alpha'^2)$). However, these instabilities will only appear at higher orders in the $\alpha'$ expansion; by adiabaticity, we expect the bubble of nothing instability should still be present, and at the very least the $t=0$ section of our euclidean solution embeds as a honest zero mass configuration (i.e. that satisfies the Hamiltonian constraint) of the full theory. Once this initial condition nucleates somehow, we expect it to expand with uniform acceleration as usual.

 One might try to construct a bubble of nothing for the supersymmetric solution by setting $H=0$ and turning on gauge fields to compensate for $\text{tr}(R^2)$, as usual in standard embedding heterotic compactifications. Although this is doomed to fail, it is interesting to see how exactly it does. We may now take $H=0$ exactly, and the gauge field equations of motion will be solved if we take the connection to be self-dual, which we can do at least in the inner layer of our model. The Yang-Mills term evaluated on the Levi-Civita connection is proportional to the Riemman tensor squared,\begin{equation} \text{tr} (F\wedge *F)= \frac12 dV R_{MNRS} R^{MNRS},\end{equation}
and as a result the first two equations in \eq{eq:HetEOM} still have the same form as in the Einstein-dilaton model, with the Gauss-Bonnet term set to zero. This is a consistent truncation of the supersymmetric lagrangian, so we cannot construct our bubble solutions in this case; the effective stress-energy tensor satisfies the dominant energy condition.

One could also ask if Witten's bubble could somehow be embedded in our model. The naive answer is no, because we chose periodic boundary conditions along the three cycles of $T^3$. But due to the presence of torsion in the gravitino connection as in \eq{ctr}, parallel transport along one of the cycles of $T^3$ rotates the gravitino,
\begin{equation}\psi_M(y^{\bar \alpha})=\exp\left(\frac{\rmi}{2} y^{\bar \alpha}  H_{\bar \alpha \bar \beta \bar \delta}\gamma^{\bar \alpha}\gamma^{\bar\delta}  \right) \psi_M(0),\end{equation}
where $y^{\bar \alpha}$ are coordinates on the $T^3$.
It is not clear to us whether by ``playing around'' with gauge instantons (which generate $H$ flux) one could embed Witten's bubble by having a gauge bundle over the disk $\cD$ such that the boundary conditions for the fermions become antiperiodic at the core. A situation like this one was explored in \cite{Blanco-Pillado:2016xvf}, where it was found that such a thing is sometimes possible with a background gauge connection. It is certainly possible for the $H$ field to behave as a $\text{Spin}^c$ connection after dimensional reduction, as exemplified by the $T$-dual of IIB on $\text{AdS}_5\times S^5$, which contains fermions on a fluxed $\mathbb{CP}^2$. In any case, these issues are absent in the dual type I model, which we present below.

Interestingly, we can get new bubbles from the one we constructed by string dualities; for instance, by $T$-dualizing along one of the $T^3$ three-cycles, the $H$ flux becomes geometric flux (a twisted torus compactification \cite{Wecht:2007wu}), and we have just constructed a bubble of nothing for this geometry involving winding modes (but perturbative in $g_s$), even for completely periodic boundary conditions. We will discuss another example in a different dual frame below.

\subsubsection{Type I/IIB and bosonic duals}
We have successfully embedded our bubble of nothing into the equations of motion of perturbative heterotic string theory on $T^4$. This theory has a large moduli space, and it is interesting to look for embeddings at other weakly coupled points. Here, we will focus on the ``S-dual'' limit obtained by compactifying perturbative type I strings on a torus. By T-duality, this is straightforwardly related (and still at weak coupling) to type II configurations with orientifolds and D-branes.

We need to check that our main ingredient, the Gauss-Bonnet term, is still available. Although it vanishes in type II in 10 dimensions \cite{Gross:1986iv}, it is present in type I \cite{Tseytlin:1995bi} strings. The dilaton dependence is however different, being $e^{-\phi}$ rather than $e^{-2\phi}$ as in the heterotic case. This signals it is an open string effect. The effective Lagrangian in 10 dimensions looks like 
\begin{equation}\mathcal{L}=e^{-2\phi}\left[R+4(\nabla\phi)^2\right]-\frac{1}{12} H^2-e^{-\phi}\left[\frac{\alpha'}{4}\text{tr}(F\wedge *F)+\frac{\alpha'}{8}R_{MNRS}R^{MNRS}\right].\label{L1}\end{equation}
The differences to the heterotic case are the aforementioned different dilaton powers for the gauge field and Gauss-Bonnet terms, and also for the 2-form kinetic term, which reflects the fact that in type I it is a RR field. However, one still has a Bianchi identity, just like \eq{bianchi-1}. It is also important that the type I fermions only couple to the geometric spin connection, and do not receive extra contributions as in \eq{ctr}. In type I language, this is clear to begin with since the orientifold projects out the $B$-field (type I strings can break). This then carries over to compactifications; in the T-dual type II language we will use later on, the zero mode of the $B$-field in the internal space is projected out by the orientifold.

In $D=10$, the heterotic model model is related to ours via S-duality: after putting our solutions in Einstein frame, we can recycle the solution from the previous situation simply by flipping the sign of the dilaton.  However, this no longer works after compactification, because the effective lower-dimensional dilaton in the heterotic frame is no longer related in a simple manner to the type I dilaton. However, since the models are so similar, we expect that a close relative of our bubble can be embedded in type I as well. We see no obstruction to solving the equations of motion just as we did in Section \ref{sec:detailedBON} in the same way, but we have just not done this explicitly. Let us, though, proceed as if we had constructed the bubble solution and discuss some interesting implications for low energy EFT's of type IIB orientifold flux compactifications, as it might bring some surprises.  

For concreteness, let's say we embed our bubble on a compactification of type I string theory on $T_A^3\times T_B^3$. That means that we first reduce on the first $T_A^3$ factor, to get the action of type I in seven dimensions, and then we consider a compactification of this seven-dimensional theory on a second $T^3$ pierced by RR three-form flux. A modified version of our bubble embeds on this model and describes the decay of the $T_B^3$ factor.

One can use this model to connect to the more standard type IIB orientifold literature. By T-dualizing along two of the cycles in $T_3$, we end up on a toroidal orientifold, $(T^2/\mathbb{Z}_2)\times T^2 \times T^2)/\Omega$, with supersymmetry breaking fluxes. The 3-form flux along\footnote{We introduce coordinates $z_1,z_2,z_3$ on the three $T^2$ factors, and $z_i=x_i+iy_i$. The orientifold action reflects $z_2$ and $z_3$. } $T_B^3$, 
\begin{equation} H_3= n\, dx_2\wedge dy_2\wedge dx_3= \frac{n}{4} dz_2 \wedge d\bar{z}_2\wedge (dz_3+d\bar{z}_3)\end{equation}
contains imaginary self-dual $(2,1)$ and an imaginary anti-self dual $(1,2)$ pieces. The latter flux is known to break supersymmetry by inducing a non-vanishing F-term for the complex structure moduli \cite{Giddings:2001yu}. This is consistent with the fact that we can only have a bubble for a non-supersymmetric compactification, since our bubble wouldn't be able to eat up supersymmetry preserving pure $(2,1)$ flux. 

From this point of view however it seems one should be able to embed our bubble whenever there is identical imaginary self dual and anti self dual pieces. One is left to wonder if, for instance, our bubbles exist in the KKLT scenario where, in addition to the ISD fluxes on the Calabi-Yau, the gaugino condensate sources an IASD $(1,2)$- flux component \cite{Baumann:2010sx}. However, the KKLT AdS vacuum (before the uplift to de Sitter) is supersymmetric and, as we have emphasized, it is not possible to have a bubble of nothing in a unitary supersymmetric compactification. So, if our bubble indeed embeds in KKLT, it could be a signal of some hidden inconsistency in the procedure. To put it another way, bubbles of nothing must be absent in \emph{unitary} supersymmetric compactifications, but might be present in non-unitary ones, since the proof of positive mass in terms of the supercharges crucially uses that one has a positive-definite inner product. So if our bubble can embed in any putative supersymmetric compactification, it would mean that the theory is non-unitary. But because generally it is not expected one can get non-unitary theories from usual string compactifications\footnote{See however, \cite{Dijkgraaf:2016lym}, which uses T-duality in timelike directions. They precisely produce supersymmetric, non-unitary theories of the kind we are discussing here.}, one would have to conclude something went wrong in whatever construction is being considered.  At the moment, we are far from concluding that the bubble can embed in KKLT, but it is surely something interesting to explore in the future.

 We could also straightforwardly embed our bubble solution in the bosonic string, since it has a Gauss-Bonnet term \cite{Zwiebach:1985uq} and the Bianchi identity receives no corrections in this case. Thus, it is completely consistent to set $H=0$. In this case, the runaway potentials would arise at one-loop in the closed string coupling, and hence they would be suppressed by powers of $e^{-2\phi}$. This is better behaviour than what we have in the heterotic embedding, where the runaways are controlled by fluxes and appear already at tree level. But in this second case we would also have the usual closed-string tachyon, which brings back a $\mathcal{O}(\alpha')$ instability. On top of this, the bosonic string does not contain fermionic states in its perturbative spectrum (it may contain them at the nonperturbative level, if the duality proposed in \cite{Bergman:1997rf} is correct, since the putative 0B dual contains massive worldsheet spinors), so the relevant bordism groups might just be oriented instead of spin. In this case, one can always use Witten's original bubble. By contrast, in the heterotic embedding, we for sure have (massive) spinors, and the distinction that the spin structure is periodic becomes meaningful.

\subsubsection{Violation of the Dominant Energy Condition\label{sec:DEC}}

 The main theme of this paper is that the usual lore that a compactification is protected against bubbles of nothing due to a topological obstruction related to the spin structure is not true in general, and in particular is clearly false whenever the relevant bordism group vanishes. However, in precisely this case, one typically admits covariantly constant spinors, and therefore the compactification is still bubble-proof as long as the local energy condition in some version of the Positive Energy Theorem holds. In the vanilla case, this is just the Dominant Energy Condition, \eq{dec}. So the topological protection has been traded by a local inequality that matter must satisfy. 

It is well-known, though, that the DEC does not hold in general, although it is typically satisfied in supersymmetric models. It was first proposed because it is a reasonable property of ordinary matter \cite{Curiel:2014zba} and facilitates the proof of interesting results, such as the singularity theorems or the positive energy theorem itself.  But it is also intimately related to superluminality: Reference \cite{Barcelo:2000zf} proved that a generic violation of the DEC (or more precisely, the Null Energy Condition, which the DEC implies) leads to traversable wormholes with faraway wormhole mouths and so to causality violation\footnote{Interestingly, the wormhole construction uses a scalar field with a transplanckian field excursion. So perhaps these wormholes are in the Swampland after all!}. So the question is what is the strongest, general statement on the low-energy EFT one can make. Clearly, there has to be some statement to prevent causality violation, but the DEC is too strong.  We take the point of view that perhaps a reasonable thing to do is not to impose somehow arbitrary energy conditions, but just enforce the absence of causality-violating effects such as traversable wormholes. Reference \cite{Mehdizadeh:2015jra} claims that traversable wormholes can be constructed in the Einstein-Gauss Bonnet theory with ordinary matter, but does not take into account the effect of the dilaton, which can cause solutions that at first sight seem to be wormholes to instead ``close up''. The prototypical example is the supergravity solution of the $D(-1)$-brane, which seems to be a wormhole in the string frame, but not on the Einstein frame \cite{Bergshoeff:1998ry,Bergshoeff:2004fq}. Figuring out rigorously whether traversable wormholes exist in the Einstein-Dilaton-Gauss Bonnet model \cite{Mehdizadeh:2015jra} is an interesting question which lies outside of the scope of this work; if they do, the wormhole throat probably has a stringy size and are not to be trusted anyway.

Clearly, as our string theory embedding shows, one should not expect the DEC to be satisfied in a general, nonsupersymmetric, string compactification. In our case, the DEC is explicitly violated by the higher derivative terms, but there are plenty of other situations where the DEC does not hold, Casimir energies and AdS space being the most prominent ones \cite{Curiel:2014zba}. 
It would be interesting to understand, though, if non-supersymmetric vacua arising from string theory always violate the DEC or the corresponding energy condition at play, allowing the vacuum to decay. Notice that, depending on the spin connection of the different fermions in the compactification, the  local energy condition that needs to get violated to allow for the existence of bubbles of nothing will be different, as we discuss in the next Section.


\subsection{Generalisations of Positive Energy Theorems}\label{sec:pet2}
In Section \ref{sec:pet}, we explained how the positive energy theorem can provide a dynamical obstruction to the existence of bubbles of nothing in cases where there is no topological protection. The proof of the theorem relies on the existence of covariantly constant spinors in the manifold whose decay we are studying. Interestingly, these spinors can be charged under additional gauge fields, and the proof of the positive energy theorem still holds, but with a modified energy connection.  This can occur, for example, when having Wilson lines or fluxes in the compactified internal dimensions. In this Subsection we briefly review some of these modifications. These lead to additional obstructions to the existence of bubbles of nothing even in cases which do not admit covariantly constant ``ordinary'' spinors (spinors which are sections of the double cover of the tangent bundle).

Usual spinors on a manifold $M$ are defined by a choice of spin structure on $TM$. If instead we have a spin structure on $TM\oplus \chi$, where $\chi$, is an additional bundle, we obtain twisted spin structures \cite{Kapustin:2014dxa}. Particular examples are $\text{Spin}^c$ when $\chi$ is a line bundle, or $\text{Spin}^{\mathbb{Z}_{2n}}$ when $\chi$ is a $\mathbb{Z}_{2n}$ bundle. In other words, given a group $G$ whose center contains a $\mathbb{Z}_2$ factor, one can define spinors as Sections of a bundle whose transition functions live in
\begin{equation} \frac{\text{Spin}\times G}{\mathbb{Z}_2},\end{equation}
where the $\mathbb{Z}_2$ identifies the center of $\text{Spin}$ with the generator of the chosen $\mathbb{Z}_2$ subgroup of $G$. We will refer to this as $\text{Spin}^G$ fermions. Similarly, a manifold with a $\text{Spin}^G$ structure will define a class of modified bordism groups $\Omega_{d}^{\text{Spin}^G}$. 

In a theory with $\text{Spin}^G$ spinors, the topological obstruction to the existence of bubbles of nothing takes  values in $\Omega_{d}^{\text{Spin}^G}$. But even if $\Omega_{d}^{\text{Spin}^G}=0$ so that this topological obstruction is absent, there will be a dynamical obstruction as long as asymptotically covariantly constant $\text{Spin}^G$ spinors exist and a local energy condition holds. We have explained this in the form of a flowchart in Figure \ref{fig:flow}.

\tikzstyle{decision} = [rectangle, draw, 
    text width=10em, text badly centered, node distance=5cm, inner sep=2pt]
\tikzstyle{block} = [rectangle, draw, 
    text width=10em, text centered, rounded corners, minimum height=4em]
\tikzstyle{line} = [draw,-{Latex[length=2mm,width=2mm]}]
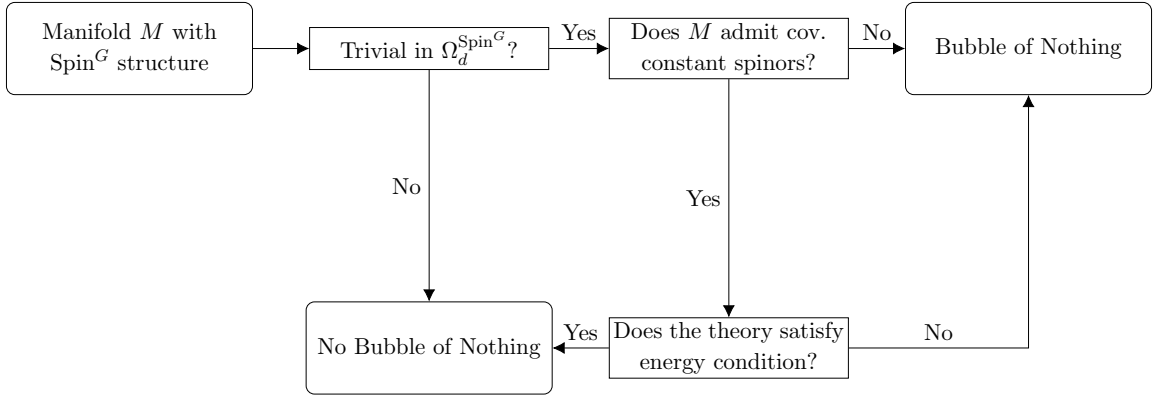
\begin{figure}
\begin{center}
\resizebox{.98\textwidth}{!}{\begin{tikzpicture}[node distance = 5cm, auto]
    \node [block] (init) {Manifold $M$ with $\text{Spin}^G$ structure};
    \node [decision, right of=init] (topobs) {Trivial in $\Omega_d^{\text{Spin}^G}$?};
    \node [decision, right of=topobs] (cccs) {Does $M$ admit cov. constant spinors?};
    \node [decision, below of=cccs] (ec) {Does the theory satisfy energy condition?};
    \node [block, below of=topobs](nobon){No Bubble of Nothing};
    \node [block,right of=cccs](bon){Bubble of Nothing};
    
    \path [line] (init) -- (topobs);
    \path [line]  (topobs) --node[anchor=south] {Yes} (cccs);
        \path [line]  (topobs) --node[anchor=east] {No} (nobon);
        \path [line]  (cccs) --node[anchor=south] {No} (bon);
        \path [line]  (cccs) --node[anchor=east] {Yes} (ec);
         \path [line]  (ec) --node[anchor=south] {Yes} (nobon);
          \path [line]  (ec) -| node[near start] {No} (bon);
\end{tikzpicture}}\end{center}

\caption{Flowchart illustrating when one can get a bubble of nothing; a more detailed version of Figure \ref{fig:flow0}. Given a manifold with bundles $M$, one first checks that there is no topological obstruction in $\Omega_{d}^{\text{Spin}^G}$. Assuming this is the case, one must make sure there are either no covariantly constant spinors in the compactification manifold, or that the relevant energy condition is violated.}
\label{fig:flow}
\end{figure}

We will now do this in detail for the case of $\text{Spin}^c$ structure. This was already worked out implicitly in \cite{Gibbons:1982jg}, which used fermions coupled to a modified background connection to obtain a positive mass theorem for black holes with charge. We believe the spinors used there are related to ours via a nonlocal field redefinition. 

The argument  in \cite{Witten:1981mf} is based on proving that there is no solution to the Dirac equation
\beq
i\slashed{D} \epsilon=0
\eeq
in which the spinor $\epsilon$ vanishes fast enough at large distances. This implies that any non-trivial configuration has non-vanishing positive ADM mass.

The proof relies on the fact that (eq. (25) of \cite{Witten:1981mf})
\begin{equation} [D_i,D_j]\epsilon= \left(\frac18 \sum_{\alpha,\beta} R_{ij\alpha\beta}\right)[\gamma^\alpha,\gamma^\beta]\epsilon \label{wittenmod}\end{equation}
which can be related to terms involving the stress-energy tensor by using Einstein's equations. In \cite{Witten:1981mf}, this then leads to a positive energy theorem provided that the matrix
\begin{equation} T_{00}\mathbf{I}+ T_{0j}\gamma^0\gamma^j\label{wittentij}\end{equation}
is positive definite, which is guaranteed by the dominant energy condition, \eq{dec}.

In the case of a Spin$^c$ structure, \eq{wittentij} receives additional contributions from the electromagnetic field when using Einstein's equations, becoming
\begin{equation} T_{00}\mathbf{I}+ T_{0j}\gamma^0\gamma^j-iq\frac{F_{ij}}{8\pi G}\gamma^i\gamma^j.\label{ustij}\end{equation}
Here, $q$ is an integer-valued parameter, the charge of the virtual fermion we are using in the proof of the theorem. We need to demand positivity of \eq{ustij} to obtain the $\text{Spin}^c$ version of the positive energy theorem\footnote{We are using an electrically charged spinor coupled canonically to the gauge field. Other choices of connection are possible. For instance, in \cite{Gibbons:1982jg}, a fermion coupled to a composite connection built out of the fieldstrength of the gauge field was used to prove a lower bound on the mass of charged black hole solutions. This leads to the same local energy condition that we have, but different global conditions. It seems that (locally), the spinor used in that reference and ours are related by a field redefinition. That means that the local energy conditions we are going to get are the same, while globally the properties might differ.}. 
We can simplify \eq{ustij} using
\begin{equation} [\gamma^i,\gamma^j]=i\epsilon^{ij}_k\gamma_5\gamma^0\gamma^k\end{equation}
to obtain $T_{00}\mathbf{I}+A$, where
\begin{equation} A\equiv  \left(T_{0k}-\frac{iq}{8\pi G} F_{ij}\epsilon^{ij}_k\gamma_5\right)\gamma^0\gamma^k.\label{ustij2}\end{equation}
Now, the matrix $A$ is hermitian, and the absolute values of the eigenvalues of $A$ are just the square root of the eigenvalues of
\begin{equation}A^\dagger A=\mathbf{I}\left( \sum_k T_{0k}T^{0k}+\sum_{i,j}\frac{q^2}{(8\pi G)^2}F_{ij}F^{ij}\right).\end{equation}
Thus, we will have a version of the positive energy theorem as long as
\begin{equation}T_{00}^2\geq  \sum_k T_{0k}T^{0k}+\sum_{i,j}\frac{q^2}{(8\pi G)^2}F_{ij}F^{ij},\label{dec2}\end{equation}
which is the same local condition as in \cite{Witten:1981mf}. Hence, this is the energy condition that needs to be violated, instead of the DEC, when the manifold has a Spin$^c$ structure. The above argument works for any value of $q$; we see now that the weakest condition is achieved for $q=1$, so this value \eq{dec2} corresponds to the weakest energy condition one needs to impose so that the positive energy theorem holds. 

The energy condition \eq{dec2} is stronger than the DEC, and it involves the gauge fields in a nontrivial way. In the model presented in \cite{Blanco-Pillado:2016xvf}, it is the local energy condition that guarantees that the vacuum is stable. As illustrated there, it is possible to violate \eq{dec2} even with classical interactions,  allowing for the existence of bubbles of nothing. This shows again that there is generically nothing obviously wrong with violating the various energy conditions related to positive energy theorems. Actually,  the situation is completely  analogous to the familiar case of false vacuum decay in supergravity theories \cite{Cvetic:1992dc,Cvetic:1992st}, where supersymmetric vacua are protected dynamically by a BPS bound, rather than the milder DEC condition.

Another example of a modified positive energy theorem takes place when we consider manifolds which do not admit covariantly constant spinors with respect to the Levi-Civita connection, but which do when the spinors are charged under an additional $\mathbb{Z}_{n}$ bundle. Since all connections associated to a discrete gauge group are flat, there is no field strength term analogous to the second term in \eq{dec2}, and so one gets the same local energy condition as in the usual positive energy theorem, i.e. the dominant energy condition. 

We could try to apply this modified version of the positive energy theorem to the $G3$ quotient of $T^3$ discussed in Section \ref{sec:detailedTopology}\footnote{If the theory includes a gauge field, we could also apply the $\text{Spin}^c$ version of the positive energy theorem, using the local energy condition \eq{dec2}.}. This manifold does not admit covariantly constant spinors with respect to the metric connection, but because the parent $T^3$ does, the quotient admits covariantly constant $\mathbb{Z}_3$-charged spinors. However, there is an obstacle: $G3$ is in a nontrivial class in $\Omega_3^{\text{Spin}}(B\mathbb{Z}_3)=\mathbb{Z}_3$ since it has a nontrivial $\eta$ invariant \cite{Baer2000DEPENDENCEOT}.  As a result, this prevents us from applying the $\mathbb{Z}_{3}$ version of the positive energy theorem, \emph{even if there are no physical $\mathbb{Z}_3$-charged spinors in the theory}. Thus, there is no energy theorem that could guarantee the absence of a bubble of nothing for $G3$ in the pure Einstein theory, which is consistent with the fact that the bubble solutions constructed in Section \ref{sec:G3decay} are smooth even when turning off the higher derivative terms that violated the DEC.
\subsection{Including fluxes}\label{sec:fluxes}
In the previous Subsection we have studied how the dynamical obstruction coming from the positive energy theorem gets modified when spinors are charged under gauge fields in the extra dimensions. Here, we will discuss how the topological obstruction is modified  in the presence of gauge fluxes, i.e. how to properly define the bordism group which is relevant for the topological construction of a bubble of nothing in compactifications with fluxes.

So far, in Sections \ref{sec:nuts}-\ref{sec:detailedBON}, we have only considered bubbles of nothing involving geometry. This means that the bordism groups that appear naturally are the spin bordism $\Omega^{\text{Spin}}(\text{pt.})$. The extension to situations with gauge bundles or abelian $p$-form fluxes is straightforward: to construct a bubble, one must have a nulbordism in which the $p$-form fluxes also extend in a smooth way in the bubble. That is, whichever nonabelian bundles or fluxes are turned on must also extend to the bordism; the corresponding bordism groups are generically denoted  $\Omega^{\text{Spin}}(B^kG)$, where $G$ denotes the relevant gauge group and $k$ is the rank of the generalized gauge symmetry under consideration; we only consider $G$ abelian for $k\geq1$. 

All that matters for the physics is that it is possible to compute these groups, and that they provide the topological obstruction to the existence of bubbles of nothing for compactifications with fluxes. These very same bordism groups arise in the study of anomalies of non-abelian gauge theories; see  \cite{Garcia-Etxebarria:2018ajm,Wan:2018bns,Wan:2019fxh} for some computations and techniques. We will briefly explain the notation for the benefit of the curious reader. In general, $\Omega_d^{\text{Spin}}(M)$ refers to equivalence classes of $d$-dimensional manifolds  equipped with a map to $M$, under bordisms where we also demand that the maps to $M$ extend to the bordism. On very general grounds, a principal $G$-bundle on a manifold $X$ is equivalent to a map from $X$ to the classifying space of the group $BG$. This is an infinite-dimensional space equipped with a $G$-bundle $\chi_{BG}$, such that any principal $G$-bundle over a manifold $X$ is the pullback $f^*\chi_{BG}$ under the map $f:X\rightarrow M$. The picture also generalizes to higher generalized symmetries, and this is the logic behind the notation $\Omega^{\text{Spin}}(B^kG)$. 

As a simple example, consider compactification of M theory to $\text{AdS}_7\times S^4$. Since $S^4$ is trivial in $\Omega^{\text{Spin}}_4=\mathbb{Z}$, it would seem there is no topological obstruction to the construction of a bubble of nothing which eats up the $S^4$ factor. However, the solution is supported by $G_4$-flux. The corresponding bordism group, $\Omega_4^{\text{Spin}}(B^3U(1))=\mathbb{Z}^2$ has two generators, and one of them is precisely a sphere with one unit of $G_4$ flux. So the solution is nontrivial in bordism and there is a topological obstruction.

We must remark again that the topological obstructions we consider in this paper are low-energy considerations. It is often the case (and in fact, \cite{McNamara:2019rup} conjectures that it is \emph{always} the case) that one can evade the topological obstruction by including some UV objects. For instance, we could have a nulbordism for $\text{AdS}_7\times S^4$ if we take a five-dimensional ball whose boundary is $S^4$ and put an appropriate number of $M5$ branes to absorb the flux. Then the topological obstruction is absent and whether a bubble of nothing exists or not becomes again a question of dynamics. We won't have a bubble in this particular case due to supersymmetry, but all bets are off in more general scenarios. Still, the bordism groups we consider provide an obstruction to constructing a bubble without involving branes or other deep UV physics.

A sufficient way to ensure that some manifold with flux is nontrivial in bordism is to exhibit a nonvanishing bordism invariant. This is some hopefully easily computable quantity that is invariant under bordisms and that vanishes on the trivial class. An easy way to obtain bordism classes is via integrals of top forms that can be constructed out of the various fieldstrengths in the theory. So for instance, 
\begin{equation} \int p_1\end{equation}
is a bordism invariant of $\Omega_4^{\text{Spin}}$, and in $\Omega_6^{\text{Spin}}(BU(1))$ we have bordism invariants
\begin{equation} \int p_1c_1,\quad \int c_1^3\end{equation}
where $c_1$ is the Chern class of the $U(1)$ bundle. In the M theory example above, 
\begin{equation} \int G_4\end{equation}
is a bordism invariant. A similar example with no bordism invariant is the IIA compactification to $\text{AdS}_4\times\mathbb{CP}^3$ discussed in \cite{Ooguri:2017njy}. There is no 6-dimensional bordism invariant one can construct involving the $G_4-$flux or the metric, and so there would seem to be no topological obstruction. Indeed, a bubble of nothing which ``unwinds'' the flux was constructed in \cite{Ooguri:2017njy}. 

This construction presents us with a puzzle in the stringy embedding of our $T^3$ bubble constructed in Section \ref{sec:string}. There we argued that the bubble is a nulbordism of a $T^3$ threaded by 12 units of $H$-flux. Yet
\begin{equation} \int H_3\end{equation}
is certainly a bordism invariant in $\Omega_3^{\text{Spin}}(B^2U(1))$. But then, it shouldn't be possible to find a bubble of nothing for a fluxed $T^3$! The answer is that due to the heterotic Bianchi identity \cite{Liu:2013dna}, the $H$-flux and the geometry mix in a nontrivial way called a ``String'' structure (see Section 4.5 of \cite{McNamara:2019rup}). The relevant bordism group is then $\Omega_{3}^{\text{String}}=\mathbb{Z}_{24}$, generated by $S^3$ with one unit of $H$-flux on top of it. We will now argue that $T^3$ with 12 units of $H$-flux is actually bordant to $S^3$ with $24$ units of $H$-flux, which is in the trivial class in $\Omega_{3}^{\text{String}}$. As a result, it is perfectly consistent to have a bubble of nothing for it in our string theory embedding (there is no topological obstruction, just as for $\Omega_3^{\text{Spin}}=0$).

\begin{figure}
\begin{center}
\includegraphics[width=0.55\textwidth]{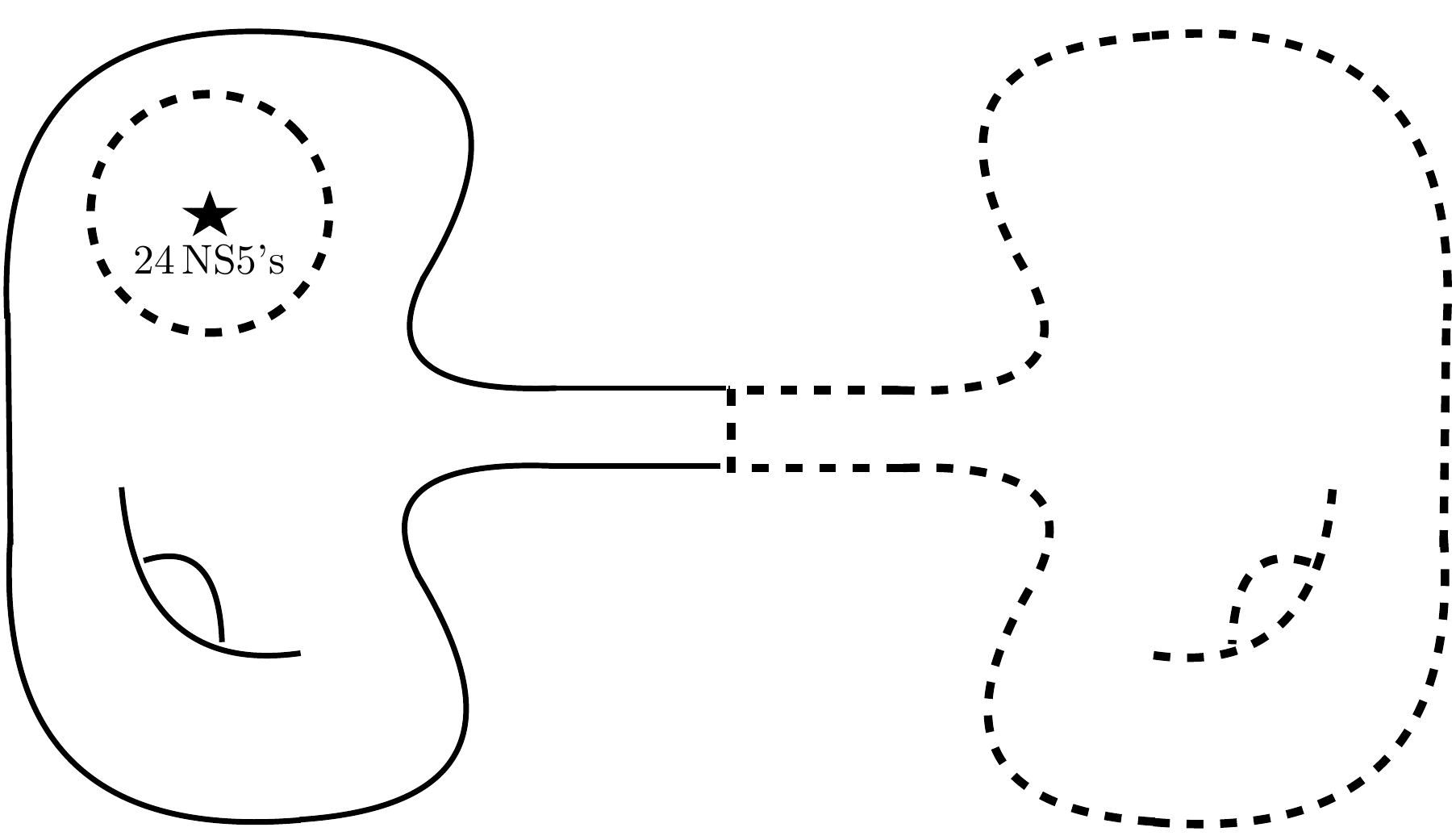}\end{center}
\caption{Construction of a bordism between a $S^3$ with 24 units of $H$-flux and a $T^3$ with 12 units of $H$ flux. One starts with K3 in the stable degeneration limit plus 24 NS5 branes, and then removes the part of the geometry encircled by dashed lines.}
\label{bordstring}
\end{figure}

To construct this bordism, consider $K3$ with $24$ NS5 branes on a point to cancel the tadpole, and deform to the stable degeneration limit where $K3$ grows an infinite tube with a $T^3$ cross Section, as in Figure \ref{bordstring}. While doing so, move all $24$ NS5-branes to one side of the tube. Then cut the geometry at the tube, and a small ball around the $24$ NS5-branes. The resulting manifold is depicted in Figure \ref{bordstring}, and it has two boundaries. One is the near-horizon region of the NS5-branes, which is an $S^3$ threaded by 24 units of $H$-flux. The other is a $T^3$ threaded by $12$ units of $H$-flux, since the part of the geometry we cut out on that side is precisely our $T^3$ nulbordism, which has $\int p_1/2=12$. The resulting manifold with $H$-flux is the bordism we wanted to construct. 

\subsection{Impact on String Phenomenology and Swampland}

Up to now, bubbles of nothing seemed to be a rare decay mode absent in typical string theory compactifications, since the original Witten's bubble is topologically forbidden unless we pick antiperiodic boundary conditions for the fermions on the shrinking circle. However, in this paper we have shown that this topological obstruction is absent for generalizations of Witten's bubble to $d$-dimensional shrinking manifolds as long as $\Omega^{\text{\text{Spin}}}_d=0$, i.e. the shrinking manifold belongs to the trivial class in the bordism group. As an example, we have explicitly constructed a new bubble of nothing for $T^3$ which is consistent even with periodic boundary conditions for the fermions, since $\Omega_3^{\text{\text{Spin}}}=0$. Hence, there may be many more bubbles of nothing in the low energy effective theories arising from string theory that one would have originally suspected. Since the topological obstruction is absent, the bubble of nothing might acquire a non-vanishing decay rate as soon as supersymmetry is broken, even if the breaking of supersymmetry occurs only at low energies and the internal space preserves some covariantly constant spinor.

Our results have clearly implications for toroidal $T^{d\geq 3}$ compactifications since topologically we can always deform it to the product $T^3\times T^{d-3}$ and construct the bubble of nothing for $T^3$. Hence, there is no topological obstruction to construct a bubble of nothing for any toroidal compactification of three or more internal dimensions. Even more interestingly, it is known that
\begin{equation}\Omega_6^{\text{Spin}}=0.\end{equation}
This means that the topological obstruction is also absent for any compactification on a six dimensional internal manifold preserving a spin structure. These are precisely the bordism groups which are relevant for type II and heterotic string compactifications to four dimensions. It should then be possible to topologically construct a bubble of nothing for any compactification of type II on a Calabi-Yau threefold for instance. Analogously, $\Omega_7^{\text{Spin}}=0$ implying that 11-dimensional M-theory compactifications to four dimensions are also not topologically protected to vacuum decay via bubbles of nothing, which includes the case of $G_2$ compactifications\footnote{Interestingly, $\Omega_{4}^{\text{Spin}}=\mathbb{Z}$, and $\Omega_8^{\text{Spin}}=\mathbb{Z}\oplus\mathbb{Z}$, so compactifications of string theory on $K3$ and F theory compactifications to four dimensions still have a geometric obstruction. Notice, though, that in \cite{Ooguri:1996me} it has been conjectured  the existence of new non-supersymmetric UV defects in these cases in order to guarantee triviality of the bordism group and the absence of global symmetries.}. We leave the topological construction of these very interesting bubbles for future work, but cannot refrain from pointing out that $CY_3$'s are conjectured to always admit a $T^3$ fibration \cite{Strominger:1996it,2014arXiv1408.6062C}, so our $T^3$ bubble might be embeddable via some sort of adiabatic argument at least in parts of the moduli space. 
 
 As explained in Section \ref{sec:fluxes}, the formalism we have introduced can be extended in a natural way to incorporate $p$-form fluxes. One only needs to consider the bordism groups of the corresponding classifying space. This means that our formalism also extends to flux compactifications: Whenever the fluxed manifold is in the trivial bordism class, there will be no topological obstruction to the existence of the bubble of nothing,  as in the string theory embedding of Section \ref{sec:heterotic} with H-flux.  

We should remark that, in this paper, we are only constructing bubbles of nothing for compactification manifolds trivial in the relevant bordism group. But what about nontrivial ones? Consider e.g. the circle with periodic spin structure, which generates $\Omega_2^{\text{Spin}}=\mathbb{Z}_2$. If we try to ``embed'' Witten's bubble in this case, the boundary conditions force us to introduce a ``spin defect'' -- a point-like defect at the origin which allows the fermions to have periodic boundary conditions around it (see e.g. \cite{Ford:1979pr}). This spin defect is clearly UV sensitive, and so, a candidate bubble can only exist in UV completions of the low-energy EFT that include such a defect as a dynamical object\footnote{We are only discussing topology; the spin defect might have a huge tension, making the bubble dynamically forbidden.}. One can incorporate the existence of these additional objects as a refinement of the bordism group which sees the UV , so that $\Omega_1^{\text{UV}}=0$ even if at low energies $\Omega_2^{\text{Spin}}\neq0$: Going to the UV has lifted the topological obstruction. Another example including fluxes might be $S^3$ with a nontrivial $U(1)$ 3-form flux on top of it, which corresponds to the generator in $\Omega_3^{\text{Spin}}(B^2U(1))$ described above. In this case, the flux can be ``eaten up'' by a brane that sources it, for instance, an NS5-brane in the case where the 3-form flux is NSNS. Again, $\Omega^{\text{UV}}=0$ while the group is non-trivial at low energies. One could expect that in string theory there always exists the adequate UV brane that can absorb the flux to guarantee triviality of the corresponding bordism group.

In fact, while completing this work, a new very interesting swampland conjecture appeared \cite{McNamara:2019rup} claiming that $\Omega^{\text{QG}}_d=0$ for any $d$. This has to be understood as the claim that the structure required for a consistent theory of quantum gravity must imply that the bordism group vanishes. In other words, cases in which $\Omega^{\#}_d\neq 0$ are not consistent compactifications in the sense that they need of additional defects to allow the bordism group to actually vanish. The reasoning underlying the conjecture is that $\Omega^{QG}_d=0$ in order to avoid the presence of global symmetries, since we can think of the different non-trivial classes of the bordism group as labelling different conserved charges and implying a global symmetry, as explained in Section \ref{sec:top_obs}. If this swampland conjecture holds, it implies that the topological obstruction to the existence of bubbles of nothing is never really there in the UV.

Of course, even if the topological obstruction is absent, there could still be some dynamical obstruction forcing the decay rate to be zero. This is what occurs in supersymmetric configurations for instance. However, once supersymmetry is broken (even at low energies) the dynamical obstruction might disappear allowing the vacuum to decay. As we have strongly remarked throughout the paper, not any breaking of supersymmetry will a priori allow for vacuum decay. If the manifold preserves an asymptotically covariantly constant spinor, then the decay rate will be non-vanishing only if the matter sector violates a certain energy condition which depends on the spin connection (e.g. the DEC \eqref{dec} for a spin structure and the modified condition \eqref{dec2} for Spin$^c$). In this paper, we have provided an example of an effective theory violating the DEC by introducing a higher derivative correction corresponding to a Gauss-Bonnet term, but depending on the context, other energy conditions involving gauge fields become relevant, as discussed in Section \ref{sec:DEC}. A violation of the DEC seems to be quite generic in QFT once quantum effects are taken into account \cite{Curiel:2014zba,Krommydas:2018fgs}\footnote{It would be interesting to see to what extent can positive energy theorems in gravity be reformulated using quantum versions of energy conditions such as the ANEC \cite{Hartman:2016lgu}. See \cite{Freivogel:2018gxj} for a proposal of such an energy condition in the gravitational context (although we do not know if this is strong enough to prove a positive energy theorem). If one finds a quantum energy condition leading to a PET, then we would demand that the corresponding quantum inequality is violated in a non-supersymmetric setup, which would be a restriction on the set of QFT's that can arise as low energy limits of quantum gravity (a Swampland constraint).}, but this is not necessarily the case for the modified energy condition \eq{dec2}. Now that the topological obstruction might always be absent in quantum gravity \cite{McNamara:2019rup}, the relevant question left for the future is whether the matter sector in a consistent, weakly coupled theory of quantum gravity necessarily violates all the relevant energy conditions when breaking supersymmetry. If this occurred,  bubbles of nothing would constitute a universal decay mode for any non-supersymmetric compactifications of string theory. This would also prove the swampland conjecture in \cite{Ooguri:2016pdq,Freivogel:2016qwc}, for which non-supersymmetric vacua must always be metastable at best. Although one might need to consider $\alpha'$ or $g_s$ corrections to explicitly see the violation of the energy condition and the consequent presence of the bubble of nothing. 

The difference between a topological and a dynamical obstruction resembles the difference between the swampland statement of not having global symmetries and the Weak Gravity Conjecture (WGC) \cite{ArkaniHamed:2006dz}. The WGC is a refinement of the former that quantifies how close we can  get to a configuration restoring a global symmetry, and therefore, necessarily constraints the dynamics of the theory. Hence, in the same way that the absence of global symmetries is intimately related to the absence of a topological obstruction for bubbles of nothing, the WGC (and similar conjectures) could then be related to the violation of the corresponding energy condition underlying the dynamical obstruction. This indeed seems to be the case for the modified energy condition \eqref{dec2} in the presence of gauge fields, which can be understood as a BPS bound as explained in \cite{Blanco-Pillado:2016xvf}. Thus, when breaking supersymmetry, a configuration violating this condition is precisely a configuration satisfying the WGC, as the latter has the rough interpretation of an anti-BPS bound at weak coupling \cite{Ooguri:2016pdq,Palti2017,Heidenreich_2019,Gendler:2020dfp}. Similarly, in \cite{Gibbons:1982jg}, \eqref{dec2} was used to prove that in general relativity any spacetime with mass $M$ and electric charge $Q$ has a mass $M\geq\vert Q \vert$. The condition \eqref{dec2} is likely violated by Schwinger pair production of WGC particles in the near-horizon geometry. We will study these connections more deeply in future work. 

Let us finally recall that this is a non-perturbative instability, so the vacuum can be very long lived. However, it can have dramatic consequences for AdS/CFT as the instability occurs instantaneously for an observer in the boundary, so the CFT is ill-defined. It could also provide a new argument to require supersymmetry preserved at some high energy scale in our universe, since otherwise it might not be sufficiently long-lived. We will explore these arguments in the future.

\section{Conclusions}\label{sec:conclus}
There is (to our knowledge) no single controlled example of an exactly stable nonsupersymmetric vacuum in string theory. And if indeed all nonsupersymetric vacua must decay \cite{Ooguri:2016pdq,Freivogel:2016qwc}, the natural question is what is the reason for this. The simple answer we have advocated in this paper is that a vacuum which carries no conserved charges should be able to pop in and out of existence, just like elementary particles are able do. The process by which a vacuum pops out of existence is a bubble of nothing\footnote{The reverse process is called a \emph{bubble from nothing} \cite{BlancoPillado:2011me}.}. Since the only charge that a vacuum can carry compatible with Poincar\'{e}  or AdS invariance is a supercharge, it would follow that any non-supersymmetric vacuum should admit a bubble of nothing. This is in line with the cobordism conjecture recently put forth in \cite{McNamara:2019rup}, but it is actually a stronger statement; while \cite{McNamara:2019rup} demands only that any vacuum can be continuously deformed to nothing, we would actually require that there is a physical process with a nonzero amplitude that allows the vacuum to decay.

In this paper we have tested this idea by trying to falsify it. A natural strategy to try and construct stable vacua is to construct vacua which have properties of supersymmetric vacua, but that  aren't actually supersymmetric.

In the context of bubbles of nothing, an old example comes from a circle compactification with the periodic spin structure for fermions \cite{Witten:1981gj}. This is the spin structure that would be required by supersymmetry, but we can have it in a nonsupersymmetric theory as well. As argued in \cite{Witten:1981gj}, the bubble of nothing that exists for the bounding spin structure is not present with periodic fermions; there is a topological obstruction to the existence of the bubble of nothing.

This seems like a robust way to engineer vacua which are at least safe from bubbles of nothing, but as we have shown, this is not really the case. All one needs to do is to add two extra compact dimensions -- their shape does not really matter --, and this topological protection is gone. Hence, once supersymmetry is broken, there might be again a bubble of nothing, this time involving the three compact extra dimensions. As a proof of concept, we have focused in a concrete example, that of compactification on $T^3$ with periodic (supersymmetry-compatible) boundary conditions along each of its one-cycles. There is a four-dimensional space with boundary (half a K3), which has $T^3$ with the right spin structure on its boundary, so unlike in the circle case, there is no topological obstruction to the existence of the bubble. Purely gravitational bubbles are classified by spin cobordism; the reason for the difference is that $\Omega_1^{\text{Spin}}=\mathbb{Z}_2$ while $\Omega_3^{\text{Spin}}=0$.

By turning on appropriate higher derivative terms (concretely, a supersymmetry-violating Gauss-Bonnet term), as well as nontrivial profiles for a dilaton in higher dimensions, we have been able to construct a metric on this half K3 with the appropriate boundary conditions as to serve as a true bubble of nothing for $\mathbb{R}^{D-3}\times T^3$ with finite action, given by  
\be
S_{BON}(\alpha) \propto \frac{1}{8 \pi G_{D-3}} \Big( \frac{24 \pi^2}{\cV_{T^3}}\alpha\Big)^{-(D-5)}.
\label{eq:actionG3-conclusions}
\ee
where $G_{D-3}$ is the lower-dimensional Newton's constant, $\alpha$ is the coefficient in front of of the Gauss-Bonnet term, and we have omitted some numerical factors that can be found in the main text. The solution has been constructed via a layered analysis involving an analytic approximation near the core of the bubble, coupled with numerical integration of Einstein's equations in the far away region, and a suitable gluing. We were able to exhibit our bubble as the zero mass solution of a one-parameter family of valid initial conditions to Einstein's equations which have arbitrarily negative mass. This also guarantees that the instability we found is genuine, and cannot be removed by any small corrections to our approximate solution to Einstein's equations. This is, to our knowledge, the first example of a bubble of nothing with supersymmetry-preserving boundary conditions, and where fermions are not charged under any gauge interaction.

Notice that the action of the bubble goes to infinity as the supersymmetry-breaking Gauss-Bonnet term is switched off, and thus the decay rate vanishes. This is what must happen, because when the GB term is switched off, the classical gravitational Lagrangian we are using has no way of knowing if it is actually a bosonic truncation of an underlying supersymmetric theory, and in this case, the decay would be impossible! From a purely gravitational point of view, what happens is that when the GB is switched off, the remaining matter obeys the Dominant Energy Condition. In this case, there is a Positive Energy Theorem that guarantees that decay of the vacuum is impossible. In this limit, therefore, there is a dynamical obstruction to the existence of the bubble.

The bubbles constructed here represent the minimal scenario where the topological obstruction is characterised by the spin bordism $\Omega^{\text{Spin}}_d$. However, in the presence of fluxes, or when the fermions are charged the spin bordism group must be generalised accordingly, which in some cases can lead to further possibilities to evade the topological constraint. The bubble of nothing constructed in \cite{Blanco-Pillado:2016xvf} is an example of this, where fermions are coupled to a $U(1)$ gauge field and the appropriate bordism is $\Omega^{\text{Spin}^c}$. As a consequence, while the model is compatible with supersymmetric fermions, there is no topological protection because  $\Omega^{\text{Spin}^c}_1=0$ is trivial. 
Interestingly, the bubbles  constructed there exhibit the same behaviour as those presented here: the compactification is unstable to decay to nothing if SUSY is broken, but in the supersymmetric limit the stability of the compactification is enforced dynamically, via a Coleman-DeLuccia type of mechanism.

In our model we get a decay of a $T^3$ with supersymmetric boundary conditions because we violate the DEC. The point is that a violation of the DEC can be quite generic whenever there is no supersymmetry: quantum corrections, higher derivative terms, and even a negative vacuum energy can all violate the DEC. So also in this example we find that the only thing that seems to guarantee vacuum stability in a robust way is supersymmetry. Incidentally, the DEC is not the only energy condition that can lead to a Positive Energy Theorem; in theories with a $\text{Spin}^c$ structure, there is a modified energy condition, involving the $U(1)$ gauge field, which takes the form of a BPS bound  in the the case of \cite{Blanco-Pillado:2016xvf}. The Positive Energy Theorem associated to the $\text{Spin}^c$ structure also guarantees stability of charge black hole solutions in the classical gravity theory \cite{Gibbons:1982jg}, and hence it has a natural relation to the Weak Gravity Conjecture; it is possible that theories with WGC states are precisely those that violate this modified energy condition. This opens a new avenue to explore in the future, namely whether the WGC and similar swampland constraints precisely imply that the relevant energy conditions underlying the dynamical obstruction to the construction of the bubble are violated in quantum gravity, allowing non-supersymmetry vacua to always decay to nothing.

Incidentally, with a slight modification of our bubble we have been able to provide the missing bubble of nothing for the last class in the classification of \cite{Acharya:2019mcu}. This reference classified all nonsupersymmetric Ricci-flat quotients of $T^3$, being able to find a bubble of nothing for each of them except for one. Our results fill in this gap, and are again aligned with the idea that nonsupersymmetric vacua always admit dynamical (i.e. they are actual solutions to the equations of motion) bubbles of nothing.

In order to make sure that the effects we are observing are not some artifact of the particular EFT we chose, we have embedded our solution into heterotic/type I string theory. The supersymmetry breaking effects are related in this case to the turning on of NS-NS flux on $T^3$. From the dual type II Calabi-Yau perspective, these correspond to both IASD and ISD fluxes, so that supersymmetry is broken. A natural question is then to what extent can one generalize the results of our toy model to more interesting and complicated flux compactifications. Since the bordism groups $\Omega_{6,7}^{\text{Spin}}=0$, one should worry about this phenomenon in any nonsupersymmetric string  or M theory compactification to four dimensions. For instance, whatever internal manifold one uses in the KKLT construction, there will be (topologically) a bubble of nothing. The IASD fluxes sourced by the gaugino condensate are of the same kind that lead to a bubble in our toy model. Of course, this would just be a non-perturbative instability, leading to a very long lifetime (see also \cite{Dine:2004uw,deAlwis:2013gka} for related discussions).

Our results can lead to more dramatic implications in the realm of holography. A non-perturbative instability is a killer in AdS/CFT, since it will be triggered near the boundary and then reach the center of AdS in a finite amount of time \cite{Harlow:2010az,Ooguri:2016pdq}. Our results suggest  a very specific strategy to construct bubble instabilities in nonsupersymmetric AdS solutions. Now, the fact that $\Omega_{5,6,7}^{\text{Spin}}=0$ suggests that any nonsupersymmetric $\text{AdS}_{4,5}$ solution in string theory might admit a bubble of nothing.  The construction in \cite{Ooguri:2017njy} is an example of this. In some of these cases, to construct a bubble of nothing, one must also kill a flux, which forces the introduction of additional branes; this is controlled by the existence of bordism invariants in the supergravity theory. 

Bubbles of nothing are a universal instability, but usually not the leading one (although they can be \cite{Ooguri:2017njy}). When present, perturbative and nonperturbative brane instabilities are usually less suppressed. If there is a bubble of nothing in the real world,  it might be highly suppressed, as it seems reasonable that its action will be at least as large as the size of the internal manifold in Planck units. But all these other instabilities are very model dependent and sometimes can be hard to detect from a low energy EFT perspective, while bubbles of nothing are, at the moment, the best candidates to constitute a universal instability of any non-supersymmetric vacua coming from a higher dimensional compactification of quantum gravity. 

In this work we have argued that, due to the general connection between bordisms, positive energy theorems and instabilities in quantum gravity, bubbles of nothing are far more common than what was previously thought and are in fact  ``lurking around the corner'', ready to hit us as soon as supersymmetry is broken in the slightest. Hence, we advise the reader to enjoy life as if there was no tomorrow, because nothing is certain in string compactifications.

\subsubsection*{Acknowledgements}

We thank Gary Horowitz, Cumrun Vafa, Thomas Hertog, Bobby Acharya, Jose Juan Blanco-Pillado, Fernando Marchesano, Tom\'as Ort\'in,  Pavel Krtous, Martin Zofka,  and especially Patrick Draper. We thank everyone who helped us with choosing a title. We also thank the ``String Theory and the Hidden Universe'' program at the Aspen Center for Theoretical Physics and the String Swampland 2020 program at KITP, where parts of this work were carried out. This research was supported in part by the National Science Foundation under Grant No. PHY-1748958. KS is supported by the Czech science foundation GA\v CR grant (19-01850S). IGE is partially supported by STFC consolidated grant ST/P000371/1. During an earlier stage of this project, MM was supported by an FWO Postdoctoral fellowship at KU Leuven and IV was supported by the Simons Foundation ``Origins of the Universe'' program at Cornell University. MM and IV are currently supported by Grant 602883 from the Simons Foundation.

\appendix

\section{BON decay of a non-supersymmetric flux compactification}
\label{app:fluxBON}
In this appendix we consider a generalisation of the model presented in Section \ref{sec:nuts} which incorporates a scalar potential to fix the three torus  volume $T^3$ of the vacuum \eqref{vacuum}. We will show that our bubble is compatible with this deformation, and also that  in this setting the BON nucleation radius is fixed as a consequence of the ingredients inducing the moduli potential ($H-$flux on $T^3$ and a cosmological constant).  The model we consider is characterised by the following action in the string frame
\be
S = -\frac{g_s^2}{16 \pi G_D} \int \sqrt{g} \rme^{-2 \phi} \left[ R + 4 (\nabla \phi)^2 - \ft{1}{12} H^2  + \ft{\alpha' \beta}{8} R_{GB}^2 -2 \rme^{-\frac{4}{D-4} \phi} \Lambda \right]
\label{eq:fluxModel}
\ee
with the Bianchi identitiy for $H_{MNP}$ given by
\be
dH = \frac{\alpha'}{4} \text{tr} R \wedge R.
\label{eq:bianchiH}
\ee 
Note that the action reduces to a consistent truncation of  heterotic string  theory with first  order $\alpha'$ corrections with the choice $\beta=1$,  and setting the cosmological constant to zero $\Lambda=0$. The equations of motion read
\bea
R_{MN} &=& -2 \nabla_M \nabla_N \phi + \ft14 H^2_{MN} +\ft{2}{D-4} \Lambda \rme^{-\frac{4}{D-4} \phi} g_{MN} \nonumber \\
&&- \ft{\alpha' \beta}{4} \[R_{MRST} R_{N}^{\phantom{N} RST} -2 R_{MSNT} R^{ST} -2 R_{MS}R_N^{\phantom{N}S} + R R_{MN}\],\\
\nabla_{(D)}^2 \phi &=& 2 (\pd \phi)^2 - \ft1{12} H^2 +\ft{\alpha' \beta}{16} R_{GB}^2-\ft{2}{D-4} \Lambda \rme^{-\frac{4}{D-4}\phi }
\eea
and 
\be
\pd_M(\rme^{-2 \phi} \sqrt{g} H^{MRS})=0.
\ee
\subsection*{Flux vacuum.}  Setting $\alpha'=0$, the  previous model admits a vacuum solution of the form $\mathrm{AdS}_{D-3} \times T^3$ 
\be
ds^2 = ds^2_{\mathrm{AdS_{D-3}}} + h_{\bar \alpha \bar \beta}^{T^3} dy^{\bar \alpha} dy^{\bar \beta},
\label{eq:AdSvacuum}
\ee
with constant dilaton $\phi= \phi_{\infty}$, and $h_{\bar \alpha \bar \beta}^{T^3}$ the metric on $T^3$.  To obtain this vacuum we turn on a flux $m=\int_{T^3} H$ on the three torus $H_{\bar \alpha \bar \beta \bar \gamma} = \frac{m}{\cV_{T^3}} \, \,  \epsilon_{\bar \alpha \bar \beta \bar \gamma}$ where $\epsilon_{\bar \alpha \bar \beta \bar \gamma}$ is the totally antisymmetric tensor on $T^3$, and $\cV_{T^3}$ is the three torus volume on the vacuum. 

From the Einstein equation  on $T^3$ it can be seen that  
the combined effect of the cosmological constant and the flux induces  an effective potential for the dilaton and the $T^3$ volume modulus, which imposes the relation
\be
\cV_{T^3}^{-2}= -\frac{4}{(D-4)} \frac{\Lambda}{m^2} \rme^{-\frac{4}{D-4}  \phi_\infty } >0.
\label{eq:fluxConstraint1}
\ee
Regarding the $\text{AdS}_{D-3}$ components the corresponding line element can be expressed as follows using a deSitter slicing parametrisation
\be
ds^2_{\text{AdS}} = L^2 \sinh^2(\rho/L) (-dt^2 + \cosh^2 (\rho)d\Omega_{D-4}^2) + d \rho^2.
\ee
 The   scalar curvature of $\mathrm{AdS}_{D-3}$ is then given by $R_{(D-3)} = -(D-3) (D-4)/L^2$,  where  $L$ is the AdS scale. The scale $L$ can be  determined from the Einstein equation on the non-compact directions, which implies
\be
L^{-2} = -\frac{2}{(D-4)^2}\Lambda \, \rme^{-\frac{4}{D-4}\phi_\infty} >0.
\label{eq:AdSscale}
\ee
Then,  the expectation value of the dilaton controls both AdS scale and the volume of the $T^3$ compact space. Since the non-compact directions of the decaying vacuum  \eqref{eq:AdSvacuum} are now AdS instead of Minkowski, the BON solutions we will now construct should have different asymptotic behaviour. In particular,  using the BON ansatz \eqref{gralBONansatz} we will need to impose
\be
\lim_{\rho\to \infty }L \, \pd_\rho W/W = 1
\label{eq:asymptoticAdS}
\ee
instead of \eqref{eq:BONbcs1}. Nevertheless, as we shall see below, when the AdS radius greatly exceeds that of the bubble $L \gg \cR$, the BON solution presented of the main text represents a good characterisation of the instanton  mediating the decay of \eqref{eq:AdSvacuum} in the region $\rho \ll L$.
\subsection*{Construction of the BON solution}

The construction of the BON solution proceeds in complete analogy to our discussion in Section \ref{sec:detailedBON}, and therefore here we will only outline the main differences with that analysis. First, the presence of an $H-$flux in the $T^3$ and the Bianchi identity \eqref{eq:bianchiH} imply that the presence of a non-zero three form on the BON solution. However, its energy momentum tensor scales as $\cO(\alpha'{}^2)$ (see Section \ref{sec:string}), and therefore its  back reaction on the geometry can be safely neglected in the construction of the inner-bubble geometry (region {\bf II.}), where we consider only  $\cO(\alpha')$ terms. Similarly, we can also tune the expectation value of the dilaton $\phi_\infty\gg1$ so that the cosmological constant produces small $\Lambda \rme^{-\frac{4}{D-4}\phi_\infty}=\cO(\alpha'{}^2)$ effects in the inner-bubble region (or alternatively tune $\Lambda\ll1$). Therefore, in the region ${\bf II.}$ we find that  the warp factor of the sphere  is unperturbed to first order $W^{(1)} = 0$, the perturbation of the volume modulus $\varphi^{(1)}$  determined by an equation analogous to \eqref{eq:poissonGB} 
\be
 \nabla^2 \varphi^{(1)} =\beta\ft{\alpha'}{16} R_{GB}^2,
\ee
with the variation of the dilaton also given by $\phi^{(1)} = \varphi^{(1)}$. Regarding the three form flux the appropriate ansatz is
\be
\text{Region {\bf II.}}:\qquad  H_{\alpha \beta \gamma} = -2 \beta^{-1}\epsilon_{\alpha \beta \gamma}^{\phantom{\alpha \beta \gamma}\delta } \, \pd_\delta \varphi^{(1)}, \qquad H_{\mu\nu\rho}=H_{\mu\alpha\beta}=H_{\mu\nu\alpha}=0,
\ee 
where $\epsilon_{\alpha \beta \gamma \delta}$ is the totally antisymmetric tensor associated to the  background Calabi-Yau geometry on the bordism $\cB$. It is easy to check that this ansatz satisfies both the Euler-Lagrange equation and the Bianchi identity $H$ to order $\cO(\alpha')$.

In the outer-bubble regime (region {\bf I.}), we can also proceed similarly to our construction in the main text. We use line element \eqref{eq:outerAnsatz} for the bubble geometry, with  $C(\rho) = R_{\text{kk}} \rme^{\varphi - \varphi_\infty}$, and  we impose the ansatz $\varphi(\rho) = \phi(\rho) - \phi_\infty + \varphi_\infty$. In this region the appropriate ansatz for the three-form  is 
\be
\text{Region {\bf I.}}:\qquad H_{\bar \alpha \bar \beta \bar \gamma} = 2 \beta^{-1}\, \epsilon_{\bar \alpha \bar \beta \bar \gamma} \, \varphi'(0) \, \rme^{\varphi(\rho)- \varphi(0)}
\ee
and all other components vanishing..  It can be  checked easily  that the form $H$ is continuous across common boundary of the regions {\bf II.} and  {\bf I.} (which we have set at $\rho_*=0$),  and that it  satisfies the equations of motion and the Bianchi identity \eqref{eq:bianchiH} regardless of the form of $\varphi(\rho)$. The value of $\varphi'(0)$ can be obtained from the matching conditions between the inner and outer bubble regimes similar to \eqref{T3matching}, which in the present model give
\be
\varphi'(0) =\beta \frac{24 \pi^2 \alpha'}{\cV_{T^3}} \rme^{3 \Delta \varphi}, \qquad \text{with} \qquad  \Delta \varphi = \varphi_\infty - \varphi(0).
\label{eq:fluxConstraint2}
\ee
Requiring  the three form to match the asymptotic configuration of the vacuum we find the relations 
\be
\frac{m}{\cV_{T^3}^*}= 2 \beta^{-1} \varphi'(0)\qquad \Longrightarrow \qquad  m = 48 \pi^2 \alpha'.
\label{eq:fluxConstraint3}
\ee
\begin{figure}[t]
\centering \hspace{-.5cm}\includegraphics[width=0.55\textwidth]{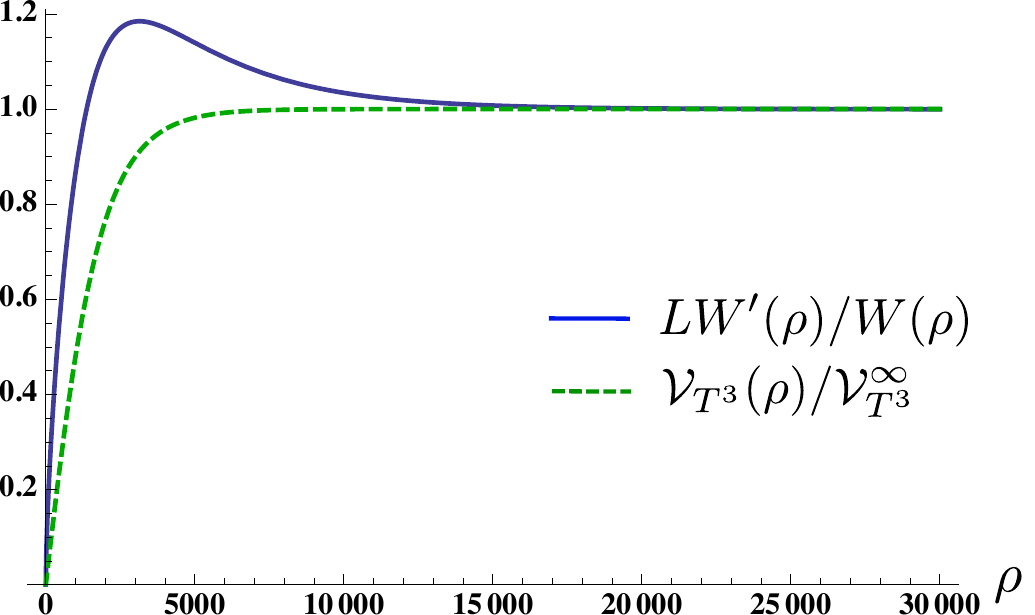} 
\caption{Outer-bubble geometry for the model \eqref{eq:fluxModel} with $D=7$, $\Delta \beta =8.61 \cdot 10^{-5}$, AdS scale $L = 6.78 \cdot10^{3}$, and the scale of  $\alpha'$ corrections given by $\frac{24 \pi^2\alpha'}{\cV_{T^3}} =1.8 \cdot 10^{-4}$. The solid line represents the warp factor on the sphere $L W'/W$, and the dashed line the three torus volume $\cV_{T^3}(\rho)/\cV_{T^3}^\infty$. The BON radius is $\cR = 5.46 \cdot 10^3$,  just slightly smaller that the AdS radius $L$. The plot displays how the non-compact space approaches to the $\mathrm{AdS}_{4}$ for large $\rho\to\infty$, i.e. $L W'/W\to1$.  
 } 
  \label{fig:AdSBON1}
\end{figure} 
The remaining fields $\varphi(\rho)$ and $W(\rho)$ should satisfy the system of equations
\bea
\hat \varphi''(\rho) + (D-4) \frac{W'}{W} \hat \varphi' + \hat \varphi'{}^2 + 2 \beta^{-2} \hat \varphi'(0)^2 \left (\rme^{-6 \hat \varphi} - \lambda \rme^{-\frac{4}{D-4} \hat \varphi} \right) &=&0,\nonumber \\
\frac{W''}{W} - (D-5) \frac{(\cR^{-2}- W'{}^2)}{W^2} + \frac{W'}{W} \hat \varphi' - 2 \beta^{-2}  \varphi'(0)^2 \lambda \rme^{-\frac{4}{D-4} \hat \varphi} =0,
\label{eq:AdSbonSys}
 \eea
where we have defined $\hat \varphi\equiv \varphi- \varphi(0)$ and the parameter $\lambda \equiv \beta^2 \frac{|\Lambda|}{\phi'(0)^2 (D-4)} \rme^{-\frac{4}{D-4} (\phi_\infty -\Delta \varphi)} $. Using the relations  \eqref{eq:fluxConstraint1} and \eqref{eq:fluxConstraint2} it follows that  the parameter $\lambda$ can be rewritten as
\be
\lambda = 
 \rme^{-\frac{2(3D-14)}{D-4} \Delta \varphi}.
\label{eq:lambdaConstraint}
\ee
The previous equations should be solved subject to the boundary conditions 
\bea
\hat \varphi(0) =0, &\qquad& \hat \varphi'(0) = \varphi'(0) = \beta \frac{24 \pi^2 \alpha'}{\cV_{T^3}} \rme^{3 \Delta \varphi},\nonumber \\
 \qquad W(0) = 1, &\qquad& W'(0)=0,
\eea
at  $\rho =0$, and  requiring that for $\rho\to \infty$ we have $\hat \varphi \to \varphi_\infty- \varphi(0) = \Delta \varphi$, which in turn automatically guarantees \eqref{eq:asymptoticAdS}, as this implies that the $T^3$ volume  and the dilaton satisfy the in the vacuum relation \eqref{eq:fluxConstraint1}. To meet these conditions we have at our disposal the  parameters $\Delta \varphi$ and the BON radius $\cR$, which we can freely vary a priori. Therefore the conditions presented above  are no sufficient to fix completely the boundary value problem  what would lead, without further input, to families of BON solutions parametrised by the nucleation radius $\cR$, similar to those  discussed in the main text.
To see how the presence of fluxes and a cosmological constant determine the BON radius we need to consider  the $\rho-\rho$ component of the Einstein's equations.  While this equation is trivially it is satisfied by construction to order\footnote{The boundary conditions at $\rho=0$ are obtained via the matching procedure  from a solution to the \emph{complete} set of Einstein's equations to order $\cO(\alpha')$ in the inner bubble region. In particular the matching guarantees that the $\rho-\rho$ Einstein's equation  is satisfied at $\rho=0$  to order $\cO(\alpha')$. } $\cO(\alpha')$, when considered to order $\cO(\alpha'{}^2)$ it leads  to an additional constraint on the outer-bubble  configuration. The resulting equation is   
\bea
\hspace{-.2cm}\frac{(D-4) (D-5)}{\cR^2 W^2}&=&  (D-4)(D-5) \frac{W'{}^2}{W^2}+2 (D-4)\frac{W'}{W} \varphi'- 2 \hat \varphi'{}^2 \nonumber\\ &&+2 \beta^{-2}  \hat \varphi'(0)^2\left(  \rme^{-6 \hat \varphi} -  (D-4)  \lambda \rme^{-\frac{4}{D-4} \hat \varphi} \right),  
\eea
which evaluated on $\rho=0$ gives an expression analogous to \eqref{eq:BONradiusScaling} after using \eqref{eq:fluxConstraint2}
\be
\cR= \rme^{-3 \Delta \varphi}\,  \sqrt{\frac{(D-4) (D-5)}{2( 1-\beta^2 -  (D-4)  \lambda )}}  \left(\frac{24 \pi^2 \alpha'}{\cV_{T^3}} \right)^{-1},
\ee
with $\lambda$ given by \eqref{eq:lambdaConstraint}.  With this additional constraint the boundary value problem becomes completely determined, and  thus it only remains to find the value of $\Delta \varphi$ (or equivalently $\cR$), what we can do solving the equations  with numerical methods. The result of such computation is  displayed in Figure \ref{fig:AdSBON1}, where we show a BON with nucleation radius just smaller than the AdS scale $\cR \lesssim L$. This solution  illustrates how the higher order $\alpha'$ effects (not considered in the main text) might fix the nucleation radius of the bubble in terms of the parameters of the compactification.

\begin{figure}[t]
\centering \hspace{-.5cm}\includegraphics[width=0.55\textwidth]{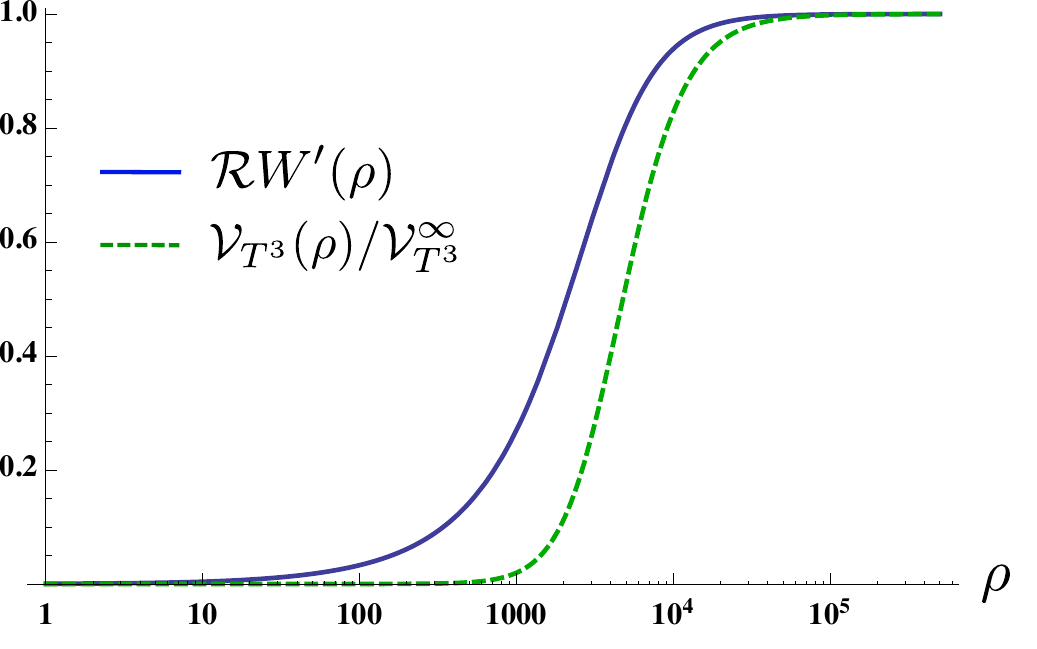} 
\caption{ Outer-bubble geometry  for the model \eqref{eq:fluxModel} with $D=7$ in the limit $\beta \to 1$ with $\cR/L=8.05\cdot 10^{-5}$. The AdS scale is $L = 4.57 \cdot10^{7}$, and the scale of  $\alpha'$ corrections is given by $\frac{24 \pi^2\alpha'}{\cV_{T^3}} =2.68 \cdot 10^{-8}$. The solid line represents the warp factor on the sphere,  expressed in terms of  $\cR W'(\rho)$, and the dashed line is the three torus volume $\cV_{T^3}(\rho)/\cV_{T^3}^\infty$. The BON nucleation radius $\cR = 3.68 \cdot 10^3\ll L$ is much smaller that the AdS scale.   In the regime $\cR\ll \rho\ll L$ the bubble spacetime approaches a configuration $\mathbb{M}_4\times T^3$, i.e. $W'(\rho)\to\cR^{-1}$, mimicking the Minkowski bubbles we discussed in the main text. } 
  \label{fig:AdSBON2}
\end{figure} 

Combining the previous equation  with \eqref{eq:fluxConstraint3}, \eqref{eq:fluxConstraint1} and \eqref{eq:AdSscale} we obtain an alternative expression for the nucleation radius in terms of the AdS scale
\be
\cR=  \sqrt{\frac{(D-5)}{( 1-\beta^2 -  (D-4)  \lambda )}} \; \rme^{-3 \Delta \varphi} \, L.
\ee
In particular we can see that the embedding of the solution presented in the main text, which asymptotes to Minkowski spacetime instead of AdS, can be achieved in the limit $\cR\ll L$, that is, when the AdS scale is far larger than the BON radius. In this limit the curvature of the AdS space is negligible near the bubble core, and then the BON spacetime in this region is expected to be similar to a bubble of nothing for a vacuum where the non-compact space is Minkowski.  To find such  solution we note that  the limit $\cR \ll L$ can be achieved provided we tune the parameter $\beta\to 1^-$, and simultaneously  $ \rme^{-6 \Delta\varphi}/(1-\beta)\to0^+$, so that the expression for the bubble radius (written in the form of  \eqref{eq:BONradiusScaling} and \eqref{eq:T3RadiusDlarge}) reduces to
\be
\lim_{\beta\to1}\cR/L= \sqrt{\frac{(D-5)}{2 \Delta \beta}} \; \rme^{-3 \Delta \varphi} \ll1 \quad \Longrightarrow \quad \cR\to \frac{\rme^{-3 \Delta \varphi}}{2}\,  \sqrt{\frac{(D-4) (D-5)}{\Delta \beta }}  \left(\frac{24 \pi^2 \alpha'}{\cV_{T^3}} \right)^{-1}. 
\ee
where $\Delta \beta\equiv 1-\beta$. Moreover, comparing this result with the expression for the radius  for the  asymptotically flat bubble  in limit $\Delta \varphi \to \infty$ \eqref{eq:T3RadiusDlarge},   we find the relations 
\be
\lim_{\Delta \varphi\to \infty} \Delta \varphi = -\ft12 \log\Big(\frac{\Delta \beta (D-5)}{D-4}\Big)\gg1,\qquad  \lim_{\beta\to1}\cR/L = \frac{(D-5)^2}{\sqrt{2} (D-4)^{3/2}} \Delta \beta.
\ee
As we anticipated in the main text, in this case the bubble radius is fixed by the higher $\alpha'$ corrections. 
  To confirm the existence of this branch of solutions we resort again to numerical methods to solve \eqref{eq:AdSbonSys}, and we find the BON configuration presented in Figure \ref{fig:AdSBON2}, which has $\cR/L=8.05\cdot 10^{-5}$. This plot shows the bubble configuration well inside the AdS radius  $\rho\ll L$. It can be observed  that the non-compact component of spacetime is indistinguishable from flat space outside the bubble core $\cR\ll \rho$. Actually the profile functions plotted in Figure \ref{fig:AdSBON2} match perfectly well the BON solution \eqref{eq:T3limitDinf} discussed in the main text, which describes the decay of the $\mathrm{M}_{D-3} \times T^3$ vacuum. 
  
\begin{figure}[t]
\centering \includegraphics[width=0.45\textwidth]{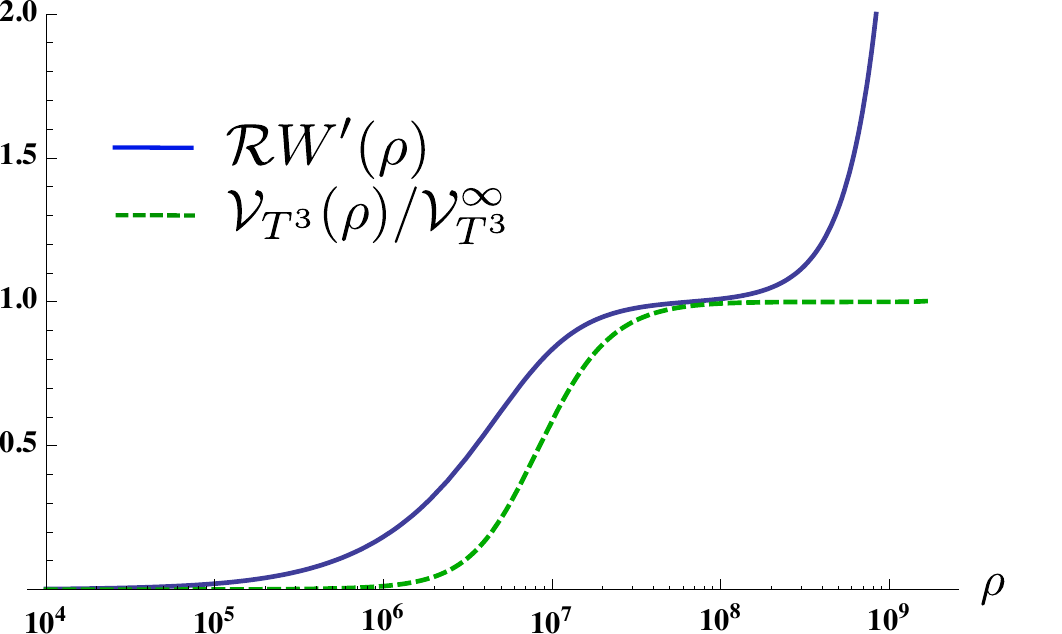}
\centering\includegraphics[width=0.45\textwidth]{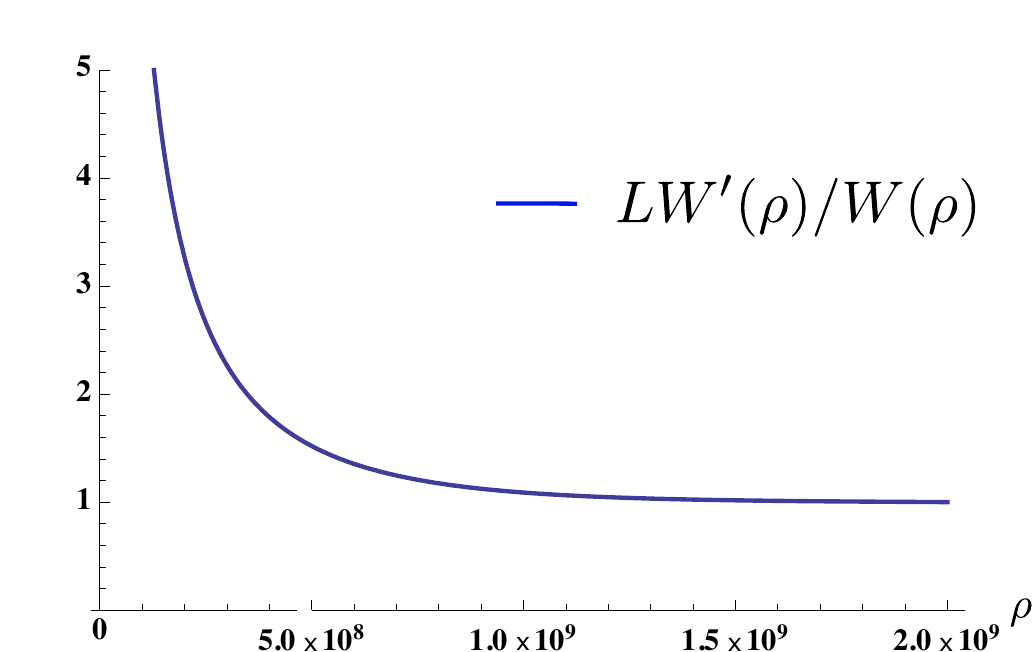} 
\caption{ Outer-bubble geometry for the model \eqref{eq:fluxModel} with $D=7$ in the intermediate regime $\cR/L=1.13\cdot 10^{-2}$. The AdS scale is $L = 6.34 \cdot10^{8}$, and the scale of  $\alpha'$ corrections is given by $\frac{24 \pi^2\alpha'}{\cV_{T^3}} =1.93 \cdot 10^{-9}$. LEFT: The solid line represents the warp factor on the sphere  $\cR W'(\rho)$, and the dashed line the three-torus volume $\cV_{T^3}(\rho)/\cV_{T^3}^\infty$. Just outside the bubble core, $\cR \lesssim\rho\ll L$ the non-compact space is almost $\mathbb{M}_{4}$ ($W' \approx \cR^{-1}$), and approaches an $\text{AdS}_{4}$  in the limit $\rho\to\infty$.  RIGHT: Warp factor on the $S^3$ sphere expressed in terms of $L W'/W$ far from the bubble $\cR \ll \rho\sim L$.  The plot shows the onset of the AdS geometry far form the bubble, i.e.  $L W'/W\to 1$. }   \label{fig:AdSBON3}
\end{figure}

For completeness in Figure \ref{fig:AdSBON3} we have also displayed an intermediate regime with smaller ratio $\cR/L=1.13\cdot 10^{-2}$. In the plots we can see two regimes of the bubble geometry:  in the left plot of Fig. \ref{fig:AdSBON3}  we see that just outside the bubble core  $\cR\lesssim\rho\ll L$  the spacetime is approximately flat, and the configuration is similar to the bubbles discussed in the main text;  in the right plot we can see that far from the bubble core $\cR\ll \rho$ the non-compact component becomes $\mathrm{AdS}_{D-3}$.


\subsection*{Step-by-step decay to Nothing}
Above and in the main text, we have described a decay process where a single instanton destroys all the flux, and the asymptotic $T^3$ geometry in a single step. 

 As shown in \eq{eq:fluxConstraint1}, the volume of the internal $T^3$ is controlled by the $H$ flux, and vanishes as $m\rightarrow0$. Since a $T^3$ of zero size is morally ``nothing'', this suggests the existence of  Euclidean solutions that  would allow one to discharge one unit of $H$ flux  at a time, arriving to the picture in \cite{Brown:2011gt}, where the bubble of nothing is described precisely as the limiting  transition where all the flux disappears in one go. 
 
 A simple way to realize this picture is to modify \eq{eq:fluxConstraint3} to consider a $T^3$ with a flux $m=48\pi^2\alpha' k$, for $k$ an arbitrary integer. This construction allows the flux to become arbitrarily large, and thus to have an internal $T^3$ volume which is also arbitrarily large.   At least topologically, it is possible to obtain a  bordism $\cB_4^{(k)}$ describing the decay to nothing of this compactification  as follows. First consider  $k$ copies of the nulbordism for $T^3$, the half K3 $\cB_4^{(1)}$,  and remove a small region $T^2\times \mathcal{D}$ in the neighbourhood of the origin from $k-1$ of the copies.    Each of these modified $k-1$ copies have now two boundaries, and both of them are topologically a  trivial $T^2\times S^1$ fibration. Thus, the nulbordism $\cB_4^{(k)}$ can be obtained  gluing in sequence these $k-1$ copies by identifying their $T^2\times S^1$ boundaries, and finally attaching the unmodified half K3 to one of the two ends. The resulting (topological) bubble would mediate the direct decay to nothing; while the modified bordism $\cB^{(1)}$,  with the region $T^2\times \mathcal{D}$ removed,  describes the topology of an instanton mediating the decay from a configuration with  $m=48\pi^2\alpha' k$ to $m=48\pi^2\alpha' (k-1)$. Hence, in a very literal way, the bubble of nothing is the limit of the small bubbles when the flux changes by $k$ units. 
 
In the above, the flux jumps are always a multiple of $48\pi ^2\alpha'$. This is twelve times the fundamental flux quantum.  We can make a similar construction where the flux  changes by a quantity which is not a multiple of 12, if we allow topology change in the process. The idea is to split the $N=12$ degenerations in our bordism between a set of $N'<N$ and $N-N'$ degenerations, and ``hide'' $N-N'$ of them inside the disk $\mathcal{D}$ on the base that we subsequently remove. The resulting manifold will be a nontrivial torus fibration over $S^1$ (for instance, for $N'=8$ it can be the $G3$ manifold as discussed in Subsection \ref{sec:detailedTopology}), and it will have $4\pi ^2\alpha'\, N'$ units of flux threading it. We can now lower $N'$ one step at a time, to arrive at a similar picture as above. Other options include using NS5 branes or nontrivial gauge bundles to change the asymptotic flux using the Bianchi identity without changing the topology of the bordism, but these do not relate to the bubble of nothing as straightforwardly as the configurations we described.

\section{Einstein frame action and equations of motion.}
\label{app:Eframe}

In this Appendix we discuss how to write down our results in Einstein frame. The action of our model in the Einstein frame has the form 
\be
S_{\text{E}} = -\frac{1}{16 \pi G_{{D-3}} \cV_{\cC_3}} \int_{\cM_D} d^Dx \sqrt{-g} \Big[ R -\frac{4}{D-2} (\nabla \phi)^2+\frac{\alpha}{8} \,  \rme^{-\frac{4 \phi}{D-2}} R^2_{GB}\Big].
\ee
When changing the conformal frame  we have assumed that the background is Ricci-flat and with a constant dilaton $\phi=\phi_0$, what simplifies greatly the transformation of the Gauss-Bonnet term (see e.g. \cite{Carneiro:2004rt}). The Einstein's equations in this frame read
\bea
G_{MN} &=& \frac{4}{D-2} \Big(\nabla_M \phi\nabla_N\phi - \frac{1}{2} (\nabla \phi)^2 g_{MN}\Big) +\frac{\tilde \alpha}{16}g_{MN} R_{GB}^2 - \frac{\tilde \alpha}{4}\Big[ R_{MRST} R_{N}^{\phantom{N}RST}\nonumber \\
&&-2 R_{MSNT} R^{ST} - 2 R_{MS} R_{N}^{\phantom{N}S} + R R_{MN}\Big],
\eea
and the dilaton equation is 
\be
\nabla^2_{(D)} \phi = \frac{\tilde \alpha}{16} R_{GB}^2.
\label{eq:appDilatonEom}
\ee
where we defined $\tilde \alpha\equiv \alpha  \rme^{-\frac{4\phi}{D-2}}$.
The ansatz (with Minkowski signature) for the bubble is the same as before
\be
ds^2 = W^2(y) \cR^2 \mathring g_{\mu\nu} dx^\mu dx^\nu+ h_{\alpha\beta}^\cB(y)  dy^\alpha dy^\beta,  \qquad \phi=\phi(y),
\ee
where $\mathring g_{\mu\nu}$ is the unit metric on  $dS_{D-4}$. Asymptotically $\rho \to \infty$ the line element should behave as
\be
ds^2 \to W^2(\rho)  \cR^2 \mathring g_{\mu\nu} dx^\mu dx^\nu +d\rho^2 +  h_{\alpha\beta}^\cC(\bar y)  dy^{\bar \alpha} dy^{\bar \beta},  \qquad \phi\to\phi(\rho).\label{outBBEF}
\ee
with $W(\rho) \to \rho/\cR$.
The Ricci tensor reads
\bea
R_{\mu\nu} &=&\Big[ - W^{-1} \nabla^2 W + W^{-2} (D-5)\[\cR^{-2} - (\nabla W)^2\] \Big]g_{\mu\nu}\nonumber \\
R_{\alpha \beta} &=& R_{\alpha\beta}^\cB - (D-4) W^{-1} \nabla_\alpha \nabla_\beta W
\eea
where $\nabla$ is the Levi-Civita connection on the bordism, and  the curvature scalar
\be
R = -2 (D-4) W^{-1} \nabla^2 W + (D-5)(D-4) \[\cR^{-2} - (\nabla W)^2 \] W^{-2} + R_{\cB}.
\ee

\paragraph{The Hamiltonian constraint.}  With this at hand we can already write down the dynamical constraint, i.e. the $t-t$ component of  Einstein's equations
\be
(D-5) W^{-1} \nabla^2 W -\frac{1}{2} (D-5) (D-6) \[\cR^{-2} - (\nabla W)^2\] W^{-2} = \frac{1}{2} R_\cB -\frac{2}{D-2} (\nabla\phi)^2 + \frac{\tilde \alpha}{16} R_{GB}^2,\label{Hconsww}
\ee
where the Gauss-Bonnet term is evaluated on the Ricci flat background $\mathbb{M}_{D-4} \times \cB_4$, and $\cR$ is the bubble radius in Einstein frame.

\section{Negative mass states and bubbles of nothing}
\label{app:negativeStates}
In this Appendix we show how 
one can construct a family of solutions of the model \eqref{eq:action} with the same topology as the bubbles of nothing constructed in the main text, but with arbitrarily large negative masses -- the Hamiltonian of the theory is unbounded from below --.  An example of such a family for Witten's bubble of nothing can be found in \cite{Brill:1991qe}, which employs special features of that solution and the four-dimensional Einstein-Maxwell theory. Here, we want to argue that such a family always exists, and is independent of the details of the bubble solution constructed in the main text.  The analysis in this Appendix is for spacetime dimension $\geq4$ and Minkowski asymptotics; in three dimensions, the fact that the mass corresponds to a deficit angle at infinity formally requires a different treatment. 

We will show the existence of these negative mass states in two ways: First, as a consequence of the Hamiltonian constraint equation \eq{Hconsww}. And secondly, we will outline a proof that these negative mass states are always present in a wide class of models whenever one can construct a bubble of nothing solution self-consistently in perturbation theory, irrespectively of the details of the construction. Thus, once one establishes the existence of a bubble of nothing in a truncation of the theory, negative mass states are unavoidable.

\paragraph{A family of states with negative ADM energy.}    To solve the Hamiltonian constraint \eq{Hconsww} we will assume that the manifold $\cB_4$ (with $N=12$ degenerations) and the dilaton are in their background configurations $R_\cB=\nabla\phi=0$, what leads to 
\be
(D-5) W^{-1} \nabla^2 W -\frac{1}{2} (D-5) (D-6) \[\cR^{-2} - (\nabla W)^2\] W^{-2} = \frac{\tilde \alpha}{16} R_{GB}^2,
\ee
Next, proceeding  as in the main text,  we consider the regime $\tilde \alpha\to 0$ ($\cR\to \infty$) so that there is a well defined  inner-bubble region  ({\bf II.}) where the Gauss-Bonnet is dominant.  In this layer of the BON spacetime we solve the linearised Hamiltonian constraint  for $W = 1 + \tilde \alpha W^{(1)}$
\be
\hspace{-0.9cm}\text{Inner-bubble region, $\cB_{\bf II}$:} \qquad  \nabla^2 W= \frac{\tilde \alpha}{16(D-5)} R_{GB}^2 + \cO(\tilde \alpha^2).
\label{eq:appInnerH}
\ee
Then, we consider the  outer-bubble region ({\bf I.}), where the Gauss-Bonnet term is approximately zero, the subdominant $\cO(\cR^{-1})$ terms become important (i.e. the $S^{D-4}$ curvature), and the $T^3$ KK modes have decayed, so that $W = W(\rho)$. In this layer, with the  ansatz \eqref{outBBEF} and considering the case with $N=12$ degenerations,  the Hamiltonian constraint reduces to 
\be
\hspace{1.3cm} \text{Outer-bubble region, $\cB_{\bf I}$:} \qquad W'' -\frac{1}{2} (D-6) \[\cR^{-2} -  W'{}^2\] W^{-1} =0,
\ee

There are no obstructions for solving the equation for the inner bubble region\footnote{Actually, the solution is given by $W = 1 +\hat \epsilon \varphi^{(1)}/(D-5)$, (with the substitution $\alpha \to \tilde \alpha$) where $\varphi^{(1)}$ is the first order variation of the volume modulus in the solution constructed in the main text, i.e. the solution to \eqref{eq:poissonGB}.} \eqref{eq:appInnerH}, as it is just a Poisson equation on $\cB_4$. We find that at the boundary between the two regions we must have
\be
W|_{\cB_{\bf I}} =1+\tilde \alpha W^{(1)}|_{\pd \cB_{\bf II}}+\cO(\tilde \alpha^2), \qquad \nabla_n W|_{\pd \cB_{\bf II}} = \frac{24 \pi^2 \tilde \alpha}{(D-5) \cV_{T^3}}+\cO(\tilde \alpha^2),
\ee
where $\cV_{T^3}$ the $T^3$ volume is measured in the Einstein frame, and $n$ is the unit normal vector to the hypersurface $\pd\cB_{\bf II}$.
At the outer-bubble region, we note that the equation admits the first integral
\be
W' = \cR^{-1} (1 + \lambda \, W^{-(D-6)})^{1/2},
\ee
where $\lambda$ is an integration constant. Then, imposing the matching conditions for the metric, (continuity of $W$ and $\nabla_n W$) we find that the Einstein-frame bubble radius is
\be
\cR^{-1} = \frac{24 \pi^2 \tilde \alpha}{(D-5) \cV_{T^3} \sqrt{1+\lambda}} + \cO(\tilde \alpha^2).
\ee
Let us now discuss what the outer bubble geometry \eq{outBBEF} represents in this case. If we make the change of variables $r = W(\rho) \cR$, the line element restricted to the Cauchy  surface $\Sigma$ at  $t=0$ reads
\be
ds^2_\Sigma|_{\cB_{\bf I}} = r^2 d\Omega_{D-5}^2   + (1+\frac{\lambda  \cR^{D-6}}{r^{D-6}})^{-1} dr^2 + ds_{T^3}^2.
\ee
This looks exactly as a spatial slice of a Schwarzschild blackhole in $D-3$ dimensions with mass parameter $-\lambda \cR^{D-6}$. Since the extra dimensions in this geometry  are inert, the  ADM energy is just proportional to the mass parameter
\be
E_{\text{ADM}} \propto -\lambda \cR^{D-6} =-\lambda  (1 + \lambda)^{\frac{D-6}{2}} \Big(\frac{24 \pi^2\tilde \alpha}{(D-5)\cV_{T^3} }\Big)^{-(D-6)},
\ee
which is negative  provided $\lambda >0$, and arbitrarily large in absolute value (even with fixed $\tilde \alpha$). Actually, the energy decreases for large values of the radius $\cR$.

Note that the bubble radius $\cR>\cR_{\text{min}}$, has a minimum value within this family of negative mass states. Since $\lambda>0$ we have 
\be
\cR_{\text{min}} = (D-5) \Big(\frac{24 \pi^2 \tilde \alpha}{\cV_{T^3}}\Big)^{-1},
\ee
where $\cR_{\text{min}}$ corresponds to $\lambda=0$, and thus a state with zero ADM energy. When we take the limit $\tilde \alpha\to0$, the radius of the bubble in these negative mass states diverges, regardless of how close to zero is their energy. So in this limit we expect  the rate of decay to these states  to  be suppressed by the Coleman-DeLuccia mechanism.

\paragraph{Negative mass states for a generic bubble.} We will now outline how to construct negative mass states more generally. Before getting into the details of the construction, let us discuss the main idea in a simpler model. Consider a field theory (no gravity) in which there is a false vacuum parametrized by a scalar field that can decay via bubble nucleation, a la Coleman-DeLuccia \cite{Coleman:1977py}. These bubbles nucleate, and then expand. The mass of the bubble must always be equal to zero, due to conservation of energy, but how does the energy balance work?  In the thin-wall approximation, one has
\begin{equation} 0=M=M_{\text{kinetic}}+ S_{d-1}TR^{d-1}- V_{d}(\Delta V)^d.\label{samis}\end{equation}
Here, $T$ is the tension of the domain wall bounding the bubble, which has radius $R$, and $\Delta V<0$ is the difference in energies between false and true vacua. $d$ is the spatial dimension and $S_{d-1}$, $V_{d}$ are the area and volume of a unit radius sphere in $\mathbb{R}^d$, respectively. $M_{\text{kinetic}}$ is the kinetic energy of the scalar field, which exactly balances out the potential and tension contribution.

In this setup, it is clear what to do to produce negative-mass solutions; just take a bubble with a supercritical radius, such that the sum of the second and third terms in \eq{samis} is negative, and switch off the time derivatives of the fields, so that $M_{\text{kinetic}}=0$. The difference in vacuum energies then overcomes the tension and produces a negative mass solution. By taking $R$ arbitrarily large, this can be made as negative as one desires.

We will do the same thing for a bubble of nothing. We will keep the discussion as general as possible, and only later particularize to the bubbles discussed in the main text. We begin with the generic euclidean metric for a bubble of nothing far away from the core of the bubble, in $d+1$ dimensions. This can always be put in the form
\begin{equation} ds^2= e^{2\varphi(r)}\left[ (dr^2+ r^2d\Omega_{d-1})+ e^{2\phi(r)} ds_B^2\right]=e^{2\varphi}[g_0+e^{2\phi}g_B],\end{equation}
where the functions $e^{\varphi},e^{2\phi}$ asymptote to a constant at infinity as a power law, as discussed elsewhere in the paper. 
 Upon continuation to Lorentzian signature of the azimuthal coordinate of the sphere $\theta$ (see Section \ref{sec:WittenBON}), and the change of variables 
\begin{equation} x= r\cosh \theta,\quad \tau = r \sinh \theta,\end{equation}
one ends up with the time-dependent Lorentzian metric
\begin{equation} ds^2= e^{2\varphi(\sqrt{x^2-\tau^2})} \left[-d\tau^2+dx^2 +x^2d\Omega_{d-1}+e^{2\phi(\sqrt{x^2-\tau^2})} ds_B^2\right].\label{ww0}\end{equation}
This describes propagation of a bubble with uniform acceleration. The metric can also be described in ADM formalism \cite{Wald:1984rg} in a simple way,
\begin{equation} ds^2=-\alpha^2d\tau^2+ \gamma_{ij} dx^i dx^j,\end{equation}
where $\alpha$ is the lapse function, and $\gamma_{ij} dx^i dx^j$ is the spatial part of the metric (projection onto constant time hypersurfaces).

Einstein's equations are evolution equations for the pair\footnote{More generally, we would also have a lapse vector $\beta^i$, but in our setup it can be consistently truncated to zero.} $\alpha,\gamma_{ij}$. However, we cannot pick just any functions $(\alpha,\gamma_{ij})$; consistent set of initial data must satisfy the Hamiltonian constraint,
\begin{equation}\mathcal{H}=-\frac{1}{2e^{2\varphi}}[K_{ij}K^{ij}-K^2-R^\gamma]=\frac{T^{00}}{2},\label{hcons}\end{equation}
as well as the momentum constraint,
\begin{equation} \nabla_j(K^{ij}-K\gamma^{ij})=8\pi G \gamma^{ij}T_{0j}.\label{mconst}\end{equation}
Here, $K_{ij}$ is the extrinsic curvature tensor, and $K\equiv K_{ij}\gamma^{ij}$ its its trace. In our setup, the expression for $K_{ij}$ is very simple, 
\begin{equation} K_{ij}=\frac{\partial_\tau\gamma_{ij}}{2e^{\varphi}}\label{cvatac}.\end{equation}
Therefore, using \eq{ww0},
\begin{equation} K_{ij}dx^i dx^j=\frac{\tau}{\sqrt{x^2-\tau^2}}\left[e^\varphi \varphi' g_s +e^{2\phi+\varphi}(\varphi'+\phi') g_B\right],\end{equation}
and we have
\begin{equation}K_{ij}K^{ij}=\frac{\tau^2}{x^2-\tau^2}e^{-2\varphi}\left[d(\varphi')^2+(\varphi'+\phi')^2k\right],\end{equation}
as well as
\begin{equation}K=\frac{\tau}{\sqrt{x^2-\tau^2}}e^{-\varphi}\left[d(\varphi')+k(\varphi'+\phi')\right],\end{equation}
which means
\begin{equation} K^2-K_{ij} K^{ij} = \frac{\tau^2e^{-2\varphi}}{x^2-\tau^2}e^{-2 \varphi } \left[\left((d+k) \varphi '+k \phi '\right)^2-d (\varphi ')^2-k \left(\varphi '+\phi '\right)^2\right].\end{equation}

\begin{figure}[!htb]
\begin{center}
\includegraphics[width=0.75\textwidth]{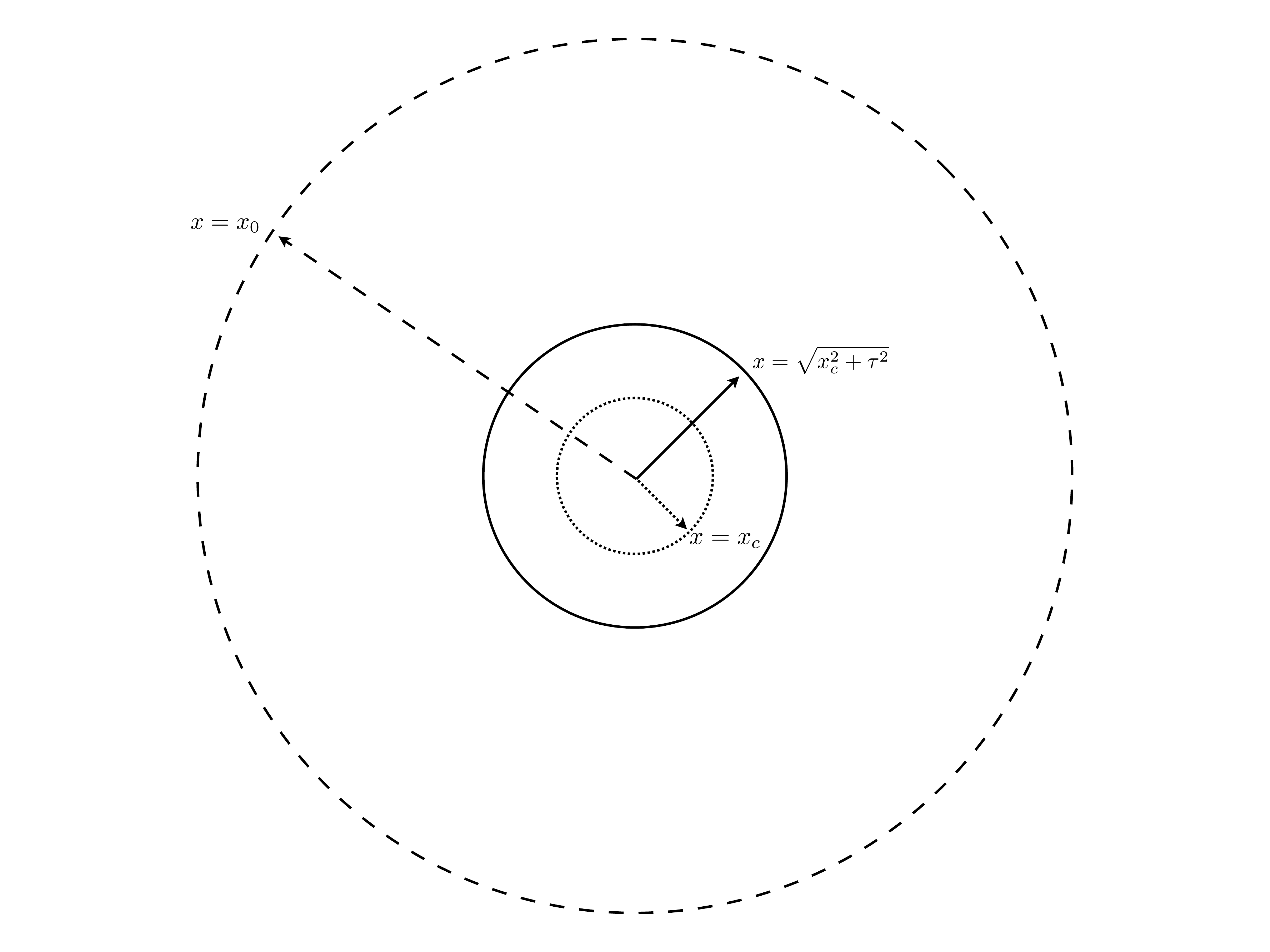}
\end{center}
\caption{Schematic construction of our bubble solutions with negative mass. The innermost circle, in the core, with radius $x\sim x_c$, corresponds to the radius of the critical bubble -- the radius of the bubble of nothing in the instant it nucleates. The circle at the solid line corresponds to the physical radius of the bubble at some time $\tau>0$, when it has been expanding for a while. The starting point of our construction is to take this $\tau>0$ ``snapshot'' of the bubble and use it as an initial condition for GR evolution, while modifying the time derivative of the metric according to \eq{b00} outside of the dashed circle at $x=x_0$, far away from the core of the bubble. This asymptotic modification is designed to take away some positive energy from the bubble, and therefore the resulting object has negative mass. As proven in the main text, this can be made arbitrarily negative, thereby establishing the instability of the vacuum even if the original bubble was not an exact (only approximate) solution to Einstein's equations.}
\label{appf1}
\end{figure}

As discussed, the bubble has zero mass, and due to energy conservation, this is true for any $\tau$; however, the balance between ``kinetic'' and ``potential energy'' changes. For $\tau=0$, the time of nucleation of the bubble, the configuration is momentarily static (the time derivatives of $\gamma_{ij}$ vanish), but for any $\tau>0$, kinetic energy (measured by the extrinsic curvature terms in \eq{hcons}) exactly balances out a negative contribution coming from the spatial curvature of the metric, just as in the field theory example. We will construct negative mass solutions by  switching off the kinetic energy from our bubble solution. More specifically, we will consider a modified initial condition where the spatial part of the metric is (a small modification of) \eq{ww0} evaluated at a generic $\tau$, but the time derivative is switched off far away from the core of the bubble (see Figure \ref{appf1}):
 \begin{equation}\partial_\tau \gamma_{ij}\vert_{\text{initial time slice},x\geq \sqrt{r^2_0+\tau^2}}\,=0.
 \label{b00}\end{equation}

Due to \eq{cvatac}, this means that the extrinsic curvature terms in the Hamiltonian constraint vanishes, and the momentum constraint is automatically satisfied. However, now that the extrinsic curvature is no longer present, we need to do something else in order to solve Hamiltonian constraint. Inspired by the fact that we somehow want to fix this by having a massive solution, we will consider the family of spatial metrics 
\begin{equation}\tilde{\gamma}_\tau= e^{2\varphi(\sqrt{x^2-\tau^2})} \left[\frac{dx^2}{1-\frac{2M(x)}{x^{d-2}}} +x^2d\Omega_{d-1}+ e^{2\phi(\sqrt{x^2-\tau^2})}  ds_B^2\right].\label{ww}\end{equation}
For $M(x)=0$ and $\tau=0$, this is just the initial condition at  of our bubble. Equation \eq{ww} is just a minor modification of our bubble solution including a Schwarzschild-like mass term, similar to what is done to describe stelar interiors \cite{Wald:1984rg}. This term is $x$-dependent, and $M(x)$ is morally the contribution to the ADM mass of the shell at radius $x$.  Taking \eq{ww} for arbitrary $\tau$ as the initial condition for a metric with vanishing $K_{ij}K^{ij}-K^2$, the Hamiltonian constraint will give a differential equation for $M(x)$.

For this, we just need to compute the Ricci scalar of \eq{ww}. It turns out that\footnote{One can compute the Ricci scalar by repeatedly using the formula for the Ricci scalar of a fibration found in \cite{o1983semi}, together with the change of the Ricci scalar under a conformal change of coordinates.} 
\begin{equation} R^{\gamma_\tau}=a\dot{M}(x)+b M(x)+c,\label{eotc}\end{equation}
where $a,b,c$ are functions of $x$ as well as $\varphi,\phi$ and their derivatives up to second order, and the dot denotes a derivative with respect to $x$. The coefficients are given explicitly as
\begin{align}a&\equiv\frac{2 e^{-2 \psi} \left(r ^2+\tau ^2\right)^{\frac{1}{2}-\frac{d}{2}} \left(\left(r ^2+\tau ^2\right) \left((d+k) e^{\psi} \psi'+k \phi'\right)+(d-1) r \right)}{r },\nonumber\\
r^3b&\equiv 2 e^{-2 \psi} \left(r ^2+\tau ^2\right)^{1-\frac{d}{2}} \left(r  (d+k) \left(r ^2+\tau ^2\right) e^{\psi} \psi'^2 \left((d+k-1) e^{\psi}+2\right)\right.\nonumber\\&\left.+(d+k) e^{\psi} \left(d r ^2-2 \tau ^2\right) \psi'+2 r  \left(r ^2+\tau ^2\right) \left((d+k) e^{\psi} \psi''+k \phi''\right)\right.\nonumber\\&\left.+k \left(d r ^2-2 \tau ^2\right) \phi'+k (k+1) r  \left(r ^2+\tau ^2\right) \phi'^2\right),\nonumber\\
r^3c&\equiv e^{-2 \psi} \left(-r  (d+k) \left(r ^2+\tau ^2\right) e^{\psi} \psi'^2 \left((d+k-1) e^{\psi}+2\right)\right.\nonumber\\&\left.-2 (d+k) e^{\psi} \left((d-1) r ^2-\tau ^2\right) \psi'-2 r  \left(r ^2+\tau ^2\right) \left((d+k) e^{\psi} \psi''+k \phi''\right)\right.\nonumber\\&\left.+2 k \left(\tau ^2-(d-1) r ^2\right) \phi'-k (k+1) r  \left(r ^2+\tau ^2\right) \phi'^2\right)
.\label{tfaf}\end{align}
Here, $r^2=x^2-\tau^2$, the natural euclidean variable, and primes denote derivatives with respect to $r$.  The Hamiltonian constraint $R^{\gamma_\tau}=0$ then becomes a first-order linear ODE for $M(x)$,
\begin{equation} \dot{M}(x)=fM(x)+g,\end{equation}
where the dot denotes a derivative with respect to $x$ and
\begin{equation}f\equiv -\frac{b}{a},\quad g\equiv -\frac{c}{a}.\label{defg}\end{equation}
 with boundary condition $M(x_0)=0$, where
 \begin{equation}x_0=\sqrt{r_0^2+\tau^2}.\end{equation}
 This corresponds to having the extrinsic curvature and stress-energy tensor terms in \eq{hcons} switched on until a radius $x_0$ (so that until that radius the solution is identical to the bubble of nothing, thus having $M=0$) and then switching them off from $x_0$ onwards (see Figure \ref{appf1}). 
 The general solution to this equation is then
\begin{equation} M(x)=\int_{x_0}^xg(x')\,\exp\left(-\int_{x'}^{x} f(x'')\,dx''\right) \,dx'\end{equation}
from which we get an expression for the ADM mass of the family of metrics 
\begin{equation} M=\int_{x_0}^\infty g(x')\,\exp\left(-\int_{x'}^\infty f(x'')\,dx''\right) \,dx'.\label{madm3}\end{equation}
This is a function of the time $\tau$ chosen in \eq{ww}; the larger $\tau$, the bigger the bubble, since the minimum value of the $x$ coordinate is  $x_0=\sqrt{\tau^2+r_0^2}$. 

It is clear from \eq{madm3} that we want $g<0$ at least somewhere to get negative mass. The  Hamiltonian constraint for the original bubble \eq{ww0} involves \eq{eotc} at $M=$0, thus becoming
\begin{equation} e^{2\varphi}T^{00}-c +K_{ij} K^{ij}-K^2=0.\end{equation}
This means that at least one of $e^{2\varphi}T^{00}$, $-c$ or $K_{ij} K^{ij}-K^2$ must be negative. The general idea is to use whichever is negative as the source for $g$, and switch off the other two terms\footnote{In cases where $K_{ij} K^{ij}-K^2$ is the negative term, one will need additional modifications to satisfy the momentum constraint \eq{mconst}. This is not the case for our bubbles.}. In the present case, $c>0$ as we will see, so we switch off $T^{00}$ and the extrinsic curvature terms. 

Since we will take $r_0$ in \eq{b00} large compared to other scales of the problem except for $\tau$, we just need the asymptotic behaviour of the coefficients $f,g$. This can be obtained from \eq{tfaf} provided that we know the asymptotic behaviour of the functions $\varphi,\phi$ in our particular bubble. As discussed elsewhere in the text, for the particular case of the $T^3$ bubble we get
\begin{equation} \varphi(r)\rightarrow \frac{-2f_1}{D-2}{r^{D-5}},\quad e^{\phi(r)}\rightarrow 1+\frac{f_1}{r^{D-5}}.\label{asymptotics}\end{equation}
Here, $D=d+k+1$ is the total dimension of spacetime, and the constant $f_1$ is
\be
f_1 = -\frac{\rme^{2 \Delta \varphi}}{(D-5)} \frac{24 \pi^2 \alpha}{\cV_{T^3}}  \cR^{D-4}
\ee
One can then obtain explicit (if cumbersome) expressions for $f$ and $g$, and compute the mass explicitly according to \eq{madm3}. The integral decomposes in two regions, according to whether $r\lesssim\tau$ or $r\gtrsim\tau$. For $r\ll\tau$ but large enough so as to trust \eq{asymptotics}, one has (for $k=3$, the case of interest)
\begin{equation}f\sim\frac{2(d+1)}{r^2},\quad g\sim -\frac{(d+1)\tau^{d-1}}{r^2},\label{asimpt1}\end{equation}
while for $r\gg\tau$, one has
\begin{equation} f\sim\frac{\alpha_d f_1}{r^d},\quad g\sim \frac{\beta_d f_1}{r^2}.\end{equation}
The $\alpha_d,\beta_d$ are dimension-dependent coefficients that can be computed explicitly on a case-by-case basis -- for instance, for $d=3$, $\alpha_3=-9/5,\beta_3=3/5,$ --. One can check that with the numerical coefficients $f_1,g_1$ we used to compute the euclidean action of the bubble, $g$ is indeed negative, as advertised. 

In any case, since $f$ is always smaller than $1/r_0^2$, for $r_0$ large enough we are entitled to drop the exponential term in \eq{madm3} and the result is a simple integral over $g$. For large $\tau$, this integral is furthermore dominated by the $r\leq\tau$ region, with asymptotics \eq{asimpt1}. One gets
\begin{equation} M\sim-(d+1)\tau^{d-1}\int_{r_0}^\tau\, \frac{dr}{r^2}\frac{r}{\sqrt{r^2+\tau^2}}\propto -\tau^{d-2}\log\left(\frac{\tau}{r_0}\right),\label{massf}\end{equation}
so we indeed get a family of bubbles whose mass is as negative as one wants. 

In this construction we have contented ourselves with stripping away the kinetic energy of the gravitational field far away from the bubble. It would be interesting to figure out what is the largest scaling one can get with $\tau$, and whether it is area ($\tau^{d-1}$) or volume ($\tau^d$) scaling. In the latter case, the coefficient in front of the $\tau^d$ term would constitute the energy density of the bubbles. Perhaps it would make sense to identify it with some sort of ``zero-point'' energy of the KK vacuum, which can then be removed by the bubble. If so, ``nothing'' seems to be the less energetic state. Perhaps  supersymmetric theories are precisely those in which ``nothing'' is degenerate with or has higher energy density than the vacuum. It would be interesting to extend and apply the formalism in this Appendix to other bubbles and see if the above ideas can be made more precise. At present, we only know that we know nothing.

\bibliographystyle{packages/JHEP}
\bibliography{bonrefs}

\end{document}